\documentclass[11pt,english]{article}
\usepackage{mathtools}

\usepackage{import}
\usepackage[utf8]{inputenc}
\usepackage[a4paper,       %% set the margins
            bindingoffset=0.0in,
            left=0.75in,
            right=0.75in,
            top=0.75in,
            bottom=0.75in,
            footskip=.25in]{geometry}
\usepackage{xcolor}
\usepackage{amsmath}
\usepackage{centernot}  
\usepackage{amsthm}    
\usepackage{amssymb}  
\usepackage{bm}       
\usepackage{moreverb}
\usepackage{accents}
\usepackage{epstopdf}
\usepackage{float}
\restylefloat{table}
\usepackage{breakcites}
\usepackage{blindtext}
\usepackage[parfill]{parskip}
\usepackage{multirow} % For \multirow command
\usepackage{booktabs} % For \toprule, \midrule, \bottomrule, \cmidrule
\usepackage{threeparttable} % For tablenotes environment
\usepackage{array}          % Optional but helpful for column formatting
\usepackage{hyperref}
\usepackage{authblk}
\usepackage[backend=bibtex,style=numeric, sorting=none]{biblatex} %% referencing 
\usepackage{breqn}
\usepackage{soul}
\usepackage{xcolor}
\usepackage{algorithm}
\usepackage{algpseudocode}
\usepackage{enumitem} 

\usepackage{makecell}

\usepackage{caption}

\usepackage{tcolorbox}
\usepackage{listings}
\newcounter{codebox}
\renewcommand{\thecodebox}{\arabic{codebox}}

\usepackage{titlesec}
\titleformat{\paragraph}
{\normalfont\normalsize\bfseries}{\theparagraph}{1em}{}
\titlespacing*{\paragraph}
{0pt}{3.25ex plus 1ex minus .2ex}{1.5ex plus .2ex}

\title{\LARGE{Ordinal regression for meta-analysis of test accuracy: a flexible approach for utilizing all threshold data}}

\author[1]{Enzo Cerullo}
\author[2]{Klaus Linde}
\author[3]{Hayley E. Jones}
\author[3]{Efthymia Derezea}
\author[1]{Tim Lucas}
\author[1]{Nicola J. Cooper}
\author[1]{Alex J. Sutton }

\affil[1]{
Biostatistics Research Group, 
Division of Public Health \& Epidemiology, 
School of Medical Sciences,
University of Leicester, 
Leicester,
UK}
\affil[2]{
Institute of General Practice and Health Services Research,
Department of Clinical Medicine,
School of Medicine and Health,
TUM University Hospital Rechts der Isar,
Technical University Munich,
Munich,
Germany
}
\affil[3]{
Population Health Sciences,
Bristol Medical School,
University of Bristol,
UK}

\date{}

\begin{document}
%%%%
%% \onehalfspacing  % for 1.5 line-spacing
%%%%

\maketitle

%%%%%%%%%%%%%%%%%%%%%%%%%%%%%%%%%%%%%%%%%%%%%%%%%%%%%%%%%%%%%%%%%%%%%%%%%%%%%%%%%%%%%%%%%%%%%%%%%%%%%%%%%%%%%%
\section*{\large{ Abstract}}
%%%%%%%%%%%%%%%%%%%%%%%%%%%%%%%%%%%%%%%%%%%%%%%%%%%%%%%%%%%%%%%%%%%%%%%%%%%%%%%%%%%%%%%%%%%%%%%%%%%%%%%%%%%%%%
%%%%
%%%% ----------------------------------------------------------------------------------------------------------------
% Standard (network) meta-analysis methods for medical test accuracy evaluation analyse the data separately
% for each test threshold - wasting data - 
% unless every study reports all thresholds.
% %%
% Previously proposed "multiple threshold" models either make overly restrictive assumptions -
% especially for ordinal tests -
% or fail to provide threshold-specific summary estimates.
%%
Standard (network) meta-analysis methods for medical test accuracy evaluation analyse the data
separately for each test threshold -
wasting data -
unless every study reports all thresholds.
Previously proposed "multiple threshold" models either fail to provide threshold-specific summary estimates,
or they assume that ordinal tests - such as questionnaires - are continuous.

%%%%
%%%% ----------------------------------------------------------------------------------------------------------------
% %%
% We propose two ordinal regression models, ordinal-bivariate and ordinal-HSROC,
% using an induced-Dirichlet framework for cutpoint parameters, 
% enabling intuitive prior specification and both fixed-effects and random-effects formulations. 
% %%
% Furthermore, unlike continuous-assumption approaches,
% our models properly respect the ordinal nature of screening 
% questionnaires and similar instruments.
%%
We propose two ordinal regression models, ordinal-bivariate and ordinal-HSROC, 
using an induced-Dirichlet framework for cutpoint parameters, 
enabling intuitive priors and both fixed-effects and random-effects cutpoints.

%%%%
%%%% ----------------------------------------------------------------------------------------------------------------
% Additionally, we conducted a simulation study to evaluate the performance of our proposed models, 
% with the simulated data being based on real anxiety screening data spanning
% $7$, $22$, and $64$ ordinal categories, with $15\%$, $40\%$ and $55\%$ missing threshold data.
% %%
% Overall, our proposed ordinal-bivariate model with fixed-effect cutpoints performed the best, 
% consistently achieved the lowest RMSE and bias across scenarios,
% including when data was generated from a recently proposed continuous-assumption model.
% %%
% For instance - even with $64$ categories -
% continuous-assumption models performed $\sim 10\% - 30\%$ worse than our ordinal approaches,
% contradicting the common assumption that many categories justify treating ordinal tests as continuous.
% %%
% Furthermore, the standard stratified-bivariate approach showed particularly poor performance,
% particularly for tests with higher missingness ($\sim 25\% - 50\%$ worse RMSE).
%%
Additionally, we conducted a simulation study to evaluate the performance of our proposed models,
with the simulated data being based on real anxiety screening data spanning $7$, $22$, and $64$ ordinal categories,
with $15\%$, $40\%$ and $55\%$ missing threshold data.

%%%%
%%%% ---------------------------------------------------------------------------------------------------------------
Our proposed ordinal-bivariate model with fixed-effect cutpoints tended to obtain the best RMSE and bias,
including when data was generated from a recently proposed continuous-assumption model.
For instance - even with 64 categories -
continuous models performed $10\% - 30\%$ worse than our models,
contradicting the common assumption that many categories justify treating ordinal tests as continuous.
Furthermore, the standard stratified-bivariate approach showed worse performance,
especially for tests with higher missingness ($25\% - 50\%$ worse RMSE).

%%%%
%%%% ---------------------------------------------------------------------------------------------------------------
% Furthermore, we implemented these models in the MetaOrdDTA R package
% (\url{https://github.com/CerulloE1996/MetaOrdDTA}),
% which provides a wide array of features, such as: 
% Stan-based estimation, K-fold cross-validation for model selection, meta-regression,
% network meta-analysis extensions, and comprehensive visualization tools including sROC plots with
% credible and prediction regions.
%%
We also implemented the models in the MetaOrdDTA R package
(\url{https://github.com/CerulloE1996/MetaOrdDTA}), 
which provides features such as: 
Stan estimation, 
K-fold cross-validation for model selection, 
meta-regression, 
network meta-analysis extensions,
and visualisation tools including sROC plots with credible/prediction regions.

% %%%%
% %%%% -------------------------------------------------------------------------------------------------------------
% Overall, our simulation study
% %% based on real-world anxiety screening data -
% suggests that our proposed models may obtain notably better accuracy estimates than previous approaches
% for tests which are ordinal - such as screening questionnaires - 
% even when the number of ordinal categories is very high.
% % %%%%
% % %%%% ---------------------------------------------------------------------------------------------------------------
% % Our ordinal regression framework establishes a more rigorous approach for synthesizing medical test 
% % accuracy data from ordinal tests, better serving researchers and clinical decision-makers.
%%
Overall, our simulation study suggests that our proposed models may obtain better accuracy estimates 
than previous approaches for ordinal tests, 
even when the number of ordinal categories is very high.

%%%%%%%%%%%%%%%%%%%%%%%%%%%%%%
\section*{\large{Keywords:}}
%%%%%%%%%%%%%%%%%%%%%%%%%%%%%%
test accuracy, 
meta-analysis,
network meta-analysis, 
missing thresholds, 
multiple thresholds, 
simulation study. 
%%%%
%%%%

*Corresponding Author

Email address: enzo.cerullo@bath.edu

%%%%%%%%%%%%%%%%%%%%%%%%%%%%%%%%%%%%%%%%%%%%%%%%%%%%%%%%%%%%%%%%%%%%%%%%%%%%%%%%%%%%%%%%%%%%%%%%%%%%%%%%%%%%
% %%%%%%%%%%%%%%%%%%%%%%%%%%%%%%%%%%%%%%%%%%%%%%%%%%%%%%
% \section*{Highlights}
% \label{section_highlights}
% %%%%%%%%%%%%%%%%%%%%%%%%%%%%%%%%%%%%%%%%%%%%%%%%%%%%%%
% \newpage 
%%%%%%%%%%%%%%%%%%%%%%%%%%%%%%%%%%%%%%%%%%%%%%%%%%%%%%%%%%%%%%%%%%%%%%%%%%%%%%%%%%%%%%%%%%%%%%%%%%%%%%%%%%%%

%%%%%%%%%%%%%%%%%%%%%%%%%%%%%%%%%%%%%%%%%%%%%%%%%%%%%%%%%%%%%%%%%%%%%%%%%%%%%%%%
\newpage
\section{Introduction \& Background}
\label{section_introduction}
%%%%%%%%%%%%%%%%%%%%%%%%%%%%%%%%%%%%%%%%%%%%%%%%%%%%%%%%%%%%%%%%%%%%%%%%%%%%%%%%
%%%%
%%%% ---------------------------------------------------------------------------------------------------------------
Diagnostic and screening tests with ordinal outcomes have become indispensable tools in modern healthcare.
A major example is for mental health screening, where symptom severity exists along a continuum.
Instruments such as:
the Generalized Anxiety Disorder scales
(i.e., the GAD-2 \supercite{Kroenke_2007_GAD_2}, GAD-7\supercite{Spitzer_2006_GAD_7}),
the Hospital Anxiety and Depression Scale (HADS\supercite{Zigmond_1983_HADS}),
and the Patient Health Questionnaire (e.g., the PHQ-2\supercite{phq_2} and PHQ-9\supercite{phq_9})
employ a graduated scoring systems that enable clinicians to assess symptom severity and guide treatment decisions.
For example, the PHQ-9\supercite{phq_9} 
for depression screening uses a $0 - 27$ scale with established cutpoints at: 
5 (mild), 
10 (moderate),
15 (moderately severe), and
20 (severe depression),
yet different studies often evaluate and report accuracy at different subsets of these thresholds.

%%%%
%%%% ---------------------------------------------------------------------------------------------------------------
Current methods for meta-analysing diagnostic tests with multiple thresholds suffer from significant limitations,
which could potentially limit their validity and clinical utility,
particularly when the underlying test accuracy data is ordinal.
The standard "bivariate" model (Reitsma et al., 2005\supercite{Reitsma2005})
and hierarchical summary receiver operating characteristic
(HSROC; Rutter \& Gatsonis, 2001\supercite{Rutter2001})
models can accommodate only one sensitivity-specificity pair per study,
forcing researchers to either arbitrarily select a single threshold,
or conduct separate meta-analyses for each cutpoint.
These widely-used approaches waste valuable data and may introduce selection bias toward certain thresholds,
which are more likely to be used in clinical practice or thought to be useful a priori.
Jones et al. (Jones et al, 2019\supercite{Jones2019}) developed a multinomial meta-analysis model.
This model can incorporate multiple thresholds per study;
however, it assumes that test results follow a continuous logistic (or normal)
distribution after a Box-Cox transformation -
an assumption that may introduce some bias when applied to truly ordinal tests with discrete,
integer-only values (such as psychiatric questionnaires for depression, anxiety, ADHD, etc).
Recent alternatives - including for instance time-to-event-based models
(i.e., Hoyer et al., 2018\supercite{Hoyer_et_al_2018_mult_thr};
Zapf et al., 2024\supercite{Zapf_et_al_2024_mult_thr})
similarly to Jones et al\supercite{Jones2019} -
also rely on continuous latent variable assumptions,
merely employing different mathematical machinery to handle the threshold data.

%%%%
%%%% ---------------------------------------------------------------------------------------------------------------
Furthermore,
whilst Dukic and Gatsonis (Dukic \& Gatsonis, 2003\supercite{dukic})
pioneered the use of ordinal regression with cutpoints as explicit parameters in test accuracy meta-analysis,
their model was designed for a fundamentally different purpose and makes substantially different assumptions.
More specifically, their framework addresses situations where diagnostic categories are non-nested,
and may not align between studies -
for instance,
when different imaging modalities use incompatible grading scales that cannot be directly mapped to each other.
In contrast,
our setting assumes coherent thresholds -
where each cutpoint k has consistent clinical meaning across all studies -
for instance, a PHQ-9 score $\ge 10$ means the same thing, regardless of study.
Most critically, although both approaches can incorporate random-effects thresholds,
the Dukic and Gatsonis model\supercite{dukic} cannot produce valid pooled sensitivity/specificity estimates in this setting.
% %%
% However, our approach, which uses the induced-Dirichlet\supercite{Betancourt2019_ordinal, cerullo_meta_ord} framework - 
% used as a prior for fixed-effects models and as a hierarchical distribution for random-effects models - 
% not only provides much more intuitive prior specification,
% but also enables computation of unbiased (less effected by the skew arising from multidimensional cutpoint constraints)
% summary estimates - through posterior medians of study-specific cutpoints.

%%%%
%%%% ---------------------------------------------------------------------------------------------------------------
We propose a novel ordinal-bivariate and ordinal-HSROC models - both MA and NMA extensions -
that address these fundamental limitations,
through a principled ordinal regression framework.
Our methods, which uses the factorised multinomial from Jones et al\supercite{Jones2019} - 
are also based on full ordinal regression 
(McCullagh et al, 1980\supercite{ord_regression_original_1st_ref_McCullagh_et_al_1980};
Winship et al, 1984\supercite{ord_regression_original_2nd_ref_Winship_et_al_1984};
Betancourt, 2019\supercite{Betancourt2019_ordinal})
with cutpoints as explicit parameters, 
eliminating the need for potentially inappropriate or overly restrictive continuous distribution assumptions.
A key methodological advance is implementing an induced-Dirichlet framework 
(Betancourt, 2019\supercite{Betancourt2019_ordinal}; 
Cerullo et al, 2022\supercite{cerullo_meta_ord})
for both fixed-effects priors and random-effects hierarchical distributions.
This allows researchers to specify intuitive prior beliefs about category probabilities,
rather than directly parameterising abstract latent cutpoints.
For random-effects cutpoint models specifically,
we introduce a novel approach using medians of posterior study-specific cutpoint distributions
to compute unbiased summary estimates.
This improves upon our previous work (Cerullo et al., 2022\supercite{cerullo_meta_ord}),
where we discovered that population-level cutpoint parameters from complex non-normal hierarchical models
can exhibit bias/skew due to multidimensional monotonicity constraints -
particularly with $>5$ categories, which our previous work did not examine or take into account, 
especially since in that paper \supercite{cerullo_meta_ord} that ordinal test only had three categories.

%%%%
%%%% ---------------------------------------------------------------------------------------------------------------
Furthermore, to facilitate widespread adoption of these methods, 
we have developed MetaOrdDTA (Cerullo et al, 2025\supercite{Cerullo_MetaOrdDTA_2025}),
a comprehensive R package for meta-analysis and network meta-analysis (NMA) 
of ordinal screening and/or diagnostic test data. 
The R package implements both our proposed ordinal-bivariate model and 
ordinal-HSROC model, 
alongside the Jones et al (Jones et al, 2019\supercite{Jones2019}) 
continuous threshold model.
Additionally, MetaOrdDTA leverages Stan's 
(Carpenter et al, 2017\supercite{Carpenter2017}) 
state-of-the-art Hamiltonian Monte Carlo algorithm
(based on NUTS-HMC; Hoffman et al, 2014\supercite{Hoffman_and_Gelman_2014_NUTS_paper})
for robust Bayesian estimation, 
providing flexible prior specification that allows incorporation of domain expertise.
The package also offers comprehensive functionality, 
including extraction of MCMC summary estimates for all parameters and derived quantities, 
automatic generation of summary sROC plots with credible and prediction regions, 
and support for both fixed-effects and random-effects cutpoint models, 
using the induced-Dirichlet \supercite{Betancourt2019_ordinal, cerullo_meta_ord} framework. 
% %%
% Future developments will incorporate covariate adjustment for meta-regression and implement faster estimation through novel state-of-the-art SNAPER-HMC and/or ChessR-HMC based algorithms \supercite{sountsov_and_Hoffman_2022_focusing_SNAPER_HMC_and_ChEESR},
% via integration with our previously developed BayesMVP R package (Cerullo, 2025 \supercite{Cerullo_BayesMVP_2025}).

%%%%
%%%% ---------------------------------------------------------------------------------------------------------------
We further extend our proposed models to NMA settings, 
enabling simultaneous comparison of multiple diagnostic tests, whilst maintaining the ordinal framework.
These NMA extensions follow the arm-based approach of 
Nyaga et al. (Nyaga et al, 2018 \supercite{Nyaga2018}) -
but adapted to our ordinal regression based approach - 
allowing researchers to compare tests even when different studies evaluate different test combinations, 
and report at different thresholds. 
Through comprehensive simulation studies and real-world application,
we demonstrate that proper ordinal modelling provides superior performance compared to methods assuming continuous latent distributions,
with benefits extending beyond missing data handling to include improved efficiency
and reduced bias -
even when threshold reporting is low and when there are a very large number of ordinal categories ($> 50$).

%%%%
%%%% ---------------------------------------------------------------------------------------------------------------
The remainder of this paper is organized as follows. 
In section \ref{section_model_specs}, 
we present the statistical framework for our ordinal-bivariate and ordinal-HSROC models,
including the within-study likelihood specification, between-study heterogeneity modeling,
and the induced-Dirichlet approach for cutpoint parameters. 
Section \ref{section_model_specs_NMA_extensions}
extends these models to the network meta-analysis setting. 
In section \ref{Sim_study},
we describe our comprehensive simulation study based on real-world anxiety and depression screening data, 
comparing our methods against the Jones et al. \supercite{Jones2019} approach -
and also to the more standard stratified-bivariate \supercite{Reitsma2005} approach - 
across scenarios with varying numbers of categories ($7$, $22$, and $64$) 
and different levels of missing threshold data ($15\%$, $40\%$, and $55\%$). 
Section \ref{MetaOrdDTA} 
provides implementation details and examples using the MetaOrdDTA R package. 
Finally, Section \ref{section_discussion} 
discusses the implications of our findings, limitations, and directions for future research.

% %%%%%%%%%%%%%%%%%%%%%%%%%%%%%%%%%%%%%%%%%%%%%%%%%%%%%%%%%%%%%%%%%%%%%%%%%%%%%%%%
% \documentclass[./Main.tex]{subfiles}
% \begin{document}
% %%%%%%%%%%%%%%%%%%%%%%%%%%%%%%%%%%%%%%%%%%%%%%%%%%%%%%%%%%%%%%%%%%%%%%%%%%%%%%%%
%%%%%%%%%%%%%%%%%%%%%%%%%%%%%%%%%%%%%%%%%%%%%%%%%%%%%%%%%%%%%%%%%%%%%%%%%%%%%%%%
\setcounter{secnumdepth}{4}  % To number these deep sections
\setcounter{tocdepth}{4}     % To include them in the table of contents
\numberwithin{equation}{section} %% number equations / appendix equations / equation numbering / bookmark
%%%%%%%%%%%%%%%%%%%%%%%%%%%%%%%%%%%%%%%%%%%%%%%%%%%%%%%%%%%%%%%%%%%%%%%%%%%%%%%%
% % Define a fourth level section command
% \titleclass{\subsubsubsection}{straight}[\subsubsection]
% \newcounter{subsubsubsection}[subsubsection]
% \renewcommand\thesubsubsubsection{\thesubsubsection.\arabic{subsubsubsection}}
% \titleformat{\subsubsubsection}
%   {\normalfont\normalsize\bfseries}{\thesubsubsubsection}{1em}{}
% \titlespacing*{\subsubsubsection}
%   {0pt}{3.25ex plus 1ex minus .2ex}{1.5ex plus .2ex}

% % Define a fifth level section command  
% \titleclass{\subsubsubparagraph}{straight}[\subsubsubsection]
% \newcounter{subsubsubparagraph}[subsubsubsection]
% \renewcommand\thesubsubsubparagraph{\thesubsubsubsection.\arabic{subsubsubparagraph}}
% \titleformat{\subsubsubparagraph}
%   {\normalfont\normalsize\itshape}{\thesubsubsubparagraph}{1em}{}
% \titlespacing*{\subsubsubparagraph}
%   {0pt}{3.25ex plus 1ex minus .2ex}{1.5ex plus .2ex}
  
% % Make them available in the document structure  
% \setcounter{secnumdepth}{5}
% \setcounter{tocdepth}{5}

%%%%%%%%%%%%%%%%%%%%%%%%%%%%%%%%%%%%%%%%%%%%%%%%%%%%%%%%%%%%%%%%%%%%%%%%%%%%%%%%%%%%%%%%%%%%%%%%%%%%%%%%%%%%%%%%%%%%%%%%%%%%%%%%%%%%%%%%%%%%%
\newpage
\section{Model specifications}
\label{section_model_specs}
%%%%%%%%%%%%%%%%%%%%%%%%%%%%%%%%%%%%%%%%%%%%%%%%%%%%%%%%%%%%%%%%%%%%%%%%%%%%%%%%%%%%%%%%%%%%%%%%%%%%%%%%%%%%%%%%%%%%%%%%%%%%%%%%%%%%%%%%%%%%%
%%%%
%%%%
%%%%%%%%%%%%%%%%%%%%%%%%%%%%%%%%%%%%%%%%%%%%%%%%%%%%%%%%%%%%%%%%%%%%%%%%%%%%%%%%
\subsection{DTA-MA model I: The "ordinal-bivariate" model}
\label{section_model_specs_Cerullo_bivariate_Reitsma_extension}
%%%%%%%%%%%%%%%%%%%%%%%%%%%%%%%%%%%%%%%%%%%%%%%%%%%%%%%%%%%%%%%%%%%%%%%%%%%%%%%%
%%%%
%%%% -----------------------------------------------------------------------------------------------------------------------
This model can be thought of as an ordinal-regression extension of the widely used,
bivariate DTA-MA model (Reitsma et al, 2005 \supercite{Reitsma2005}).
As we mentioned in the introduction (see section \ref{section_introduction}),
the benefit of this is that it allows analysts to make use of all the aggregate data,
regardless of which thresholds study authors decide to report test accuracy at, 
but whilst making less assumptions compared to other approaches which have explored this issue 
(e.g. Steinhauser et al, \supercite{Steinhauser2016}; Jones et al, 2019 \supercite{Jones2019}). 
% %%
% Furthermore, the within-study likelihoods are based on the multinomial factorisation method
% (as a series of conditionally independent binomials) proposed by Jones et al\supercite{Jones2019}.

%%%%
%%%% -----------------------------------------------------------------------------------------------------------------------
This model can also be thought of as an extension -
but at the same time a simpler version -
of the model for single studies proposed by Xu et al (Xu et al, 2013\supercite{Xu2013}),
with the following key differences:
\begin{itemize}
    \item[(i)] In this paper, we are assuming a perfect gold standard,
    whereas the model from Xu et al\supercite{Xu2013} is a latent class model - 
    and hence does not assume any of the tests are perfect (i.e., $100\%$ sensitivity and specificity).
    \item[(ii)] The model from Xu et al\supercite{Xu2013} models the (within-study) conditional dependence,
    using an ordinal multivariate probit model\supercite{Chib_Greenberg_MVP_1998} -
    which we have previously extended to meta-analysis (Cerullo et al\supercite{cerullo_meta_ord},
    and also implemented in a separate R package -
    BayesMVP\supercite{Cerullo_BayesMVP_2025}).
    On the other hand,
    for this paper, our within-study likelihood function has the same form as the one proposed in
    Jones et al (Jones et al, 2019\supercite{Jones2019}) - more specifically,
    it is based on the binomial factorization of the multinational distribution and univariate ordinal regression -
    as opposed to multivariate - 
    as we are not modelling the conditional dependence (i.e. the within-study correlations) between tests, 
    since we are assuming a perfect gold standard.
    \item[(iii)] Our model can handle any number of test thresholds, 
    whereas the model from Xu et al\supercite{Xu2013} is just for tests with "intermediate" results (i.e. 2 thresholds).
    \item[(iv)] Our model is for meta-analysis, 
    whereas the model from Xu et al\supercite{Xu2013} is for single studies.
\end{itemize}

%%%%
%%%% -----------------------------------------------------------------------------------------------------------------------
In addition, another way to think of our proposed ordinal-bivariate model is a variation
of the model proposed by Jones et al \supercite{Jones2019}.
More specifically, key differences between our proposed model and the model in Jones et al\supercite{Jones2019} are:
(i) we estimate the cutpoints as parameters 
(which is the standard approach in ordinal regression modeling), 
whereas Jones et al\supercite{Jones2019} use explicit log-transformed or box-cox-transformed thresholds; and:
(ii) we do not estimate (or need to\supercite{Xu2013}) any scale parameters, 
whereas Jones et al\supercite{Jones2019} do - in both the diseased and non-diseased groups.

%%%%
%%%% -----------------------------------------------------------------------------------------------------------------------
Furthermore, the key differences between this model and the "HSROC" model extension 
(which we will refer to as the "ordinal-HSROC" model) - 
which we propose in section \ref{section_model_specs_Cerullo_R_and_G_HSROC}) - 
is that in the ordinal-bivariate model:
\begin{itemize}
    \item 
    There are no scale parameters (they are fixed to 1 in both the non-diseased and diseased groups);
    \item 
    The location parameters are freely estimated in both groups,
    whereas in the ordinal-HSROC model both groups share the same location parameters;
    \item 
    Each group has it's own set of cutpoints,
    whereas in the ordinal-HSROC model both groups share the same set of cutpoints; and:
    \item 
    The between-study correlation is modelled \textbf{explicitly} in this model, 
    using a bivariate-normal distribution
    (hence "bivariate" model);
    which is in contrast to the ordinal-HSROC model,
    which models the between-study correlation \textbf{implicitly} by imposing a structural relationship 
    (which we discuss in detail in section
    \ref{section_model_specs_relationship_between_Cerullo_bivariate_and_Cerullo_R_and_G_HSROC_MA_models}).
\end{itemize}
%%%%
%%%%
%%%%%%%%%%%%%%%%%%%%%%%%%%%%%%%%%%%%%%%%%%%%%%
\subsubsection{Within-study model}   
\label{section_model_specs_Cerullo_bivariate_Reitsma_extension_within_study_model}
%%%%%%%%%%%%%%%%%%%%%%%%%%%%%%%%%%%%%%%%%%%%%%
%%%% -------------------------------------------------------------------------------------------------------------------------------------
In the ordinal-bivariate model, the test accuracy measures at the $k$th cutpoint in study $s$ are represented as:
\begin{equation}
\begin{aligned}
Sp_{s, k} & =     \Phi  \left( C_{s, k}^{[d-]} - \beta^{[d-]}_{s}\right)   \\
Se_{s, k} & = 1 - \Phi  \left( C_{s, k}^{[d+]} - \beta^{[d+]}_{s}\right)
\end{aligned}
\end{equation}
Where:
$\Phi(\cdot)$ is the CDF of a standard normal,
$\beta^{[d-]}_{s}$ and $\beta^{[d+]}_{s}$ are the location parameters in the non-diseased and diseased groups,
respectively, and:
$C_{k}^{[d-]}$ and $C_{k}^{[d+]}$ are the latent cutpoint parameters, 
corresponding to test threshold $k$ in study $s$ in the non-diseased 
(denoted by the superscript $[d+]$) and the diseased (denoted by $[d-]$) group.

%%%%
%%%% -------------------------------------------------------------------------------------------------------------------------------------
Note that we show how to incorporate study-level covariates in 
section \ref{section_model_specs_NMA_extensions}, 
sub-section \ref{section_model_specs_NMA_ordinal_bivariate_between_study_model}.

%%%%
%%%% -------------------------------------------------------------------------------------------------------------------------------------
Furthermore, the \textbf{within-study log-likelihood}
for study $s$ has the essentially the same form as that described in Jones et al \supercite{Jones2019}.
To be more specific, we will be using a multinomial likelihood -
but expressed as a series of conditionally independent binomial likelihoods.
As discussed in Jones et al \supercite{Jones2019},
this allows us to use the data expressed in terms of \textit{cumulative} counts -
as opposed to ordered categorical counts
(i.e. the number of patients falling within each category).
The benefit of this is that it makes it easier to incorporate "missing thresholds" -
i.e. we can include studies regardless of which (and what number) thresholds are reported,
which allows us to include all studies in the meta-analysis.
This is in contrast to standard methods 
(Reitsma et al, 2005 \supercite{Reitsma2005};
Rutter \& Gatsonis, 2001 \supercite{Rutter2001}),
which cause data wastage as we can only include studies which report at exactly the same threshold.

%%%%%%%%%%%%%%%%%%%%%%%%%%%%%%%%%%%%%%%%%%%%%%%%%%%%%%%%%%%
\paragraph{\underline{Ordinal log-likelihood}}
%%%%%%%%%%%%%%%%%%%%%%%%%%%%%%%%%%%%%%%%%%%%%%%%%%%%%%%%%%%
%%%%
%%%%%%%%%%%%%%%%%%%%%%%%%%%%%%%%%%%%%%%%%%%%%%%%%%%%%%%%%%%
\subparagraph{\underline{Ordinal probabilities} \newline} 
%%%%%%%%%%%%%%%%%%%%%%%%%%%%%%%%%%%%%%%%%%%%%%%%%%%%%%%%%%%
%%%% -------------------------------------------------------------------------------------------------------------------------------------
The likelihood for univariate ordinal regression
\supercite{Betancourt2019_ordinal, 
ord_regression_original_1st_ref_McCullagh_et_al_1980, 
ord_regression_original_2nd_ref_Winship_et_al_1984}
is based on the \textbf{ordinal probabilities}, which for our model are written
(assuming no individual-level covariates) as:
%%%%
\begin{equation}
\begin{aligned}
P^{ord[d]}_{1} & =  \Phi\left(   C_{s, 1}^{[d]}  -  \beta_{s}^{[d]}  \right) \\
P^{ord[d]}_{k} & =  \Phi\left(   C_{s, k}^{[d]}  -  \beta_{s}^{[d]}  \right)  -   \Phi\left(   C_{s, k - 1}^{[d]}  -  \beta_{s}^{[d]} \right) ~ , ~ \forall k \in \{2, \hdots, K - 1 \} \\
P^{ord[d]}_{K} & =  1 - \Phi\left(   C_{s, K - 1}^{[d]}  -  \beta_{s}^{[d]}  \right)
\end{aligned}
\label{equation_ordinal_bivariate_model_ordinal_probs_defn_for_log_likelihood}
\end{equation}
%%%%
Where: 
$ d \in \{ 0, 1 \} $ is an indicator for disease group, and
$ p_{k}^{[d]} $ is the probability of falling within test category $k$, conditional on disease status $d$.
%%%%

%%%%%%%%%%%%%%%%%%%%%%%%%%%%%%%%%%%%%%%%%%%%%%%%%%%%%%%%%%%
\subparagraph{\underline{Log-likelihood} \newline} 
%%%%%%%%%%%%%%%%%%%%%%%%%%%%%%%%%%%%%%%%%%%%%%%%%%%%%%%%%%%
%%%% -------------------------------------------------------------------------------------------------------------------------------------
Due to the inclusion of individual-level covariates, the likelihood for standard ordinal regression is conventionally 
\supercite{ord_regression_original_1st_ref_McCullagh_et_al_1980, ord_regression_original_2nd_ref_Winship_et_al_1984, Betancourt2019_ordinal}
incremented at the \textbf{individual-level}.
%%%%
Specifically for our ordinal-bivariate model, the log-likelihood for study $s$ conditional on disease group $d$ would be written 
in terms of the ordinal probabilities defined in 
\ref{equation_ordinal_bivariate_model_ordinal_probs_defn_for_log_likelihood}
as:
%%%%
\begin{equation}
\begin{aligned}
\text{log_lik} \left(  \beta_{s}^{[d]} , \underline{C}_{s}^{[d]} \mid \underline{Y} \right) & = 
         \sum_{n = 1}^{N_{s}^{[d]}} \left[ \text{I}\left[ Y_{n} = k \right]   \cdot    \log\left[     \Phi\left(   C_{s, k}^{[d]}  -  \beta_{s}^{[d]} \right)  -   \Phi\left(   C_{s, k - 1}^{[d]}  -  \beta_{s}^{[d]} \right) \right]  \right] \\ 
     & = \sum_{n = 1}^{N_{s}^{[d]}} \left[ \text{I}\left[ Y_{n} = k \right]   \cdot    \log\left(  P_{k}^{ord[d]}    \right)  \right]      
\end{aligned}
\label{equation_ordinal_bivariate_model_ordinal_regression_within_study_likelihood_defn}
\end{equation}
%%%%
Where:
$d \in \{ 0, 1 \} $ is an indicator for the disease group, 
$\underline{C}_{s}^{[d]}$ is the vector of cutpoint parameters for study s (i.e., $\underline{C}_{s}^{[d]} = \{ C_{s, 1}^{[d]}, \hdots,C_{s, K - 1}^{[d]} \} $),
$ N_{s}^{[d]} $ is the number of individuals in study $s$ in disease group $d$, and
$ \text{I}\left[ Y_{n} = k \right]$ is an indicator that individual $n$ has a test score equal to $k$.
%%%%

%%%%%%%%%%%%%%%%%%%%%%%%%%%%%%%%%%%%%%%%%%%%%%%%%%%%%%%%%%%
\paragraph{\underline{Conditional probability log-likelihood}}
%%%%%%%%%%%%%%%%%%%%%%%%%%%%%%%%%%%%%%%%%%%%%%%%%%%%%%%%%%%
%%%%
%%%%%%%%%%%%%%%%%%%%%%%%%%%%%%%%%%%%%%%%%%%%%%%%%%%%%%%%%%%
\subparagraph{\underline{Cumulative probabilities} \newline} 
%%%%%%%%%%%%%%%%%%%%%%%%%%%%%%%%%%%%%%%%%%%%%%%%%%%%%%%%%%%
%%%% -------------------------------------------------------------------------------------------------------------------------------------
Before defining the conditional probabilities, we will first define the \textbf{cumulative probabilities}. 
For our model, these probabilities are equal to:
%%%%
\begin{equation}
\begin{aligned}
P_{s, k}^{cumul[d]} & = Pr(Y_{s, n} \le k)   ~ , ~ \forall k \in \{ 1, \hdots, K - 1 \} \\            
                    & = \Phi\left(C_{s, k}^{[d]} -\beta_{s}^{[d]}\right)  ~ , ~ \forall k \in \{ 1, \hdots, K - 1 \} \\
P_{s, K}^{cumul[d]} & = Pr(Y_{s, n} \le K) = 1.
\end{aligned}
\label{equation_ordinal_bivariate_model_CUMULATIVE_probs_defn_for_log_likelihood}
\end{equation}
%%%%
I.e., $ P_{k}^{cumul[d]} $ represents the probability of an individual having a score equal to or below $k$. 
%%%%
Hence, $P_{k}^{cumul[d]}$ are \textbf{increasing} with $k$ (i.e., $ P_{k} \ge P_{k - 1} ~ \forall k$ - 
as is the standard convention in ordinal regression modeling - 
see e.g., McCullagh et al, 1980; \supercite{ord_regression_original_1st_ref_McCullagh_et_al_1980}; 
Windship et al, 1984\supercite{ord_regression_original_2nd_ref_Winship_et_al_1984}).

%%%%%%%%%%%%%%%%%%%%%%%%%%%%%%%%%%%%%%%%%%%%%%%%%%%%%%%%%%%
\subparagraph{\underline{Survival probabilities} \newline} 
%%%%%%%%%%%%%%%%%%%%%%%%%%%%%%%%%%%%%%%%%%%%%%%%%%%%%%%%%%%
%%%% -------------------------------------------------------------------------------------------------------------------------------------
Next, we will define the (ordinal) \textbf{"survival" probabilities}:
%%%%
\begin{equation}
\begin{aligned}
\textbf{}
%%%% P_{s, 1}^{surv[d]}  & = Pr( Y_{s, n} > 1 ) = 1. \\
P_{s, k}^{surv[d]}  & = Pr( Y_{s, n} > k )   ~ , ~ \forall k \in \{ 1, \hdots, K - 1 \} \\       
  & = Pr( Y_{s, n} \ge k + 1 )   \\ %%   ~ , ~ \forall k \in \{ 1, \hdots, K - 1 \} \\  
  & = 1 - Pr( Y_{s, n} \le k  )  \\ %%    ~ , ~ \forall k \in \{ 1, \hdots, K - 1 \} \\  
\implies P_{s, k}^{surv[d]}  & = 1 - Pr_{s, k}^{cumul[d]}  ~ , ~ \forall k \in \{ 1, \hdots, K - 1 \} \\  
%% \implies P_{s, k}^{surv[d]}  & = 1 - \Phi\left(     C_{s, k}^{[d]}  - \beta_{s}^{[d]}       \right)   \\ %%  ~ , ~ \forall k \in \{ 1, \hdots, K - 1 \} \\
\implies P_{s, k}^{surv[d]}  & = \Phi\left(     \beta_{s}^{[d]}   -  C_{s, k}^{[d]}    \right)   ~ , ~ \forall k \in \{ 1, \hdots, K - 1 \} \\
\end{aligned}
\label{equation_ordinal_bivariate_model_SURVIVAL_probs_defn_for_log_likelihood}
\end{equation}
%%%%
I.e., $ P_{k}^{surv[d]} $ represents the probability of an individual having a score above $k$.
%%%%
Hence, $P_{k}^{surv[d]}$ are \textbf{decreasing} with increasing $k$ (i.e., $ P_{k} \le P_{k - 1} ~ \forall k$. 
%%%%
These are the same probabilities that Jones et al \supercite{Jones2019} uses to define their model,
and in this paper, we will be using the same approach for the within-study likelihood.

%%%%%%%%%%%%%%%%%%%%%%%%%%%%%%%%%%%%%%%%%%%%%%%%%%%%%%%%%%%
\subparagraph{\underline{Conditional probabilities} \newline} 
%%%%%%%%%%%%%%%%%%%%%%%%%%%%%%%%%%%%%%%%%%%%%%%%%%%%%%%%%%%
%%%% -------------------------------------------------------------------------------------------------------------------------------------
Next, we can define the \textbf{conditional probabilities} - 
which we will write in terms of the survival probabilities defined in 
equation \ref{equation_ordinal_bivariate_model_SURVIVAL_probs_defn_for_log_likelihood} above:
%%%%
\begin{equation}
\begin{aligned}
P_{s, k}^{cond[d]}    & = Pr\left( Y \ge k + 1 \mid  Y \ge k \right)  ~ , ~ \forall k \in \{ 2, \hdots, K - 1 \} \\
                      & = \frac{   Pr(Y_{s, n} \ge k + 1)     }{    Pr(Y_{s, n} \ge k)     } \\ 
\implies P_{s, k}^{cond[d]}     & = \frac{   P_{s, k}^{surv[d]}    }{    P_{s, k - 1}^{surv[d]}      }  ~ , ~ \forall k \in \{ 2, \hdots, K - 1 \} \\
%% \implies P_{s, k}^{cond[d]}  & = \frac{    \Phi\left(     \beta_{s}^{[d]}   -  C_{s, k}^{[d]}    \right)      }{     \Phi\left(     \beta_{s}^{[d]}   -  C_{s, k - 1}^{[d]}    \right)    } \\                     
P_{s, 1}^{cond[d]}    & = Pr\left( Y \ge 2 \mid  Y \ge 1 \right)  \\
\implies P_{s, 1}^{cond[d]}    & = P_{s, 1}^{surv[d]},     ~ ~ \text{since:} ~ Pr(Y \ge 1) = 1. \\
\end{aligned}
\label{equation_ordinal_bivariate_model_CONDITIONAL_probs_IN_TERMS_OF_CUMUL_PROBS_defn_for_log_likelihood}
\end{equation}
%%%%
Hence, we are modelling the probability that a test result falls in categories $1$ through $k + 1$,
\textit{\textbf{given that}} it falls in categories $1$ through $k$.
%%%%
In other words, we are essentially asking:
"out of all of the individuals with results $ \ge k $, what fraction have results $ \ge k + 1$?".

%%%%
%%%% -------------------------------------------------------------------------------------------------------------------------------------
We can write these conditional probabilities 
(defined in \ref{equation_ordinal_bivariate_model_CONDITIONAL_probs_IN_TERMS_OF_CUMUL_PROBS_defn_for_log_likelihood})
directly in terms of the parameters of our ordinal-bivariate model parameters as:
%%%%
\begin{equation}
P_{s, k}^{cond[d]} = 
    \begin{dcases}
            \Phi\left(   \beta_{s}^{[d]}  -   C_{s, 1}^{[d]}     \right), & \text{if} ~ k = 1 \\
        \frac{    \Phi\left(  \beta_{s}^{[d]}  -   C_{s, k}^{[d]}  \right)    }{    \Phi\left(    \beta_{s}^{[d]}  - C_{s, k - 1}^{[d]}    \right)    }, & \text{if} ~ k \in \{2, \dots, K - 1\}
    \end{dcases}
\label{equation_ordinal_bivariate_model_CONDITIONAL_probs_defn_for_log_likelihood}
\end{equation}

%%%%%%%%%%%%%%%%%%%%%%%%%%%%%%%%%%%%%%%%%%%%%%%%%%%%%%%%%%%
\subparagraph{\underline{Log-likelihood} \newline} 
%%%%%%%%%%%%%%%%%%%%%%%%%%%%%%%%%%%%%%%%%%%%%%%%%%%%%%%%%%%
%%%% -------------------------------------------------------------------------------------------------------------------------------------
Then, instead of using the ordinal probabilities $ P_{k}^{ord[d]} $ 
(defined in equation \ref{equation_ordinal_bivariate_model_ordinal_probs_defn_for_log_likelihood}), 
for the ordinal-regression log-likelihood 
(as we defined in equation \ref{equation_ordinal_bivariate_model_ordinal_regression_within_study_likelihood_defn}), 
we can now re-parameterize this log-likelihood and write it in terms of the \textbf{conditional probabilities} which we just defined 
(see equations 
\ref{equation_ordinal_bivariate_model_CONDITIONAL_probs_IN_TERMS_OF_CUMUL_PROBS_defn_for_log_likelihood}  and
\ref{equation_ordinal_bivariate_model_CONDITIONAL_probs_defn_for_log_likelihood}).
%%%%
In other words, the \textbf{within-study log-likelihood} for study $s$ is implemented using a multinomial likelihood expressed as a series of conditionally independent binomial likelihoods. 
%%%%
This is the same approach as used in Jones et al, 2019 \supercite{Jones2019}, 
and which we will now define for our ordinal-bivariate model.

%%%%
%%%% -------------------------------------------------------------------------------------------------------------------------------------
Let $N^{[d]}_{s}$ denote the total number of individuals with disease status $d$ (where $d=0$ for diseased and $d=1$ for diseased) in study $s$. 
%%%%
Furthermore, let $x^{[d]}_{s, k}$ denote the number of individuals with test results falling at or above threshold $k$.
%%%%
Then, using the binomial factorization of the multinomial distribution (i.e., the approach used in Jones et al \supercite{Jones2019}), 
and using the conditional probabilities we defined in \ref{equation_ordinal_bivariate_model_CONDITIONAL_probs_defn_for_log_likelihood}, 
the factorized binomial likelihood for disease status group $d$ in study $s$ can be written as:
%%%%
\begin{equation}
\begin{aligned}
%% x^{[d]}_{s, k} & \sim \text{Binomial} \left(  N^{[d]}_{s}, P^{[d]}_{s,1}  \right) \\
% \left[ x^{[d]}_{s, k + 1} \mid x^{[d]}_{s, k}  \right]
% & \sim \text{Binomial}\left(  x^{[d]}_{s, k}, \frac{  P^{cumul [d]}_{s, k}  }{  P^{cumul [d]}_{s, k + 1}  }  \right),  ~ ~  k = \{2, \ldots, K \} \\
\left[ x^{[d]}_{s, 1}  \right]
& \sim \text{Binomial}\left(  N^{[d]}_{s}, ~ P^{cond[d]}_{s, 1}   \right) \\
\left[ x^{[d]}_{s, k + 1} \mid x^{[d]}_{s, k}  \right]
& \sim \text{Binomial}\left(  x^{[d]}_{s, k}, ~ P^{cond[d]}_{s, k + 1}   \right),  \quad k = \{2, \ldots, K - 1 \}
\end{aligned}
\label{equation_ordinal_bivariate_model_FACTORISED_MULTINOMIAL_BINOMIAL_defn_for_likelihood}
\end{equation}
% %%%%
% Where $xijt$ represents the number of individuals in group $j$ of study $i$ with test results at or below threshold $t$. This parameterization allows us to directly model the cumulative counts while maintaining the constraint that, as threshold increases, the probability of a test result falling below the threshold must increase.
%  \begin{equation*} \text{Binomial}(n~|~N,\theta) = \binom{N}{n}
% \theta^n (1 - \theta)^{N - n}. \end{equation*}
And hence, the \textbf{within-study log-likelihood} for study $s$ can be written as (using the binomial promability mass function): 
%%%%
% \begin{eqnarray*}
% \log \text{Binomial}(n~|~N, P) & = & \log
% \Gamma(N+1) - \log \Gamma(n + 1) - \log \Gamma(N- n + 1) \\[4pt] & & {
% } + n \log(P) + (N - n) \log(1 - P), 
% \end{eqnarray*}
\begin{equation}
\begin{aligned}
% L(x^{[d]}_{s,1:K} | \beta_{s}^{[d]}, C^{[d]}) &= \prod_{k=1}^{K-1} \binom{x_{s,k+1}^{[d]}}{x_{s,k}^{[d]}} \left(P_{s,k}^{cond[d]}\right)^{x_{s,k}^{[d]}} \left(1-P_{s,k}^{cond[d]}\right)^{x_{s,k+1}^{[d]} - x_{s,k}^{[d]}} \\
\text{log\_lik}\left(   \underline{x}^{[d]}_{s, 1:K} \mid \beta_{s}^{[d]}, \underline{C}^{[d]}_{[s]}  \right)  \propto  ~ 
& x_{s, 1}^{[d]}  \log\left(    P_{s, 1}^{cond[d]}  \right)   +   \left(   N_{s}^{[d]} - x_{s, 1}^{[d]}  \right) \log\left(  1 - P_{s, 1}^{cond[d]}  \right) ~ + \\
& \sum_{k = 2}^{K - 1}
\left[ x_{s, k}^{[d]} \log\left(  P_{s, k}^{cond[d]}  \right) + \left(  x_{s, k + 1}^{[d]} - x_{s, k}^{[d]}  \right) \log\left(  1 - P_{s, k}^{cond[d]}  \right) \right]
\end{aligned}
\label{equation_ordinal_bivariate_model_WITHIN_STUDY_LOG_LIKELIHOOD}
\end{equation}

%%%%
%%%% ---------------------------------------------------------------------------------------------------------------------------------
This factorization offers several advantages:
\begin{itemize}
    \item
    It allows modelling of cumulative negative test probabilities.
    \item
    It naturally accommodates studies reporting at different or missing thresholds.
    \item
    The conditional independence structure simplifies computation.
    (i.e., it does not require full individual-level test response data).
    \item
    It correctly maintains the ordering constraint that specificity
    and (1 - sensitivity) must change monotonically with threshold.
\end{itemize}

%%%%
%%%% ---------------------------------------------------------------------------------------------------------------------------------
Using this approach, we can make efficient use of all available threshold data, 
avoiding the data wastage that would occur when restricting analysis to studies reporting at the same thresholds.

%%%%%%%%%%%%%%%%%%%%%%%%%%%%%%%%%%%%%%%%%%%%%%
\subsubsection{Between-study heterogeneity}
\label{section_model_specs_Cerullo_bivariate_Reitsma_extension_between_study_heterogenity_bs}
%%%%%%%%%%%%%%%%%%%%%%%%%%%%%%%%%%%%%%%%%%%%%%
%%%%
%%%% ---------------------------------------------------------------------------------------------------------------------------------
Then, we model the between-study heterogeneity and between-study correlation \textbf{explicitly}, 
using a bivariate-normal distribution:
\begin{equation}
      \begin{bmatrix}    
                  {\beta_{s}}^{[d-]} \\
                  {\beta_{s}}^{[d+]}   
      \end{bmatrix} 
\sim
\text{bivariate\_normal}  \left( \underline{\mu_{\beta}} =
            \begin{bmatrix} 
                  \beta_{\bullet}^{[d-]} \\
                  \beta_{\bullet}^{[d+]} 
            \end{bmatrix}, ~~
            \boldsymbol\Sigma_{\beta} = 
            \begin{bmatrix} 
                 \ \left(\sigma^{[d-]}\right)^{2} &
                 \rho_{\beta}  \sigma_{\beta}^{[d-]}  \sigma_{\beta}^{[d+]} \\
                 \rho_{\beta}  \sigma_{\beta}^{[d-]}  \sigma_{\beta}^{[d+]} & 
                \ \left(\sigma^{[d+]}\right)^{2}
             \end{bmatrix}
 \right),
\label{equation_between_study_model_Cerullo_bivariate_Reitsma}
\end{equation}
Where:
We use the sub-script bullet ($\bullet$) notation to denote the summary (pooled) estimates
(i.e. $ \beta_{\bullet}^{[d]} $) -
as opposed to the study-specific estimates
which use an $s$-subscript (i.e. $ \beta_{s}^{[d]}, ~ C^{[d]}_{s, k} $).
$ \rho_{\beta} $ is the between-study correlation for the location parameters, 
between the diseased and non-diseased groups.
In other words, $ \rho_{\beta} $ models the correlation between $ {\beta^{[d-]}_{s}} $ and $ {\beta^{[d+]}_{s}} $ .
$ \sigma_{\beta}^{[d-]}$ and $ \sigma_{\beta}^{[d+]}$ are the standard deviations for the $ {\beta^{[d-]}_{s}} $ and $ {\beta^{[d+]}_{s}} $
(i.e., the location) parameters, which model the between-study heterogeneity.

% %%%%
% %%%% ------------------------------------------------------------------------------------------------------------------------------
% Furthermore, unlike the standard and routinely used bivariate model (Reitsma et al, 2005\supercite{Reitsma2005}) - 
% and hence our extension of it here (see section \ref{section_model_specs_Cerullo_bivariate_Reitsma_extension}) - 
% the between-study correlation is not modelled explicitly.
% Instead, it is modelled * implicitly * due to the relationship between the location and scale parameters, 
% in the non-diseased and diseased g roups. 

%%%%
%%%% ------------------------------------------------------------------------------------------------------------------------------
For the cutpoint parameters ($ C_{s, k}^{[d]} $), in this paper we consider 2 cases: 
%%%%
(i) using a fixed-effects between-study cutpoint model (i.e. $ C^{[d]}_{s, k} = C^{[d]}_{k} ~ \forall k$),
using the Induced-Dirichlet distribution \supercite{Betancourt2019_ordinal, cerullo_meta_ord} as a prior -
which we discuss in more detail in section \ref{section_model_specs_induced_Dirichlet_fixed_cutpoints_definition}); or:
%%%%
(ii) using a random-effects between-study cutpoint model, again using the Induced-Dirichlet distribution -
but as part of the observational model/likelihood itself - 
as opposed to only using it as a prior.
We define and discuss this in more detail in
section \ref{section_model_specs_induced_Dirichlet_random_cutpoints_definition}).

%%%%%%%%%%%%%%%%%%%%%%%%%%%%%%%%%%%%%%%%%%%%%%
\subsubsection{Summary (pooled) estimates}   
\label{section_model_specs_Cerullo_R_and_G_HSROC_summary_pooled_estimates}
%%%%%%%%%%%%%%%%%%%%%%%%%%%%%%%%%%%%%%%%%%%%%%
%%%% ------------------------------------------------------------------------------------------------------------------------------
The \textbf{summary estimates} for the ordinal-bivariate model are given by:
\begin{equation}
\begin{aligned}
Sp_{k}  &=      \Phi\left(      C^{[d-]}_{\bullet, k} - \beta_{\bullet}^{[d-]}      \right) \\
Se_{k}  &=  1 - \Phi\left(      C^{[d+]}_{\bullet, k} - \beta_{\bullet}^{[d+]}      \right)
\end{aligned}
\label{summary_estimates_Se_Sp_for_ordinal_bivariate_model}
\end{equation}
Where we use $C^{[d]}_{\bullet, k}$ to denote the summary-level cutpoints -
which could be pooled posterior median estimates (if random-effect/partially pooled cutpoints),
or complete pooling if using fixed-effect cutpoints,
in which case:
$C^{[d]}_{\bullet, k}$ = $C^{[d]}_{k}$ = $C^{[d]}_{s, k}$.
%%
% Where we use the sub-script bullet ($\bullet$) notation to denote the summary (pooled) estimates
% (i.e. $ \beta_{\bullet}^{[d]}, ~ C^{[d]}_{\bullet, k} $) -
% as opposed to the study-specific estimates
% which use an $s$-subscript (i.e. $ \beta_{s}^{[d]}, ~ C^{[d]}_{s, k} $).

%%%%
%%%% ------------------------------------------------------------------------------------------------------------------------------
Furthermore, in equation \ref{summary_estimates_Se_Sp_for_ordinal_bivariate_model},
the summary-level location parameters
(i.e., $ \beta_{\bullet}^{[d+]}, \beta_{\bullet}^{[d-]} $)
are obtained immediately from the bivariate-normal hierarchical model
(defined previously in
section \ref{section_model_specs_Cerullo_bivariate_Reitsma_extension_between_study_heterogenity_bs},
equation \ref{equation_between_study_model_Cerullo_bivariate_Reitsma}).
%%
% However, the summary-level cutpoint parameters 
% (i.e., $ C^{[d-]}_{\bullet, k},  C^{[d+]}_{\bullet, k} $) are only obtained immediately 
% if we are using a \textbf{fixed-effects cutpoints} version of the model,
% (i.e., if $C^{[d]}_{\bullet, k} = C^{[d]}_{s, k} = C^{[d]}_{k} ~ \forall k $)
%%
Additionally, we will discuss how to obtain the summary-level cutpoints for the more
general random-effects-cutpoints version of the model later in
section \ref{section_model_specs_induced_Dirichlet_fixed_cutpoints_definition}.

%%%%%%%%%%%%%%%%%%%%%%%%%%%%%%%%%%%%%%%%%%%%%%%%%%%%%%%%%%%%%%%%%%%%%%%%%%%%%%%%
\subsection{DTA-MA model II: The "Ordinal-HSROC" model}
\label{section_model_specs_Cerullo_R_and_G_HSROC}
%%%%%%%%%%%%%%%%%%%%%%%%%%%%%%%%%%%%%%%%%%%%%%%%%%%%%%%%%%%%%%%%%%%%%%%%%%%%%%%%
%%%%
%%%% --------------------------------------------------------------------------------------------------------------------------------
For the meta-analysis of a single ordinal diagnostic or screening test,
the measures of test accuracy at the $k$th cutpoint in study $s$ can be represented as:
\begin{equation}
    \begin{aligned}
    Sp_{s, k} &=
    \Phi\left( \frac{\left( C_{s, k} - (-1) \beta_{s}\right)}{\text{f}(  (-1) \gamma_{s}) }  \right) \\
    Se_{s, k} &=
    1 - \Phi\left( \frac{\left( C_{s, k} - (+1) \beta_{s}\right)}{\text{f}(  (+1) \gamma_{s}) }  \right)
    \end{aligned}
\end{equation}
Where:
$\Phi(\cdot)$ is the CDF of a standard normal,
and $f(\cdot)$ is some positive-constrained function.
For $f(\cdot)$, one could assume the "raw" scale parameters ($\gamma_{s}$) in study $s$ are log-normal so that:
$f(x) = \exp(x)$, in which case the scale parameters
are given by:
$\text{exp}(  1 \cdot \gamma) $ and:
$\text{exp}( -1 \cdot \gamma) $,
in the diseased and non-diseased groups, respectively.
Alternatively, one could use the softplus function to mitigate any non-linearity issues caused by using a log-normal transformation.
%%%%
% However, for the remainder of this paper, we will just use $f(\cdot) = exp(\cdot)$.
% This is because in pilot runs we did not find any issues caused from non-linearity 
% This is in contrast to the issue with the random-effect-cutpoint models, which require taking the pooled median estimate - 
% we discuss this non-linearity/skew issue in section .............
%%%%
% In this paper, we will consider both of these choices for $f$:
% %%%%
% \begin{equation}
% \begin{aligned}
% f \left( x \right) & = \text{softplus} \left( x \right) = \log\left(1 + \exp\left(x\right) \right), \text{or:} \\
% f \left( x \right) & = \text{exp} \left( x \right).
% \end{aligned}
% \end{equation}
% %%%%
% $f(\cdot) = \exp(\cdot)$ is less computationally expensive than the softplus function, 
% and is also more convenient for settings priors for the scale parameters; 
% for instance, using a prior of:
% $ \gamma \sim \text{normal}\left(0, 1\right) $
% translates to a 95\% prior interval of:

% \textcolor{red}{INSERT PRIOR HERE}

% and perhaps consider using the mean of the log-normal distribution
% (i.e. $\exp(\mu + 0.5  \cdot \sigma)$) rather than the median $\exp(\mu)$.

%%%%
%%%% -------------------------------------------------------------------------------------------------------------------------------------
The \textbf{within-study} model has the same form as that described in Jones et al. 
That is, a factorised multinomial likelihood - 
in other words, instead of using the to the \textbf{ordinal probabilities}: 
\begin{equation}
    \begin{aligned}
    P^{[d]ord}_{k} = \Phi\left(  \frac{ \left(C_{k} - (-1)^{(d + 1)} \beta_{s} \right)}{ (-1)^{(d + 1)}  f(\gamma_{s})} \right) - 
                     \Phi\left(  \frac{ \left(C_{k - 1}  - (-1)^{(d + 1)} \beta_{s} \right)}{ (-1)^{(d + 1)} f(\gamma_{s})} \right)
    \end{aligned}
\end{equation}
we instead use the \textbf{conditional probabilities} 
(i.e. ratio of successive \textbf{cumulative probabilities}), 
by using a series of conditionally independent binomial log-likelihoods.
Furthermore note that:
\begin{itemize}
    %%%%
    \item Any given $ C_{s, k} $ for test $t$ may be missing in some studies. 
    %%%%
    \item For the \textbf{fixed-effects-cutpoints} version of the model, we use the \textbf{induced-Dirichlet prior} for the cutpoints 
    $ C_{k} $ by mapping them to ordinal probabilities $ p^{*}_{k} = \Phi(C_{k} - \phi) - \Phi(C_{k - 1} - \phi) $, 
    and implementing the necessary Jacobian adjustment (e.g., in Stan).
    %% where $ \phi $ is an arbitrary anchor point which we set to zero. 
    %%%%
    % \item For the \textbf{random-effects cutpoints} version of the mode, we use the \textbf{induced-dirichlet model} for the study-specific cutpoints 
    % $ C_{s, k} $.
    %%%%
\end{itemize}
%%%%
%%%%%%%%%%%%%%%%%%%%%%%%%%%%%%%%%%%%%%%%%%%%%%
\subsubsection{Between-study heterogeneity}
\label{section_model_specs_Cerullo_R_and_G_HSROC_between_study_heterogenity_bs}
%%%%%%%%%%%%%%%%%%%%%%%%%%%%%%%%%%%%%%%%%%%%%%
%%%%
%%%% -------------------------------------------------------------------------------------------------------------------------------------
For the proposed Ordinal-HSROC model, \textbf{heterogeneity} (i.e., between-study variation)
is modelled with independent normal distributions, for both the location and scale parameters:
%%%%
\begin{equation}
\begin{aligned}
\beta_{s}  \sim \text{normal}\left( \mu_{\beta},  \sigma_{\beta} \right) \\
\gamma_{s} \sim \text{normal}\left( \mu_{\gamma}, \sigma_{\gamma} \right)
\end{aligned}
\end{equation}
%%%%
Furthermore, note that unlike the "bivariate" model (Reitsma et al, \supercite{Reitsma2005}) -
and hence our extension of it (see section \ref{section_model_specs_Cerullo_bivariate_Reitsma_extension}) -
in this model the between-study correlation is not modelled explicitly. 
Instead, it is modelled implicitly, due to the relationship between the location and scale parameters 
in the non-diseased and diseased groups.

%%%%%%%%%%%%%%%%%%%%%%%%%%%%%%%%%%%%%%%%%%%%%%
\subsubsection{Summary (pooled) estimates}
\label{section_model_specs_Cerullo_R_and_G_HSROC_summary_pooled_estimates}
%%%%%%%%%%%%%%%%%%%%%%%%%%%%%%%%%%%%%%%%%%%%%%
%%%%
%%%% -------------------------------------------------------------------------------------------------------------------------------------
The \textbf{summary estimates} for the Ordinal-HSROC model are given by:
%%%
\begin{equation}
\begin{aligned}
Se_{\bullet, k}  &=  
1 - \Phi\left( \frac{\left( C_{\bullet, k} - (-1) {\beta_{\bullet}} \right)}{\text{f}(  (-1) {\gamma_{\bullet}}) }  \right) \\
Sp_{\bullet, k}  &=     
\Phi\left( \frac{\left( C_{\bullet, k}  - (+1) {\beta_{\bullet}} \right)}{\text{f}(  (+1) {\gamma_{\bullet}}) }  \right)
\end{aligned}
\end{equation}
%%%%
%%%%
%%
% Sp_{s, k} &=
% \Phi\left( \frac{\left( C_{s, k} - (-1) \beta_{s}\right)}{\text{f}(  (-1) \gamma_{s}) }  \right) \\
% Se_{s, k} &=
% 1 - \Phi\left( \frac{\left( C_{s, k} - (+1) \beta_{s}\right)}{\text{f}(  (+1) \gamma_{s}) }  \right)
%%
%%%%%%%%%%%%%%%%%%%%%%%%%%%%%%%%%%%%%%%%%%%%%%%%%%%%%%%%%%%%%%%%%%%%%%%%%%%%%%%%
\subsection{ Modelling cutpoints using the induced-Dirichlet distribution}
\label{section_model_specs_induced_Dirichlet_distribution}
%%%%%%%%%%%%%%%%%%%%%%%%%%%%%%%%%%%%%%%%%%%%%%%%%%%%%%%%%%%%%%%%%%%%%%%%%%%%%%%%
%%%%
%%%% -------------------------------------------------------------------------------------------------------------------------
For either the ordinal-HSROC (see section \ref{section_model_specs_Cerullo_R_and_G_HSROC}),
or the ordinal-bivariate
(see section \ref{section_model_specs_Cerullo_bivariate_Reitsma_extension}) models,
defining sensible and coherent priors for the cutpoints is very challenging,
due to their ordering constraint,
and also due to the fact that they are defined in the latent space.

%%%%
%%%% -------------------------------------------------------------------------------------------------------------------------
This section will be dedicated to defining and presenting a solution to this problem -
which essentially boils down to modelling ordinal probabilities using a Dirichlet distribution -
rather than attempting to directly model the latent cutpoint parameters.
We will refer to this Dirichlet distribution as the "induced\_Dirichlet" distribution,
using the same terminology as Betancourt, 2019 \supercite{Betancourt2019_ordinal}.

%%%%%%%%%%%%%%%%%%%%%%%%%%%%%%%%%%%%%%%%%%%%%%%%%%%%%%%%%%%%%%%%%%%%%%%%%%%%%%%%
\subsubsection{The induced-Dirichlet distribution}
\label{section_model_specs_induced_Dirichlet_distribution_definition}
%%%%%%%%%%%%%%%%%%%%%%%%%%%%%%%%%%%%%%%%%%%%%%%%%%%%%%%%%%%%%%%%%%%%%%%%%%%%%%%%
%%%%
%%%% -------------------------------------------------------------------------------------------------------------------------
We can mitigate this difficulty by transforming the $K - 1$ cutpoint parameters
($\mathbf{C} = \{C_1, \hdots, C_{K - 1} \}$)
into $K$ ordinal probabilities
($\mathbf{P}^{ord} = \{P^{ord}_{1}, \hdots, P^{ord}_{K} \}$) -
which only depend on the cutpoints and some arbitrary constant $ \phi $.
%%%%
In other words, we define the ordinal probabilities to be functions such that:
$ \mathbf{P}^{ord} = \mathbf{P}^{ord}( \mathbf{C}, \phi ) $,
and these induced ordinal probabilities are defined as:
%%%%
\begin{equation}
\begin{aligned}
P^{ord}_{1} & = \Phi\left(  C_{1} - \phi  \right) \\
P^{ord}_{k} & = \Phi\left(  C_{k} - \phi  \right)  - \Phi\left(  C_{k - 1} - \phi  \right),
~ \text{for} ~ k \in \{2, \hdots, K - 1 \}  \\
P^{ord}_{K} & = 1 - \Phi\left(  C_{K - 1} - \phi  \right)
\end{aligned}
\label{equation_induced_Dirichlet_definition_induced_ordinal_probs_Po}
\end{equation}
%%%%
%%%%
Then, rather than putting a prior directly on the abstract and awkward cutpoints $\mathbf{C}$,
we can instead put a prior directly on the ordinal probabilities $\mathbf{P}^{ord}$.

%%%%
%%%% --------------------------------------------------------------------------------------------------------------------------
We can achieve this by putting a Dirichlet prior on $\mathbf{P}^{ord}$,
and then making sure to include the Jacobian adjustment for the transformation $ \mathbf{C} \mapsto \mathbf{P}^{ord} $.

%%%%
%%%% --------------------------------------------------------------------------------------------------------------------------
Note that we need a Jacobian adjustment since our real model parameters (i.e., Stan parameters block)
are $\mathbf{C}$ - not $\mathbf{P}^{ord}$ -
which are "transformed parameters" (i.e., defined in the Stan "transformed parameters").

%%%%
%%%% --------------------------------------------------------------------------------------------------------------------------
We can define this "Induced-Dirichlet" density more formally
(see Betancourt, 2019 \supercite{Betancourt2019_ordinal})
by writing:
%%%%
%%%%
\begin{equation}
\begin{aligned}
\text{Induced-Dir}\left( \mathbf{C} \mid \boldsymbol\alpha, \phi\right)
=
\text{Dir}\left( \mathbf{P}^{ord}( \mathbf{C}, \phi) \mid \boldsymbol\alpha \right)
  \cdot 
  \left| \mathbf{J}_{  \mathbf{C} \mapsto \mathbf{P}^{ord}  } \right|,
\end{aligned}
\label{induced_dirichlet_density_definition}
\end{equation}
Where:
\begin{itemize}
    \item
    $\mathbf{C}$ is the vector of $K - 1$ cutpoints (as previously defined).
    \item
    $ \mathbf{P}^{ord} $ are the $K$ "induced" ordinal probabilities
    (as previously defined in equation
    \ref{equation_induced_Dirichlet_definition_induced_ordinal_probs_Po}).
    \item
    $ \boldsymbol{\alpha} = (\alpha_{1}, \hdots, \alpha_{K}) $,
    denotes the set of $K$ Dirichlet "concentration" parameters.
    \item 
    $ \phi $ denotes the Induced-Dirichlet "anchor point"
    (which is typically set to zero).
\end{itemize}

%%%%
%%%% ----------------------------------------------------------------------------------------------------------------------------
Furthermore, $ \left| \mathbf{J}_{\mathbf{C} \mapsto \mathbf{P}^{ord} } \right| $
in equation \ref{induced_dirichlet_density_definition} denotes the Jacobian 
(i.e. the absolute derivative) of the transformation of going from cutpoints to the induced-Dirichlet ordinal probabilities. 
%%
%% This is needed because our actual model parameters are the cutpoints - not the ordinal probabilities
%% (they are "transformed parameters").#
%%
To compute this, we need to find the elements of the matrix  
$  \left| \mathbf{J}_{\mathbf{C}}  \right| 
= \left| \mathbf{J}_{\mathbf{C} \mapsto \mathbf{P}^{ord} } \right| $,
which are the partial derivatives of the transformation 
(i.e., each $ J_{i, j} =   \frac{\partial P^{ord}_{i} }{\partial C_{j} } $)
and then compute it's determinant. 
For more details on the Induced-Dirichlet distribution (including the Jacobian adjustment), 
please see Michael Betancourt's case study 
(Betancourt, 2019 \supercite{Betancourt2019_ordinal}).

%%%%%%%%%%%%%%%%%%%%%%%%%%%%%%%%%%%%%%%%%%%%%%%%%%%%%%%%%%%%%%%%%%%%%%%%%%%%%%%%
\subsubsection{Fixed-effects cutpoints}
\label{section_model_specs_induced_Dirichlet_fixed_cutpoints_definition}
%%%%%%%%%%%%%%%%%%%%%%%%%%%%%%%%%%%%%%%%%%%%%%%%%%%%%%%%%%%%%%%%%%%%%%%%%%%%%%%%
%%%%
%%%% ------------------------------------------------------------------------------------------------------------------------------
We can use the Induced-Dirichlet distribution, which we defined in
section \ref{section_model_specs_induced_Dirichlet_distribution_definition}
to define a fixed-effects cutpoint model for the cutpoint parameters
$ \mathbf{C} =  (C_{1}, \hdots, C_{K - 1}) $.
This can be done for both the ordinal-bivariate
(see section \ref{section_model_specs_Cerullo_bivariate_Reitsma_extension})
and ordinal-HSROC
(see section \ref{section_model_specs_Cerullo_R_and_G_HSROC})
models.

%%%%
%%%% ------------------------------------------------------------------------------------------------------------------------------
This allows us to set priors directly on the induced ordinal probabilities,
namely by using a Dirichlet prior for the Dirichlet concentrations
$ \boldsymbol{\alpha} = (\alpha_{1}, \hdots, \alpha_{K})  $.

%%%%
%%%% ----------------------------------------------------------------------------------------------------------------------------
This allows us to easily set either informative priors or a "flat" (i.e., uniform) prior on the induced ordinal
probabilities (and hence the cutpoints).
For example, we can set a flat prior by specifying the Dirichlet concentrations to be a ($K - 1$)-length vector of ones, i.e.:
$$ \boldsymbol\alpha^{\text{prior}} = (1, \hdots, 1) $$

\subsubsection{ Random-effects cutpoints}
\label{section_model_specs_induced_Dirichlet_random_cutpoints_definition}
%%%%%%%%%%%%%%%%%%%%%%%%%%%%%%%%%%%%%%%%%%%%%%%%%%%%%%%%%%%%%%%%%%%%%%%%%%%%%%%%
%%%%
%%%% --------------------------------------------------------------------------------------------------------------------------
For random-effects models, we use the induced-Dirichlet distribution to specify
a hierarchical partial-pooling observational model rather than merely as a prior.
For each study $s$, the study-specific cutpoint parameters
$ \mathbf{C}_{s}^{[d]} = \{C_{s, 1}^{[d]}, \hdots, C_{s, K - 1}^{[d]} \} $
are modelled as:
\begin{equation}
\begin{aligned}
\mathbf{C}_{s}^{[d]}    & \sim \text{induced\_Dirichlet}\left( \boldsymbol\alpha^{[d]} \right), \text{ where:} \\
\boldsymbol\alpha^{[d]} & = 0.01 + \boldsymbol\phi^{[d]} \cdot \kappa^{[d]}, \\
\boldsymbol\phi^{[d]}   & \sim \text{Dirichlet}\left( \boldsymbol\alpha^{\text{prior}} \right), \\
\log(\kappa^{[d]})      & \sim \text{Student-}t\left( \nu^{[d]}, \mu_{\kappa}^{[d]}, \sigma_{\kappa}^{[d]} \right)
\end{aligned}
\label{equation_induced_Dirichlet_kappa_parameterization}
\end{equation}
where:
\begin{itemize}
    \item
    $\boldsymbol\phi^{[d]}$ is a $K$ -
    simplex representing the mean category probabilities.
    \item
    $\kappa^{[d]} \ge 1$ controls the concentration (precision) of the distribution.
    \item
    The small constant $0.01$ ensures numerical stability.
    \item 
    The Student-$t$ prior on $\log(\kappa^{[d]})$ includes degrees of freedom $\nu^{[d]}$,
    location $\mu_{\kappa}^{[d]}$, 
    and scale $\sigma_{\kappa}^{[d]}$.
    \item 
    A Jacobian adjustment of $-\log(\kappa^{[d]})$ is applied for the log-scale parameterization. 
    See below for details.
\end{itemize}
%%%%
%%%% --------------------------------------------------------------------------------------------------------------------------
%%%%%%%%%%%%%%%%%%%%%%%%%%%%%%%%%%%%%%%%%%%%%%%%%%%%%%%%%%%
\paragraph{\underline{ Jacobian adjustment (for prior on $log(\kappa^{[d]})$): }}
%%%%%%%%%%%%%%%%%%%%%%%%%%%%%%%%%%%%%%%%%%%%%%%%%%%%%%%%%%%
%%%%
%%%% --------------------------------------------------------------------------------------------------------------------------
Since the prior is directly on
$\log(\kappa^{[d]})$, to obtain $\log|J|$,
we need to find the following derivative:
$\frac{d\log(\kappa)}{d\kappa}$
(see equation \ref{log_det_J_kappa} below).
\begin{equation}
\begin{aligned}
    % g: \kappa \mapsto \log(\kappa) \\ 
    % g^{-1}: \log(\kappa) \mapsto \kappa = \exp(\log(\kappa)) \\
    p(\kappa) = \pi(\log(\kappa)) \cdot \left| \frac{d\log(\kappa)}{d\kappa} \right| \\
    \frac{d\log(\kappa)}{d\kappa} = \frac{1}{\kappa} \\
    |J| = \frac{1}{\kappa} \\
    \log|J| = \log\left(\frac{1}{\kappa}\right) = -\log(\kappa)
\label{log_det_J_kappa}
\end{aligned}
\end{equation}
Note that we are not differentiating w.r.t the unconstrained parameter ($\kappa^{\text{unc}}$) -
since we are using Stan -
which allows you to declare parameters with constraints and does the necessary Jacobian adjustments behind the scenes
(in C++).
However, if we were writing our own sampler, 
we would also need to multiply the Jacobian adjustment in equation \ref{log_det_J_kappa} by the derivative of:
$\kappa$ w.r.t $\kappa^{\text{unc}}$.
%%
% $$$$ %% \textbf{Forward transform} (from parameter to prior space):
% $$$$ %% \textbf{Reverse transform} (from prior space to parameter):
% %%
% For change of variables in probability densities, 
% if we specify a prior on $\log(\kappa)$ but sample $\kappa$, we need:
% %%
% $$$$
% $$$$ %% \textbf{The derivative} (forward transform):
% $$$$ %% \textbf{Jacobian} (absolute value):
% $$$$ %% \textbf{Log-Jacobian} (what appears in Stan):
%%
%%%%%%%%%%%%%%%%%%%%%%%%%%%%%%%%%%%%%%%%%%%%%%%%%%%%%%%%%%%
\paragraph{\underline{ Computing summary cutpoints}}
%%%%%%%%%%%%%%%%%%%%%%%%%%%%%%%%%%%%%%%%%%%%%%%%%%%%%%%%%%%
%%%%
%%%% --------------------------------------------------------------------------------------------------------------------------
Our previous work (Cerullo et al, 2022\supercite{cerullo_meta_ord})
derived summary cutpoint estimates from the population-level concentration parameters $\boldsymbol\alpha^{[d]}$
by computing expected category probabilities
$P^{ord}_{\bullet, k} = \alpha_{k} / \alpha_{0}$
and transforming these back to the cutpoint scale.

%%%%
%%%% --------------------------------------------------------------------------------------------------------------------------
However, we subsequently discovered that this approach produces biased summary estimates
when $K > 5$ categories due to skew introduced by the complex multidimensional ordering constraints
inherent to hierarchical ordinal regression models.
This limitation was not detectable in our earlier work because we only examined tests with $K \le 5$ categories.

%%%%
%%%% --------------------------------------------------------------------------------------------------------------------------
To address this issue, we now compute summary cutpoints as the \textbf{medians} 
of the posterior distributions of study-specific cutpoints:
\begin{equation}
C_{\bullet, k}^{[d]} = \text{median}\left( \{ C_{1, k}^{[d]}, C_{2, k}^{[d]}, \hdots, C_{N_{\text{studies}}, k}^{[d]} \} \right)
\label{equation_summary_cutpoints_via_medians}
\end{equation}
The median provides a robust summary statistic that is unaffected by the geometric constraints
and asymmetry of the posterior distribution,
yielding unbiased pooled sensitivity and specificity estimates,
even for tests with many ordinal categories.
We verified this was the case in pilot studies 
(results not shown here).
This methodological improvement represents a key advance over our previous framework,
enabling valid inference for complex screening instruments such as the BAI ($64$ categories).

% %%%%%%%%%%%%%%%%%%%%%%%%%%%%%%%%%%%%%%%%%%%%%%%%%%%%%%%%%%%
% \paragraph{\underline{ Computing summary cutpoints:  }}
% %%%%%%%%%%%%%%%%%%%%%%%%%%%%%%%%%%%%%%%%%%%%%%%%%%%%%%%%%%%
% %%%%
% %%%% --------------------------------------------------------------------------------------------------------------------------
% %%
% Unlike population-level approaches that derive summary estimates from $\boldsymbol\alpha^{[d]}$,
% as in our previous work
% (Cerullo et al, 2022\supercite{cerullo_meta_ord}).
% %%
% we compute summary cutpoints as the \textbf{medians} of the posterior distributions
% of study-specific cutpoints:
% %%
% \begin{equation}
% C_{\bullet, k}^{[d]} = \text{median}\left( \{ C_{1, k}^{[d]}, C_{2, k}^{[d]}, \hdots, C_{N_{\text{studies}}, k}^{[d]} \} \right)
% \label{equation_summary_cutpoints_via_medians}
% \end{equation}
% %%
% This approach avoids bias/skew arising from the complex multidimensional ordering constraints
% on cutpoint parameters in hierarchical models,
% particularly when $K > 5$ categories.
% %%
% The median provides a robust summary that is less affected by these geometric constraints
% than population-level parameter estimates.

%%%%%%%%%%%%%%%%%%%%%%%%%%%%%%%%%%%%%%%%%%%%%%%%%%%%%%%%%%%%%%%%%%%%%%%%%%%%%%%%
\paragraph{\underline{ The ordinal-bivariate model}}
\label{section_model_specs_induced_Dirichlet_random_cutpoints_for_ordinal_bivariate}
%%%%%%%%%%%%%%%%%%%%%%%%%%%%%%%%%%%%%%%%%%%%%%%%%%%%%%%%%%%%%%%%%%%%%%%%%%%%%%%%
%%%%
%%%% --------------------------------------------------------------------------------------------------------------------------
For the ordinal-bivariate model
(section \ref{section_model_specs_Cerullo_bivariate_Reitsma_extension}),
we have two independent sets of cutpoints for non-diseased ($d = 0$) and diseased ($d = 1$) populations:
$ \{C_{s, 1}^{[0]}, \hdots, C_{s, K - 1}^{[0]} \} $
and
$ \{C_{s, 1}^{[1]}, \hdots, C_{s, K - 1}^{[1]} \} $.
The between-study model for each disease group is:
\begin{equation}
\begin{aligned}
{\mathbf{P}^{[d]}_{s}}^{ord}( {\mathbf{C}_{s}}^{[d]}, \phi = 0) &
\sim \text{Dir}\left( \boldsymbol\alpha^{[d]} \right) \circ
\left| \mathbf{J}_{ {\mathbf{C}_{s}}^{[d]} \mapsto {\mathbf{P}^{[d]}_{s}}^{ord} } \right|, \\
\text{where: } \boldsymbol\alpha^{[d]} & = 0.01 + \boldsymbol\phi^{[d]} \cdot \kappa^{[d]}, \\
\boldsymbol\phi^{[d]} & \sim \text{Dir}\left( \boldsymbol\alpha^{\text{prior}, [d]} \right), \\
\log(\kappa^{[d]}) & \sim \text{Student-}t\left( \nu^{[d]}, \mu_{\kappa}^{[d]}, \sigma_{\kappa}^{[d]} \right), \\
\text{and } d & \in \{0, 1\}.
\end{aligned}
\label{equation_induced_Dirichlet_RE_ordinal_bivariate_kappa}
\end{equation}
The Jacobian term accounts for the transformation from cutpoints to ordinal probabilities.
Summary cutpoints for computing pooled Se/Sp are obtained via 
equation \ref{equation_summary_cutpoints_via_medians}.
%%
%%
%%%%%%%%%%%%%%%%%%%%%%%%%%%%%%%%%%%%%%%%%%%%%%%%%%%%%%%%%%%%%%%%%%%%%%%%%%%%%%%%
\paragraph{\underline{ The ordinal-HSROC model}}
\label{section_model_specs_induced_Dirichlet_random_cutpoints_for_ordinal_HSROC}
%%%%%%%%%%%%%%%%%%%%%%%%%%%%%%%%%%%%%%%%%%%%%%%%%%%%%%%%%%%%%%%%%%%%%%%%%%%%%%%%
%%%%
%%%% -------------------------------------------------------------------------------------------------------------------------
%%
For the ordinal-HSROC model
(section \ref{section_model_specs_Cerullo_R_and_G_HSROC}) -
unlike the ordinal-bivariate model -
cutpoints are shared across disease groups; however, they vary by study.
More specifically:

%%%%
%%%% -------------------------------------------------------------------------------------------------------------------------
Let $\mathbf{C}_{s} = \{C_{s, 1}, \hdots, C_{s, K - 1}\}$ denote 
study $s$'s cutpoint vector.
The between-study model is:
%%%%
\begin{equation}
\begin{aligned}
{\mathbf{P}_{s}}^{ord}( \mathbf{C}_{s}, \phi = 0) & 
\sim \text{Dir}\left( \boldsymbol\alpha \right) \circ 
\left| \mathbf{J}_{ \mathbf{C}_{s} \mapsto {\mathbf{P}_{s}}^{ord} } \right|, \\
\text{where: } \boldsymbol\alpha & = 0.01 + \boldsymbol\phi \cdot \kappa, \\
\boldsymbol\phi & \sim \text{Dir}\left( \boldsymbol\alpha^{\text{prior}} \right), \\
\log(\kappa) & \sim \text{Student-}t\left( \nu, \mu_{\kappa}, \sigma_{\kappa} \right).
\end{aligned}
\label{equation_induced_Dirichlet_RE_ordinal_HSROC_kappa}
\end{equation}
%%%%
Since cutpoints are shared, we use a single induced-Dirichlet hierarchy 
(unlike the ordinal-bivariate model which requires two independent hierarchies).
Summary cutpoints are computed via equation \ref{equation_summary_cutpoints_via_medians},
then used with the HSROC parameterization 
(section \ref{section_model_specs_Cerullo_R_and_G_HSROC})
to compute pooled Se/Sp at specified $\Theta$ and $\Lambda$ values.

\subsection{Quantifying between-study heterogeneity \& correlation}
\label{section_model_specs_quantifying_between_study_heterogenity}
%%%%%%%%%%%%%%%%%%%%%%%%%%%%%%%%%%%%%%%%%%%%%%%%%%%%%%%%%%%%%%%%%%%%%%%%%%%%%%%%
%%%%
%%%%
%%%%%%%%%%%%%%%%%%%%%%%%%%%%%%%%%%%%%%%%%%%%%%%%%%%%%%%%%%%%%%%%%%%%%%%%%%%%%%%%
\subsubsection{ Fixed-effects cutpoint models}
\label{section_model_specs_quantifying_between_study_heterogenity_with_fixed_cutpoints}
%%%%%%%%%%%%%%%%%%%%%%%%%%%%%%%%%%%%%%%%%%%%%%%%%%%%%%%%%%%%%%%%%%%%%%%%%%%%%%%%
%%%%
%%%% ------------------------------------------------------------------------------------------------------------------------------
As we mentioned in section \ref{section_model_specs_Cerullo_bivariate_Reitsma_extension}, 
for the ordinal-bivariate model,
the between-study heterogeneity in the 
study-specific sensitivity and specificities is quantified by the between-study heterogeneity of the location parameters
($ \beta_{s}^{[d+]}, \beta_{s}^{[d-]}$), 
which are \textbf{directly} modelled and quantified by the standard deviations of the bivariate normal between-study model,
that is, by
$ \sigma_{\beta}^{[d-]} $ and $ \sigma_{\beta}^{[d+]} $ in the non-diseased and diseased groups, respectively.

%%%%
%%%% ------------------------------------------------------------------------------------------------------------------------------
On the other hand, for the ordinal-HSROC model, 
since we do not model the between-study heterogeneity of the probit-transformed sensitivities and specificities as directly as we do
in the ordinal-bivariate model, we do not immediately obtain parameters which quantify this heterogeneity. 
%%%%
However, we can obtain equivalent "bivariate-like" quantities from the ordinal-HSROC
model by using the relationships we found in section \ref{section_model_specs_relationship_between_Cerullo_bivariate_and_Cerullo_R_and_G_HSROC_MA_models}.

%%%%
%%%% ------------------------------------------------------------------------------------------------------------------------------
In other words, for the ordinal-HSROC model, 
we can quantify the between-study heterogeneity in the probit-transformed sensitivities and specificities by computing:
$$ \sigma_{\beta}^{[d]} =  \exp\left( (-1)^d \mu_{\gamma}^H  \right) \cdot  \sqrt{ \left[ \left(  \sigma_{\beta}^H \right)^2 ~ + ~  \left(  \mu_{\beta}^H \cdot \sigma_{\gamma}^H  \right)^2  \right] } $$
%  $$ \sigma_{\beta}^{[d-]} =  \exp\left(   \mu_{\gamma}^H  \right)   \sqrt{ \cdot \left[ \left(  \sigma_{\beta}^H \right)^2 ~ + ~  \left(  \mu_{\beta}^H \cdot \sigma_{\gamma}^H  \right)^2  \right] } $$#
Where $ d $ is an indicator for disease status 
(i.e., $ d = 1 $ for the diseased population, 
and $ d = 0 $ for the non-diseased population).

%%%%%%%%%%%%%%%%%%%%%%%%%%%%%%%%%%%%%%%%%%%%%%%%%%%%%%%%%%%%%%%%%%%%%%%%%%%%%%%%
\subsubsection{ Random-effects cutpoint models}
\label{section_model_specs_quantifying_between_study_heterogenity_with_random_cutpoints}
%%%%%%%%%%%%%%%%%%%%%%%%%%%%%%%%%%%%%%%%%%%%%%%%%%%%%%%%%%%%%%%%%%%%%%%%%%%%%%%%
%%%%
%%%% ------------------------------------------------------------------------------------------------------------------------------
It is very important to note that $ \sigma_{\beta}^{[d-]} $ and $ \sigma_{\beta}^{[d+]} $ 
can only explain $100\%$ of the between-study heterogeneity of the (probit-transformed) 
sensitivities and specificities if we are using a \textbf{fixed-effects} 
between-study model for the cutpoint parameters -
that is, only if:
$ C_{s, k}^{[d+]} = C_{k}^{[d+]} $ and $ C_{s, k}^{[d-]} = C_{k}^{[d-]}  $  $ ~ \forall s $ 
if we are using the ordinal-bivariate model, and only if: 
$ C_{s, k}^H = C_{k}^H ~ \forall s $
if we are using the ordinal-HSROC model.

%%%%
%%%% ------------------------------------------------------------------------------------------------------------------------------
Now we need to work out how to quantify between-study heterogeneity in the ordinal-bivariate and ordinal-HSROC models
if we use a random-effects induced-Dirichlet between-study model for the cutpoints.

%%%%%%%%%%%%%%%%%%%%%%%%%%%%%%%%%%%%%%%%%%%%%%%%%%%%%%%%%%%%%%%%%%%%%%%%%%%%%%%%
\subsection{ The relationship between the proposed ordinal-bivariate and the ordinal-HSROC models }
\label{section_model_specs_relationship_between_Cerullo_bivariate_and_Cerullo_R_and_G_HSROC_MA_models}
%%%%%%%%%%%%%%%%%%%%%%%%%%%%%%%%%%%%%%%%%%%%%%%%%%%%%%%%%%%%%%%%%%%%%%%%%%%%%%%%
%%%%
%%%% -----------------------------------------------------------------------------------------------------------------------------
In appendix
\ref{appendix_model_specs_relationship_between_Cerullo_bivariate_and_Cerullo_R_and_G_HSROC_MA_models},
we show how the two models we proposed in sections
\ref{section_model_specs_Cerullo_R_and_G_HSROC} and
\ref{section_model_specs_Cerullo_bivariate_Reitsma_extension}
are very closely related -
and that the ordinal-HSROC model can in fact be thought of a more restricted,
test-accuracy-specific version of the ordinal-bivariate model.

%%%%%%%%%%%%%%%%%%%%%%%%%%%%%%%%%%%%%%%%%%%%%%%%%%%%%%%%%%%%%%%%%%%%%%%%%%%%%%%%
\setcounter{secnumdepth}{4}  % To number these deep sections
\setcounter{tocdepth}{4}     % To include them in the table of contents
\numberwithin{equation}{section} %% number equations / appendix equations / equation numbering / bookmark
%%%%%%%%%%%%%%%%%%%%%%%%%%%%%%%%%%%%%%%%%%%%%%%%%%%%%%%%%%%%%%%%%%%%%%%%%%%%%%%%

%%%%%%%%%%%%%%%%%%%%%%%%%%%%%%%%%%%%%%%%%%%%%%%%%%%%%%%%%%%%%%%%%%%%%%%%%%%%%%%%%%%%%%%%%%%%%%%%%%%%%%%%%%%%%%%%%%%%%%%%%%%%%%%%%%%%%%%
\newpage
\section{Network meta-analysis (NMA) extensions}
\label{section_model_specs_NMA_extensions}
%%%%%%%%%%%%%%%%%%%%%%%%%%%%%%%%%%%%%%%%%%%%%%%%%%%%%%%%%%%%%%%%%%%%%%%%%%%%%%%%%%%%%%%%%%%%%%%%%%%%%%%%%%%%%%%%%%%%%%%%%%%%%%%%%%%%%%%
%%%%
%%%% ------------------------------------------------------------------------------------------------------------------------------
We now extend our previously defined ordinal regression models
(see section \ref{section_model_specs})
for DTA-MA to the network meta-analysis framework (DTA-NMA),
allowing for the inclusion of studies that assess multiple index tests, 
and enabling comparative analysis between tests.
These extensions follow the arm-based approach of Nyaga et al.\supercite{Nyaga2018},
but tailored to our ordinal regression framework.
%%%%
%%%%
%%%%%%%%%%%%%%%%%%%%%%%%%%%%%%%%%%%%%%%%%%%%%%%%%%%%%%%%%%%%%%%%%%%%%%%%%%%%%%%%
\subsection{Background: The Nyaga et al. Arm-Based NMA Model for Diagnostic Test Accuracy}
\label{section_model_specs_Nyaga_NMA_background}
%%%%%%%%%%%%%%%%%%%%%%%%%%%%%%%%%%%%%%%%%%%%%%%%%%%%%%%%%%%%%%%%%%%%%%%%%%%%%%%%
%%%%
%%%% ------------------------------------------------------------------------------------------------------------------------------
Before presenting our network meta-analysis (NMA) extensions of the ordinal regression DTA models, 
we first review the arm-based NMA approach for screening/diagnostic test accuracy, 
as proposed by Nyaga et al \supercite{Nyaga2018}.

Consider a setting with $T$ tests and $S$ studies, where not all tests are evaluated in each study. 
For a certain study $s$, 
let $(Y^{[1]}_{s, t}, Y^{[0]}_{s, t})$ 
denote the true positives and true negatives, 
$(N^{[1]}_{s, t}, N^{[0]}_{s, t})$ the diseased and healthy individuals, and 
$(\pi^{[1]}_{s, t}, \pi^{[0]}_{s, t})$ 
the "unobserved" sensitivity and specificity respectively with test $t$ in study $s$. 
The study serves as a block where all diagnostic accuracy tests are hypothetically evaluated, though some may be missing.

Given study-specific sensitivity and specificity, the distribution of true positives and true negatives among the diseased and the healthy individuals follows:
%%%%
\begin{equation}
Y^{[d]}_{s, t} \mid \pi^{[d]}_{s, t} \sim \text{Binomial}(\pi^{[d]}_{s, t}, N^{[d]}_{s, t}), ~ s = \{1, \ldots, S\}, ~ t = \{1, \ldots, T\}.
\end{equation}
%%%%
\noindent where $d = 1$ refers to the diseased individuals and $d = 0$ to the healthy individuals.

The arm-based single-factor design with repeated measures model is written as follows:
%%%%
\begin{equation}
\pi^{[d]}_{s, t} = \Phi\left( \mu^{[d]}_{t} + \eta^{[d]}_{s} + \delta^{[d]}_{s, t} \right)
\end{equation}
%%%%
\begin{equation}
\begin{pmatrix} \eta^{[0]}_{s} \\ \eta^{[1]}_{s} \end{pmatrix} \sim N \left( \begin{pmatrix} 0 \\ 0 \end{pmatrix}, \boldsymbol{\Sigma} \right)
\end{equation}
%%%%
\begin{equation}
\boldsymbol{\Sigma} = \begin{bmatrix} \sigma^2_1 & \rho\sigma_1\sigma_2 \\ \rho\sigma_1\sigma_2 & \sigma^2_2 \end{bmatrix}
\end{equation}
%%%%
\begin{equation}
(\delta^{[d]}_{s, 1}, \delta^{[d]}_{s, 1}, \ldots \delta^{[d]}_{s, T}) \sim N(\mathbf{0}, ~ \text{diag}( {  (\tau^{[d]}})^2) )
\end{equation}
%%%%
\noindent where:
$\mu^{[1]}_{t}$ and $\mu^{[0]}_{t}$ are the mean sensitivity and specificity in a hypothetical study with random-effects equal to zero respectively. 
$\eta^{[d]}_{s}$ is the study effect for healthy individuals ($d = 1$) or diseased individuals ($d = 2$) and represents the deviation of a particular study $s$ from the mean sensitivity ($d = 1$) or specificity ($d = 0$), inducing between-study correlation. 
The study effects are assumed to be a random sample from a population of such effects.

The between-study variability of sensitivity and specificity and the correlation thereof is captured by the parameters $\sigma^2_1$, $\sigma^2_2$, and $\rho$ respectively. 
$\delta^{[d]}_{s, t}$ is the error associated with the sensitivity ($d = 1$) or specificity ($d = 2$) of test $t$ in the $s^{th}$ study. 
Conditional on study $s$, the repeated measurements are independent with variance constant across studies such that:
$ (\tau^{[d]})^2 = (( \tau^{[d]}_{1})^2, \ldots, (\tau^{[d]}_{T})^2)$ is a $T$ dimensional vector of homogeneous variances.

In essence, the model separates the variation in the studies into two components: the within-study variation 
$\text{diag}(  (\tau^{[d]})^2 )$, 
referring to the variation in the repeated sampling of the study results if they were replicated, 
and the between-study variation 
$\boldsymbol{\Sigma}$, 
referring to variation in the studies' true underlying effects.

\subsection{ NMA Extension of the ordinal-bivariate model}
\label{section_model_specs_NMA_ordinal_bivariate}
%%%%%%%%%%%%%%%%%%%%%%%%%%%%%%%%%%%%%%%%%%%%%%%%%%%%%%%%%%%%%%%%%%%%%%%%%%%%%%%%
%%%%
%%%% -------------------------------------------------------------------------------------------------------------------------------------
In this section, we extend the proposed ordinal-bivariate model 
(see section \ref{section_model_specs_Cerullo_bivariate_Reitsma_extension}) 
model to the network meta-analysis (NMA) framework, 
using the NMA structure based on Nyaga et al, 2018\supercite{Nyaga2018}.
For each test $t$ in study $s$,
the test accuracy measures at the $k$th cutpoint are represented as:
\begin{equation}
\begin{aligned}
Sp_{s, t, k} & =     \Phi  \left( C_{s, t, k}^{[d-]} - \beta^{[d-]}_{s, t}\right) \\
Se_{s, t, k} & = 1 - \Phi  \left( C_{s, t, k}^{[d+]} - \beta^{[d+]}_{s, t}\right)
\end{aligned}
\end{equation}
Where:
$\beta^{[d-]}_{s, t}$ and $\beta^{[d+]}_{s, t}$ are the location parameters for test $t$ in study $s$, 
in the non-diseased ($[d-]$) and diseased ($[d+]$) groups, respectively, and:
$C_{s, t, k}^{[d-]}$ and $C_{s, t, k}^{[d+]}$ are the latent cutpoint parameters for study $s$, test $t$ at cutpoint $k$, 
in the non-diseased ($[d-]$) and diseased ($[d+]$) groups, respectively.
%%%%
%%%%
%%%%%%%%%%%%%%%%%%%%%%%%%%%%%%%%%%%%%%%%%%%%%%%%%%%%%%%%%%%%%%%%%%%%%%%%%%%%%%%%
\subsubsection{ Within-study model}
\label{section_model_specs_NMA_ordinal_bivariate_within_study_model}
%%%%%%%%%%%%%%%%%%%%%%%%%%%%%%%%%%%%%%%%%%%%%%%%%%%%%%%%%%%%%%%%%%%%%%%%%%%%%%%%
%%%%
%%%% -----------------------------------------------------------------------------------------------------------------------------------
The within-study models are defined exactly as they were for the non-NMA 
(i.e., MA/single test only) 
version of this model; 
see section \ref{section_model_specs_Cerullo_bivariate_Reitsma_extension_within_study_model}
for more information.
The within-study models are defined exactly this way for every test $t$.
%%%%
%%%%
%%%%%%%%%%%%%%%%%%%%%%%%%%%%%%%%%%%%%%%%%%%%%%%%%%%%%%%%%%%%%%%%%%%%%%%%%%%%%%%%
\subsubsection{ Hierarchical Structure and between-study model}
\label{section_model_specs_NMA_ordinal_bivariate_between_study_model}
%%%%%%%%%%%%%%%%%%%%%%%%%%%%%%%%%%%%%%%%%%%%%%%%%%%%%%%%%%%%%%%%%%%%%%%%%%%%%%%%
%%%%
%%%% -------------------------------------------------------------------------------------------------------------------------------------
Following the Nyaga et al.\supercite{Nyaga2018} framework and using similar notation
(see section \ref{section_model_specs_Nyaga_NMA_background}), 
we decompose the location parameters using a two-level hierarchical structure:
\begin{equation}
\begin{aligned}
\beta_{s,t}^{[d]}  &= \beta_{\bullet,t}^{[d]}  + \eta_{\beta, s}^{[d]}    + \delta_{\beta, s, t}^{[d]}
\end{aligned}
\label{equation_NMA_ordinal_bivariate_two_level_structure_for_locations}
\end{equation}
Where:
\begin{itemize}
    \item $\beta_{\bullet, t}^{[d]}$ are the mean location parameters for test $t$ in disease group $d$.
    \item $\eta_{\beta, s}^{[d]}$ are study-level random effects shared across all tests within study $s$, disease group $d$.
    \item $\delta_{\beta, s, t}^{[d]}$ are test-specific deviations for test $t$ in study $s$, disease group $d$.
\end{itemize}
%%%%
%%%% -------------------------------------------------------------------------------------------------------------------------------------
The study-level random effects 
(i.e., the $\eta_{\beta, s}^{[d]}$ in equation \ref{equation_NMA_ordinal_bivariate_two_level_structure_for_locations} above) - 
which are shared across tests -
follow a bivariate normal distribution:
\begin{equation}
    \begin{pmatrix} \eta_{\beta, s}^{[d-]} \\ \eta_{\beta, s}^{[d+]} \end{pmatrix}
    \sim \text{bivariate\_normal}
    \left( \begin{bmatrix} 0 \\ 0 \end{bmatrix}, ~ \boldsymbol{\Sigma}^{NMA}_\beta \right)
\end{equation}
Where:
\begin{equation}
    \boldsymbol{\Sigma}^{NMA}_\beta = 
    \begin{bmatrix} 
    (\sigma_{\beta}^{[d-]})^2 & \rho_\beta \sigma_{\beta}^{[d-]} \sigma_{\beta}^{[d+]} \\ 
    \rho_\beta \sigma_{\beta}^{[d-]} \sigma_{\beta}^{[d+]} & (\sigma_{\beta}^{[d+]})^2 
    \end{bmatrix}
\end{equation}
The test-specific deviations 
(i.e., the $\delta_{\beta, s, t}^{[d]}$ in equation \ref{equation_NMA_ordinal_bivariate_two_level_structure_for_locations} above)
are modeled independently:
\begin{equation}
    \begin{aligned}
    \delta_{\beta, s, t}^{[d]} &\sim \text{normal}\left( 0, ~  \left(\tau_{\beta, t}^{[d]} \right)^2 \right)
    \end{aligned}
\end{equation}
%%%%
%%%%
%%%%%%%%%%%%%%%%%%%%%%%%%%%%%%%%%%%%%%%%%%%%%%%%%%%%%%%%%%%%%%%%%%%%%%%%%%%%%%%%
\subsubsection{ Incorporating Covariates}
\label{section_model_specs_NMA_ordinal_bivariate_between_study_model}
%%%%%%%%%%%%%%%%%%%%%%%%%%%%%%%%%%%%%%%%%%%%%%%%%%%%%%%%%%%%%%%%%%%%%%%%%%%%%%%%
%%%%
%%%% -------------------------------------------------------------------------------------------------------------------------------------
The model accommodates study-level covariates through the mean parameters. For test $t$:
\begin{equation}
\begin{aligned}
\left( X\beta \right)_{\bullet,t}^{[d-]} &= \mathbf{X}^{[d                                                                            
         -]}_{t} \cdot \boldsymbol{\beta}_{\bullet,t}^{[d-]} \\
\left( X\beta \right)_{\bullet,t}^{[d+]} &= \mathbf{X}^{[d+]}_{t} \cdot \boldsymbol{\beta}_{\bullet,t}^{[d+]}
\end{aligned}
\end{equation}
Where:
$\mathbf{X}^{[d-]}_{t}$ and $\mathbf{X}^{[d+]}_{t}$ 
are the design matrices for covariates in the non-diseased and diseased groups for test $t$, and:
$\boldsymbol{\beta}_{\bullet, t}^{[d]}$ are the corresponding coefficient vectors.
%%%%
%%%%
%%%%%%%%%%%%%%%%%%%%%%%%%%%%%%%%%%%%%%%%%%%%%%%%%%%%%%%%%%%%%%%%%%%%%%%%%%%%%%%%
\subsubsection{ Summary Estimates}
%%%%%%%%%%%%%%%%%%%%%%%%%%%%%%%%%%%%%%%%%%%%%%%%%%%%%%%%%%%%%%%%%%%%%%%%%%%%%%%%
%%%%
%%%% -------------------------------------------------------------------------------------------------------------------------------------
The summary accuracy measures for test $t$ at threshold $k$ are obtained by evaluating at the mean parameter values with 
baseline covariate values:
\begin{equation}
\begin{aligned}
    \text{Sp}_{\bullet,t,k} &= 
    \Phi\left(   C^{[d-]}_{\bullet, t, k} - \left( X\beta \right)_{\bullet,t}^{[d-],\text{baseline}}  \right) \\
    \text{Se}_{\bullet,t,k} &= 
    1 - \Phi\left(    C^{[d+]}_{\bullet, t, k} - \left( X\beta \right)_{\bullet,t}^{[d+],\text{baseline}}  \right) \\
\end{aligned}
\end{equation}
Where:
\begin{equation}
    \begin{aligned}
    \left( X\beta \right)_{\bullet,t}^{[d],\text{baseline}}  &=
    \mathbf{X}_{\text{baseline}}^T \cdot \boldsymbol{\beta}_{\bullet, t}^{[d]}
    \end{aligned}
\end{equation}
%%%%
%%%%
%%%%%%%%%%%%%%%%%%%%%%%%%%%%%%%%%%%%%%%%%%%%%%%%%%%%%%%%%%%%%%%%%%%%%%%%%%%%%%%%
\subsubsection{ Variance Decomposition}
%%%%%%%%%%%%%%%%%%%%%%%%%%%%%%%%%%%%%%%%%%%%%%%%%%%%%%%%%%%%%%%%%%%%%%%%%%%%%%%%
%%%%
%%%% ---------------------------------------------------------------------------------------------------------------------------------
Following Nyaga et al \supercite{Nyaga2018}, 
the total between-study variance for each test can be decomposed as:
\begin{equation}
\text{Var}[\beta_{s,t}^{[d]}] = (\sigma_{\beta}^{[d]})^2 + (\tau_{\beta,t}^{[d]})^2
\end{equation}
The proportion of variance explained by test-specific heterogeneity is:
\begin{equation}
\text{Prop}_{\beta,t}^{[d]} = \frac{(\tau_{\beta,t}^{[d]})^2}{(\sigma_{\beta}^{[d]})^2 + (\tau_{\beta,t}^{[d]})^2}
\end{equation}
This decomposition allows assessment of how much heterogeneity is due to general study effects versus test-specific variations.
%%%%
%%%% 
%%%%%%%%%%%%%%%%%%%%%%%%%%%%%%%%%%%%%%%%%%%%%%%%%%%%%%%%%%%%%%%%%%%%%%%%%%%%%%%%
\subsection{ NMA Extension of the ordinal-HSROC model}
\label{section_model_specs_NMA_ordinal_HSROC}
%%%%%%%%%%%%%%%%%%%%%%%%%%%%%%%%%%%%%%%%%%%%%%%%%%%%%%%%%%%%%%%%%%%%%%%%%%%%%%%%
%%%%
%%%% -------------------------------------------------------------------------------------------------------------------------------------
In this section, we extend the second model we proposed - 
the ordinal-HSROC model 
(see section \ref{section_model_specs_Cerullo_R_and_G_HSROC}) - 
NMA framework, again using the NMA structure based on Nyaga et al, 2018\supercite{Nyaga2018}.
For each test $t$ in study $s$, the test accuracy measures at the $k$th cutpoint are represented as:
\begin{equation}
\begin{aligned}
Sp_{s, t, k} & = \Phi\left( \frac{C_{s, t, k} - (-1)\beta_{s, t}}{\exp((-1)\gamma_{s, t})}\right) \\
Se_{s, t, k} & = 1 - \Phi\left( \frac{C_{s, t, k} - (+1)\beta_{s, t}}{\exp((+1)\gamma_{s, t})}\right)
\end{aligned}
\end{equation}
Where:
$\beta^{[d-]}_{s, t}$ and $\beta^{[d+]}_{s, t}$ are the location parameters for test $t$ in study $s$, 
in the non-diseased ($[d-]$) and diseased ($[d+]$) groups, respectively,
$\gamma^{[d-]}_{s, t}$ and $\gamma^{[d+]}_{s, t}$ are the raw scale parameters for test $t$ in study $s$, 
in the non-diseased and diseased groups, respectively, and
$C_{s, t, k}$ are the latent cutpoint parameters for study $s$, test $t$ at cutpoint $k$ 
(note: these are shared between disease groups - unlike the ordinal-HSROC model).
%%%%
%%%%
%%%%%%%%%%%%%%%%%%%%%%%%%%%%%%%%%%%%%%%%%%%%%%%%%%%%%%%%%%%%%%%%%%%%%%%%%%%%%%%%
\subsubsection{ Within-study model}
\label{section_model_specs_NMA_ordinal_HSROC_within_study_model}
%%%%%%%%%%%%%%%%%%%%%%%%%%%%%%%%%%%%%%%%%%%%%%%%%%%%%%%%%%%%%%%%%%%%%%%%%%%%%%%%
%%%%
%%%% -----------------------------------------------------------------------------------------------------------------------------------
The within-study models are defined exactly as they were for the non-NMA (i.e., MA/single stest only) 
version of this model; 
see section \ref{section_model_specs_Cerullo_R_and_G_HSROC}
for more information.
The within-study models are defined exactly this way for every test $t$.
%%%%
%%%%
%%%%%%%%%%%%%%%%%%%%%%%%%%%%%%%%%%%%%%%%%%%%%%%%%%%%%%%%%%%%%%%%%%%%%%%%%%%%%%%%
\subsubsection{ Hierarchical Structure and between-study model}
\label{section_model_specs_NMA_ordinal_HSROC_between_study_model}
%%%%%%%%%%%%%%%%%%%%%%%%%%%%%%%%%%%%%%%%%%%%%%%%%%%%%%%%%%%%%%%%%%%%%%%%%%%%%%%%
%%%%
%%%% -------------------------------------------------------------------------------------------------------------------------------------
Following the Nyaga et al\supercite{Nyaga2018} framework, and using similar notation
(see section \ref{section_model_specs_Nyaga_NMA_background}),
we decompose both the location and scale parameters using a two-level hierarchical structure.
\begin{equation}
\begin{aligned}
\beta_{s,t}  &= \beta_{\bullet,t} + \eta_{\beta, s} + \delta_{\beta, s, t} \\
\gamma_{s,t} &= \gamma_{\bullet,t} + \eta_{\gamma, s} + \delta_{\gamma, s, t}
\end{aligned}
\label{equation_NMA_ordinal_HSROC_two_level_structure}
\end{equation}
Where:
\begin{itemize}
    \item $\beta_{\bullet, t}^{[d]}$ and $\gamma_{\bullet, t}^{[d]}$ 
    are the mean location and raw scale parameters parameters (respectively) for test $t$ in disease group $d$.
    \item $\eta_{\beta, s}^{[d]}$ and $\eta_{\gamma, s}^{[d]}$ 
    are study-level random effects shared across all tests within study $s$, disease group $d$.
    \item $\delta_{\beta, s, t}^{[d]}$ and $\delta_{\gamma, s, t}^{[d]}$ 
    are test-specific deviations for test $t$ in study $s$, disease group $d$.
\end{itemize}
%%%%
%%%% -------------------------------------------------------------------------------------------------------------------------------------
The study-level random effects for locations 
($\eta_{\beta, s}$ in equation \ref{equation_NMA_ordinal_HSROC_two_level_structure}) 
and scales 
($\eta_{\gamma, s}$ in equation \ref{equation_NMA_ordinal_HSROC_two_level_structure})
are modeled independently:
\begin{equation}
\begin{aligned}
\eta_{\beta, s}^{[d]}  &\sim \text{normal}\left(0, (\sigma_{\beta}^{[d]})^2\right) \\
\eta_{\gamma, s}^{[d]} &\sim \text{normal}\left(0, (\sigma_{\gamma}^{[d]})^2\right)
\end{aligned}
\end{equation}
Note that, unlike the ordinal-bivariate NMA model 
(see section \ref{section_model_specs_NMA_ordinal_bivariate_between_study_model}),
the HSROC formulation does not explicitly model the between-study correlation 
between disease groups at the study level using a bivariate normal distribution.
Instead, the between-study correlation between disease groups is modeled implicitly.
Furthermore, the test-specific deviations are also modeled independently:
\begin{equation}
\begin{aligned}
\delta_{\beta, s, t}^{[d]}  &\sim \text{normal}\left(0, (\tau_{\beta, t}^{[d]})^2\right) \\
\delta_{\gamma, s, t}^{[d]} &\sim \text{normal}\left(0, (\tau_{\gamma, t}^{[d]})^2\right)
\end{aligned}
\end{equation}
%%%%
%%%%
%%%%%%%%%%%%%%%%%%%%%%%%%%%%%%%%%%%%%%%%%%%%%%%%%%%%%%%%%%%%%%%%%%%%%%%%%%%%%%%%
\subsubsection{ Incorporating Covariates}
\label{section_model_specs_NMA_ordinal_HSROC_covariates}
%%%%%%%%%%%%%%%%%%%%%%%%%%%%%%%%%%%%%%%%%%%%%%%%%%%%%%%%%%%%%%%%%%%%%%%%%%%%%%%%
%%%%
%%%% -------------------------------------------------------------------------------------------------------------------------------------
The model accommodates study-level covariates through the mean parameters. 
For test $t$:
\begin{equation}
\begin{aligned}
(X\beta)_{\bullet,t}  &= \mathbf{X}_{t} \cdot \boldsymbol{\beta}_{\bullet,t} \\
(X\gamma)_{\bullet,t} &= \mathbf{X}_{t} \cdot \boldsymbol{\gamma}_{\bullet,t}
\end{aligned}
\end{equation}
Where:
$\mathbf{X}_{t}$ is the design matrix for covariates for test $t$, and
$\boldsymbol{\beta}_{\bullet, t}$ and $\boldsymbol{\gamma}_{\bullet, t}$ are the corresponding coefficient vectors 
for location and raw scale parameters.
%%%%
%%%%
%%%%%%%%%%%%%%%%%%%%%%%%%%%%%%%%%%%%%%%%%%%%%%%%%%%%%%%%%%%%%%%%%%%%%%%%%%%%%%%%
\subsubsection{ Summary Estimates}
\label{section_model_specs_NMA_ordinal_HSROC_summary_estimates}
%%%%%%%%%%%%%%%%%%%%%%%%%%%%%%%%%%%%%%%%%%%%%%%%%%%%%%%%%%%%%%%%%%%%%%%%%%%%%%%%
%%%%
%%%% -------------------------------------------------------------------------------------------------------------------------------------
The summary accuracy measures for test $t$ at threshold $k$ are obtained by evaluating at the mean parameter values with 
baseline covariate values:
\begin{equation}
\begin{aligned}
\text{Sp}_{\bullet,t,k} &= \Phi\left(\frac{C_{\bullet, t, k} - (-1)(X\beta)_{\bullet,t}^{\text{baseline}}}{\exp((-1)(X\gamma)_{\bullet,t}^{\text{baseline}})}\right) \\
\text{Se}_{\bullet,t,k} &= 1 - \Phi\left(\frac{C_{\bullet, t, k} - (+1)(X\beta)_{\bullet,t}^{\text{baseline}}}{\exp((+1)(X\gamma)_{\bullet,t}^{\text{baseline}})}\right)
\end{aligned}
\end{equation}
Where:
\begin{equation}
\begin{aligned}
(X\beta)_{\bullet,t}^{\text{baseline}}  &= \mathbf{X}_{\text{baseline}}^T \cdot \boldsymbol{\beta}_{\bullet, t} \\
(X\gamma)_{\bullet,t}^{\text{baseline}} &= \mathbf{X}_{\text{baseline}}^T \cdot \boldsymbol{\gamma}_{\bullet, t}
\end{aligned}
\end{equation}
% %%
% \begin{equation}
% \begin{aligned}
% (X\beta)_{\bullet,t}^{[d],\text{baseline}} &= \text{mult}^{[d]} \times \mathbf{X}_{\text{baseline}}^T \boldsymbol{\beta}_{\bullet, t} \\
% \gamma_{\bullet,t}^{[d],\text{baseline}} &= \exp\left(\text{mult}^{[d]} \times \mathbf{X}_{\text{baseline}}^T \boldsymbol{\gamma}_{\bullet, t}\right)
% \end{aligned}
% \end{equation}
% %%
% with $\text{mult}^{[d-]} = -1$ and $\text{mult}^{[d+]} = +1$.
%%%%
%%%%
%%%%%%%%%%%%%%%%%%%%%%%%%%%%%%%%%%%%%%%%%%%%%%%%%%%%%%%%%%%%%%%%%%%%%%%%%%%%%%%%
\subsubsection{ Variance Decomposition}
\label{section_model_specs_NMA_ordinal_HSROC_variance_decomp}
%%%%%%%%%%%%%%%%%%%%%%%%%%%%%%%%%%%%%%%%%%%%%%%%%%%%%%%%%%%%%%%%%%%%%%%%%%%%%%%%
%%%%
%%%% ---------------------------------------------------------------------------------------------------------------------------------
Following Nyaga et al \supercite{Nyaga2018}, 
the total between-study variance for each test can be decomposed for both parameters:
\begin{equation}
\begin{aligned}
\text{Var}[\beta_{s,t}]  &= \sigma_{\beta}^2 + \tau_{\beta,t}^2 \\
\text{Var}[\gamma_{s,t}] &= \sigma_{\gamma}^2 + \tau_{\gamma,t}^2
\end{aligned}
\end{equation}
The proportion of variance explained by test-specific heterogeneity is:
\begin{equation}
\begin{aligned}
\text{Prop}_{\beta,t}^{[d]}  &= \frac{(\tau_{\beta,t}^{[d]})^2}{(\sigma_{\beta}^{[d]})^2 + (\tau_{\beta,t}^{[d]})^2} \\
\text{Prop}_{\gamma,t}^{[d]} &= \frac{(\tau_{\gamma,t}^{[d]})^2}{(\sigma_{\gamma}^{[d]})^2 + (\tau_{\gamma,t}^{[d]})^2}
\end{aligned}
\end{equation}
This decomposition allows assessment of how much heterogeneity is due to general study effects versus test-specific variations,
for both location and scale parameters.
%%%%
%%%%
%%%%
%%%%
%%%%%%%%%%%%%%%%%%%%%%%%%%%%%%%%%%%%%%%%%%%%%%%%%%%%%%%%%%%%%%%%%%%%%%%%%%%%%%%%
\setcounter{secnumdepth}{4}  % To number these deep sections
\setcounter{tocdepth}{4}     % To include them in the table of contents
\numberwithin{equation}{section} %% number equations / appendix equations / equation numbering / bookmark
%%%%%%%%%%%%%%%%%%%%%%%%%%%%%%%%%%%%%%%%%%%%%%%%%%%%%%%%%%%%%%%%%%%%%%%%%%%%%%%%
%%%%%%%%%%%%%%%%%%%%%%%%%%%%%%%%%%%%%%%%%%%%%%%%%%%%%%%%%%%%%%%%%%%%%%%%%%%%%%%%%%%%%%%%%%%%%%%%%%%%%%%%%%%%%%%%%%%%%%%%%%%%%%%%%%%%%%
\newpage
\section{Simulation study}
\label{Sim_study}
%%%%%%%%%%%%%%%%%%%%%%%%%%%%%%%%%%%%%%%%%%%%%%%%%%%%%%%%%%%%%%%%%%%%%%%%%%%%%%%%%%%%%%%%%%%%%%%%%%%%%%%%%%%%%%%%%%%%%%%%%%%%%%%%%%%%%%
%%%%
%%%% ---------------------------------------------------------------------------------------------------------------------------------
In this section,
we will present and discuss the results of a simulation study we conducted to compare the following models
(as well as variations of them) to one another:
\begin{itemize}
    \item[(i)]
    The continuous-data model proposed by Jones et al\supercite{Jones2019}
    (i.e., the model which our proposed models build upon).
    \item[(ii)]
    The ordinal-bivariate model,
    which we proposed in section \ref{section_model_specs_Cerullo_bivariate_Reitsma_extension}.
    \item[(iii)]
    The ordinal-HSROC model,
    which we proposed in section \ref{section_model_specs_Cerullo_R_and_G_HSROC}.
\end{itemize}
%%
%%%%
%%%%
%%%%%%%%%%%%%%%%%%%%%%%%%%%%%%%%%%%%%%%%%%%%%%%%%%%%%%%%%%%%%%%%%%%%%%%%%%%%%%%%%%%%%%%%%%%%%%%%%%%%%%%%%%%%%%%%%%%%
\subsection{Design}
\label{Sim_study_design}
%%%%%%%%%%%%%%%%%%%%%%%%%%%%%%%%%%%%%%%%%%%%%%%%%%%%%%%%%%%%%%%%%%%%%%%%%%%%%%%%%%%%%%%%%%%%%%%%%%%%%%%%%%%%%%%%%%%%
%%%%
%%%% -------------------------------------------------------------------------------------------------------------------------------
We followed the ADEMP framework \supercite{morris_using_2019} to design and report our simulation study.
%%%%
%%%%
%%%%%%%%%%%%%%%%%%%%%%%%%%%%%%%%%%%%%%%%%%%%%%%%%%%%%%%%%%%%%%%%%%%%%%%%%%%%%%%%
\subsubsection{ Aims}
\label{Sim_study_design_aims}
%%%%%%%%%%%%%%%%%%%%%%%%%%%%%%%%%%%%%%%%%%%%%%%%%%%%%%%%%%%%%%%%%%%%%%%%%%%%%%%%
%%%%
%%%% -------------------------------------------------------------------------------------------------------------------------------
To evaluate the performance of ordinal-bivariate and ordinal-HSROC models compared to existing approaches
(Jones continuous "multiple thresholds" model \supercite{Jones2019}
and the stratified "standard" bivariate analysis \supercite{Reitsma2005})
for meta-analysing screening/diagnostic accuracy studies with ordinal test data, 
under realistic conditions of missing threshold reporting.
%%%%
%%%%
%%%%%%%%%%%%%%%%%%%%%%%%%%%%%%%%%%%%%%%%%%%%%%%%%%%%%%%%%%%%%%%%%%%%%%%%%%%%%%%%
\subsubsection{ Data-generating mechanisms}
\label{Sim_study_design_DGMs}
%%%%%%%%%%%%%%%%%%%%%%%%%%%%%%%%%%%%%%%%%%%%%%%%%%%%%%%%%%%%%%%%%%%%%%%%%%%%%%%%
%%%%
%%%% -------------------------------------------------------------------------------------------
The DGM's were based on the following two things:
\begin{enumerate}
    \item
    Three anxiety screening instruments,
    based on real-life meta-analytic data we obtained:
    \begin{itemize}
        \item[(1)] The Generalized Anxiety Disorder 2-item scale
        (GAD-2\supercite{Kroenke_2007_GAD_2}).
        This has a total of $7$ categories ($6$ thresholds).
        For this test,
        we assumed a small amount of missing data ($\sim 15\%$).
        \item[(2)] The HADS\supercite{Zigmond_1983_HADS}.
        This test has a total of $22$ categories ($21$ thresholds).
        For this test,
        we assumed a large amount of missing threshold data ($\sim 40\%$).
        \item[(3)] The BAI\supercite{Beck_1988_BAI},
        which has a total of $64$ categories ($63$ thresholds).
        For this test,
        we assumed a large amount of missing threshold data ($\sim 55\%$).
        \item For each of the three screening tests mentioned above
        (i.e. the GAD-2, HADS, and the BAI),
        the proportion of missing threshold data
        (which was $15\%$, $40\%$, and $55\%$, respectively)
        was consistent with the real-life data we had.
    \end{itemize}
    \item The kind of model to simulate from,
    which was one of the following:
    \begin{enumerate}
        \item \textbf{"Jones"}:
        The model proposed in Jones et al, 2019\supercite{Jones2019}.
        \item \textbf{"O-bivariate-FC" (or "O-biv-FC"):}
        The ordinal-bivariate model we proposed in this paper
        (see section \ref{section_model_specs_Cerullo_bivariate_Reitsma_extension}),
        with fixed-effects cutpoint parameters
        (see section \ref{section_model_specs_induced_Dirichlet_distribution},
        sub-section \ref{section_model_specs_induced_Dirichlet_fixed_cutpoints_definition}).
        \item \textbf{"O-bivariate-RC" (or "O-biv-RC"):}
        The ordinal-bivariate model we proposed in this paper
        (see section \ref{section_model_specs_Cerullo_bivariate_Reitsma_extension}),
        with random-effects cutpoint parameters
        (see section \ref{section_model_specs_induced_Dirichlet_distribution},
        sub-section \ref{section_model_specs_induced_Dirichlet_random_cutpoints_definition}).
        \item \textbf{"O-HSROC-RC":}
        The ordinal-HSROC model we proposed in this paper
        (see section \ref{section_model_specs_Cerullo_R_and_G_HSROC}),
        with random-effects cutpoint parameters
        (see section \ref{section_model_specs_induced_Dirichlet_distribution},
        sub-section \ref{section_model_specs_induced_Dirichlet_random_cutpoints_definition}).
        Note: we did not use the fixed-effects cutpoint variant in this simulation study as pilot runs showed very bad estimates
        (this isn't surprising given the nature of the model).
    \end{enumerate}
\end{enumerate}
Hence, overall, there are: $3$ (\# of tests) x $4$ (\# of DGM's per test) = $12$ scenarios per model.
\paragraph*{ Cutpoint heterogeneity}
%%%%%%%%%%%%%%%%%%%%%%%%%%%%%%%%%%%%%%%%%%%%%%%%%%%%%%%%%%%%%%%%%%%%%%%%%%%%%%%
% %%%% -------------------------------------------------------------------------------------------
% %%%%
% \begin{table}[H]
% \centering
% \caption{Mean cutpoint heterogeneity (SD on probability scale) - by test and DGM}
% \small
% \begin{tabular}{lcccccc}
% \toprule
% \textbf{Test} & \textbf{Jones-FC} & \textbf{O-biv-FC} & \multicolumn{2}{c}{\textbf{O-biv-RC}} & \textbf{O-HSROC-RC} \\
% & \textbf{(DGM \#1)} & \textbf{(DGM \#2)} & \multicolumn{2}{c}{\textbf{(DGM \#3)}} & \textbf{(DGM $\#4^*$)} \\
% \cmidrule(lr){4-5}
% & & & ND & D & \\
% %%
% \midrule
% %%
% GAD-2 & $0$ & $0$ & $0.036$ & $0.068$ & $0.069$ \\
% HADS  & $0$ & $0$ & $0.010$ & $0.011$ & $0.010$ \\
% BAI   & $0$ & $0$ & $0.005$ & $0.003$ & $0.003$ \\
% %%
% \bottomrule
% \end{tabular}
% \begin{tablenotes}
% \footnotesize
% \item
% FC = Fixed cutpoints (zero heterogeneity by definition).
% %%
% \item
% RC = Random cutpoints;
% $D-$ = Non-diseased, $D+$ = Diseased.
% %%
% \item 
% Values represent mean SD of cutpoint parameters on probability scale.
% %%
% \item 
% *Note: For DGM $\#4$, recall that cutpoints are the same in ND and D groups \\
% (hence one heterogeneity value per test, rather than 2).
% %%
% \end{tablenotes}
% \label{table:cutpoint_heterogeneity_by_dgm}
% \end{table}
\begin{table}[H]
\centering
\caption{Mean cutpoint heterogeneity (SD on probability scale) by test and DGM}
\small
\begin{tabular}{lccccccc}
\toprule
& & & \multicolumn{5}{c}{\textbf{Data Generating Mechanism (DGM)}} \\
\cmidrule(lr){4-8}
\textbf{Test} & \textbf{ } & \textbf{$K-1$} & \textbf{Jones-FC} & \textbf{O-biv-FC} & \multicolumn{2}{c}{\textbf{O-biv-RC}} & \textbf{O-HSROC-RC} \\
& \textbf{$N_{studies}$} & \textbf{cutpoints} & \textbf{(DGM \#1)} & \textbf{(DGM \#2)} & \multicolumn{2}{c}{\textbf{(DGM \#3)}} & \textbf{(DGM $\#4^*$)} \\
\cmidrule(lr){6-7}
& & & & & $D-$ & $D+$ & \\
\midrule
GAD-2 & $29$ & $6$  & $0$ & $0$ & $0.036$ & $0.068$ & $0.069$ \\
HADS  & $36$ & $21$ & $0$ & $0$ & $0.010$ & $0.011$ & $0.010$ \\
BAI   & $7$  & $63$ & $0$ & $0$ & $0.005$ & $0.003$ & $0.003$ \\
\bottomrule
\end{tabular}
\begin{tablenotes}
\footnotesize
\item $N_{studies}$ = number of studies in real dataset used to calibrate simulation parameters.
\item $K-1$ cutpoints = number of thresholds ($K$ = number of ordinal categories).
\item FC = Fixed cutpoints (zero heterogeneity by definition).
\item RC = Random cutpoints; $D-$ = Non-diseased, $D+$ = Diseased.
\item Values represent mean SD of cutpoint parameters on probability scale.
\item *Note: For DGM \#4 (O-HSROC-RC), cutpoints are shared between disease groups, 
hence one heterogeneity value per test.
\end{tablenotes}
\label{table:cutpoint_heterogeneity_by_dgm}
\end{table}
%%%%
%%%%
%%%% -------------------------------------------------------------------------------------------
%%
%% ---- For GAD-2:
%%
Table \ref{table:cutpoint_heterogeneity_by_dgm} shows the
true values of the cutpoint heterogeneity (i.e., the SD) -
on the probability scale.
Recall that the O-HSROC-RC model only has one SD -
not two like the O-biv-RC model -
% because this model shares the same set of cutpoints between
% the two groups - non-diseased ($D-$) and diseased ($D+$).
since it has one set of cutpoints, 
shared between both the diseased and non-diseased groups.

%%%%
%%%% -------------------------------------------------------------------------------------------
As we can see (table \ref{table:cutpoint_heterogeneity_by_dgm}),
the GAD-2 (with $6$ cutpoints) has somewhat notable cutpoint heterogeneity -
for both the O-biv-RC model (SD = $0.036$ and SD = $0.068$ in the D- and D+ groups, respectively),
as well as the O-HSROC-RC model (SD = $0.069$).
The fact that the cutpoint heterogeneity is approximately $2$-fold greater
in the D+ group for the O-biv-RC model ($0.068$ vs. $0.036$) is not surprising,
and in many cases might be expected.

%%%% -------------------------------------------------------------------------------------------
%%%%
%%
%% ---- For HADS:
%%
Unlike the GAD-2, the real dataset (not shown) for the HADS ($21$ cutpoints)
suggests that the cutpoint heterogeneity
is essentially the same in both the D+ and D- groups -
being $0.010$ and $0.011$ for the O-biv-RC in the D- and D+ groups, respectively.
Furthermore - for the O-HSROC-RC model - it is essentially the same magnitude ($SD = 0.010$)
as it is for the O-biv-RC model.
Also notice that the magnitude of cutpoint heterogeneity is substantially lower
for the HADS ($0.010 - 0.011$) compared to the GAD-2 ($0.036 - 0.068$).

%%%% -------------------------------------------------------------------------------------------
%%%%
%%
%% ---- For BAI:
%%
For the BAI ($63$ cutpoints), the real dataset (not shown) suggested low cutpoint heterogeneity -
being $0.005$ and $0.003$ in the non-diseased and diseased groups, respectively,
for the O-biv-RC model.
For the O-HSROC-RC model, the cutpoint heterogeneity was $0.003$.
It is also interesting to note that here the cutpoint heterogeneity (for the O-biv-RC model)
was actually greater in the non-diseased group compared to the diseased group;
however, for the BAI (real dataset) we only had $7$ studies - 
whereas we had $29$ and $36$ studies for the real GAD-2 and HADS datasets, respectively.

%%%%%%%%%%%%%%%%%%%%%%%%%%%%%%%%%%%%%%%%%%%%%%%%%%%%%%%%%%%%%%%%%%%%%%%%%%%%%%%
\paragraph*{ More information on test I: GAD-2}
%%%%%%%%%%%%%%%%%%%%%%%%%%%%%%%%%%%%%%%%%%%%%%%%%%%%%%%%%%%%%%%%%%%%%%%%%%%%%%%
%%%%
%%%% ------------------------------------------------------------------------------------------------------------
The Generalized Anxiety Disorder 2-item scale (GAD-2) is an ultra-brief screening tool derived from the GAD-7,
consisting of its first two items that assess nervousness,
and inability to control worry over the past two weeks\supercite{Kroenke_2007_GAD_2}.
With scores ranging from $0-6$ and a widely used standard clinical cut-off of $\geq 3$,
the GAD-2 demonstrates good screening properties for generalized anxiety disorder,
and performs well for detecting panic disorder, social anxiety, and PTSD in primary care settings -
where rapid screening is essential \supercite{Spitzer_2006_GAD_7}.
For example, in 2016 a comprehensive meta-analysis
(Plummer et al, 2016 \supercite{Plummer_MA_for_GAD_2})
reported a sensitivity and specificity of
$0.76$ ($95\%$ CI = $(0.55, 0.89)$),
and 
$0.81$ ($95\%$ CI = $(0.60, 0.92)$).
%%%%
%%%%            
%%%%%%%%%%%%%%%%%%%%%%%%%%%%%%%%%%%%%%%%%%%%%%%%%%%%%%%%%%%%%%%%%%%%%%%%%%%%%%%%
\paragraph*{ More information on test II: HADS }
%%%%%%%%%%%%%%%%%%%%%%%%%%%%%%%%%%%%%%%%%%%%%%%%%%%%%%%%%%%%%%%%%%%%%%%%%%%%%%%%
%%%%
%%%% ------------------------------------------------------------------------------------------------------------
The Hospital Anxiety and Depression Scale (HADS) is a $14$-item self-report measure with
7 items each for anxiety (HADS-A) and depression (HADS-D),
originally designed for medical outpatients by deliberately excluding somatic symptoms
that might confound assessment in physically ill populations \supercite{Zigmond_1983_HADS}.
Each item scores 0-3 yielding sub-scale scores of $0 - 21$, with clinical cutoffs of $0 - 7$ (normal),
8-10 (borderline), and $\geq 11$ (probable case),
where the optimal screening threshold of 8 provides is thought to provide approximately $80\%$
sensitivity and specificity across diverse settings\supercite{Bjelland_2002_HADS}.
However, despite being widely used for anxiety screening,
little evidence exists evaluating this test.
%%%%
%%%%
%%%%%%%%%%%%%%%%%%%%%%%%%%%%%%%%%%%%%%%%%%%%%%%%%%%%%%%%%%%%%%%%%%%%%%%%%%%%%%%%
\paragraph*{ More information on test III: BAI}
%%%%%%%%%%%%%%%%%%%%%%%%%%%%%%%%%%%%%%%%%%%%%%%%%%%%%%%%%%%%%%%%%%%%%%%%%%%%%%%%
%%%%
%%%% ------------------------------------------------------------------------------------------------------------
The Beck Anxiety Inventory (BAI) is a $21$-item self-report measure assessing anxiety severity over the past week,
with scores ranging from $0 - 63$ and particular emphasis on physiological symptoms -
15 of $21$ items focus on somatic manifestations like numbness,
sweating, and dizziness \supercite{Beck_1988_BAI}.
This somatic focus helps differentiate anxiety from depression
(achieving the scale's design goal)
but may inflate scores in medically ill populations,
with clinical ranges of:
$0-7$ (minimal),
$8-15$ (mild),
$16-25$ (moderate), and
$26-63$ (severe anxiety).
%%%%
%%%%
%%%%%%%%%%%%%%%%%%%%%%%%%%%%%%%%%%%%%%%%%%%%%%%%%%%%%%%%%%%%%%%%%%%%%%%%%%%%%%%%
\subsubsection{ Estimands}
\label{Sim_study_design_estimands}
%%%%%%%%%%%%%%%%%%%%%%%%%%%%%%%%%%%%%%%%%%%%%%%%%%%%%%%%%%%%%%%%%%%%%%%%%%%%%%%%
%%%%
%%%% ------------------------------------------------------------------------------------------------------------
The estimates for this simulation study are the average sensitivity and specificity,
for each of the three tests and each DGM,
with "average" meaning the mean of the threshold-specific sensitivity and specificity at each test threshold.
More specifically, 

%%%%
%%%%
%%%%%%%%%%%%%%%%%%%%%%%%%%%%%%%%%%%%%%%%%%%%%%%%%%%%%%%%%%%%%%%%%%%%%%%%%%%%%%%%
\subsubsection{ Methods}
\label{Sim_study_design_methods}
%%%%%%%%%%%%%%%%%%%%%%%%%%%%%%%%%%%%%%%%%%%%%%%%%%%%%%%%%%%%%%%%%%%%%%%%%%%%%%%%
%%%%
%%%% ------------------------------------------------------------------------------------------------------------
Five models were compared:
\begin{enumerate}
    \item The stratified-bivariate model
    (standard method;
    Reitsma et al, 2005 \supercite{Reitsma2005}).
    \item Jones continuous "multiple thresholds" model
    (Jones et al, 2019\supercite{Jones2019}).
    \item The ordinal-bivariate model with fixed cutpoints
    (O-bivariate-FC;
    proposed in section \ref{section_model_specs_Cerullo_bivariate_Reitsma_extension}).
    \item The ordinal-bivariate with random cutpoints
    (O-bivariate-RC;
    proposed in section \ref{section_model_specs_Cerullo_bivariate_Reitsma_extension}).
    \item The ordinal-HSROC with random cutpoints
    (O-HSROC-RC;
    proposed in section \ref{section_model_specs_Cerullo_R_and_G_HSROC}).
\end{enumerate}
%%%%
%%%%

Note that we did not include the sixth model, 
which would have been the Ordinal-HSROC with fixed-effects cutpoints.
This is because pilot runs (results not shown) 
suggested relatively very poor performance and time was limited.
%%%%
%%%%
%%%%%%%%%%%%%%%%%%%%%%%%%%%%%%%%%%%%%%%%%%%%%%%%%%%%%%%%%%%%%%%%%%%%%%%%%%%%%%%%
\subsubsection{ Performance measures}
\label{Sim_study_design_performance_measures}
%%%%%%%%%%%%%%%%%%%%%%%%%%%%%%%%%%%%%%%%%%%%%%%%%%%%%%%%%%%%%%%%%%%%%%%%%%%%%%%%
%%%%
%%%% ------------------------------------------------------------------------------------------------------------
We investigated the following measures of model performance:
%%%% 
% \begin{itemize}
%     \item 
%     Primary measure:    \textbf{Root Mean Square Error (RMSE):} Overall accuracy combining bias and variance.
%     \item 
%     Secondary measure:  \textbf{Bias:} Mean absolute percentage deviation from true values.
%     \item 
%     Additional measure: \textbf{Coverage:} Proportion of $95\%$ credible intervals containing the true values.
%     \item 
%     Additional measure: \textbf{Precision:} Mean width of $95\%$ credible intervals.
% \end{itemize}
% %%%%
% %%%%
% We assess the following performance measures as suggested in Morris et al\supercite{morris_using_2019}, 
% using RMSE of accuracy (sensitivity and specificities) as primary measures:
% Since the targets are estimands, we will measure: 
% \textbf{bias}, \textbf{precision} (i.e., SE) and \textbf{coverage} of 95\% posterior (or ``credible") intervals, 
% as suggested in Morris et al\supercite{morris_using_2019}. 
% %%
% These are defined as:
%%%%
%%%%
\begin{enumerate}
   \item 
   \textbf{\underline{ Primary measure: }}
   \textbf{Root Mean Square Error (RMSE):} 
   Overall accuracy combining bias and variance
   $\left( \sqrt{\mathbb{E}\left[(\hat\theta - \theta)^2\right]} \right) $:
   \begin{itemize} 
     \item Estimated by: 
              $\sqrt{\frac{1}{N_{sim}} \sum_{i=1}^{N_{sim}} (\hat{\theta}_{i} - \theta)^2}$
     \item Monte Carlo SE is estimated by:
             $\frac{\widehat{\text{RMSE}}}{\sqrt{2 \cdot N_{sim}}}$
   \end{itemize}
   \item 
   \textbf{\underline{ Secondary measure: }}
   \textbf{Bias:} 
   Mean absolute percentage deviation from true values
   $\left( \mathbb{E}\left[\hat\theta\right] - \theta \right)$  
   \begin{itemize} 
     \item Estimated by: 
              $\frac{1}{N_{sim}} \sum_{i=1}^{N_{sim}} \hat{\theta}_{i} - \theta$
     \item Monte Carlo SE: as shown above
   \end{itemize}
   \item 
   \textbf{\underline{ Additional measure: }}
   \textbf{Coverage:} 
   Proportion of $95\%$ credible intervals containing the true values.
   $ \left( \text{Prob} \left( \hat\theta_{\text{low}} \le \theta \le  \hat\theta_{\text{upper}}  \right) \right)$
   \begin{itemize} 
     \item Estimated by: 
              $\frac{1}{N_{sim}} \sum_{i=1}^{N_{sim}}
              \mathbb{I} \left(  \hat\theta_{\text{low}} \le \theta \le  \hat\theta_{\text{upper}}  \right)$
     \item Monte Carlo SE: as shown above
   \end{itemize}
   \item 
   \textbf{\underline{ Additional measure: }}
   \textbf{Interval Width (precision):} Mean width of $95\%$ credible intervals.
   Mean width of $95\%$ credible intervals
   \begin{itemize} 
     \item Estimated by: 
              $\frac{1}{N_{sim}} \sum_{i=1}^{N_{sim}} (\hat\theta_{\text{upper},i} - \hat\theta_{\text{low},i})$
   \end{itemize}
\end{enumerate}
%%%%
%%%%
% %%%%%%%%%%%%%%%%%%%%%%%%%%%%%%%%%%%%%%%%%%%%%%%%%%%%%%%%%%%%%%%%%%%%%%%%%%%%%%%%
% \subsubsection{ Sample size determination}
% \label{Sim_study_design_sample_size_n_sims}
% %%%%%%%%%%%%%%%%%%%%%%%%%%%%%%%%%%%%%%%%%%%%%%%%%%%%%%%%%%%%%%%%%%%%%%%%%%%%%%%%
%%%%
%%%% ------------------------------------------------------------------------------------------------------------
% Pilot simulations determined minimum iterations needed for stable Monte Carlo standard errors (MCSE < 5\% of estimate). 
% %%%%
% Final simulations used 200-2,200 iterations depending on model complexity and convergence rates, 
% with scenarios of $N_{studies} \in \{10, 50\}$ studies and average study size $N = 500$ (with $SD(N) = 500$).
%%%%
%%%%
%%%%%%%%%%%%%%%%%%%%%%%%%%%%%%%%%%%%%%%%%%%%%%%%%%%%%%%%%%%%%%%%%%%%%%%%%%%%%%%%
\subsubsection{ Implementation}
\label{Sim_study_design_implementation_MCMC}
%%%%%%%%%%%%%%%%%%%%%%%%%%%%%%%%%%%%%%%%%%%%%%%%%%%%%%%%%%%%%%%%%%%%%%%%%%%%%%%%
%%%%
%%%% ----------------------------------------------------------------------------------------
The models were implemented using the MetaOrdDTA R 
package\supercite{Cerullo_MetaOrdDTA_2025} we developed
(on GitHub: \url{https://github.com/CerulloE1996/MetaOrdDTA}).
This R package uses Stan\supercite{Carpenter2017} 2.36.0,
using the default adaptive NUTS-HMC\supercite{Hoffman_and_Gelman_2014_NUTS_paper} 
MCMC algorithm for sampling.

%%%%
%%%% ----------------------------------------------------------------------------------------
All models will be run using 4 chains (in parallel) on a local HPC/server
computer with 96 cores (AMD EPYC 9654; 96 cores/192 threads),
and 384GB of DDR5 RAM.
Furthermore, we will run 24 models at once 
(hence 96 parallel chains total).

%%%%
%%%% ----------------------------------------------------------------------------------------
All code and supplementary materials which we used to run this simulation study are available at:
\url{https://github.com/CerulloE1996/MetaOrdDTA/tree/main/inst/examples}

%%%%%%%%%%%%%%%%%%%%%%%%%%%%%%%%%%%%%
\paragraph{ Adaptive Simulation Size ($N_{sim}$)}
\label{section_adaptive_N_sim}
%%%%%%%%%%%%%%%%%%%%%%%%%%%%%%%%%%%%%
%%%%
%%%% ----------------------------------------------------------------------------------------
Rather than fixing $N_{sim}$ a priori, 
we implemented an adaptive stopping rule that terminated simulations 
when the maximum Monte Carlo standard error for RMSE(Se) fell below $0.125\%$.
This ensured adequate precision while avoiding unnecessary computation.
The algorithm checked convergence after every simulation.
The mean MCSE of the sensitivity RMSE and specificitiy RMSE 
was used as the convergence criterion.

%%%%
%%%%
%%%%%%%%%%%%%%%%%%%%%%%%%%%%%%%%%%%%%%%%%%%%%%%%%%%%%%%%%%%%%%%%%%%%%%%%%%%%%%%%%%%%%%%%%%%%%%%%%%%%%%%%%%%%%%%%%%%%
\subsection{Results: Test I (GAD-2)}
\label{Sim_study_GAD_2}
%%%%%%%%%%%%%%%%%%%%%%%%%%%%%%%%%%%%%%%%%%%%%%%%%%%%%%%%%%%%%%%%%%%%%%%%%%%%%%%%%%%%%%%%%%%%%%%%%%%%%%%%%%%%%%%%%%%%
%%%%
%%%%
\begin{table}[H]
\centering
\caption{
Overall model performance for all four DGMs - GAD-2
(4-group classification based on statistical and practical significance)
}
\footnotesize
\setlength{\tabcolsep}{4pt}
\begin{tabular}{c|l|l}
\toprule
\textbf{DGM} & $\mathbf{N_{studies} = 10}$ & $\mathbf{N_{studies} = 50}$ \\
\midrule
%%
%% ---- DGM = Jones:
%%
Jones & \makecell[l]{\underline{Group 1 - Best:} \\
                Jones [RMSE: $9.75$, Bias: $2.08$]. \\
                \\
                \underline{Group 2 - Stat only:} \\
                O-biv-FC [RMSE: $10.25$/$5.2\%$, Bias: $1.61$]. \\
                \\
                \underline{Group 3 - Practical only:} \\
                None. \\
                \\
                \underline{Group 4 - Worse:} \\
                Strat-biv [RMSE: $11.76$/$20.6\%$, Bias: $3.03$].}
  & \makecell[l]{\underline{Group 1 - Best:} \\
                Jones [RMSE: $4.55$, Bias: $0.32$], \\
                O-biv-FC [RMSE: $4.81$/$5.8\%$, Bias: $0.64$]. \\
                \\
                \underline{Group 2 - Stat only:} \\
                None. \\
                \\
                \underline{Group 3 - Practical only:} \\
                None. \\
                \\
                \underline{Group 4 - Worse:} \\
                Strat-biv [RMSE: $5.32$/$17.0\%$, Bias: $0.36$].} \\
\midrule
%%
%% ---- DGM: O-biv-FC:
%%
O-biv-FC & \makecell[l]{\underline{Group 1 - Best:} \\
               O-biv-FC [RMSE: $8.21$, Bias: $0.76$], \\
               Jones [RMSE: $8.23$/$0.2\%$, Bias: $2.25$]. \\
               \\
               \underline{Group 2 - Stat only:} \\
               None. \\
               \\
               \underline{Group 3 - Practical only:} \\
               None. \\
               \\
               \underline{Group 4 - Worse:} \\
               Strat-biv [RMSE: $9.39$/$14.3\%$, Bias: $2.06$].}
  & \makecell[l]{\underline{Group 1 - Best:} \\
                O-biv-FC [RMSE: $3.84$, Bias: $0.17$]. \\
                \\
                \underline{Group 2 - Stat only:} \\
                Strat-biv [RMSE: $4.19$/$9.2\%$, Bias: $0.39$]. \\
                \\
                \underline{Group 3 - Practical only:} \\
                None. \\
                \\
                \underline{Group 4 - Worse:} \\ 
                Jones [RMSE: $4.40$/$14.6\%$, Bias: $1.87$].} \\
\midrule
%%
%% ---- DGM: O-biv-RC:
%%
O-biv-RC & \makecell[l]{\underline{Group 1 - Best:} \\
               O-biv-RC [RMSE: $8.23$, Bias: $1.08$], \\
               O-biv-FC [RMSE: $8.41$/$2.1\%$, Bias: $1.20$], \\
               Jones [RMSE: $8.45$/$2.7\%$, Bias: $2.44$]. \\
               \\
               \underline{Group 2 - Stat only:} \\
               O-HSROC-RC [RMSE: $8.65$/$5.1\%$, Bias: $1.45$], \\
               Strat-biv [RMSE: $8.86$/$7.6\%$, Bias: $1.58$]. \\
               \\
               \underline{Group 3 - Practical only:} \\
               None. \\
               \\
               \underline{Group 4 - Worse:} \\
               None.}
  & \makecell[l]{\underline{Group 1 - Best:} \\
                O-biv-FC [RMSE: $3.69$, Bias: $0.49$], \\
                O-biv-RC [RMSE: $3.85$/$4.4\%$, Bias: $0.38$], \\
                O-HSROC-RC [RMSE: $4.04$/$9.5\%$, Bias: $1.23$]. \\
                \\
                \underline{Group 2 - Stat only:} \\
                None. \\
                \\
                \underline{Group 3 - Practical only:} \\
                None. \\
                \\
                \underline{Group 4 - Worse:} \\
                Strat-biv [RMSE: $4.10$/$11.0\%$, Bias: $0.49$], \\
                Jones [RMSE: $4.29$/$16.1\%$, Bias: $1.95$].} \\
\midrule
%%
%% ---- DGM: O-HSROC-RC:
%%
O-HSROC-RC & \makecell[l]{\underline{Group 1 - Best:} \\
               O-biv-RC [RMSE: $8.22$, Bias: $0.92$], \\
               Jones [RMSE: $8.33$/$1.4\%$, Bias: $2.59$], \\
               O-HSROC-RC [RMSE: $8.38$/$2.0\%$, Bias: $0.87$], \\
               O-biv-FC [RMSE: $8.45$/$2.8\%$, Bias: $1.33$]. \\
               \\
               \underline{Group 2 - Stat only:} \\
               None. \\
               \\
               \underline{Group 3 - Practical only:} \\
               None. \\
               \\
               \underline{Group 4 - Worse:} \\
               Strat-biv [RMSE: $9.08$/$10.5\%$, Bias: $1.39$].}
  & \makecell[l]{\underline{Group 1 - Best:} \\
                O-HSROC-RC [RMSE: $3.85$, Bias: $0.58$], \\
                O-biv-RC [RMSE: $3.88$/$0.9\%$, Bias: $0.74$], \\
                Strat-biv [RMSE: $4.12$/$7.2\%$, Bias: $0.77$], \\
                O-biv-FC [RMSE: $4.14$/$7.5\%$, Bias: $0.85$]. \\
                \\
                \underline{Group 2 - Stat only:} \\
                None. \\
                \\
                \underline{Group 3 - Practical only:} \\
                None. \\
                \\
                \underline{Group 4 - Worse:} \\ 
                Jones [RMSE: $4.24$/$10.3\%$, $1.89$].} \\ 
\bottomrule
\end{tabular}
\label{Table:summary_table_overall_GAD_2}
\end{table}
%%%%
%%%%
%%%%
%%%%
\begin{table}[H]
\centering
\caption{
Overall model performance for all four DGMs - GAD-2
(4-group classification ordered by BIAS performance)
}
\footnotesize
\setlength{\tabcolsep}{4pt}
\begin{tabular}{c|l|l}
\toprule
\textbf{DGM} & $\mathbf{N_{studies} = 10}$ & $\mathbf{N_{studies} = 50}$ \\
\midrule
%%
%% ---- DGM = Jones (Bias ordering):
%%
Jones & \makecell[l]{\underline{Group 1 - Best (Bias):} \\
                O-biv-FC [Bias: $1.61$, RMSE: $10.25$]. \\
                \\
                \underline{Group 2 - Stat only:} \\
                None. \\
                \\
                \underline{Group 3 - Practical only:} \\
                Jones [Bias: $2.08$/$29.0\%$, RMSE: $9.75$]. \\
                \\
                \underline{Group 4 - Worse:} \\
                Strat-biv [Bias: $3.03$/$88.2\%$, RMSE: $11.76$].}
  & \makecell[l]{\underline{Group 1 - Best (Bias):} \\
                Jones [Bias: $0.32$, RMSE: $4.55$]. \\
                \\
                \underline{Group 2 - Stat only:} \\
                None. \\
                \\
                \underline{Group 3 - Practical only:} \\
                Strat-biv [Bias: $0.36$/$12.1\%$, RMSE: $5.32$], \\
                O-biv-FC [Bias: $0.64$/$101.8\%$, RMSE: $4.81$]. \\
                \\
                \underline{Group 4 - Worse:} \\
                None.} \\
\midrule
%%
%% ---- DGM = O-biv-FC (Bias ordering):
%%
O-biv-FC & \makecell[l]{\underline{Group 1 - Best (Bias):} \\
               O-biv-FC [Bias: $0.76$, RMSE: $8.21$]. \\
               \\
               \underline{Group 2 - Stat only:} \\
               None. \\
               \\
               \underline{Group 3 - Practical only:} \\
               None. \\
               \\
               \underline{Group 4 - Worse:} \\
               Strat-biv [Bias: $2.06$/$171.1\%$, RMSE: $9.39$], \\
               Jones [Bias: $2.25$/$197.2\%$, RMSE: $8.23$].}
  & \makecell[l]{\underline{Group 1 - Best (Bias):} \\
                O-biv-FC [Bias: $0.17$, RMSE: $3.84$]. \\
                \\
                \underline{Group 2 - Stat only:} \\
                None. \\
                \\
                \underline{Group 3 - Practical only:} \\
                Strat-biv [Bias: $0.39$/$127.8\%$, RMSE: $4.19$]. \\
                \\
                \underline{Group 4 - Worse:} \\ 
                Jones [Bias: $1.87$/$989.8\%$, RMSE: $4.40$].} \\
\midrule
%%
%% ---- DGM = O-biv-RC (Bias ordering):
%%
O-biv-RC & \makecell[l]{\underline{Group 1 - Best (Bias):} \\
               O-biv-RC [Bias: $1.08$, RMSE: $8.23$]. \\
               \\
               \underline{Group 2 - Stat only:} \\
               None. \\
               \\
               \underline{Group 3 - Practical only:} \\
               O-biv-FC [Bias: $1.20$/$10.6\%$, RMSE: $8.41$], \\
               O-HSROC-RC [Bias: $1.45$/$33.6\%$, RMSE: $8.65$], \\
               Strat-biv [Bias: $1.58$/$45.4\%$, RMSE: $8.86$]. \\
               \\
               \underline{Group 4 - Worse:} \\
               Jones [Bias: $2.44$/$125.3\%$, RMSE: $8.45$].}
  & \makecell[l]{\underline{Group 1 - Best (Bias):} \\
                O-biv-RC [Bias: $0.38$, RMSE: $3.85$]. \\
                \\
                \underline{Group 2 - Stat only:} \\
                None. \\
                \\
                \underline{Group 3 - Practical only:} \\
                O-biv-FC [Bias: $0.49$/$28.1\%$, RMSE: $3.69$], \\
                Strat-biv [Bias: $0.49$/$28.3\%$, RMSE: $4.10$]. \\
                \\
                \underline{Group 4 - Worse:} \\
                O-HSROC-RC [Bias: $1.23$/$223.6\%$, RMSE: $4.04$], \\
                Jones [Bias: $1.95$/$413.3\%$, RMSE: $4.29$].} \\
\midrule
%%
%% ---- DGM = O-HSROC-RC (Bias ordering):
%%
O-HSROC-RC & \makecell[l]{\underline{Group 1 - Best (Bias):} \\
               O-HSROC-RC [Bias: $0.87$, RMSE: $8.38$], \\
               O-biv-RC [Bias: $0.92$, RMSE: $8.22$]. \\
               \\
               \underline{Group 2 - Stat only:} \\  
               None. \\
               \\
               \underline{Group 3 - Practical only:} \\
               O-biv-FC [Bias: $1.33$/$51.9\%$, RMSE: $8.45$], \\
               Strat-biv [Bias: $1.39$/$58.8\%$, RMSE: $9.08$]. \\
               \\
               \underline{Group 4 - Worse:} \\
               Jones [Bias: $2.59$/$196.3\%$, RMSE: $8.33$].}
  & \makecell[l]{\underline{Group 1 - Best (Bias):} \\ 
                O-HSROC-RC [Bias: $0.58$, RMSE: $3.85$]. \\
                \\
                \underline{Group 2 - Stat only:} \\
                None. \\
                \\
                \underline{Group 3 - Practical only:} \\
                O-biv-RC [Bias: $0.74$/$26.6\%$, RMSE: $3.88$], \\
                Strat-biv [Bias: $0.77$/$31.9\%$, RMSE: $4.12$], \\
                O-biv-FC [Bias: $0.85$/$46.5\%$, RMSE: $4.14$]. \\
                \\
                \underline{Group 4 - Worse:} \\ 
                Jones [Bias: $1.89$/$225.4\%$, RMSE: $4.24$].} \\ 
\bottomrule
\end{tabular}
\label{Table:summary_table_overall_GAD_2_bias}
\end{table}
%%%%
%%%%
%%%% --------------------------------------------------------------------------------------------------------------------
Table \ref{Table:summary_table_overall_GAD_2} shows the results for the GAD-2 test based DGMs
(recall that the GAD-2 data has $7$ ordinal categories,
and approximately $\sim 15\%$ missing data),
for all four DGM's.
Furthermore, in the appendix we have additional figures; namely:
A figure showing the RMSE results
(for all four DGMs;
see figure  \ref{Sim_study_RMSE_GAD_2},
in appendix \ref{appendix_B_RMSE_plots},
section \ref{appendix_B_RMSE_plots_GAD_2_test_I}),
an analogous figure but for the bias
(see figure \ref{Sim_study_Bias_GAD_2},
in appendix \ref{appendix_C_Bias_plots}, 
section \ref{appendix_C_Bias_plots_GAD_2_test_I}),
and finally a figure showing the coverage for all four DGMs for the GAD-2 data
(see figure \ref{Sim_study_Coverage_GAD_2},
in appendix \ref{appendix_D_Coverage_plots},
section \ref{appendix_D_Coverage_plots_GAD_2_test_I}).
Additionally, appendix \ref{appendix_E_detailed_results_tables} shows detailed results -
which includes the RMSE(Se), RMSE(Sp), Bias(Se), Bias(Sp), Coverage(Se), Coverage(Sp), 
and also the interval widths of Se and Sp - 
for the GAD-2, these are specifically in 
appendix section \ref{appendix_E_detailed_results_tables_GAD_2_test_I}.
%%%%
%%%%
%%%%%%%%%%%%%%%%%%%%%%%%%%%%%%%%%%%%%%
\subsubsection{GAD-2: DGM \#1 (Jones)}
\label{Sim_study_GAD_2_Jones_DGM_1}
%%%%%%%%%%%%%%%%%%%%%%%%%%%%%%%%%%%%%%
%%%%
%%%% --------------------------------------------------------------------------------------------------------------------
%%
%% ---- 10 studies; RMSE:
%%
For the first DGM (Jones), we can see from the top-left of table \ref{Table:summary_table_overall_GAD_2}
that the Jones model is the only model that made it into the "best" group (RMSE = $9.75$),
and we found it was statistically significantly better than the O-biv-FC model (RMSE = $10.25$);
however, the magnitude of the relative difference in RMSE was not practically significant ($5.2\% < 10\%$), 
hence why the O-biv-FC model made it into the "worse (stat only)" group rather than the "worse" group.
Furthermore, the Jones model was notably better than the strat-biv model
(RMSE(strat-biv) = $11.76$, $20.6\%$ worse than the Jones model),
hence why the strat-biv model made it into the "worse" group.

%%%%
%%%% --------------------------------------------------------------------------------------------------------------------
%%
%% ---- 10 studies; Bias:
%%
If we instead focus on bias alone (see table \ref{Table:summary_table_overall_GAD_2_bias}) -
which ignores precision, which RMSE also takes into account -
we obtain quite different results.
Namely, it is now the O-biv-FC model in the lead and the only model in the "best" group 
(bias = $1.61$), with the Jones model now being in the "worse (practical only)" group, 
with a bias of $2.08$ ($29.0\%$ worse than the O-biv-FC model).
Furthermore, the strat-biv model still ends up being the only model in the "worse" group, 
with it's bias being nearly 2-fold as bad as the O-biv-FC's bias
(bias(strat-biv) = $3.03$, relative difference vs. O-biv-FC = $88.2\%$).

%%%%
%%%% --------------------------------------------------------------------------------------------------------------------
%%
%% ---- 50 studies; RMSE:
%%
On the other hand, when we have $50$ studies, both the Jones (RMSE = $4.55$) and the O-biv-FC (RMSE = $4.81$)
models are in the "best" group, with the O-biv-FC's difference in RMSE not being either statistically significantly
(RMSE diff = $0.26 < 0.50$) nor practically significantly worse (rel. diff = $5.8\% < 10\%$) than the Jones model.
However - similarly to the $10$ studies case - 
the strat-biv model performs statistically significantly
(as well as practically) worse, with an RMSE of $5.32$ (vs. $4.55 - 4.81$ for the Jones and O-biv-FC models),
being $17.0\%$ worse compared to the Jones model.

%%%%
%%%% --------------------------------------------------------------------------------------------------------------------
%%
%% ---- 50 studies; Bias:
%%
Additionally, the bias of the strat-biv model is quite good ($0.36$) -
being in between the Jones and O-biv-FC models
($0.32$ and $0.64$ for Jones and O-biv-FC, respectively);
hence, it ending up in the "worse (practical only)" group
(see table \ref{Table:summary_table_overall_GAD_2_bias}) - 
when we instead rank model performance by bias (as opposed to RMSE).

%%%%%%%%%%%%%%%%%%%%%%%%%%%%%%%%%%%%%%
\subsubsection{GAD-2: DGM \#2 (O-biv-FC)}
\label{Sim_study_GAD_2_O_biv_FC_DGM_2}
%%%%%%%%%%%%%%%%%%%%%%%%%%%%%%%%%%%%%%
%%%%
%%%% --------------------------------------------------------------------------------------------------------------------
%%
%% ---- 10 studies; RMSE:
%%
For the second DGM (O-biv-FC; left of second quadrant of table \ref{Table:summary_table_overall_GAD_2}),
both the O-biv-FC and Jones models obtain essentially the same RMSE
($8.21$ and $8.23$ for O-biv and Jones models, respectively).
Hence, both make it into the "best" group - as they are well within 1 MCSE (1 MCSE here = $0.125$) of one another.
However, the strat-biv model performs notably worse (RMSE = $9.39$ vs. $8.21 - 8.23$ for the other two models),
with a relative difference in RMSE of $14.3\%$ compared to the leading O-biv-FC model.

%%%%
%%%% --------------------------------------------------------------------------------------------------------------------
%%
%% ---- 10 studies; Bias:
%%
On the other hand, the bias (see table \ref{Table:summary_table_overall_GAD_2_bias}
- the secondary measure for this study - is substantially better ($\sim 1.5\%$ better) 
for the O-biv-FC model compared to the Jones model
($0.76$ vs. $2.25$ for O-biv-FC and Jones models, respectively).
Furthermore, regarding the strat-biv model - 
this actually achieves slightly better bias than the Jones model here (but not significant), despite 
notably worse RMSE (bias/RMSE = $2.06/9.39$ for strat-biv vs. $2.25/8.23$ for Jones model).
This is because RMSE takes the variance - as well as the bias itself - into account;
hence this shows that the strat-biv model performs worse than Jones because its precision is worse 
(i.e., it is less consistent).

%%%%
%%%% --------------------------------------------------------------------------------------------------------------------
%%
%% ---- 50 studies; RMSE:
%%
When we have $50$ studies, the pattern changes 
(see right of second quadrant of table \ref{Table:summary_table_overall_GAD_2}).
More specifically, we can see that - unlike the $N_{studies} = 10$ case -
the Jones model is no longer in the "best" group,
and we now instead have the O-biv-FC model (RMSE = $3.84$) alone in the "best" group.
Furthermore, we can see that the strat-biv model is in the "worse (stat only)" group -
since despite it being statistically significantly worse than the O-biv-FC model (RMSE diff = $0.35$),
the relative difference in RMSE does not quite reach our threshold (rel. diff. in RMSE = $9.2\% < 10\%$).

%%%%
%%%% --------------------------------------------------------------------------------------------------------------------
%%
%% ---- 50 studies; Bias:
%%
On the other hand, the Jones model obtains a statistically significantly worse as well as practically worse
(rel. diff = $14.6\% > 10\%$) RMSE than the O-biv-FC model
(RMSE = $4.40$ vs. $3.84$ for Jones and O-biv-FC models, respectively).
%% -- bookmark: ** maybe move to discussion ** 
Furthermore, if we focus on bias instead
(see table \ref{Table:summary_table_overall_GAD_2_bias}),
the difference between the two "multiple threshold" models is much more substantial,
with the O-biv-FC having very low bias ($0.17$ for O-biv-FC vs. $1.87$ for Jones).
However, for the strat-biv model, the bias is actually quite good (bias = $0.39$).

%%%%%%%%%%%%%%%%%%%%%%%%%%%%%%%%%%%%%%
\subsubsection{A note on DGMs \#1 and \#2 (fixed-cutpoint DGMs)}
\label{Sim_study_GAD_2_note_on_DGMs_1_and_2}
%%%%%%%%%%%%%%%%%%%%%%%%%%%%%%%%%%%%%%
%%%%
%%%% --------------------------------------------------------------------------------------------------------------------
Note that, for the first two fixed-cutpoint DGMs (i.e., the Jones and O-biv-FC DGMs), we only ran 2 out of 4 models 
(Jones and O-biv-FC). 
This is because:
(i) the random-effect-cutpoint models - when run on these data generated by fixed-effect models - 
have very inconsistent and often poor ESS
(this is because $ \kappa \Rightarrow \infty $ when heterogeneity approaches zero which creates a tendency to cause numerical instability).
(ii) they're not realistic (even though perhaps a good approximation); in reality,
there is typically notable between-study heterogeneity in the cutpoints,
even if (as our simulation study shows) this does not need to be explicitly modeled to recover Se/Sp well/optimally.
Importantly, the scientifically relevant comparison is whether fixed-cutpoint models 
remain robust when applied to random-cutpoint DGMs (which we demonstrate they do), 
rather than whether random-cutpoint models can handle the unrealistic scenario of zero heterogeneity.

%%%%%%%%%%%%%%%%%%%%%%%%%%%%%%%%%%%%%%
\subsubsection{GAD-2: DGM \#3 (O-biv-FC)}
\label{Sim_study_GAD_2_O_biv_RC_DGM_3}
%%%%%%%%%%%%%%%%%%%%%%%%%%%%%%%%%%%%%%
%%%%
%%%% --------------------------------------------------------------------------------------------------------------------
%%
%% ---- 10 studies; RMSE:
%%
For the first random-cutpoints DGM with $10$ studies 
(O-biv-RC; see left of third quadrant of table \ref{Table:summary_table_overall_GAD_2}), 
we can see that three out of four of the "multiple threshold" models make it into the "best" group - all except O-HSROC-RC -
with the RMSE range across these three models being $8.23 - 8.45$, 
and the RMSE of the O-HSROC-RC model being statistically significantly (but not practically)
worse than the leading O-biv-RC model
(RMSE(O-HSROC-RC) = $8.65$, rel. diff. = $5.1\%$).
Furthermore, the strat-biv model also joins the O-HSROC-RC model in the "stat only" group
(RMSE(strat-biv) = $8.86$, rel. diff. = $7.6\%$).

%%%%
%%%% --------------------------------------------------------------------------------------------------------------------
%%
%% ---- 10 studies; Bias:
%%
Additionally, there are notable differences in bias between some of the "multiple thresholds" models
(see table \ref{Table:summary_table_overall_GAD_2_bias}).
More specifically, we can see that both of the O-biv models have essentially the same bias - 
around just $1\%$ (bias range: $1.08 - 1.20$) -
with the Jones model coming in third, with worse bias ($2.44$);
finally, the O-HSROC-RC model comes in last place out of these four models (recall that order is ranked by RMSE); 
however, the bias is quite good (bias = $1.45$).
Furthermore, despite having the worst RMSE ($8.86$), the strat-biv model obtains relatively good bias ($1.58$) - 
better than the Jones model ($2.44$) - despite having a worse RMSE
(RMSE was $8.86$ and $8.45$ for strat-biv and Jones models, respectively). 
%%
%% ---- ** bookmark: might move to discussion ** ----
%%
As mentioned above, this discrepancy between bias and RMSE rankings is because the strat-biv model is inconsistent (high variance).

%%%%
%%%% --------------------------------------------------------------------------------------------------------------------
%%
%% ---- 50 studies; RMSE:
%%
When we move up to $50$ studies (right of third quadrant of table \ref{Table:summary_table_overall_GAD_2}), 
we have a different pattern than the $10$-study case. 
More specifically, we have three models in the "Best" group, but a different set of models than before: 
the O-biv-FC model leading with the lowest RMSE (RMSE = $3.69$) - 
despite the true DGM being that of the O-biv-RC model - 
which is then followed by the O-biv-RC model (RMSE = $3.85$, rel. diff. = $4.4\% < 10\%$),
and finally last place in the "best" group is the O-HSROC-RC model (RMSE = $4.04$, rel. diff. = $9.5\% < 10\%$).
In contrast, both the strat-biv and the Jones model are in the "worse" group, 
with both of these models having notably worse RMSE magnitudes than the leading O-biv-FC model
(rel. diff. in RMSE is $11.0\%$ and $16.1\%$ for strat-biv and Jones models, respectively), 
as well as both being statistically significantly worse than the leading O-biv-FC model.

%%%%
%%%% --------------------------------------------------------------------------------------------------------------------
%%
%% ---- 50 studies; Bias:
%%
Furthermore, both O-biv models also have effectively the same bias  
(see table \ref{Table:summary_table_overall_GAD_2_bias}); 
$0.49$ and $0.38$ for the FC and RC O-biv models, respectively).
However, the O-HSROC-RC model has substantially worse (but still good) bias compared to these two models 
($1.23$ for O-HSROC-RC vs. $0.38 - 0.49$ for the O-biv models). 
Additionally, the Jones model also has notably worse bias than the FC/RC O-biv models
(bias = $1.95$ for Jones vs. $0.38 - 0.49$ for the O-biv models), 
and slightly worse than the O-HSROC-RC model 
(bias = $1.95$ vs $1.23$ for Jones and O-HSROC-RC models, respectively). 
We can also see that the strat-biv model actually obtains very good bias here ($0.49$) - 
better than both the Jones ($1.95$) and O-HSROC-RC ($1.23$) models - 
being the same as the bias of the O-biv-FC model ($0.49$).
This is despite it having a notably (as well as statistically significantly) worse RMSE 
($4.10$ vs. $3.69 - 3.85$ for strat-biv and the O-biv models, respectively).

%%%%%%%%%%%%%%%%%%%%%%%%%%%%%%%%%%%%%%
\subsubsection{GAD-2: DGM \#4 (O-HSROC-RC)}
\label{Sim_study_GAD_2_O_HSROC_RC_DGM_4}
%%%%%%%%%%%%%%%%%%%%%%%%%%%%%%%%%%%%%%
%%%%
%%%% --------------------------------------------------------------------------------------------------------------------
%%
%% ---- 10 studies; RMSE:
%%
Finally, for the second random cutpoints DGM
(O-HSROC-RC; bottom-left of table \ref{Table:summary_table_overall_GAD_2}) -
all four "multiple thresholds" models make it into the "best" group, and there is very
little variation in the RMSE across all four models ($8.22 - 8.45$).
However, we can see that the strat-biv model ends up in the "worse" group,
with an RMSE of $9.08$, which is statistically significantly worse than the leading model
(O-biv-RC; RMSE = $8.22$), as well as practically significant ($10.5\% > 10\%$).

%%%%
%%%% --------------------------------------------------------------------------------------------------------------------
%%
%% ---- 10 studies; Bias:
%%
On the other hand, if we focus on bias instead 
(\ref{Table:summary_table_overall_GAD_2})
the Jones model has notably worse bias here
($2.59$ vs. $0.92 - 1.33$ for the Jones model and the other three "multiple threshold" models, respectively).
Furthermore, despite statistically significantly worse RMSE -
$9.08$ for the strat-biv vs. $8.22 - 8.45$ for the other four models -
the strat-biv model actually obtains good bias ($1.39$),
being better than the Jones model ($2.59$), despite notably worse RMSE
(RMSE = $9.08$ and $8.33$ for the strat-biv and Jones models, respectively).

%%%%
%%%% --------------------------------------------------------------------------------------------------------------------
%%
%% ---- 50 studies; RMSE:
%%
The pattern changes somewhat with with five times more data 
($50$ studies; see bottom-right of table \ref{Table:summary_table_overall_GAD_2}).
Unlike the $N_{studies} = 10$ case, the strat-biv model is now in the "best" group (not "worse" group), 
and the Jones model is now in the "worse" group - 
being both statistically significantly worse than the leading O-HSROC-RC model 
(RMSE diff = $0.39$), as well as practically worse in magnitude (rel. diff.  = $10.3\% > 10\%$).

%%%%
%%%% --------------------------------------------------------------------------------------------------------------------
%%
%% ---- 50 studies; Bias:
%%
Furthermore, all four models - except for the Jones model - have a rather narrow bias range
(see table \ref{Table:summary_table_overall_GAD_2_bias}), 
with the Jones model having substantially worse bias
($1.89$ and $0.58 - 0.85$ for the Jones model and the other four models, respectively).

%%%%%%%%%%%%%%%%%%%%%%%%%%%%%%%%%%%%%%%%%%%%%%%%%%%%%%%%%%%%%%%%%%%%%%%%%%%%%%%%%%%%%%%%%%%%%%%%%%%%%%%%%%%%%%%%%%%%
\subsection{Results: Test II (HADS)}
\label{Sim_study_HADS}
%%%%%%%%%%%%%%%%%%%%%%%%%%%%%%%%%%%%%%%%%%%%%%%%%%%%%%%%%%%%%%%%%%%%%%%%%%%%%%%%%%%%%%%%%%%%%%%%%%%%%%%%%%%%%%%%%%%%
%%%%
%%%%
\begin{table}[H]
\centering
\caption{
Overall model performance for all four DGMs - HADS
(4-group classification based on statistical and practical significance)
}
\footnotesize
\setlength{\tabcolsep}{4pt}
\begin{tabular}{c|l|l}
\toprule
\textbf{DGM} & $\mathbf{N_{studies} = 10}$ & $\mathbf{N_{studies} = 50}$ \\
\midrule
%%
%% ---- DGM = Jones:
%%
Jones & \makecell[l]{\underline{Group 1 - Best:} \\
                Jones [RMSE: $5.59$, Bias: $2.99$], \\
                O-biv-FC [RMSE: $5.84$/$4.4\%$, Bias: $3.19$]. \\
                \\
                \underline{Group 2 - Worse (stat. signif. only):} \\
                None. \\
                \\
                \underline{Group 3 - Worse (practical only):} \\
                None. \\
                \\
                \underline{Group 4 - Worse:} \\
                Strat-biv [RMSE: $6.98$/$24.8\%$, Bias: $3.40$].}
  & \makecell[l]{\underline{Group 1 - Best:} \\
                Jones [RMSE: $3.22$, Bias: $2.52$], \\
                O-biv-FC [RMSE: $3.29$/$2.4\%$, Bias: $2.52$]. \\
                \\
                \underline{Group 2 - Worse (stat. signif. only):} \\
                None. \\
                \\
                \underline{Group 3 - Worse (practical only):} \\
                None. \\
                \\
                \underline{Group 4 - Worse:} \\
                Strat-biv [RMSE: $3.56$/$10.8\%$, Bias: $2.50$].} \\
\midrule
%%
%% ---- DGM = O-biv-FC:
%%
O-biv-FC & \makecell[l]{\underline{Group 1 - Best:} \\
               O-biv-FC [RMSE: $4.60$, Bias: $0.51$], \\
               Jones [RMSE: $4.66$/$0.06\%$, Bias: $0.95$]. \\
               \\
               \underline{Group 2 -  Worse (stat. signif. only):} \\
               None. \\
               \\
               \underline{Group 3 - Worse (practical only):} \\
               None. \\
               \\
               \underline{Group 4 - Worse:} \\
               Strat-biv [RMSE: $6.34$/$37.8\%$, Bias: $1.45$].}
  & \makecell[l]{\underline{Group 1 - Best:} \\
                O-biv-FC [RMSE: $1.85$, Bias: $0.11$], \\
                Jones [RMSE: $2.02$/$9.2\%$, Bias: $0.84$]. \\
                \\
                \underline{Group 2 - Worse (stat. signif. only):} \\
                None. \\
                \\
                \underline{Group 3 - Worse (practical only):} \\
                None. \\
                \\
                \underline{Group 4 - Worse:} \\ 
                Strat-biv [RMSE: $2.33$/$26.0\%$, Bias: $0.33$].} \\
\midrule
%%
%% ---- DGM = O-biv-RC:
%%
O-biv-RC & \makecell[l]{\underline{Group 1 - Best:} \\
               O-biv-RC [RMSE: $4.96$, Bias: $0.85$], \\
               O-biv-FC [RMSE: $4.99$/$0.7\%$, Bias: $0.57$], \\
               Jones [RMSE: $5.04$/$1.8\%$, Bias: $1.39$]. \\
               \\
               \underline{Group 2 - Worse (stat. signif. only):} \\
               None. \\
               \\
               \underline{Group 3 - Worse (practical only):} \\
               None. \\
               \\
               \underline{Group 4 - Worse:} \\
               O-HSROC-RC [RMSE: $5.77$/$16.3\%$, Bias: $2.49$], \\
               Strat-biv [RMSE: $6.43$/$29.8\%$, Bias: $1.55$].}
  & \makecell[l]{\underline{Group 1 - Best:} \\
                O-biv-FC [RMSE: $2.03$, Bias: $0.24$], \\
                O-biv-RC [RMSE: $2.08$/$2.2\%$, Bias: $0.48$]. \\
                \\
                \underline{Group 2 - Worse (stat. signif. only):} \\
                None. \\
                \\
                \underline{Group 3 - Worse (practical only):} \\
                None. \\
                \\
                \underline{Group 4 - Worse:} \\
                Jones [RMSE: $2.55$/$25.5\%$, Bias: $1.43$], \\
                Strat-biv [RMSE: $2.59$/$27.4\%$, Bias: $0.33$], \\
                O-HSROC-RC [RMSE: $3.23$/$58.9\%$, Bias: $2.22$].} \\
\midrule
%%
%% ---- DGM = O-HSROC-RC:
%%
O-HSROC-RC & \makecell[l]{\underline{Group 1 - Best:} \\
               O-HSROC-RC [RMSE: $3.54$, Bias: $0.43$], \\
               Jones [RMSE: $3.58$/$1.5\%$, Bias: $0.69$], \\
               O-biv-RC [RMSE: $3.72$/$5.0\%$, Bias: $0.72$], \\
               O-biv-FC [RMSE: $3.80$/$7.2\%$, Bias: $0.57$]. \\
               \\
               \underline{Group 2 - Worse (stat. signif. only):} \\  
               None. \\
               \\
               \underline{Group 3 - Worse (practical only):} \\
               None. \\
               \\
               \underline{Group 4 - Worse:} \\
               Strat-biv [RMSE: $4.70$/$32.7\%$, Bias: $0.68$].}
  & \makecell[l]{\underline{Group 1 - Best:} \\ 
                O-HSROC-RC [RMSE: $1.52$, Bias: $0.39$], \\
                Jones [RMSE: $1.52$/$0.1\%$, Bias: $0.61$], \\
                O-biv-FC [RMSE: $1.59$/$4.4\%$, Bias: $0.43$], \\
                O-biv-RC [RMSE: $1.66$/$9.3\%$, Bias: $0.44$]. \\
                \\
                \underline{Group 2 - Worse (stat. signif. only):} \\
                None. \\
                \\
                \underline{Group 3 - Worse (practical only):} \\
                None. \\
                \\
                \underline{Group 4 - Worse:} \\ 
                Strat-biv [RMSE: $1.99$/$30.7\%$, Bias: $0.36$].} \\ 
\bottomrule
\end{tabular}
\label{Table:summary_table_overall_HADS}
\end{table}
%%%%
%%%%
%%%%
%%%%
\begin{table}[H]
\centering
\caption{
Overall model performance for all four DGMs - HADS
(4-group classification ordered by BIAS performance)
}
\footnotesize
\setlength{\tabcolsep}{4pt}
\begin{tabular}{c|l|l}
\toprule
\textbf{DGM} & $\mathbf{N_{studies} = 10}$ & $\mathbf{N_{studies} = 50}$ \\
\midrule
%%
%% ---- DGM = Jones (Bias ordering):
%%
Jones & \makecell[l]{\underline{Group 1 - Best (Bias):} \\
                Jones [$2.99$, RMSE: $5.59$], \\
                O-biv-FC [$3.19$, RMSE: $5.84$]. \\
                \\
                \underline{Group 2 - Worse (stat. signif. only):} \\
                None. \\
                \\
                \underline{Group 3 - Worse (practical only):} \\
                Strat-biv [$3.40$/$13.8\%$, RMSE: $6.98$]. \\
                \\
                \underline{Group 4 - Worse:} \\
                None.}
  & \makecell[l]{\underline{Group 1 - Best (Bias):} \\
                Strat-biv [$2.50$, RMSE: $3.56$], \\
                O-biv-FC [$2.52$, RMSE: $3.29$], \\
                Jones [$2.52$, RMSE: $3.22$]. \\
                \\
                \underline{Group 2-4:} All in Group 1.} \\
\midrule
%%
%% ---- DGM = O-biv-FC (Bias ordering):
%%
O-biv-FC & \makecell[l]{\underline{Group 1 - Best (Bias):} \\
               O-biv-FC [$0.51$, RMSE: $4.60$]. \\
               \\
               \underline{Group 2 - Worse (stat. signif. only):} \\
               None. \\
               \\
               \underline{Group 3 - Worse (practical only):} \\
               Jones [$0.95$/$86.3\%$, RMSE: $4.66$]. \\
               \\
               \underline{Group 4 - Worse:} \\
               Strat-biv [$1.45$/$185.9\%$, RMSE: $6.34$].}
  & \makecell[l]{\underline{Group 1 - Best (Bias):} \\
                O-biv-FC [$0.11$, RMSE: $1.85$]. \\
                \\
                \underline{Group 2 - Worse (stat. signif. only):} \\
                None. \\
                \\
                \underline{Group 3 - Worse (practical only):} \\
                Strat-biv [$0.33$/$201.4\%$, RMSE: $2.33$]. \\
                \\
                \underline{Group 4 - Worse:} \\ 
                Jones [$0.84$/$665.9\%$, RMSE: $2.02$].} \\
\midrule
%%
%% ---- DGM = O-biv-RC (Bias ordering):
%%
O-biv-RC & \makecell[l]{\underline{Group 1 - Best (Bias):} \\
               O-biv-FC [$0.57$, RMSE: $4.99$]. \\
               \\
               \underline{Group 2 - Worse (stat. signif. only):} \\
               None. \\
               \\
               \underline{Group 3 - Worse (practical only):} \\
               O-biv-RC [$0.85$/$48.0\%$, RMSE: $4.96$]. \\
               \\
               \underline{Group 4 - Worse:} \\
               Jones [$1.39$/$143.5\%$, RMSE: $5.04$], \\
               Strat-biv [$1.55$/$170.7\%$, RMSE: $6.43$], \\
               O-HSROC-RC [$2.49$/$336.4\%$, RMSE: $5.77$].}
  & \makecell[l]{\underline{Group 1 - Best (Bias):} \\
                O-biv-FC [$0.24$, RMSE: $2.03$]. \\
                \\
                \underline{Group 2 - Worse (stat. signif. only):} \\
                None. \\
                \\
                \underline{Group 3 - Worse (practical only):} \\
                Strat-biv [$0.33$/$40.8\%$, RMSE: $2.59$], \\
                O-biv-RC [$0.48$/$104.9\%$, RMSE: $2.08$]. \\
                \\
                \underline{Group 4 - Worse:} \\
                Jones [$1.43$/$506.8\%$, RMSE: $2.55$], \\
                O-HSROC-RC [$2.22$/$843.0\%$, RMSE: $3.23$].} \\
\midrule
%%
%% ---- DGM = O-HSROC-RC (Bias ordering):
%%
O-HSROC-RC & \makecell[l]{\underline{Group 1 - Best (Bias):} \\
               O-HSROC-RC [$0.43$, RMSE: $3.54$]. \\
               \\
               \underline{Group 2 - Worse (stat. signif. only):} \\  
               None. \\
               \\
               \underline{Group 3 - Worse (practical only):} \\
               O-biv-FC [$0.57$/$32.3\%$, RMSE: $3.80$], \\
               Strat-biv [$0.68$/$58.0\%$, RMSE: $4.70$], \\
               Jones [$0.69$/$60.1\%$, RMSE: $3.58$], \\
               O-biv-RC [$0.72$/$68.0\%$, RMSE: $3.72$]. \\
               \\
               \underline{Group 4 - Worse:} \\
               None.}
  & \makecell[l]{\underline{Group 1 - Best (Bias):} \\ 
                Strat-biv [$0.36$, RMSE: $1.99$], \\
                O-HSROC-RC [$0.39$, RMSE: $1.52$]. \\
                \\
                \underline{Group 2 - Worse (stat. signif. only):} \\
                None. \\
                \\
                \underline{Group 3 - Worse (practical only):} \\
                O-biv-FC [$0.43$/$18.9\%$, RMSE: $1.59$], \\
                O-biv-RC [$0.44$/$21.5\%$, RMSE: $1.66$], \\
                Jones [$0.61$/$69.4\%$, RMSE: $1.52$]. \\
                \\
                \underline{Group 4 - Worse:} \\ 
                None.} \\ 
\bottomrule
\end{tabular}
\label{Table:summary_table_overall_HADS_bias}
\end{table}
%%%%
%%%% ------------------------------------------------------------------------------------------------------------
Table \ref{Table:summary_table_overall_HADS} shows the results for the HADS test
(which has $22$ ordinal categories, and approximately $\sim 40\%$ missing data),
for all four DGM's.
Furthermore, in the appendix we have additional figures; namely:
A figure showing the RMSE results
(for all four DGMs;
see figure  \ref{Sim_study_RMSE_HADS},
in appendix \ref{appendix_B_RMSE_plots},
section \ref{appendix_B_RMSE_plots_HADS_test_II}),
an analogous figure but for the bias
(see figure \ref{Sim_study_Bias_HADS},
in appendix \ref{appendix_C_Bias_plots}, 
section \ref{appendix_C_Bias_plots_HADS_test_II}),
and finally a figure showing the coverage for all four DGMs for the HADS data
(see figure \ref{Sim_study_Coverage_HADS},
in appendix \ref{appendix_D_Coverage_plots},
section \ref{appendix_D_Coverage_plots_HADS_test_II}).
Additionally, appendix \ref{appendix_E_detailed_results_tables} shows detailed results -
which includes the RMSE(Se), RMSE(Sp), Bias(Se), Bias(Sp), Coverage(Se), Coverage(Sp), 
and also the interval widths of Se and Sp - 
for the HADS, these are specifically in 
appendix section \ref{appendix_E_detailed_results_tables_HADS_test_II}.
%%%%
%%%%
%%%%%%%%%%%%%%%%%%%%%%%%%%%%%%%%%%%%%%
\subsubsection{HADS: DGM \#1 (Jones)}
\label{Sim_study_HADS_Jones_DGM_1}
%%%%%%%%%%%%%%%%%%%%%%%%%%%%%%%%%%%%%%
%%%%
%%%% ------------------------------------------------------------------------------------------------------------
%%
%% ---- 10 studies; RMSE:
%%
For the Jones DGM and for $N_{studies} = 10$
(see top-left quadrant of table \ref{Table:summary_table_overall_HADS}),
we can see that the Jones and O-biv-FC models perform very similarly
(relative improvement of Jones model over O-biv-FC = $4.4\%$),
and are equivalent form a statistical significance perspective
(RMSE = $5.59$ and $5.84$ for Jones and O-biv-FC models, respectively).
Furthermore, the strat-biv model is the only model is the "worse" group,
performing statistically significantly worse than the leading model (Jones),
with an RMSE of $6.98$ (vs. $5.59$ for Jones),
as well as practically worse
(relative change in magnitude of RMSE vs. Jones model = $24.8\%$).

%%%%
%%%% ------------------------------------------------------------------------------------------------------------
%%
%% ---- 10 studies; Bias:
%%
Additionally, if we look at bias alone
(not RMSE; see table \ref{Table:summary_table_overall_HADS_bias}),
we can see that we obtain almost the same pattern -
the only difference being that the strat-biv model is now in the "worse (practical only)" group -
as opposed to the more general "worse" group.
In other words, even though the strat-biv models' bias ($3.40$) is worse than the leading Jones model
from a practical standpoint (rel. change = $13.8\% > 10\%$), this did not quite make statistical significance.
% both "multiple threshold" models achieve very similar total absolute bias
% ($2.99$ and $3.19$ for the Jones and O-biv-FC models, respectively).
% Also, the strat-biv model does not actually do much worse when looking at bias
% (bias = $3.40$ and $2.99 - 3.19$ for the strat-biv model and the other two models, respectively).

%%%%
%%%% ------------------------------------------------------------------------------------------------------------
%%
%% ---- 50 studies; RMSE:
%%
The pattern is similar when we move up to $50$ studies
(top-right quadrant of table \ref{Table:summary_table_overall_HADS});
More specifically, we can see that - despite not performing as badly as it did for $10$ studies -
the strat-biv model still makes it into the "worse" group,
because the relative difference in RMSE (vs. the Jones model) is over $10\%$ ($10.8\%$).
Furthermore, the differences between the Jones and O-biv-FC models, in terms of RMSE,
is both statistically (RMSE diff = $0.07$) and practically (rel. diff. in RMSE = $2.4\%$) not significant.

%%%%
%%%% ------------------------------------------------------------------------------------------------------------
%%
%% ---- 50 studies; Bias:
%%
Regarding bias,
(not RMSE; see table \ref{Table:summary_table_overall_HADS_bias}),
we can see that - unlike when we had only $10$ studies -
we now essentially obtain identical bias between all three models (bias range: $2.50 - 2.52$),
hence they all end up in the "best" group when only bias is considered.
%%
%% ---- note: could move this to discussion:
%%
This discrepancy between rankings/groupings when using RMSE (see table \ref{Table:summary_table_overall_HADS}), 
vs. bias (see table \ref{Table:summary_table_overall_HADS_bias}) 
suggests that the reason why the strat-biv model is clearly worse 
(both practically and statistically) than the leading Jones model is due to poorer precision, 
as opposed to being driven by bias.

%%%%%%%%%%%%%%%%%%%%%%%%%%%%%%%%%%%%%%
\subsubsection{HADS: DGM \#2 (O-biv-FC)}
\label{Sim_study_HADS_O_biv_FC_DGM_2}
%%%%%%%%%%%%%%%%%%%%%%%%%%%%%%%%%%%%%%
%%%%
%%%% ------------------------------------------------------------------------------------------------------------
%%
%% ---- 10 studies; RMSE:
%%
For the O-biv-FC DGM and $N_{studies} = 10$
(see left, second quadrant of table \ref{Table:summary_table_overall_HADS}),
the Jones and O-biv-FC model again essentially perform the same in terms of RMSE
(RMSE: $4.60 - 4.66$ for Jones and O-biv-FC, respectively - well within MCSE of one another).
We can also see that the strat-biv - just like the previous DGM we discussed above (Jones model DGM) -
is the only model in the "worse" group, obtaining an RMSE of $6.34$ (vs. $4.60 - 4.66$ for the other two models), 
and a very substantial relative difference in RMSE (vs. the O-biv-FC model) of $37.8\%$.

%%%%
%%%% ------------------------------------------------------------------------------------------------------------
%%
%% ---- 10 studies; Bias:
%%
When we focus on bias alone (see table \ref{Table:summary_table_overall_HADS_bias}),
we obtained a notably smaller bias for the O-biv-FC model ($0.51$) compared to the Jones model ($0.95$) - 
hence, the Jones model makes it into the "worse (practical only)" group here -
as opposed to "best" group when using RMSE. 
However, the strat-biv model still remains in the "worse" group, with statistically significantly
worse bias ($1.45$) as well as practically worse (relative change vs. O-biv-FC model = $185.9\%$). 
% %%
% In addition, the strat-biv model also obtains notably worse bias than both of the "multiple threshold" models, 
% mirroring the pattern seen when using RMSE 
% (bias = $1.45$ for strat-biv vs $0.51 - 0.95$ for the other two models, respectively). 

%%%%
%%%% ------------------------------------------------------------------------------------------------------------
%%
%% ---- 50 studies; RMSE:
%%
For $50$ studies (see right, second quadrant of table \ref{Table:summary_table_overall_HADS}),
we obtain a similar pattern when looking at RMSE; however, the difference between the O-biv-FC and Jones model 
goes from practically zero (when $N_{studies} = 10$) to almost practically significant 
(RMSE: $1.85$ and $2.02$ for the O-biv-FC and Jones models, respectively), 
with a $9.2\%$ ($< 10\%$) relative difference. 
Furthermore - similar to the $10$-study case - 
the strat-biv model is again the only model in the "worse" group,
obtaining an RMSE of $2.33$ (vs. $1.85 - 2.02$ for the other two models), 
and a substantial relative RMSE difference (vs. O-biv-FC model) of $26.0\%$.

%%%%
%%%% ------------------------------------------------------------------------------------------------------------
%%
%% ---- 50 studies; Bias:
%%
If we look at bias alone (ignoring variance/precision), the pattern changes
(see table \ref{Table:summary_table_overall_HADS_bias}).
Most notably, the Jones model goes from being in the "best" group to the "worse" group, 
with relative bias change of $665.9\%$. 
This shows that despite relatively much worse bias (but still good in absolute terms, being $< 1\%$), 
the Jones model performs very well in terms of precision, hence much better RMSE ranking. 
% Regarding bias, the O-biv-FC model obtains exceptional bias ($0.11$), followed by the strat-biv model ($0.33$), 
% and finally the Jones model obtains the worst bias ($0.84$) - however this is still very good ($ < 1\%$). 

%%%%%%%%%%%%%%%%%%%%%%%%%%%%%%%%%%%%%%
\subsubsection{HADS: DGM \#3 (O-biv-RC)}
\label{Sim_study_HADS_O_biv_RC_DGM_3}
%%%%%%%%%%%%%%%%%%%%%%%%%%%%%%%%%%%%%%
%%%%
%%%% ------------------------------------------------------------------------------------------------------------
%%
%% ---- 10 studies; RMSE:
%%
For the O-biv-RC DGM and $N_{studies} = 10$
(see left, third quadrant of table \ref{Table:summary_table_overall_HADS}),
the O-biv-FC, O-biv-RC, and Jones model all get essentially the same RMSE (RMSE range: $4.96 - 5.04$).
However, the O-HSROC-RC model made it into the "worse" group,
because it has a statistically significantly worse RMSE than the leading model in the "Best" group
(RMSE = $5.77$ for the O-HSROC-RC model vs. $4.96 - 5.04$ for the other three "multiple threshold" models),
this difference is not substantial, but still somewhat notable ($\sim 0.75\%$ difference in RMSE).
We can also see that the strat-biv model performs the worst - having an RMSE of $6.43$ -
around $0.65\%$ worse than the O-HSROC-RC model (RMSE = $5.77$).

%%%%
%%%% ------------------------------------------------------------------------------------------------------------
%%
%% ---- 10 studies; Bias:
%%
However, if we focus on bias instead (secondary measure; see table \ref{Table:summary_table_overall_HADS_bias}),
the model hierarchy is now:
O-biv-FC ($0.57$) $>$ O-biv-RC ($0.85$) $>$ Jones ($1.39$) / strat-biv ($1.55$) $>$ O-HSROC-RC ($2.49$).
Notably, we can see that now (unlike when we used RMSE) the O-biv-FC model is the only model which makes 
it into the "best" group - despite the DGM itself being the O-biv-RC model.
More specifically, we can see that the RC variant is now in the "worse (practical only)" group, 
since its bias is notably worse in relative terms 
($0.57$ vs. $0.85$, giving a relative change of $48.0\%$). 
Furthermore, we can see that the Jones model moves from the "best" group (when using RMSE) to the "worse" group, 
due to its poorer bias performance here (bias(Jones) = $1.39$, relative change vs. O-biv-FC = $143.5\%$). 
% that the strat-biv model is actually better
% than the O-HSROC-RC model in terms of bias (bias = $2.41$ and $1.65$ for the O-HSROC-RC and strat-biv models, respectfully).

%%%%
%%%% ------------------------------------------------------------------------------------------------------------
%%
%% ---- 50 studies; RMSE:
%%
For $50$ studies (see right, third quadrant of table \ref{Table:summary_table_overall_HADS}),
both of the O-biv models essentially perform the same in terms of RMSE (RMSE: $2.03 - 2.08$),
and are the only two models that make it into the "best" group, being well within MCSE of one another.
On the other hand, the Jones, strat-biv, as well as the O-HSROC-RC models are all in the "worse" group here
(RMSE = $2.55$, $2.59$ and $3.23$ for the Jones, strat-biv, and the O-HSROC-RC models, respectively),
with the O-HSROC-RC model (RMSE = $3.23$, relative change vs. O-biv-FC = $58.9\%$) 
performing notably worse than both the 
Jones model (RMSE = $2.55$, relative change vs. O-biv-FC = $25.5\%$)
and even the strat-biv model (RMSE = $2.59$, relative change vs. O-biv-FC = $27.4\%$).

%%%%
%%%% ------------------------------------------------------------------------------------------------------------
%%
%% ---- 50 studies; Bias:
%%
Focusing just on the bias (see table \ref{Table:summary_table_overall_HADS_bias}),
the pattern is quite different to when also taking precision into account (i.e., RMSE).
More specifically, we can see that now the O-biv-FC model is the only model making it into the "best" group
(bias = $0.24$), 
with the O-biv-RC model moving from the "best" group (when using RMSE) to the "worse (practical only)" group, 
since despite not reaching statistical significance, it's bias is substantially worse in relative terms 
($0.24$ vs. $0.48$ - relative change of $104.9\%$). 
Furthermore, we can see that the strat-biv model is also in the "worse (practical only)" group, 
as opposed to the "worse" group, since it's bias is not notably worse in absolute terms 
($0.33$ vs. $0.24$ for O-biv-FC), however it is in relative terms ($40.8\%$ worse than O-biv-FC's bias).
% Furthermore, the bias of the O-HSROC-RC model is consistent with the aforementioned RMSE patterns;
% more specifically, we can see that the O-HSROC-RC model obtains a substantial bias of $2.22$,
% whereas the Jones model obtains $1.43$ and the strat-biv obtains just $0.33$.
% Interestingly, we can see that the strat-biv performs extremely well in terms of bias ($0.33$) -
% but not RMSE (which is the primary measure here) - this is likely due to worse coverage and/or precision.
%%
% however, in terms of absolute bias, the difference is more notable,
% with the FC model paradoxically doing better than the RC one, even though the DGM itself is the RC model
% ($0.24$ vs. $0.48$ for the FC and RC O-biv models, respectively).

%%%%%%%%%%%%%%%%%%%%%%%%%%%%%%%%%%%%%%
\subsubsection{HADS: DGM \#4 (O-HSROC-RC)}
\label{Sim_study_HADS_O_HSROC_RC_DGM_4}
%%%%%%%%%%%%%%%%%%%%%%%%%%%%%%%%%%%%%%
%%%%
%%%% ------------------------------------------------------------------------------------------------------------
%%
%% ---- 10 studies; RMSE:
%%
Finally, for the O-HSROC-RC DGM (bottom-left quadrant of table \ref{Table:summary_table_overall_HADS}),
for $N_{studies} = 10$ we can see that
all four "multiple threshold" models make it into the "best" group (RMSE range: $3.54 - 3.80$).
On the other hand, the strat-biv model is in the "worse" group, obtaining a substantially worse 
RMSE of $4.70$.

%%%%
%%%% ------------------------------------------------------------------------------------------------------------
%%
%% ---- 10 studies; Bias:
%%
When looking at the bias in isolation (see table \ref{Table:summary_table_overall_HADS_bias}),
the pattern is now very different, with the O-HSROC-RC model being the only model
making it into the "best" group (bias = $0.43$), 
and all four of the other models making it into the worse (practical only) group, 
having bias ranges between $0.57 - 0.72$ and relative magnitude bias changes (vs. O-HSROC-RC)
between $32.3\% - 68.0\%$.
However, it's important to note that despite being much worse in relative terms, the bias for all five models here
is still below $1\%$ (i.e., very good). 
%%
%% ---- MAYBE for discussion:
%%
This discrepancy between model rankings when usuing RMSE as opposed to bias shows that when the variance/precision
is also taken into account (as well as the bias - since RMSE = bias$^2$ + variance), 
all "multiple threshold" models do in fact perform very similarly, 
with the strat-biv model clearly being worse due to less precise estimates.

%%%%
%%%% ------------------------------------------------------------------------------------------------------------
%%
%% ---- 50 studies; RMSE:
%%
When we move up to $50$ studies (bottom-right quadrant of table \ref{Table:summary_table_overall_HADS}), 
the pattern is very similar as it was for $N_{studies} = 10$, 
with all four "multiple threshold" models again making it into the "best" group (RMSE range: $1.52 - 1.66$).
Additionally, the strat-biv model again is the only model in the "worse" group (RMSE = $1.99$), 
with a relative change in RMSE (vs. the leading O-HSROC-RC model) of $30.7\%$ ($> 10\%$).

%%%%
%%%% ------------------------------------------------------------------------------------------------------------
%%
%% ---- 50 studies; RMSE:
%%
When we look at bias alone instead of RMSE (see table \ref{Table:summary_table_overall_HADS_bias}),
we can see that the pattern changes.
More specifically, it is now the strat-biv and O-HSROC-RC models which are in the "best" group
(bias = $0.36 - 0.39$, with strat-biv leading with $0.36$), 
and the other three models all being in the "worse (practical only)" group, 
with bias between $0.43 - 0.61$ and relative change in bias (vs. strat-biv model) between $18.9\% - 69.4\%$.
%%
% Furthermore, the bias (\textcolor{red}{see appendix ......, table ......}) 
% is also very similar between the four "multiple threshold" models (bias range: $0.39 - 0.61$). 
% However, unlike the RMSE, the bias of the strat-biv model is actually better than all four 
% "multiple threshold" models, albeit not to any significant or notable degree 
% ($0.36$ vs. $0.39 - 0.61$ for the strat-biv and the four "multiple threshold" models, respectfully). 

%%%%%%%%%%%%%%%%%%%%%%%%%%%%%%%%%%%%%%%%%%%%%%%%%%%%%%%%%%%%%%%%%%%%%%%%%%%%%%%%%%%%%%%%%%%%%%%%%%%%%%%%%%%%%%%%%%%%
\subsection{Results: Test III (BAI)}
\label{Sim_study_BAI}
%%%%%%%%%%%%%%%%%%%%%%%%%%%%%%%%%%%%%%%%%%%%%%%%%%%%%%%%%%%%%%%%%%%%%%%%%%%%%%%%%%%%%%%%%%%%%%%%%%%%%%%%%%%%%%%%%%%%
%%%%
%%%%
\begin{table}[H]
\centering
\caption{
Overall model performance for all four DGMs - BAI
(4-group classification based on statistical and practical significance)
}
\footnotesize
\setlength{\tabcolsep}{4pt}
\begin{tabular}{c|l|l}
\toprule
\textbf{DGM} & $\mathbf{N_{studies} = 10}$ & $\mathbf{N_{studies} = 50}$ \\
\midrule
%%
%% ---- DGM = Jones:
%%
Jones & \makecell[l]{\underline{Group 1 - Best:} \\
                Jones [RMSE: $6.14$, Bias: $0.60$]. \\
                \\
                \underline{Group 2 - Worse (stat. signif. only):} \\
                O-biv-FC [RMSE: $6.59$/$7.4\%$, Bias: $1.45$]. \\
                \\
                \underline{Group 3 - Worse (practical only):} \\
                None. \\
                \\
                \underline{Group 4 - Worse:} \\
                Strat-biv [RMSE: $8.42$/$37.2\%$, Bias: $1.64$].}
  & \makecell[l]{\underline{Group 1 - Best:} \\
                Jones [RMSE: $2.64$, Bias: $0.42$], \\
                O-biv-FC [RMSE: $2.87$/$8.8\%$, Bias: $0.90$]. \\
                \\
                \underline{Group 2 - Worse (stat. signif. only):} \\
                None. \\
                \\
                \underline{Group 3 - Worse (practical only):} \\
                None. \\
                \\
                \underline{Group 4 - Worse:} \\
                Strat-biv [RMSE: $4.03$/$52.6\%$, Bias: $0.99$].} \\
\midrule
%%
%% ---- DGM = O-biv-FC:
%%
O-biv-FC & \makecell[l]{\underline{Group 1 - Best:} \\
               O-biv-FC [RMSE: $5.75$, Bias: $1.17$], \\
               Jones [RMSE: $5.81$/$1.0\%$, Bias: $1.66$]. \\
               \\
               \underline{Group 2 - Worse (stat. signif. only):} \\
               None. \\
               \\
               \underline{Group 3 - Worse (practical only):} \\
               None. \\
               \\
               \underline{Group 4 - Worse:} \\
               Strat-biv [RMSE: $7.70$/$33.9\%$, Bias: $1.53$].}
  & \makecell[l]{\underline{Group 1 - Best:} \\
                O-biv-FC [RMSE: $2.32$, Bias: $0.27$]. \\
                \\
                \underline{Group 2 - Worse (stat. signif. only):} \\
                None. \\
                \\
                \underline{Group 3 - Worse (practical only):} \\
                None. \\
                \\
                \underline{Group 4 - Worse:} \\ 
                Jones [RMSE: $3.01$/$29.6\%$, Bias: $1.53$], \\
                Strat-biv [RMSE: $3.15$/$35.6\%$, Bias: $0.41$].} \\
\midrule
%%
%% ---- DGM = O-biv-RC:
%%
O-biv-RC & \makecell[l]{\underline{Group 1 - Best:} \\
               O-biv-RC [RMSE: $5.46$, Bias: $1.84$], \\
               O-biv-FC [RMSE: $5.72$/$4.8\%$, Bias: $1.70$], \\
               Jones [RMSE: $5.81$/$6.5\%$, Bias: $2.27$]. \\
               \\
               \underline{Group 2 - Worse (stat. signif. only):} \\
               O-HSROC-RC [RMSE: $5.90$/$8.2\%$, Bias: $2.04$]. \\
               \\
               \underline{Group 3 - Worse (practical only):} \\
               None. \\
               \\
               \underline{Group 4 - Worse:} \\
               Strat-biv [RMSE: $7.25$/$33.0\%$, Bias: $1.82$].}
  & \makecell[l]{\underline{Group 1 - Best:} \\
                O-biv-RC [RMSE: $3.32$, Bias: $2.38$], \\
                O-biv-FC [RMSE: $3.36$/$1.0\%$, Bias: $2.42$], \\
                O-HSROC-RC [RMSE: $3.36$/$1.0\%$, Bias: $2.07$]. \\
                \\
                \underline{Group 2 - Worse (stat. signif. only):} \\
                None. \\
                \\
                \underline{Group 3 - Worse (practical only):} \\
                None. \\
                \\
                \underline{Group 4 - Worse:} \\
                Jones [RMSE: $3.67$/$10.4\%$, Bias: $2.79$], \\
                Strat-biv [RMSE: $3.95$/$18.8\%$, Bias: $2.25$].} \\
\midrule
%%
%% ---- DGM = O-HSROC-RC:
%%
O-HSROC-RC & \makecell[l]{\underline{Group 1 - Best:} \\
               O-HSROC-RC [RMSE: $6.06$, Bias: $0.63$], \\
               O-biv-RC [RMSE: $6.31$/$4.1\%$, Bias: $1.12$], \\
               O-biv-FC [RMSE: $6.33$/$4.4\%$, Bias: $1.41$], \\
               Jones [RMSE: $6.47$/$6.7\%$, Bias: $1.72$]. \\
               \\
               \underline{Group 2 - Worse (stat. signif. only):} \\  
               None. \\
               \\
               \underline{Group 3 - Worse (practical only):} \\
               None. \\
               \\
               \underline{Group 4 - Worse:} \\
               Strat-biv [RMSE: $8.37$/$38.1\%$, Bias: $1.81$].}
  & \makecell[l]{\underline{Group 1 - Best:} \\ 
                O-HSROC-RC [RMSE: $2.73$, Bias: $0.90$], \\
                O-biv-FC [RMSE: $2.94$/$7.5\%$, Bias: $1.22$]. \\
                \\
                \underline{Group 2 - Worse (stat. signif. only):} \\
                None. \\
                \\
                \underline{Group 3 - Worse (practical only):} \\
                O-biv-RC [RMSE: $3.05$/$11.6\%$, Bias: $1.07$]. \\
                \\
                \underline{Group 4 - Worse:} \\ 
                Jones [RMSE: $3.41$/$24.9\%$, Bias: $1.73$], \\
                Strat-biv [RMSE: $3.92$/$43.5\%$, Bias: $1.16$].} \\ 
\bottomrule
\end{tabular}
\label{Table:summary_table_overall_BAI}
\end{table}
%%%%
%%%%
%%%%
%%%%
\begin{table}[H]
\centering
\caption{
Overall model performance for all four DGMs - BAI
(4-group classification ordered by BIAS performance)
}
\footnotesize
\setlength{\tabcolsep}{4pt}
\begin{tabular}{c|l|l}
\toprule
\textbf{DGM} & $\mathbf{N_{studies} = 10}$ & $\mathbf{N_{studies} = 50}$ \\
\midrule
%%
%% ---- DGM = Jones (Bias ordering):
%%
Jones & \makecell[l]{\underline{Group 1 - Best (Bias):} \\
                Jones [Bias: $0.60$, RMSE: $6.14$]. \\
                \\
                \underline{Group 2 - Worse (stat. signif only):} \\
                None. \\
                \\
                \underline{Group 3 - Worse (practical only):} \\
                None. \\
                \\
                \underline{Group 4 - Worse:} \\
                O-biv-FC [Bias: $1.45$/$141.2\%$, RMSE: $6.59$], \\
                Strat-biv [Bias: $1.64$/$173.2\%$, RMSE: $8.42$].}
  & \makecell[l]{\underline{Group 1 - Best (Bias):} \\
                Jones [Bias: $0.42$, RMSE: $2.64$]. \\
                \\
                \underline{Group 2 - Worse (stat. signif only):} \\
                None. \\
                \\
                \underline{Group 3 - Worse (practical only):} \\
                O-biv-FC [Bias: $0.90$/$115.7\%$, RMSE: $2.87$]. \\
                \\
                \underline{Group 4 - Worse:} \\
                Strat-biv [Bias: $0.99$/$137.3\%$, RMSE: $4.03$].} \\
\midrule
%%
%% ---- DGM = O-biv-FC (Bias ordering):
%%
O-biv-FC & \makecell[l]{\underline{Group 1 - Best (Bias):} \\
               O-biv-FC [Bias: $1.17$, RMSE: $5.75$]. \\
               \\
               \underline{Group 2 - Worse (stat. signif only):} \\
               None. \\
               \\
               \underline{Group 3 - Worse (practical only):} \\
               Strat-biv [Bias: $1.53$/$30.8\%$, RMSE: $7.70$]. \\
               \\
               \underline{Group 4 - Worse:} \\
               Jones [Bias: $1.66$/$41.6\%$, RMSE: $5.81$].}
  & \makecell[l]{\underline{Group 1 - Best (Bias):} \\
                O-biv-FC [Bias: $0.27$, RMSE: $2.32$]. \\
                \\
                \underline{Group 2 - Worse (stat. signif only):} \\
                None. \\
                \\
                \underline{Group 3 - Worse (practical only):} \\
                Strat-biv [Bias: $0.41$/$51.8\%$, RMSE: $3.15$]. \\
                \\
                \underline{Group 4 - Worse:} \\ 
                Jones [Bias: $1.53$/$472.1\%$, RMSE: $3.01$].} \\
\midrule
%%
%% ---- DGM = O-biv-RC (Bias ordering):
%%
O-biv-RC & \makecell[l]{\underline{Group 1 - Best (Bias):} \\
               O-biv-FC [Bias: $1.70$, RMSE: $5.72$], \\
               Strat-biv [Bias: $1.82$, RMSE: $7.25$], \\
               O-biv-RC [Bias: $1.84$, RMSE: $5.46$]. \\
               \\
               \underline{Group 2 - Worse (stat. signif only):} \\
               None. \\
               \\
               \underline{Group 3 - Worse (practical only):} \\
               O-HSROC-RC [Bias: $2.04$/$20.0\%$, RMSE: $5.90$]. \\
               \\
               \underline{Group 4 - Worse:} \\
               Jones [Bias: $2.27$/$33.5\%$, RMSE: $5.81$].}
  & \makecell[l]{\underline{Group 1 - Best (Bias):} \\
                O-HSROC-RC [Bias: $2.07$, RMSE: $3.36$], \\
                Strat-biv [Bias: $2.25$, RMSE: $3.95$]. \\
                \\
                \underline{Group 2 - Worse (stat. signif only):} \\
                None. \\
                \\
                \underline{Group 3 - Worse (practical only):} \\
                None. \\
                \\
                \underline{Group 4 - Worse:} \\
                O-biv-RC [Bias: $2.38$/$15.1\%$, RMSE: $3.32$], \\
                O-biv-FC [Bias: $2.42$/$16.9\%$, RMSE: $3.36$], \\
                Jones [Bias: $2.79$/$34.9\%$, RMSE: $3.67$].} \\
\midrule
%%
%% ---- DGM = O-HSROC-RC (Bias ordering):
%%
O-HSROC-RC & \makecell[l]{\underline{Group 1 - Best (Bias):} \\
               O-HSROC-RC [Bias: $0.63$, RMSE: $6.06$]. \\
               \\
               \underline{Group 2 - Worse (stat. signif only):} \\
               None. \\
               \\
               \underline{Group 3 - Worse (practical only):} \\
               O-biv-RC [Bias: $1.12$/$78.9\%$, RMSE: $6.31$]. \\
               \\
               \underline{Group 4 - Worse:} \\
               O-biv-FC [Bias: $1.41$/$124.9\%$, RMSE: $6.33$], \\
               Jones [Bias: $1.72$/$174.2\%$, RMSE: $6.47$], \\
               Strat-biv [Bias: $1.81$/$188.0\%$, RMSE: $8.37$].}
  & \makecell[l]{\underline{Group 1 - Best (Bias):} \\
                O-HSROC-RC [Bias: $0.90$, RMSE: $2.73$]. \\
                \\
                \underline{Group 2 - Worse (stat. signif only):} \\
                None. \\
                \\
                \underline{Group 3 - Worse (practical only):} \\
                O-biv-RC [Bias: $1.07$/$18.9\%$, RMSE: $3.05$], \\
                Strat-biv [Bias: $1.16$/$28.2\%$, RMSE: $3.92$], \\
                O-biv-FC [Bias: $1.22$/$35.7\%$, RMSE: $2.94$]. \\
                \\
                \underline{Group 4 - Worse:} \\
                Jones [Bias: $1.73$/$91.8\%$, RMSE: $3.41$].} \\
\bottomrule
\end{tabular}
\label{Table:summary_table_overall_BAI_bias}
\end{table}
%%%%
%%%% -------------------------------------------------------------------------------------------------------------------
The results for the BAI test are shown in table \ref{Table:summary_table_overall_BAI}.
This test has 63 categories, and approximately $\sim 55\%$ missing data.
Furthermore, in the appendix we have additional figures; namely:
A figure showing the RMSE results
(for all four DGMs;
see figure  \ref{Sim_study_RMSE_BAI},
in appendix \ref{appendix_B_RMSE_plots},
section \ref{appendix_B_RMSE_plots_BAI_test_III}),
an analogous figure but for the bias
(see figure \ref{Sim_study_Bias_BAI},
in appendix \ref{appendix_C_Bias_plots},
section \ref{appendix_C_Bias_plots_BAI_test_III}),
and finally a figure showing the coverage for all four DGMs for the GAD-2 data
(see figure \ref{Sim_study_Coverage_BAI},
in appendix \ref{appendix_D_Coverage_plots},
section \ref{appendix_D_Coverage_plots_BAI_test_III}).
Additionally, appendix \ref{appendix_E_detailed_results_tables} shows detailed results -
which includes the RMSE(Se), RMSE(Sp), Bias(Se), Bias(Sp), Coverage(Se), Coverage(Sp), 
and also the interval widths of Se and Sp - 
for the BAI, these are specifically in 
appendix section \ref{appendix_E_detailed_results_tables_BAI_test_III}.
%%%%
%%%%
%%%%%%%%%%%%%%%%%%%%%%%%%%%%%%%%%%%%%%
\subsubsection{BAI: DGM \#1 (Jones)}
\label{Sim_study_BAI_Jones_DGM_1}
%%%%%%%%%%%%%%%%%%%%%%%%%%%%%%%%%%%%%%
%%%%
%%%% -------------------------------------------------------------------------------------------------------------------
%%
%% ---- 10 studies:
%%
More specifically, the results for the Jones DGM for $N_{studies} = 10$ 
are shown on the top-left quadrant of table \ref{Table:summary_table_overall_BAI}.
We can see that the Jones model is the only model in the "best" group here (RMSE = $6.14$).
However, the O-biv-FC model is in the "worse (stat only)" group,
since it is statistically significantly worse than the Jones model, 
but not practically worse ($7.4\% < 10\%$ relative diff vs. Jones model),
On the other hand, we can see that the strat-biv model performs much worse, and therefore makes it into the "worse" group
(RMSE = $8.42$ $<<$ $6.14 - 6.59$, relative diff vs. Jones model = $37.2\%$).

%%%%
%%%% -------------------------------------------------------------------------------------------------------------------
%%
%% ---- 50 studies:
%%
If we instead focus on bias (see table \ref{Table:summary_table_overall_BAI_bias}), 
we can see that this time the O-biv-FC model is in the "worse" group (not stat only),
obtaining much worse bias ($1.45$) than the Jones model (relative change in bias vs. Jones = $141.2\%$),
and the strat-biv model obtaining a bias of $1.64$ (relative change vs. Jones model = $173.2\%$).

%%%%
%%%% -------------------------------------------------------------------------------------------------------------------
%%
%% ---- 50 studies:
%%
For when we have $50$ studies (see right, second quadrant of table \ref{Table:summary_table_overall_BAI}),
we can see that both the Jones and O-biv-FC model are tied (easily within MCSE of one another),
obtaining an RMSE range of $2.64 - 2.87$,
with the O-biv-FC being only $8.8\%$ ($< 10\%$) worse than the Jones model.
However, the strat-biv model does much worse, obtaining an RMSE of $4.03$ -
and therefore being $52.6\%$ worse than the leading Jones model -
hence being the only model in the "worse" group.

%%%%
%%%% -------------------------------------------------------------------------------------------------------------------
%%
%% ---- 50 studies:
%%
Furthermore, both "multiple threshold" models obtain very good bias
(under $1\%$; see table \ref{Table:summary_table_overall_BAI_bias});
namely, $0.42$ and $0.90$ for the Jones and O-biv-FC models, respectively.
Also - unlike the RMSE - the bias for the strat-biv model is also very good
(despite being worse than both of the other models in terms of RMSE),
being just under $1\%$ (bias = $0.99$).
However - the strat-biv model still ends up in the "worse" group when ranking by bias -
because in relative terms it's much worse compared to the leading Jones model
($137.3\%$ worse, with abs. bias diff of $0.57$). %% 0.99 - 0.42 = 0.57
% %%
% %% ---- bookmark: MAYBE move this to discussion:
% %%
% This discrepancy between the RMSE and bias for the O-biv-FC model
% (i.e., going from "best" group for RMSE and "worse (practical only)" when ranking by bias)
% is because 

%%%%%%%%%%%%%%%%%%%%%%%%%%%%%%%%%%%%%%
\subsubsection{BAI: DGM \#2 (O-biv-FC)}
\label{Sim_study_BAI_O_biv_FC_DGM_2}
%%%%%%%%%%%%%%%%%%%%%%%%%%%%%%%%%%%%%%
%%%%
%%%% -----------------------------------------------------------------------------------------------------------------------
%%
%% ---- 10 studies:
%%
Looking at the O-biv-FC DGM (second row, left side of table \ref{Table:summary_table_overall_BAI}),
we get effectively the same RMSE for both of the "multiple threshold" models (RMSE range: $5.75 - 5.81$). 
However, the strat-biv model - 
similarly to the previous DGM (i.e. the Jones DGM discussed above) - 
ends up being the lone model in the "worse" group, obtaining a substantially worse RMSE 
($7.70$ vs. $5.75 - 5.81$ - i.e., being $33.9\%$ worse than the leading O-biv-FC model here).

%%%%
%%%% ------------------------------------------------------------------------------------------------------------------------
%%
%% ---- 50 studies:
%%
Regarding bias (see table \ref{Table:summary_table_overall_BAI_bias}),
we can see that there is a somewhat notable difference between the two "multiple threshold" models
($1.17$ vs. $1.53$ for the O-biv and Jones models, respectively) - 
with the strat-biv model being $30.8\%$ worse than the O-biv-FC model in terms of bias, 
and the Jones model now being in the "worse" group (unlike when using RMSE),
with it's bias ($1.66$) being $41.6\%$ worse than the O-biv-FC model.
% %%
% However, very similarly to the Jones DGM we discussed in the previous paragraph, the bias for the strat-biv 
% model is actually quite good - being in between the O-biv and Jones models
% (bias = $1.57$, $1.16$, $1.62$ for the strat-biv, O-biv-FC, and Jones models, respectively).

%%%%
%%%% -------------------------------------------------------------------------------------------------------------------
%%
%% ---- 50 studies:
%%
When we move on to looking at the $50$ study case
(second row, right-hand side of table \ref{Table:summary_table_overall_BAI}),
we can see that the pattern changes quite notably.
Namely, the O-biv-FC model is now the only model in the "best" group (RMSE = $2.32$).
However, the Jones model is now statistically significantly - as well as practically -
worse than the leading O-biv-FC model,
with an RMSE of $3.01$ (i.e. around $29.6\%$ worse than the O-biv-FC model).
Furthermore, the strat-biv model is also in the "worse" group,
obtaining the worst RMSE out of all three models for this DGM
(RMSE = $3.15$, being $35.6\%$ worse than the O-biv-FC model).

%%%%
%%%% -------------------------------------------------------------------------------------------------------------------
%%
%% ---- 50 studies:
%%
Additionally, when we look at the bias (see table \ref{Table:summary_table_overall_BAI_bias}),
we can see that the O-biv-FC obtains a bias of just $0.27\%$,
whereas the Jones model obtains a notably higher bias (but still good in absolute terms) of $1.53$,
being $472.1\%$ worse than the O-biv-FC model in relative terms.
Furthermore, the strat-biv model moves from the "worse" group (when using RMSE)
to the "worse (practical only)" group.
This is because its bias difference of just $0.14$
(bias was $0.27$ and $0.41$ for O-biv-FC and strat-biv models, respectively)
is not statistically significantly worse than the leading O-biv-FC model;
however, in relative terms, its bias is $51.8\%$ worse than the O-biv-FC model. %% 0.41 - 0.27 = 0.14

%%%%%%%%%%%%%%%%%%%%%%%%%%%%%%%%%%%%%%
\subsubsection{BAI: DGM \#3 (O-biv-RC)}
\label{Sim_study_BAI_O_biv_RC_DGM_3}
%%%%%%%%%%%%%%%%%%%%%%%%%%%%%%%%%%%%%%
%%%%
%%%% -------------------------------------------------------------------------------------------------------------------
%%
%% ---- 10 studies:
%%
Moving on to the third DGM
(O-biv-RC; see third row, left side of table \ref{Table:summary_table_overall_BAI}),
we can see that all four of the "multiple threshold" models - except for the O-HSROC-RC model -
make it into the "best" group.
More specifically, we can see that all three of these models are within MCSE of the O-biv-RC model
(the leading model; RMSE range for these three models: $5.46 - 5.81$, relative diff = $4.8\% - 6.5\%$).
The O-HSROC-RC model is in the "worse (stat only)" group, since it is statistically significantly
worse than the leading O-biv-RC model; however, this difference does not reach practical significance
(relative diff = $8.2\% < 10\%$).
Furthermore, the strat-biv model performs much worse than the other four models,
obtaining an RMSE of $7.25$ (vs. $5.46 - 5.90$ for the other four models).
Additionally, it performs $33.0\%$ worse than the leading O-biv-RC model.

%%%%
%%%% -------------------------------------------------------------------------------------------------------------------
%%
%% ---- 10 studies:
%%
When we instead rank models by bias 
(secondary measure; \ref{Table:summary_table_overall_BAI_bias}),
it is now the Jones model which is the lone model in the "worse" group, 
obtaining $33.5\%$ worse bias than the leading O-biv-FC (not RC as when using RMSE) model.

%%%%
%%%% -------------------------------------------------------------------------------------------------------------------
%%
%% ---- 50 studies:
%%
Looking at the $50$-study case 
(third row, right hand side of table \ref{Table:summary_table_overall_BAI}),
we again have three models in the "best" group; however, it is now: 
O-biv-RC (RMSE = $3.32$), O-biv-FC (RMSE = $3.36$), and finally O-HSROC-RC in third place
(not Jones like when we had $10$ studies)
with an RMSE of $3.36$. 
On the other hand, the strat-biv model is worse than the leading O-biv-RC model 
(RMSE is $18.8\%$ worse for the strat-biv model vs. the leading O-biv-RC model), 
and the Jones model now also moves from the "best" group (10-study case) to the "worse" group, 
since it is both statistically significantly and practically worse than the leading O-biv-RC model
(RMSE diff = $0.35$, rel. diff = $10.4\% > 10\%$).  %% 3.32 - 3.67

%%%%
%%%% ------------------------------------------------------------------------------------------------------------------
%%
%% ---- 10 studies:
%%
Regarding the bias results
(see table 
\ref{Table:summary_table_overall_BAI_bias})
we can see that it is now the O-HSROC-RC model in the lead (bias = $2.07$), followed by
the strat-biv model (bias = $2.25$) -
these are now the only two models in the "best" group.
On the other hand, the three remaining models are all in the "worse" group, being both
statistically significantly as well as practically worse than the leading O-HSROC-RC model.
More specifically, we obtained relative bias differences of: $15.1\%$, $16.9\%$ and $34.9\%$
for the O-biv-RC, O-biv-FC, and the Jones model, respectively.
This discrepancy between the bias results and the RMSE is because the RMSE accounts for both the bias itself,
as well as variance/precision.

%%%%%%%%%%%%%%%%%%%%%%%%%%%%%%%%%%%%%%
\subsubsection{BAI: DGM \#4 (O-HSROC-RC)}
\label{Sim_study_BAI_O_biv_RC_DGM_4}
%%%%%%%%%%%%%%%%%%%%%%%%%%%%%%%%%%%%%%
%%%%
%%%% -------------------------------------------------------------------------------------------------------------------
%%
%% ---- 10 studies:
%%
Finally, for the last DGM (O-HSROC-RC; see last row, left of table \ref{Table:summary_table_overall_BAI}),
for the $10$-study case, we can see that the leading model is the O-HSROC-RC itself 
(RMSE = $6.06$), followed by the following three models (all in the "best" group):
O-biv_RC (RMSE = $6.31$), O-biv-FC (RMSE = $6.33$), Jones (RMSE = $6.47$).
Furthermore, the only model which makes it into the "worse" group is the strat-biv model, 
which obtains an RMSE ($8.37$) which is $38.1\%$ worse than the leading O-HSROC-RC model ($6.06$).

%%%%
%%%% -------------------------------------------------------------------------------------------------------------------
%%
%% ---- 10 studies:
%%
If we instead just focus on bias - our secondary measure 
(see table  \ref{Table:summary_table_overall_BAI_bias}) - 
We can see that now the only model which makes it into the "best" group is the O-HSROC-RC model (bias = 0.63), 
with the O-biv-RC model being practically - but not statistically significantly - 
worse ($78.9\%$ worse than the O-HSROC-RC model, obtaining a bias of $1.12$ - which is still good in absolute terms).
Additionally, the following three models now are all in the "worse" group:
O-biv-FC (bias = $1.41$),
Jones (bias = $1.72$), and
strat-biv (bias = $1.81$) - 
being $124.9\%$, $174.2\%$,
and $188.0\%$ worse than the O-HSROC-RC model, 
respectively.

%%%%
%%%% -------------------------------------------------------------------------------------------------------------------
%%
%% ---- 50 studies:
%%
When we move up to $N_{studies} = 50$ (see last row, right of table \ref{Table:summary_table_overall_BAI}),
the pattern changes somewhat - now we only have two models (not four) in the "best" group -
the O-HSROC-RC again in the lead (RMSE = $2.73$), followed by the O-biv-FC model
(RMSE = $2.94$), 
which is $7.5\%$ ($< 10\%$) worse than the O-HSROC-RC model).
Then we have the O-biv-RC model being practically worse than the leading O-HSROC-RC model
($11.6\% > 10\%$) - but not making statistical significance.
Furthermore, both the Jones and strat-biv models are the only two models in the "worse" group, 
being $24.9\%$ (RMSE = $3.41$) and $43.5\%$ (RMSE = $3.92$) worse than the leading O-HSROC-RC model, respectively.

%%%%
%%%% -------------------------------------------------------------------------------------------------------------------
%%
%% ---- 10 studies:
%%
When we focus on bias instead (see table \ref{Table:summary_table_overall_BAI}),
we again (like the $10$-study case) only have one model in the "best" group - the O-HSROC-RC (bias = $0.90$), 
and the following three models in the "worse (practical only)" group:
O-biv-RC  (bias = $1.07$, rel. diff vs O-HSROC-RC = $18.9\%$), 
strat-biv (bias = $1.16$, rel. diff vs O-HSROC-RC = $28.2\%$), 
O-biv-FC  (bias = $1.22$, rel. diff vs O-HSROC-RC = $35.7\%$).
Finally, the only model in the "worse" group is the Jones model, with a bias of $1.73$
($91.8\%$ worse than the leading O-HSROC-RC model).

\subsection{Coverage and interval widths (all tests)}
\label{Sim_study_coverage_and_interval_widths}
%%%%%%%%%%%%%%%%%%%%%%%%%%%%%%%%%%%%%%%%%%%%%%%%%%%%%%%%%%%%%%%%%%%%%%%%%%%%%%
%%%%
%%%% ---------------------------------------------------------------------------------------------------------------
The results for coverage and interval widths are shown in
appendix \ref{appendix_E_detailed_results_tables};
more specifically, in:
\ref{appendix_E_detailed_results_tables_GAD_2_test_I},
\ref{appendix_E_detailed_results_tables_HADS_test_II}, and
\ref{appendix_E_detailed_results_tables_BAI_test_III},
for the GAD-2, HADS and the BAI, respectively.

%%%%
%%%% ---------------------------------------------------------------------------------------------------------------
Coverage probability was generally acceptable ($> 90\%$) for correctly specified scenarios
with low threshold counts (e.g., for GAD-2: $91\% - 97\%$).
However, under-coverage emerged under two conditions:
(1) continuous-assumption models applied to ordinal data
(e.g., Jones model on HADS with 50 studies: $67.3\%$ for Se),
and
(2) high threshold counts with substantial missingness
(e.g., BAI O-Biv-RC with 50 studies: $74.6\%$ for Se).

%%%%
%%%% ---------------------------------------------------------------------------------------------------------------
Notably, under-coverage sometimes occurred despite low RMSE,
indicating accurate point estimates paired with poorly calibrated uncertainty intervals.

%%%%%%%%%%%%%%%%%%%%%%%%%%%%%%%%%%%%%%%%%%%%%%%%%%%%%%%%%%%%%%%%%%%%%%%%%%%%%%%%%%%%%%%%%%%%%%%%%%%%%%%%%%%%%%%%%%%%
\newpage
\section{MetaOrdDTA R package: A network meta-analysis (NMA) example}
\label{MetaOrdDTA}
%%%%%%%%%%%%%%%%%%%%%%%%%%%%%%%%%%%%%%%%%%%%%%%%%%%%%%%%%%%%%%%%%%%%%%%%%%%%%%%%%%%%%%%%%%%%%%%%%%%%%%%%%%%%%%%%%%%%
%%%%
%%%% --------------------------------------------------------------------------------------------------------
In this section, we demonstrate the MetaOrdDTA R package, which we created as part of this project. 
This R package is available to download on GitHub here:
\url{https://github.com/CerulloE1996/MetaOrdDTA/}
Furthermore, all of the example code and data is available in the examples folder 
(link: \url{https://github.com/CerulloE1996/MetaOrdDTA/tree/main/inst/examples}).
The files most relevant for the examples in this section are the R files whose file names start with 
"NMA\_example";
for instance, for the sROC plots (section \ref{MetaOrdDTA_results_baseline_analysis_sROC_plots}),
it is the "NMA_example_plots_sROC.R" file which contains the full R code.
%%%%
%%%%
%%%%%%%%%%%%%%%%%%%%%%%%%%%%%%%%%%%%%%%%%%%%%%%%%%%%%%%%%%%%%%%%%%%%%%%%%%%%%%%%%%%%%%%%%%%%%%%%%%%%%%%%%%%%%%%%%%%%
\subsection{The dataset}
\label{MetaOrdDTA_dataset}
%%%%%%%%%%%%%%%%%%%%%%%%%%%%%%%%%%%%%%%%%%%%%%%%%%%%%%%%%%%%%%%%%%%%%%%%%%%%%%%%%%%%%%%%%%%%%%%%%%%%%%%%%%%%%%%%%%%%
%%%%
%%%% ---------------------------------------------------------------------------------------------------------------
The results presented in this section use simulated data, generated for methodological demonstration only.
Note that these estimates should NOT be interpreted as clinical findings or used for clinical decision-making.
The true diagnostic accuracy of these instruments is reported in Linde et al
(Linde et al, 2025\supercite{protocol_Linde_et_al_2025}).
The simulated data was intentionally generated from a model that is not the best-fitting model for the real data -
to help further preserve confidentiality,
whilst still being able to demonstrate the features of the MetaOrdDTA R package.

\subsection{Baseline analysis (no covariates)}
\label{MetaOrdDTA_results_baseline_analysis}
%%%%%%%%%%%%%%%%%%%%%%%%%%%%%%%%%%%%%%%%%%%%%%%%%%%%%%%%%%%%%%%%%%%%%%%%%%%%%%%%%%%%%%%%%%%%%%%%%%%%%%%%%%%%%%%%%%%%
%%%%
%%%%
%%%%%%%%%%%%%%%%%%%%%%%%%%%%%%%%%%%%%%%%%%%%%%%%%%%%%%%%%%%%%%%%%%%%%%%%%%%%%%%%
\subsubsection{ Data}
\label{MetaOrdDTA_results_data_format}
%%%%%%%%%%%%%%%%%%%%%%%%%%%%%%%%%%%%%%%%%%%%%%%%%%%%%%%%%%%%%%%%%%%%%%%%%%%%%%%%
%%%%
%%%% -------------------------------------------------------------------------------------------------------------------
For NMA analyses, MetaOrdDTA requires data structured as a nested list (\verb|x|), with the following specifications:
\begin{itemize}
\item \textbf{Outer list}: One element per screening/diagnostic test (4 elements in our GAD example: 
GAD-2, GAD-7, HADS, BAI)
\item \textbf{Inner list}: Two matrices per test:
  \begin{itemize}
  \item First matrix: counts for non-diseased group.
  \item Second matrix: counts for diseased group.
  \end{itemize}
\item \textbf{Matrix dimensions}: 
  \begin{itemize}
  \item Number of rows: this must be equal to the total number of studies ($80$ in our example).
  \item Number of columns: this must be equal to the number of thresholds plus one - 
  (which can and often will vary by test - e.g., for our example we had: 7 for GAD-2, 22 for GAD-7 and HADS, 64 for BAI).
  \end{itemize}
\item \textbf{Missing data}: indicated by -1 values.
\end{itemize}
%%%%
%%%% -------------------------------------------------------------------------------------------------------------------
For the first column of the inner matrices, 
each row must represent the total number of non-diseased individuals in that study (for the first inner matrix), 
or the total number of diseased individuals in that study (for the second inner matrix).
%%%
For the rest of the columns (i.e., column \#2 onwards), 
the counts in each cell of the inner matrices represent the cumulative number of individuals scoring at or above each threshold
(i.e., increasing the column count represents increased disease severity required to test positive).
Hence, as we go to the right (i.e. increasing column count), the cell counts in the matrices must be monotonically decreasing.
This cumulative structure is essential for maintaining the ordinal nature of the screening/diagnostic test data.

%%%%
%%%% -------------------------------------------------------------------------------------------------------------------
Additionally, for NMA analyses, 
the \verb|indicator_index_test_in_study| argument maps which tests appear in each study. 
This is a matrix where rows represent studies and columns represent tests, 
with binary indicators (1 = test present in study, 0 = test absent in study).

%%%%
%%%% -------------------------------------------------------------------------------------------------------------------
For standard (non-NMA) meta-analyses,
the \verb|indicator_index_test_in_study| can be omitted, 
and the data structure (\verb|x|) simplifies to a single list of matrices -
rather than a nested "list of list of matrices" -
containing the two matrices for the single index test being evaluated:
the first matrix for the non-diseased group, and the second matrix for the diseased group.

%%%%%%%%%%%%%%%%%%%%%%%%%%%%%%%%%%%%%%%%%%%%%%%%%%%%%%%%%%%%%%%%%%%%%%%%%%%%%%%%
\subsubsection{ Model fitting}
\label{MetaOrdDTA_results_baseline_analysis_model_fitting}
%%%%%%%%%%%%%%%%%%%%%%%%%%%%%%%%%%%%%%%%%%%%%%%%%%%%%%%%%%%%%%%%%%%%%%%%%%%%%%%%
%%%%
%%%% -------------------------------------------------------------------------------------------------------------------
\refstepcounter{codebox} %% increment the counter
%% \begin{tcolorbox}[colback=black!75,colframe=gray!75!black,title=Code box \thecodebox: Plotting sROC plots]
\begin{tcolorbox}
[colback=gray!75,colframe=gray!75!black,colupper=black,fonttitle=\color{white}\bfseries,
title=Code box \thecodebox: Model fitting workflow (NMA example)]
\label{code_box_MetaOrdDTA_baseline_analysis_model_fitting}
%% \begin{lstlisting}[escapeinside={(*@}{@*)},basicstyle=\footnotesize\ttfamily,columns=fullflexible]
\begin{lstlisting}[escapeinside={(*@}{@*)},basicstyle=\scriptsize\ttfamily\color{black},columns=fullflexible]
(*@\textcolor{blue}{\#\# ---- Call "MetaOrd\_model\$new()" to initialise model (create a new R6 class):  }@*) 
model_prep_obj <- MetaOrdDTA::(*@\textcolor{red}{MetaOrd_model\$new}@*)(  
      x = x, (*@\textcolor{blue}{\#\# The data }@*)
      indicator_index_test_in_study = indicator_index_test_in_study, (*@\textcolor{blue}{\#\# See manuscript for details }@*)
      intercept_only = TRUE, (*@\textcolor{blue}{\#\#  If set to TRUE, then "cov_data" will be ignored }@*)
      cov_data = NULL, (*@\textcolor{blue}{\#\# Only needed if doing meta-regression (i.e. if intercept_only = FALSE) }@*)
      network = TRUE,
      compound_symmetry = TRUE, (*@\textcolor{blue}{\#\# Only relevant for NMA }@*)
      prior_only = FALSE,
      box_cox = FALSE, (*@\textcolor{blue}{\#\# Optional + only relevant for "Jones" model parameterisation }@*)
      model_parameterisation = "ord_bivariate", (*@\textcolor{blue}{\#\# For NMA, only "ord_bivariate" and "Jones" available }@*)
      random_thresholds = FALSE,
      Dirichlet_random_effects_type = NULL) (*@\textcolor{blue}{\#\# Only used if random_thresholds = TRUE. Default is "kappa" }@*)
(*@\textcolor{blue}{  \#\# ---- You can also extract (to later modify) the inits and priors: }@*)    
init_lists_per_chain <- model_prep_obj$init_lists_per_chain
priors <- model_prep_obj$priors
(*@\textcolor{blue}{ \#\# ---- E.g., if you want to change some specific prior(s) before sampling: }@*)
priors$prior_beta_sigma_SD (*@\textcolor{blue}{ \#\# default is: c(0.5, 0.5) }@*)
priors$prior_beta_sigma_SD[2] <- 1.0 (*@\textcolor{blue}{ \#\# e.g., if we think there is very high between-study heterogeneity in D+ group }@*) 
(*@\textcolor{blue}{  \#\# ---- Can also use the "resize_init_list" helper function to rezise the list to match n_chains: }@*)
n_chains <- 16 (*@\textcolor{blue}{ \#\# Set the number of chains for sampling }@*)
init_lists_per_chain <- MetaOrdDTA::resize_init_list( init_lists_per_chain = init_lists_per_chain,
                                                          n_chains_new = n_chains)
(*@\textcolor{blue}{\#\# ---- Sample model using the "\$sample()" R6 method:  }@*)
model_samples_obj <-  model_prep_obj$(*@\textcolor{red}{sample}@*)(
  n_chains = n_chains,
  n_burnin = 1000,
  n_iter   = 1000,
  algorithm = "Stan", (*@\textcolor{blue}{\#\# Default - but can use "BayesMVP" (if installed - see manuscript for details) }@*)
  priors = priors, (*@\textcolor{blue}{\#\# Optional (otherwise defaults) }@*)
  init_lists_per_chain = init_lists_per_chain (*@\textcolor{blue}{\#\# Optional (otherwise defaults) }@*)
)
(*@\textcolor{blue}{\#\# ---- Model results (including summary and traces, etc) using the "\$summary()" R6 method  }@*)
model_summary_and_trace_obj <- model_samples_obj$(*@\textcolor{red}{summary}@*)(
  compute_main_params = TRUE,
  compute_transformed_parameters = TRUE, 
  compute_generated_quantities = TRUE,
  use_BayesMVP_for_faster_summaries = TRUE) (*@\textcolor{blue}{\#\# Only works if BayesMVP installed }@*)
(*@\textcolor{blue}{\#\# ---- Extract summary tibbles using built-in R6 class methods:  }@*)
tibble_main <- model_summary_and_trace_obj$(*@\textcolor{red}{get_summary_main()}@*) %>% print(n = 200)
tibble_tp   <- model_summary_and_trace_obj$(*@\textcolor{red}{get_summary_transformed()}@*) %>% print(n = 100)
tibble_gq   <- model_summary_and_trace_obj$(*@\textcolor{red}{get_summary_generated_quantities()}@*) %>% print(n = 1000)
(*@\textcolor{blue}{\#\# ---- Extract specific params (as tibbles) using the "\$extract\_params()" R6 method (e.g., Se, Sp):  }@*)
Se <- model_summary_and_trace_obj$(*@\textcolor{red}{extract_params}@*)(params = c("Se_baseline")) %>% print(n = 20)
Sp <- model_summary_and_trace_obj$(*@\textcolor{red}{extract_params}@*)(params = c("Sp_baseline")) %>% print(n = 20)
(*@\textcolor{blue}{\#\# ---- MCMC info and diagnostics: }@*)
model_summary_and_trace_obj$(*@\textcolor{red}{get_divergences()}@*)
model_summary_and_trace_obj$(*@\textcolor{red}{get_efficiency_metrics()}@*)
model_summary_and_trace_obj$(*@\textcolor{red}{get_HMC_info()}@*)
(*@\textcolor{blue}{\#\# ---- MCMC plots: }@*)
model_summary_and_trace_obj$(*@\textcolor{red}{plot_traces()}@*)
model_summary_and_trace_obj$(*@\textcolor{red}{plot_densities()}@*)
\end{lstlisting}
\end{tcolorbox}
%%%%
%%%% -------------------------------------------------------------------------------------------------------------------------------------
Code snippet box \ref{code_box_MetaOrdDTA_baseline_analysis_model_fitting}
demonstrates the model fitting workflow for the MetaOrdDTA R package.
This is for an NMA example dataset - 
we discuss how the data should be inputted in section \ref{MetaOrdDTA_results_data_format}.
In this case, since we are not doing meta-regression
(see section \ref{MetaOrdDTA_results_meta_reg} for the meta-regression demonstration/example),
we must set \verb|intercept_only = TRUE|. 
Furthermore, note that we do not need to specify anything for the \verb|cov_data| argument; 
however, if we do (e.g., by accident or left-over from previously running meta-regression), 
then it will simply be ignored - unless \verb|intercept_only| is set to \verb|FALSE|.

%%%%
%%%% -------------------------------------------------------------------------------------------------------------------------------------
In code box \ref{code_box_MetaOrdDTA_baseline_analysis_model_fitting},
we begin by initialising our model
using the R6 class method \verb|MetaOrd_model$new()|.
We discuss what the data variables
("\verb|x|" and "\verb|indicator_index_test_in_study|")
must look like in section \ref{MetaOrdDTA_results_data_format}.
For non-NMA only, "\verb|indicator_index_test_in_study|" is redundant, 
so will be ignored if specified.

%%%%
%%%% -------------------------------------------------------------------------------------------------------------------------------------
For modelling options,
"\verb|compound_symmetry|" is only needed for NMA.
If "\verb|compound_symmetry = TRUE|",
then all of the test-specific variances
within each disease group will be set to be equal to one another.
These test-specific variances were defined in section \ref{section_model_specs_NMA_extensions}.
In our example here, we are running the ordinal-bivariate model 
(which we proposed and defined in section 
\ref{section_model_specs_Cerullo_bivariate_Reitsma_extension})
by setting: \newline
"\verb|model_parameterisation = "ord_bivariate"|",
but fixed-effect thresholds 
(hence: \newline "\verb|random_thresholds = FALSE|).

%%%%
%%%% -------------------------------------------------------------------------------------------------------------------------------------
Code box \ref{code_box_MetaOrdDTA_baseline_analysis_model_fitting} demonstrates the complete model fitting workflow for MetaOrdDTA.
This process consists of the following steps:
\begin{enumerate}
\item \textbf{Model initialization} using \verb|MetaOrd_model$new()|, with the prepared data and model specifications.
\item \textbf{(Optional) prior and/or initial value modification} - extracting and customizing priors and/or initial values before sampling.
\item \textbf{MCMC sampling} using the \verb|$sample()| method to obtain the model samples object (here named \verb|model_samples_obj|).
\item \textbf{Results extraction} using the \verb|$summary()| method to obtain the model summary and traces object
(here named \verb|model_summary_and_trace_obj|).
\item \textbf{Parameter extraction} including:
    \begin{itemize}
    \item Summary tibbles for Stan model blocks (e.g., \verb|$get_summary_main()| for the Stan \verb|parameters| block).
    \item Specific parameters using \verb|$extract_params()|.
    \item All tibbles include $ESS$ and $\hat{R}$ for convergence assessment.
    \end{itemize}
\item \textbf{MCMC diagnostics:} extracting MCMC information, such as HMC divergence information and other HMC sampler information,
and optionally plotting trace and density plots.
For this step, we recommend at least checking divergences using the \verb|$get_divergences()| method
(note that convergence can also be checked in step (5) since the tibbles include $ESS$ and $\hat{R}$ information).
\end{enumerate}
In this example, we specify the ordinal bivariate model with compound symmetry for the NMA correlation structure and fixed-effect thresholds. 
For non-NMA analyses, the \verb|indicator_index_test_in_study| argument can be omitted.

%%%%%%%%%%%%%%%%%%%%%%%%%%%%%%%%%%%%%%%%%%%%%%%%%%%%%%%%%%%%%%%%%%%%%%%%%%%%%%%%
\subsubsection{ Model selection using K-fold CV}
\label{MetaOrdDTA_results_baseline_analysis_model_selection_k_fold}
%%%%%%%%%%%%%%%%%%%%%%%%%%%%%%%%%%%%%%%%%%%%%%%%%%%%%%%%%%%%%%%%%%%%%%%%%%%%%%%%
%%%%
%%%% -------------------------------------------------------------------------------------------------------------------------------------
\refstepcounter{codebox} %% increment the counter
%% \begin{tcolorbox}[colback=black!75,colframe=gray!75!black,title=Code Snippet \thecodebox: Plotting sROC plots]
\begin{tcolorbox}
[colback=gray!75,colframe=gray!75!black,colupper=black,fonttitle=\color{white}\bfseries,
title=Code box \thecodebox: Model selection using K-fold cross-validation (CV)]
\label{code_box_MetaOrdDTA_model_selection_k_fold_cv}
%% \begin{lstlisting}[escapeinside={(*@}{@*)},basicstyle=\footnotesize\ttfamily,columns=fullflexible]
\begin{lstlisting}[escapeinside={(*@}{@*)},basicstyle=\scriptsize\ttfamily\color{black},columns=fullflexible]
(*@\textcolor{blue}{\#\# ---- Call "\$make\_folds\_k\_fold\_CV()" R6 method: }@*) 
fold_assignments <- Model_A$model_summary_and_trace_obj$(*@\textcolor{red}{make\_folds\_k\_fold\_CV}@*)( 
                                         seed = 123, (*@\textcolor{blue}{\#\# make sure always use same seed when comparing models }@*) 
                                         K = 5) (*@\textcolor{blue}{\#\# number of folds }@*) 
(*@\textcolor{blue}{\#\# ---- Next, call "\$run\_k\_fold\_CV()" R6 method: }@*) 
outs_kfold <- Model_A$model_summary_and_trace_obj$(*@\textcolor{red}{run\_k\_fold\_CV}@*)(    
                                   seed = 123,
                                   fold_assignments = fold_assignments,
                                   priors = NULL, (*@\textcolor{blue}{\#\# optional (see manuscript text for details) }@*)
                                   init_lists_per_chain = NULL, (*@\textcolor{blue}{\#\# optional (see manuscript text for details) }@*)
                                   n_burnin = 1000,
                                   n_iter = 1000,
                                   n_chains = 2,
                                   adapt_delta = 0.80,
                                   max_treedepth = 10, 
                                   parallel = TRUE, (*@\textcolor{blue}{\#\# Note: No. of threads used = K*n_workers*n_chains }@*) 
                                   n_workers = 5)
(*@\textcolor{blue}{\#\# ---- Save the results somewhere: }@*)     
saveRDS(outs_kfold, file.path("path_to_project_dir", "cv_results", "Model_A_fixed_cutpoints_CS.RDS"))
(*@\textcolor{blue}{\#\# ---- Now, suppose we have conducted K-fold on 4 models, all with the same "fold_assignments" and seed. }@*) 
(*@\textcolor{blue}{\#\# ---- We will now load all 4 of these K-fold objects, and then use the MetaOrdDTA helper functions }@*) 
(*@\textcolor{blue}{\#\# ---- to compare them to one another: }@*) 
kfold_Model_A <- readRDS(file.path("cv_results", "Model_A_fixed_cutpoints_CS.RDS"))
kfold_Model_B <- readRDS(file.path("cv_results", "Model_B_fixed_cutpoints_UN.RDS"))
kfold_Model_C <- readRDS(file.path("cv_results", "Model_C_random_cutpoints_CS.RDS"))                                                          
kfold_Model_D <- readRDS(file.path("cv_results", "Model_D_random_cutpoints_UN.RDS"))
(*@\textcolor{blue}{\#\# ---- Now we need to make a K-fold results list and a vector of model names: }@*) 
kfold_results_list <- list(kfold_Model_A, kfold_Model_B, kfold_Model_C, kfold_Model_D)
model_names <- c( "Model_A (FIXED-C + CS)", 
                  "Model_B (FIXED-C + UN)",
                  "Model_C (RAND-C + CS)", 
                  "Model_D (RAND-C + UN)")
(*@\textcolor{blue}{\#\# ---- Then, we call the "compare_models_k_fold_CV" helper function to compare model fit: }@*)      
comparison <- MetaOrdDTA::(*@\textcolor{red}{compare\_models\_k\_fold\_CV}@*)( kfold_results_list = kfold_results_list,
                                                model_names = model_names,
                                                min_ess_threshold = 100)
(*@\textcolor{blue}{\#\# ---- We can also call the "summarise_model_comparison_k_fold_CV" helper function to get a model ranking }@*) 
(*@\textcolor{blue}{\#\# ---- (in order from best -> worst) by ELPD: }@*) 
MetaOrdDTA::(*@\textcolor{red}{summarise\_model\_comparison\_k\_fold\_CV}@*)(comparison)
(*@\textcolor{blue}{\#\# ---- We can also make a LaTeX table of the K-fold comparison results: }@*)
MetaOrdDTA::(*@\textcolor{red}{make\_latex\_table\_k\_fold\_CV}@*)(comparison)
\end{lstlisting}
\end{tcolorbox}
%%%%
%%%% -------------------------------------------------------------------------------------------------------------------------------------
Code snippet box \ref{code_box_MetaOrdDTA_model_selection_k_fold_cv} 
shows us how we can conduct K-fold cross-validation using the \newline
\verb|$make_folds_k_fold_CV()| and \verb|$run_k_fold_CV()|
R6 class methods.
It also shows us how we can use the
\verb|compare_models_k_fold_CV()| 
helper function to compare model fit between any number of previous models which we have conducted K-fold CV on.
This function will also output an ESS summary across all of the models, and will discard any folds (across all models) 
which have an ESS of less than 100 (by default - ESS threshold can be changed using the \verb|min_ess_threshold| argument).
Code box \ref{code_box_MetaOrdDTA_model_selection_k_fold_cv} 
also shows us how we can use the 
\verb|summarise_model_comparison_k_fold_CV()|
helper function to create a summary which will rank and order the models by ELPD.

%%%%
%%%% --------------------------------------------------------------------------------------------------------------------------------
As mentioned in the comments in code box \ref{code_box_MetaOrdDTA_model_selection_k_fold_cv},
the \verb|priors| and \verb|init_lists_per_chain| arguments are optional here,
and if not specified (or set to \verb|NULL|), then they will simply be inherited from 
the originally fitted model; however, if there was no model fit 
(i.e., only initialized - see section \ref{MetaOrdDTA_results_baseline_analysis_model_fitting}), 
then these arguments will be set to the defaults.

%%%%
%%%% --------------------------------------------------------------------------------------------------------------------------------
Furthermore, note that the MCMC arguments here 
(i.e. \verb|n_burnin|, \verb|n_iter|, \verb|adapt_delta|, 
\verb|max_treedepth|) are all optional, 
and the defaults are as specified in code box 
\ref{code_box_MetaOrdDTA_model_selection_k_fold_cv}.
Finally, as shown in code box \ref{code_box_MetaOrdDTA_model_selection_k_fold_cv}, 
we can also use the \verb|$make_folds_k_fold_CV()| helper function 
to generate LaTeX code to make a table presenting these results, 
which will look just like table \ref{table_MetaOrdDTA_model_selection_k_fold_table}.
%%%%
%%%%
\begin{table}[H]
\centering
\caption{K-fold cross-validation model comparison}
\label{tab:kfold_comparison}
\begin{tabular}{clcc}
\hline
Rank & Model & ELPD (SE) & $\Delta$ELPD (SE) \\
\hline
1 & Model\_A (FIXED-C + CS) & $-24881$ ($1960$) & ---  \textbf{[Best]} \\
2 & Model\_B (FIXED-C + UN) & $-25338$ ($1849$) & $-457$ ($2695$) \\
3 & Model\_C (RAND-C + CS) & $-27581$ ($2392$) & $-2699$ ($3093$) \\
4 & Model\_D (RAND-C + UN) & $-28139$ ($2269$) & $-3258$ ($2998$) \\
\hline
\end{tabular}
\vspace{0.2cm}
\begin{flushleft}
\footnotesize ELPD = Expected Log Pointwise Predictive Density (higher is better). \\
\end{flushleft}
\label{table_MetaOrdDTA_model_selection_k_fold_table}
\end{table}
%%%%
%%%% --------------------------------------------------------------------------------------------------------------------------------
\textbf{Why K-fold cross-validation for model selection?}

K-fold cross-validation directly evaluates predictive performance on held-out data, 
addressing the critical question for screening/diagnostic test accuracy meta-analyses: 
how well will the model predict accuracy in new studies?
This aligns with clinical needs—practitioners require reliable predictions for their specific settings, 
not explanations of historical data.

%%%%
%%%% --------------------------------------------------------------------------------------------------------------------------------
For hierarchical NMA models, leave-one-out cross-validation (LOO-IC) fails systematically. 
The Pareto-smoothed importance sampling requires stable importance weights, 
but hierarchical models yield Pareto $\hat{k} > 0.7$ for most observations, 
indicating unreliable estimates (Vehtari et al., 2017 \supercite{Vehtari2017}). 
This occurs because removing individual studies substantially affects the between-study variance structure -
each study contributes to both test-specific and shared NMA parameters. 
With typical NMA sample sizes 
(e.g., our $N_{studies} = 80$, 
with each study having up to 4 tests), 
this influence is too large for importance sampling to overcome.
However, full K-fold CV with multiple studies per fold provides stable out-of-sample predictions 
even for complex hierarchical structures, 
making it the only viable option for NMA model comparison.

%%%%%%%%%%%%%%%%%%%%%%%%%%%%%%%%%%%%%%%%%%%%%%%%%%%%%%%%%%%%%%%%%%%%%%%%%%%%%%%%
\subsubsection{Area under the curve (AUC)}
\label{MetaOrdDTA_results_baseline_analysis_AUC}
%%%%%%%%%%%%%%%%%%%%%%%%%%%%%%%%%%%%%%%%%%%%%%%%%%%%%%%%%%%%%%%%%%%%%%%%%%%%%%%%
%%%%
%%%% -------------------------------------------------------------------------------------------------------------------------------------
\refstepcounter{codebox} %% increment the counter
\begin{tcolorbox}[colback=gray!75,colframe=gray!75!black,title=Code box \thecodebox: Extracting AUC Results]
\label{code_box_MetaOrdDTA_base_model_AUC_results_for_best_kfold_model}
%% \scriptsize
\begin{lstlisting}[escapeinside={(*@}{@*)},basicstyle=\scriptsize\ttfamily,columns=fullflexible]
(*@\textcolor{blue}{\#\# ---- Call "extract\_AUC()" R6 method:}@*)  
outs_AUC <- Base_Model_best$model_summary_and_trace_obj$(*@\textcolor{red}{extract\_AUC}@*)( test_names = test_names,
                                                                           
  make_latex_table = TRUE)
(*@\textcolor{blue}{ \#\# ---- Output tibbles: }@*) }@*) 
outs_AUC$outs_AUC$auc (*@\textcolor{blue}{\#\# tibble with AUC estimates per test: }@*)
outs_AUC$outs_AUC$auc_pred (*@\textcolor{blue}{\#\# tibble with prediction intervals and pairwise differences: }@*)
outs_AUC$outs_AUC$auc_diff (*@\textcolor{blue}{\#\# tibble with prediction pairwise differences: }@*)
(*@\textcolor{blue}{\#\# ---- Output LaTeX table with AUC results:}@*)
cat(outs_AUC$outs_AUC_latex_table)
\end{lstlisting}
\end{tcolorbox}
%%%%
%%%% 
%%%% -------------------------------------------------------------------------------------------------------------------------------------
Code snippet box \ref{code_box_MetaOrdDTA_base_model_AUC_results_for_best_kfold_model} 
shows how we can extract AUC estimates using the "\verb|$extract_AUC()|" R6 class method.
This also has an option 
("\verb|make_latex_table = TRUE|")
to generate LaTeX code presenting the results, 
which will look just like table 
\ref{table_MetaOrdDTA_base_model_AUC_results_for_best_kfold_model} - 
and works for any number of tests and test names.
%%%%
%%%%
\begin{table}[H]
\centering
\caption{AUC values, pairwise differences, and predictive intervals}
\small
\begin{tabular}{lcc}
\toprule
\multicolumn{3}{l}{\textbf{AUC estimates (95\% CI):}} \\
\midrule
GAD-2 & 84.5 & (80.8, 87.7) \\
GAD-7 & 89.2 & (86.8, 91.4) \\
HADS & 85.5 & (82.4, 88.2) \\
BAI & 86.7 & (82.1, 90.6) \\
\midrule
\multicolumn{3}{l}{\textbf{Pairwise AUC differences:}} \\
\midrule
GAD-2 vs GAD-7 & -4.8 & (-8.4, -1.3)* \\
GAD-2 vs HADS & -1.1 & (-5.2, 3.1) \\
GAD-2 vs BAI & -2.2 & (-7.6, 3.3) \\
GAD-7 vs HADS & 3.7 & (0.5, 7.0)* \\
GAD-7 vs BAI & 2.6 & (-2.0, 7.4) \\
HADS vs BAI & -1.2 & (-6.0, 4.0) \\
\midrule
\multicolumn{3}{l}{\textbf{AUC predictive intervals:}} \\
\midrule
GAD-2 & - & (58.3, 96.0) \\
GAD-7 & - & (64.6, 97.6) \\
HADS  & - & (58.3, 96.8) \\
BAI   & - & (59.7, 97.3) \\
\bottomrule
\end{tabular}
\begin{tablenotes}
\footnotesize
\item * indicates 95\% CI excludes zero (significant difference)
\item Predictive intervals incorporate between-study heterogeneity
\end{tablenotes}
\label{table_MetaOrdDTA_base_model_AUC_results_for_best_kfold_model}
\end{table}
%%%%
%%%%
%%%%
%%%%
%%%%%%%%%%%%%%%%%%%%%%%%%%%%%%%%%%%%%%%%%%%%%%%%%%%%%%%%%%%%%%%%%%%%%%%%%%%%%%%%
\subsubsection{Plots: sROC}
\label{MetaOrdDTA_results_baseline_analysis_sROC_plots}
\begin{figure}[H]
    \centering
    \includegraphics[width=12cm]{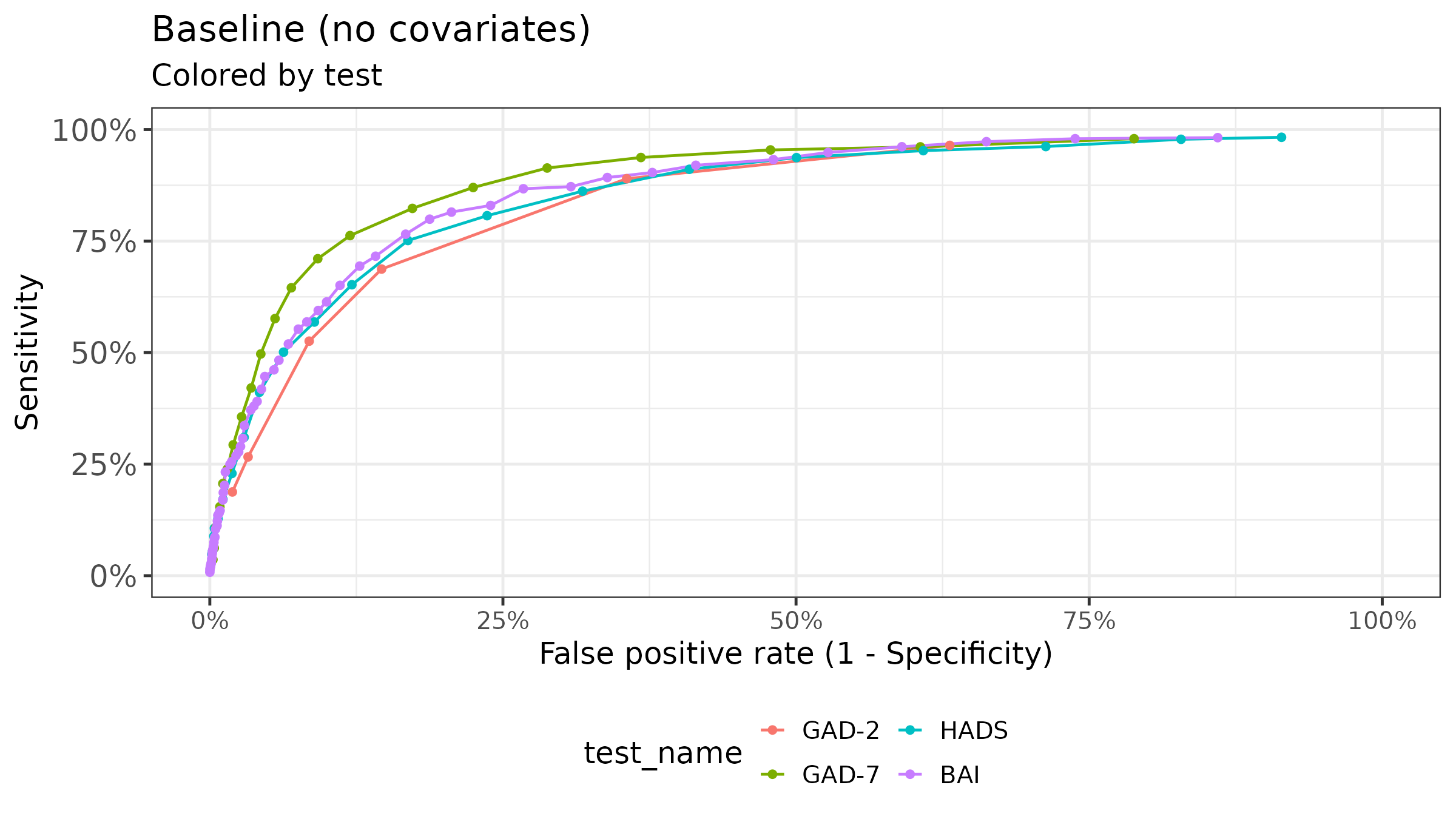}
    \caption{\footnotesize{
          sROC Plot for baseline analysis,
          for best-fitting model according to K-fold cross-validation (Model A).
    }}
    \label{Figure_MetaOrdDTA_baseline_analysis_sROC}
\end{figure}

%%%%%%%%%%%%%%%%%%%%%%%%%%%%%%%%%%%%%%%%%%%%%%%%%%%%%%%%%%%%%%%%%%%%%%%%%%%%%%%%
%%%%
%%%% -------------------------------------------------------------------------------------------------------------------------------------
\refstepcounter{codebox} %% increment the counter
%% \begin{tcolorbox}[colback=black!75,colframe=gray!75!black,title=Code Snippet \thecodebox: Plotting sROC plots]
\begin{tcolorbox}[colback=gray!75,colframe=gray!75!black,colupper=black,fonttitle=\color{white}\bfseries,title=Code box \thecodebox: sROC plots]
\label{code_box_MetaOrdDTA_base_model_sROC_plots}
%% \begin{lstlisting}[escapeinside={(*@}{@*)},basicstyle=\footnotesize\ttfamily,columns=fullflexible]
\begin{lstlisting}[escapeinside={(*@}{@*)},basicstyle=\scriptsize\ttfamily\color{black},columns=fullflexible]
(*@\textcolor{blue}{\#\# ---- Input relevant thresholds for each test (optional - otherwise shows all thresholds): }@*)  
relevant_thresholds = list( "GAD-2" = c(1:6), 
                            "GAD-7" =  c(3:18), 
                            "HADS" =  c(3:18),
                            "BAI" = c(3:38))
(*@\textcolor{blue}{\#\# ---- Call "plot\_sROC()" R6 method: }@*) 
(*@\textcolor{blue}{\#\# ---- Note: we can pass optional ggplot customization options }@*) 
(*@\textcolor{blue}{\#\# ---- for the first sROC plot using the R list "by_scenario_sROC_settings", }@*) 
(*@\textcolor{blue}{\#\# ---- and for the second sROC plot using the R list "by_scenario_grid_separate_sROC_settings" }@*)
outs_plots_sROC <- model_summary_and_trace_obj$(*@\textcolor{red}{plot\_sROC}@*)( 
                                test_names = test_names, 
                                relevant_thresholds = relevant_thresholds, (*@\textcolor{blue}{ \#\# optional }@*)
                                by_scenario_sROC_settings = NULL, (*@\textcolor{blue}{ \#\# optional }@*)  
                                by_scenario_grid_separate_sROC_settings = NULL) (*@\textcolor{blue}{ \#\# optional }@*)
(*@\textcolor{blue}{\#\# ---- sROC plot 1: shows the sROC curves for each test all on a single panel }@*) 
outs_plots_sROC$all_tests
(*@\textcolor{blue}{\#\# ---- sROC plot 2: shows the sROC curve for each test on 4 (= No. of tests) seperate }@*)
(*@\textcolor{blue}{\#\# ---- panels, together with 95\% credible and prediction regions. }@*) 
outs_plots_sROC$panels_with_regions
\end{lstlisting}
\end{tcolorbox}
%%%%
%%%% -------------------------------------------------------------------------------------------------------------------------------------
Code snippet box \ref{code_box_MetaOrdDTA_base_model_sROC_plots} 
shows how we can plot sROC plots for our fitted model, 
by using the 
"\verb|$plot_sROC()|" R6 class method. 
This will generate two different sROC plots. 
The first sROC plot (see figure 
\ref{Figure_MetaOrdDTA_baseline_analysis_sROC}) 
shows the sROC curve for each 
of the four tests on a single panel, 
with each curve coloured by test type.

\begin{figure}[H]
    \centering
    \includegraphics[width=12cm]{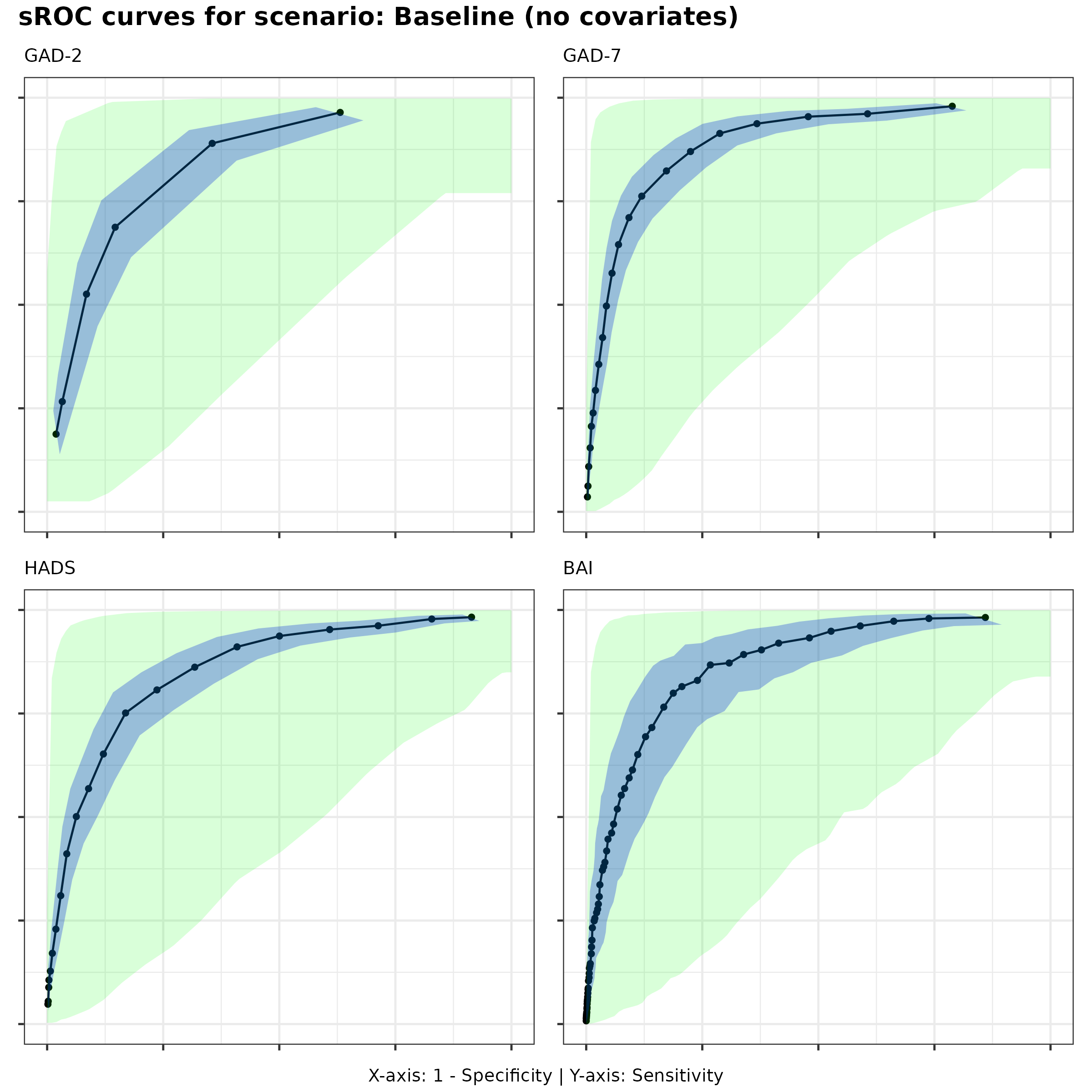}
    \caption{\footnotesize{
         sROC panel plot for baseline analysis; 
         with 95\% CrI's and 95\% PrI's;
         for best-fitting model according to K-fold cross-validation (Model A).
    }}
    \label{Figure_MetaOrdDTA_baseline_analysis_sROC_panel_w_CrI_PrI}
\end{figure}

%%%%
%%%% -------------------------------------------------------------------------------------------------------------------------------------
The second sROC plot (see figure 
\ref{Figure_MetaOrdDTA_baseline_analysis_sROC_panel_w_CrI_PrI})
shows each sROC curve for each of the four tests (in this case - as we have four tests)
on its own separate panel,
together with the $95\%$ credible regions 
(shaded in blue by default), 
and $95\%$ prediction regions 
(shaded in green by default).

%%%%
%%%% -------------------------------------------------------------------------------------------------------------------------------------
As shown from the note in code snippet box \ref{code_box_MetaOrdDTA_base_model_sROC_plots},
users can also customize the plots.
This can be achieved by creating two R lists -
one called \verb|by_scenario_sROC_settings| 
(for sROC plot \#1, see figure \ref{Figure_MetaOrdDTA_baseline_analysis_sROC}),
and another R list called \verb|by_scenario_grid_separate_sROC_settings| 
(for sROC plot \#2, see figure \ref{Figure_MetaOrdDTA_baseline_analysis_sROC_panel_w_CrI_PrI}).
More specifically, for sROC plot \#1, the R list "\verb|by_scenario_sROC_settings|" can accept the following customization options:
\verb|base_size|,
\verb|point_size|,
\verb|line_size|,
\verb|rows_in_legend|,
\verb|legend_position|,
as well as options for adding credible and prediction regions 
(\verb|add_regions|, \verb|show_conf|, \verb|show_pred|, \verb|conf_region_alpha|, \verb|pred_region_alpha|),
although we would not recommend this for this plot (figure \ref{Figure_MetaOrdDTA_baseline_analysis_sROC})
because of the overlap - unless perhaps there's only 2-3 tests being evaluated in the DTA-NMA.
Furthermore, the second R list (\verb|by_scenario_grid_separate_sROC_settings|) - 
for the plot in figure \ref{Figure_MetaOrdDTA_baseline_analysis_sROC_panel_w_CrI_PrI} -
can accept all of the above arguments, 
in addition to: 
\verb|conf_region_colour| and \verb|pred_region_colour|.
%%%% 
%%%%
%%%%
%%%%
%%%%
%%%%
%%%%%%%%%%%%%%%%%%%%%%%%%%%%%%%%%%%%%%%%%%%%%%%%%%%%%%%%%%%%%%%%%%%%%%%%%%%%%%%%
\subsubsection{Plots: Accuracy vs. threshold}
\label{MetaOrdDTA_results_baseline_analysis_accuracy_vs_thr_plots}
%%%%%%%%%%%%%%%%%%%%%%%%%%%%%%%%%%%%%%%%%%%%%%%%%%%%%%%%%%%%%%%%%%%%%%%%%%%%%%%%
%%%%
%%%% -------------------------------------------------------------------------------------------------------------------------------------
Code snippet box \ref{code_box_MetaOrdDTA_base_model_accuracy_vs_thr_plots} 
shows how we can plot accuracy vs. threshold plots for our fitted model, 
by using the "\verb|$plot_threshold()|" R6 class method.

%%%%
%%%% -------------------------------------------------------------------------------------------------------------------------------------
\refstepcounter{codebox} %% increment the counter
%% \begin{tcolorbox}[colback=black!75,colframe=gray!75!black,title=Code Snippet \thecodebox: Plotting sROC plots]
\begin{tcolorbox}
[colback=gray!75,colframe=gray!75!black,colupper=black,fonttitle=\color{white}\bfseries,title=Code box \thecodebox: Accuracy vs. threshold plots]
\label{code_box_MetaOrdDTA_base_model_accuracy_vs_thr_plots}
%% \begin{lstlisting}[escapeinside={(*@}{@*)},basicstyle=\footnotesize\ttfamily,columns=fullflexible]
\begin{lstlisting}[escapeinside={(*@}{@*)},basicstyle=\scriptsize\ttfamily\color{black},columns=fullflexible]
(*@\textcolor{blue}{\#\# ---- Input relevant thresholds for each test (optional - otherwise shows all thresholds): }@*)  
relevant_thresholds = list( "GAD-2" = c(1:6), 
                            "GAD-7" =  c(3:18), 
                            "HADS" =  c(3:18),
                            "BAI" = c(3:38))
(*@\textcolor{blue}{\#\# ---- Call "plot\_threshold()" R6 method: }@*) 
(*@\textcolor{blue}{\#\# ---- Note: we can pass optional ggplot customization options }@*) 
(*@\textcolor{blue}{\#\# ---- for the first sROC plot using the R list "by_scenario_sROC_settings", }@*) 
(*@\textcolor{blue}{\#\# ---- and for the second sROC plot using the R list "by_scenario_grid_separate_sROC_settings" }@*)
outs_plots_thr <- model_summary_and_trace_obj$(*@\textcolor{red}{plot\_threshold}@*)( 
                                test_names = test_names, 
                                plot_type = "both",  (*@\textcolor{blue}{ \#\# either "Se", "Sp", or "both" }@*)
                                show_actual_thresholds = FALSE, 
                                relevant_thresholds = relevant_thresholds, (*@\textcolor{blue}{ \#\# optional }@*)
                                by_scenario_sROC_settings = NULL, (*@\textcolor{blue}{ \#\# optional }@*)  
                                by_scenario_grid_separate_sROC_settings = NULL) (*@\textcolor{blue}{ \#\# optional }@*)
(*@\textcolor{blue}{\#\# ---- Accuracy (Se and/or Sp) vs. thr plot \#1: shows the Se and/or Sp curves for each test all on a single panel }@*) 
outs_plots_thr$all_tests_combined_Se_and_Sp
(*@\textcolor{blue}{\#\# ---- Accuracy (Se and/or Sp) vs. thr plot \#2: shows the Se and/or Sp curve for each test on 4 (= No. of tests) }@*)
(*@\textcolor{blue}{\#\# ---- seperate panels, together with 95\% credible and prediction bands. }@*) 
outs_plots_thr$panels_with_regions_Se_and_Sp
\end{lstlisting}
\end{tcolorbox}
%%%%
%%%%
%%%%
%%%% -------------------------------------------------------------------------------------------------------------------------------------

\begin{figure}[H]
    \centering
    \includegraphics[width=14cm]{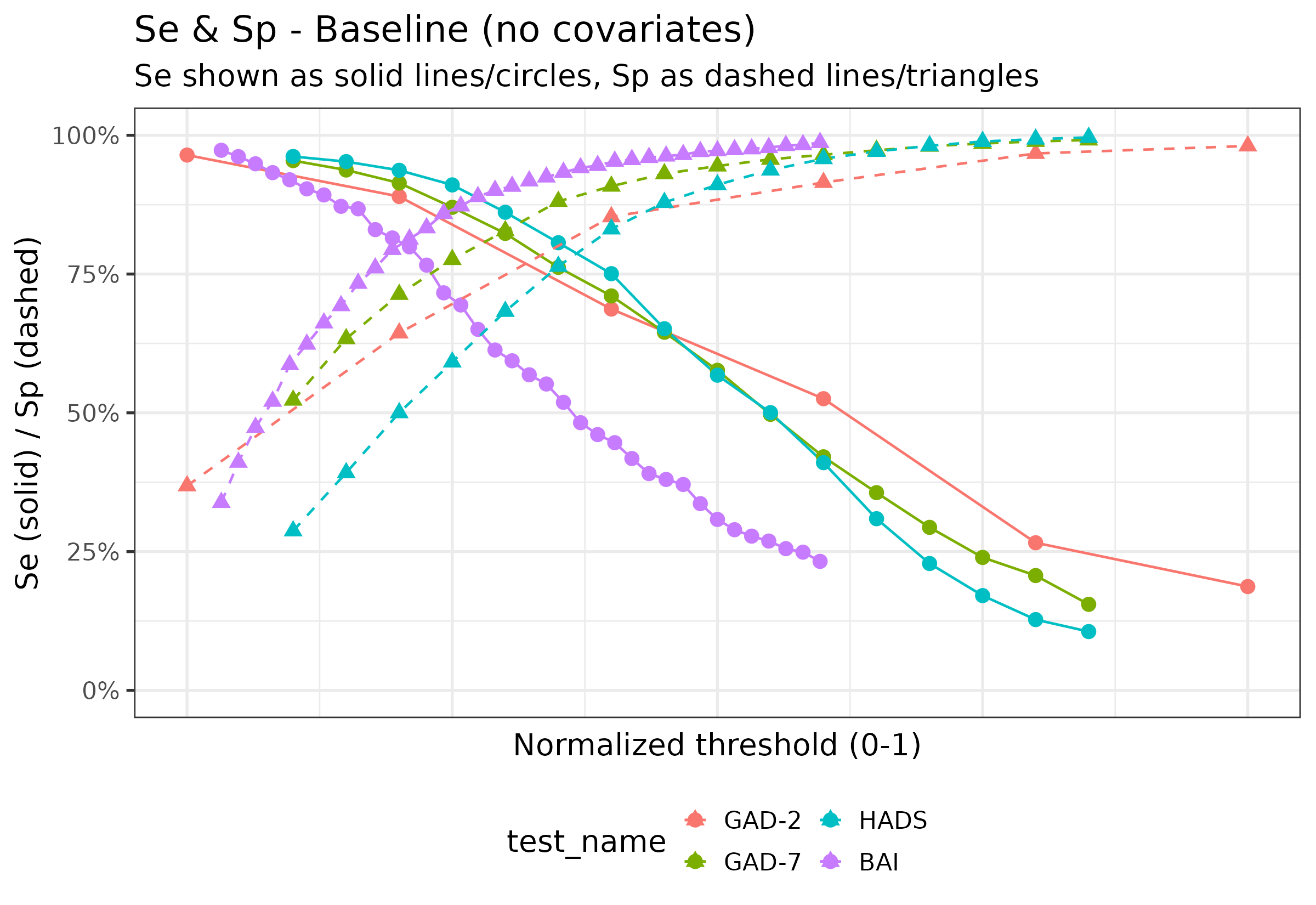}
    \caption{\footnotesize{
      Se/Sp vs. threshold plot for baseline analysis; 
      for best-fitting model according to K-fold cross-validation (Model A).
    }}
    \label{Figure_MetaOrdDTA_baseline_analysis_Se_Sp_vs_thr}
\end{figure}

%%
Similarly to the sROC plots (see section \ref{MetaOrdDTA_results_baseline_analysis_sROC_plots}), 
this will produce two plots - which besides the different type of curves -
are quite analogous to the corresponding sROC curves previously discussed. 
The first accuracy vs. threshold plot 
(see figure \ref{Figure_MetaOrdDTA_baseline_analysis_Se_Sp_vs_thr}) 
shows the curve for each 
of the four tests on a single panel, with each curve coloured by test type.

\begin{figure}[H]
    \centering
    \includegraphics[width=14cm]{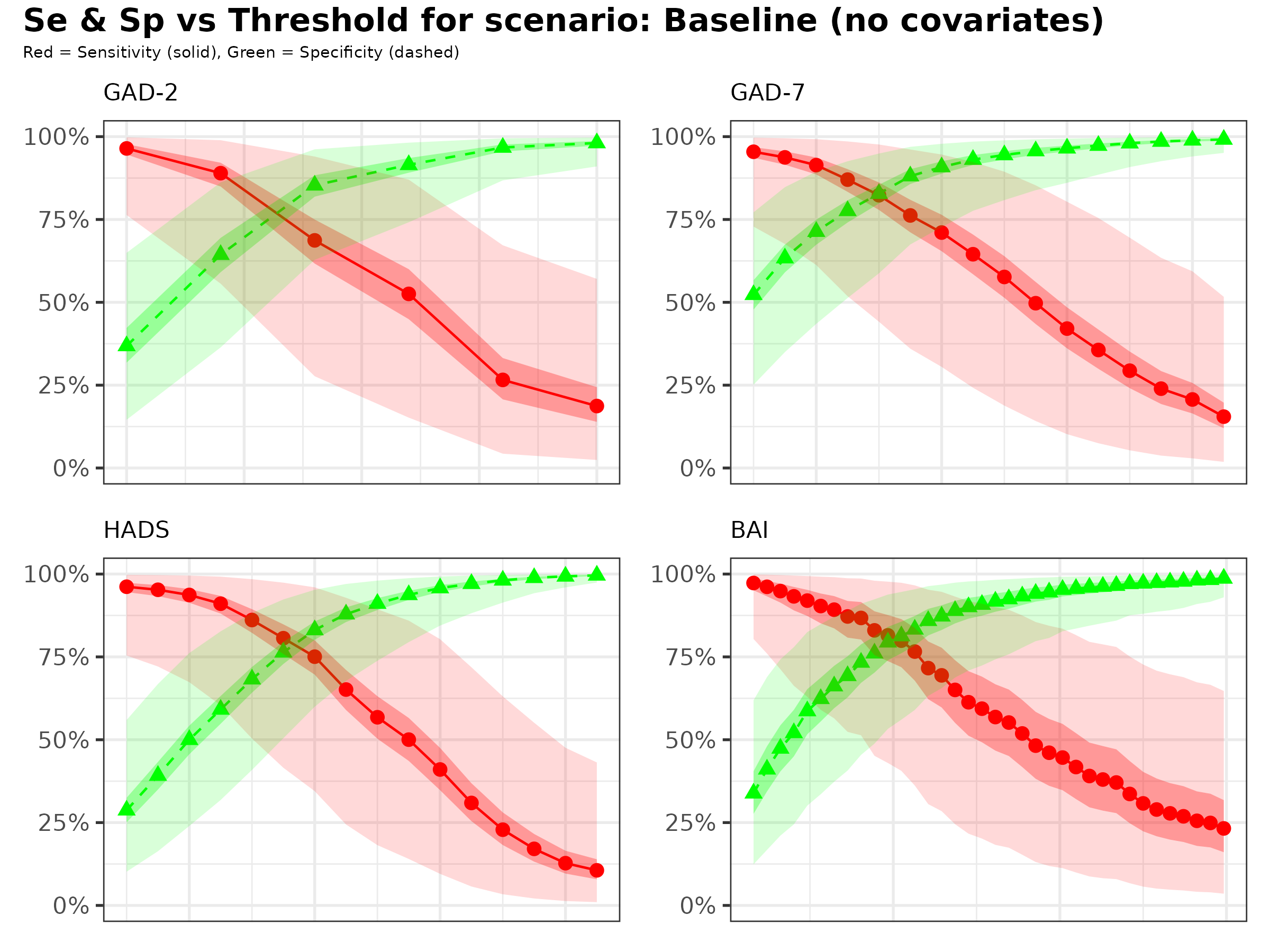}
    \caption{\footnotesize{
      Se/Sp vs. threshold panel plot for baseline analysis; with $95\%$ CrI's and $95\%$ PrI's; 
      for best-fitting model according to K-fold cross-validation (Model A).
    }}
    \label{Figure_MetaOrdDTA_baseline_analysis_Se_Sp_vs_thr_panel_w_CrI_PrI}
\end{figure}

%%%%
%%%% -------------------------------------------------------------------------------------------------------------------------------------
Again, analogously to the second sROC plot (see figure \ref{Figure_MetaOrdDTA_baseline_analysis_sROC_panel_w_CrI_PrI}), 
the second accuracy vs. threshold plot 
(see figure \ref{Figure_MetaOrdDTA_baseline_analysis_Se_Sp_vs_thr_panel_w_CrI_PrI})
shows each Se and/or Sp curve for each test on its own separate panel,
together with the $95\%$ credible bands (slightly darker shading) 
as well as the $95\%$ prediction bands (very light shading).
Furthermore, by default, all sensitivity curves and regions are shown in red, and specificity all in green.

%%%%
%%%% -------------------------------------------------------------------------------------------------------------------------------------
As shown from the note in code snippet box \ref{code_box_MetaOrdDTA_base_model_accuracy_vs_thr_plots},
users can also customize the plots.
This can be done by creating two R lists - 
one called \verb|by_scenario_threshold_settings| 
(for plot \#1, see figure \ref{Figure_MetaOrdDTA_baseline_analysis_Se_Sp_vs_thr}),
and another R list called \verb|by_scenario_grid_separate_threshold_settings| 
(for plot \#2, see figure \ref{Figure_MetaOrdDTA_baseline_analysis_Se_Sp_vs_thr_panel_w_CrI_PrI}).

%%%%
%%%% -------------------------------------------------------------------------------------------------------------------------------------
More specifically, for plot \#1, the R list \verb|by_scenario_threshold_settings| can accept the following customization options:
\verb|base_size|,
\verb|point_size|,
\verb|line_size|,
\verb|rows_in_legend|,
\verb|legend_position|,
as well as options for adding credible and prediction bands 
(\verb|add_regions|, \verb|show_conf|, \verb|show_pred|, \verb|conf_region_alpha|, \verb|pred_region_alpha|),
although we would not recommend this for this plot (figure \ref{Figure_MetaOrdDTA_baseline_analysis_Se_Sp_vs_thr})
because of the overlap - unless perhaps there's only 2-3 tests being evaluated in the DTA-NMA.
% %%
% Note that all of these aforementioned options are identical to those for sROC plot \#1. 
% However, we also have four additional arguments we can pass:
% \verb|conf_region_colour_Se| (default red),
% \verb|conf_region_colour_Sp| (default green)
% \verb|pred_region_colour_Se|
% \verb|pred_region_colour_Sp|
%%
Additionally, the second R list (\verb|by_scenario_grid_separate_threshold_settings|) - 
which is for the second plot (shown in figure \ref{Figure_MetaOrdDTA_baseline_analysis_Se_Sp_vs_thr_panel_w_CrI_PrI}) -
can accept all of the above arguments, 
in addition to: 
\verb|conf_region_colour_Se| (default red),
\verb|conf_region_colour_Sp| (default green),
\verb|pred_region_colour_Se| (default red),
and \verb|pred_region_colour_Sp| (default green).

%%%%
%%%% -------------------------------------------------------------------------------------------------------------------------------------
As mentioned in code snippet box \ref{code_box_MetaOrdDTA_base_model_accuracy_vs_thr_plots}, 
the \verb|"plot_type"| argument can be set to either "both", "Se" or "Sp". 
If "Se" (or "Sp") is specified - rather than "both" (as we have done in our example here) - 
then only sensitivity (or specificity) curves will be displayed, but otherwise the plots will look the same.
However, there is one key difference: 
in plot \#1 (figure \ref{Figure_MetaOrdDTA_baseline_analysis_Se_Sp_vs_thr}),
if "Se" (or "Sp") is specified, then users will have the option to plot $95\%$ credible and/or $95\%$ prediction bands onto the plots, 
by setting: \newline
\verb|by_scenario_threshold_settings$show_conf <- TRUE| and/or: \newline
\verb|by_scenario_threshold_settings$show_pred <- TRUE|
for credible and/or prediction bands, respectively.
%%%
%%%
%%%%
%%%%
%%%%
%%%%
%%%%%%%%%%%%%%%%%%%%%%%%%%%%%%%%%%%%%%%%%%%%%%%%%%%%%%%%%%%%%%%%%%%%%%%%%%%%%%%%
\subsubsection{Plots and tables: Pairwise accuracy differences}
\label{MetaOrdDTA_results_baseline_analysis_pairwise_diffs}
\begin{figure}[H]
    \centering
    \includegraphics[width=14cm]{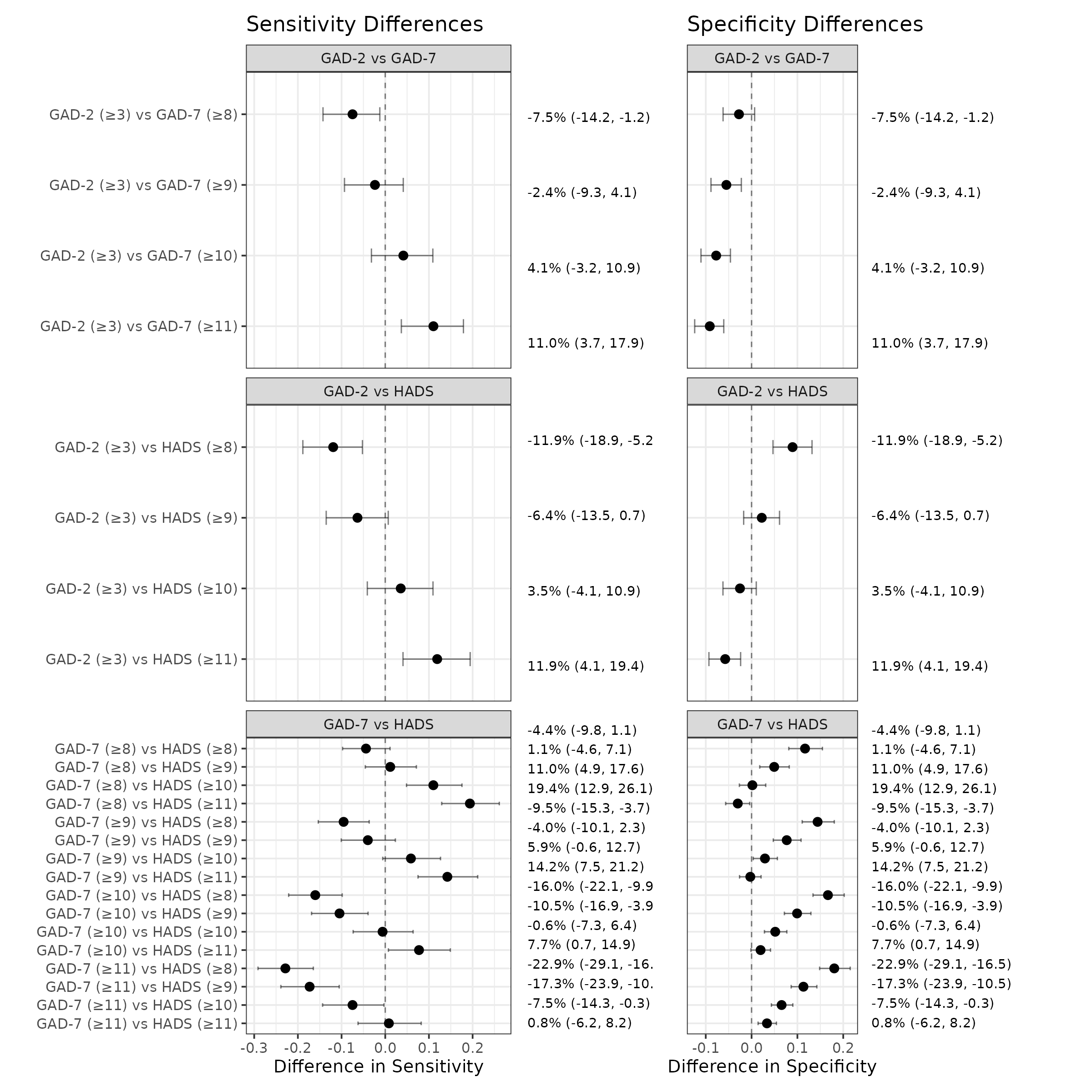}
    \caption{\footnotesize{
      Se and Sp pairwise differences with $95\%$ CrI's; 
      for best-fitting model according to K-fold cross-validation (Model A).
    }}
    \label{Figure_MetaOrdDTA_baseline_analysis_Se_and_Sp_pairwise_diffs_screening_GAD}
\end{figure}

%%%%%%%%%%%%%%%%%%%%%%%%%%%%%%%%%%%%%%%%%%%%%%%%%%%%%%%%%%%%%%%%%%%%%%%%%%%%%%%%
%%%%
%%%% -------------------------------------------------------------------------------------------------------------------------------------
Code snippet box \ref{code_box_MetaOrdDTA_base_model_pairwise_diffs_plots} 
shows how to:
(i) extract NMA pairwise comparisons (i.e. Se and Sp pairwise differences) for our fitted model,
by using the \verb|$extract_NMA_comparisons()| R6 class method, and:
(ii) plot these pairwise differences as forest plots (with optional side-by-side tables) for our fitted model, 
by using the \verb|$plot_pairwise()| R6 class method.
%%%%
%%%% -------------------------------------------------------------------------------------------------------------------------------------
\refstepcounter{codebox} %% increment the counter
\begin{tcolorbox}
[colback=gray!75,colframe=gray!75!black,colupper=black,fonttitle=\color{white}\bfseries,
title=Code box \thecodebox: Accuracy vs. threshold plots]
\label{code_box_MetaOrdDTA_base_model_pairwise_diffs_plots}
\begin{lstlisting}[escapeinside={(*@}{@*)},basicstyle=\scriptsize\ttfamily\color{black},columns=fullflexible]
(*@\textcolor{blue}{ \#\# ---- First, extract NMA comparisons using "\$extract\_NMA\_comparisons()" R6 class method: }@*)
(*@\textcolor{blue}{ \#\# ---- Note: this calls an internal Rcpp (C++) function to rapidly compute all pairwise comparisons for all test/threshold combos, }@*)
(*@\textcolor{blue}{ \#\# ---- using the full posterior Stan (or BayesMVP) traces - not just naively using posterior medians or means }@*)
outs_NMA_comp <- model_summary_and_trace_obj$(*@\textcolor{red}{extract\_NMA\_comparisons}@*)(test_names = test_names)
(*@\textcolor{blue}{ \#\# ---- Filter out BAI from tibble: }@*)
outs_NMA_comp_subset <- outs_NMA_comp %>% filter(test1_name != "BAI", test2_name != "BAI")
(*@\textcolor{blue}{ \#\# ---- Make R list w/ thresholds of interest for each test (not including BAI for this example): }@*)
thr_screening = list(
      "GAD-2" = c(3),              (*@\textcolor{blue}{ \#\# 3 is typical screening thr. for GAD }@*)
      "GAD-7" = c(8, 9, 10),       (*@\textcolor{blue}{ \#\# 10 is a typical screening thr. }@*)
      "HADS" = c(9, 10, 11)
)
(*@\textcolor{blue}{ \#\# ---- Modify plot settings: }@*)
baseline_plot_settings <- list()
baseline_plot_settings$show_values <- TRUE (*@\textcolor{blue}{ \#\# also shows posterior medians and 95\% CrI's for pairwise diffs }@*)
(*@\textcolor{blue}{ \#\# ---- Call the "\$plot\_pairwise()" R6 class method: }@*)
outs_plot_pairwise <- model_summary_and_trace_obj$(*@\textcolor{red}{plot\_pairwise}@*)( 
                                tibble_comp = outs_NMA_comp_subset,
                                comparison_type = "both", (*@\textcolor{blue}{ \#\# either "Se", "Sp", or "both" }@*) 
                                test_names = c("GAD-2", "GAD-7", "HADS"), 
                                relevant_thresholds = thr_screening,                     
                                baseline_plot_settings = baseline_plot_settings, (*@\textcolor{blue}{ \#\# optional }@*)  
                                metareg_plot_settings = NULL) (*@\textcolor{blue}{ \#\# optional - only for meta-reg }@*)
outs_plot_pairwise$Se  (*@\textcolor{blue}{ \#\# Se pairwise diffs only }@*)    
outs_plot_pairwise$Sp  (*@\textcolor{blue}{ \#\# Sp pairwise diffs only }@*)
outs_plot_pairwise$combined (*@\textcolor{blue}{ \#\# Se and Sp pairwise diffs, side-by-side }@*)
\end{lstlisting}
\end{tcolorbox}
%%%%
%%%% ------------------------------------------------------------------------------------------------------------------------------------
Figure \ref{Figure_MetaOrdDTA_baseline_analysis_Se_and_Sp_pairwise_diffs_screening_GAD}
shows the resulting plot, together with the side-by-side estimates table
(from the output in \verb|outs_plot_pairwise$combined|;
see code box \ref{code_box_MetaOrdDTA_base_model_pairwise_diffs_plots}).
As shown in code box \ref{code_box_MetaOrdDTA_base_model_pairwise_diffs_plots},
in our example we used the optional \verb|baseline_plot_settings| argument,
and set: \newline 
\verb|baseline_plot_settings$show_values <- TRUE|,
which shows the actual estimates
(posterior medians and 95\% credible intervals)
for the pairwise differences to the right-hand side of the forest plots
(see figure \ref{Figure_MetaOrdDTA_baseline_analysis_Se_and_Sp_pairwise_diffs_screening_GAD}).

%%%%
%%%% -------------------------------------------------------------------------------------------------------------------------------------
Other arguments which can be passed via the R list "\verb|baseline_plot_settings|" include the following:
\verb|base_size|,  %% yes
\verb|point_size|, %% yes
%% \verb|line_size|, %% no
\verb|error_bar_width|, %% yes
\verb|error_bar_alpha|, %% yes
%% \verb|legend_position|, %% no
%% \verb|facet_ncol|, %% no
%% \verb|facet_scales|, %% no
\verb|y_axis_text_size_multiplier|, %% yes
\verb|show_values| (as we already mentioned), %% yes
\verb|value_size_multiplier|, %% yes
\verb|ratio_plot|, %% yes
and
\verb|ratio_text|. %% yes

%%%%
%%%% -------------------------------------------------------------------------------------------------------------------------------------
%%%%
%%%%

%%%%%%%%%%%%%%%%%%%%%%%%%%%%%%%%%%%%%%%%%%%%%%%%%%%%%%%%%%%%%%%%%%%%%%%%%%%%%%%%%%%%%%%%%%%%%%%%%%%%%%%%%%%%%%%%%%%%
\newpage
\subsection{Meta-regression}
\label{MetaOrdDTA_results_meta_reg}
%%%%%%%%%%%%%%%%%%%%%%%%%%%%%%%%%%%%%%%%%%%%%%%%%%%%%%%%%%%%%%%%%%%%%%%%%%%%%%%%%%%%%%%%%%%%%%%%%%%%%%%%%%%%%%%%%%%%
%%%%
%%%%
%%%%%%%%%%%%%%%%%%%%%%%%%%%%%%%%%%%%%%%%%%%%%%%%%%%%%%%%%%%%%%%%%%%%%%%%%%%%%%%%
\subsubsection{ Data}
\label{MetaOrdDTA_results_meta_reg_data_prep}
%%%%%%%%%%%%%%%%%%%%%%%%%%%%%%%%%%%%%%%%%%%%%%%%%%%%%%%%%%%%%%%%%%%%%%%%%%%%%%%%
%%%%
%%%% ------------------------------------------------------------------------------------------------------
There are two primary ways to implement meta-regression (i.e., study-level) covariates:
\begin{enumerate}
    \item 
    The first way is directly inputting the list of lists \verb|X|
    (or just a list of matrices if doing non-NMA) -
    note that we define \verb|X| below,
    later in this section.
    \item
    The second way - which we will discuss here in this section -
    is by starting off with a tibble,
    as shown in table \ref{table:covariate_data_structure},
    where there's one row per study-test combination.
\end{enumerate}
%%
%%%%
%%%% ------------------------------------------------------------------------------------------------------
Our example here 
(see table \ref{table:covariate_data_structure})
includes a total of four covariates
(six after converting the two 3-level categorical covariates to binary indicators).
\begin{itemize}
    \item 
    A single \textbf{continuous} covariate
    (logit-transformed disease prevalence).
    \item 
    Two \textbf{categorical} covariates 
    (e.g., study setting, reference standard -
    both having 3 levels in this case).
    \item 
    One \textbf{binary} covariate (Risk of Bias [RoB]).
\end{itemize}

%%%%
%%%% ------------------------------------------------------------------------------------------------------
\begin{table}[H]
\centering
\caption{Example structure of covariate data input for meta-regression analysis}
\small
\begin{tabular}{clcccccc}
\toprule
Study & Author & Setting & prev\_GAD & Ref\_test & RoB\_low & logit\_prev \\
\midrule
1 & Author A & 3 & 0.14 & SCID & No & -1.82 \\
2 & Author B & 1 & 0.23 & MINI & No & -1.21 \\
2 & Author C & 1 & 0.06 & MINI & No & -2.75 \\
3 & Author D & 3 & 0.49 & Structured & Yes & -0.04 \\
4 & Author E & 3 & 0.11 & Structured & Yes & -2.09 \\
5 & Author F & 3 & 0.29 & SCID & No & -0.90 \\
6 & Author G & 3 & 0.06 & MINI & No & -2.75 \\
7 & Author H & 3 & 0.09 & MINI & No & -2.31 \\
8 & Author I & 2 & 0.61 & MINI & Yes & 0.45 \\
... & ... & ... & ... & ... & ... & ... \\
\bottomrule
\end{tabular}
\begin{tablenotes}
\footnotesize
\item Setting: 1 = primary care, 2 = secondary care, 3 = tertiary/specialist care
\item prev\_GAD: GAD prevalence in study population
\item Ref\_test: Reference standard used (MINI, SCID, or structured clinical interview)
\item RoB\_low: Low risk of bias according to QUADAS-2
\item logit\_prev: Logit-transformed prevalence
\item NA indicates missing covariate values
\end{tablenotes}
\label{table:covariate_data_structure}
\end{table}
%%%%
%%%%
%%%%
%%%% ------------------------------------------------------------------------------------------------------
The MetaOrdDTA R package \supercite{Cerullo_MetaOrdDTA_2025}
provides a comprehensive workflow for preparing covariates for model estimation:
\begin{enumerate}
    \item 
    \textbf{Initial covariate preparation}:
    Users provide a tibble with study-level covariates
    (note: if any study-level covariate values are missing,
    these can be can be represented as NA, NaN, Inf, or 999).
    \item \textbf{Covariate subsetting and transformation}:
    The 
    \newline \verb|R_fn_create_subsetted_covariates_for_tests()| 
    function processes the covariate tibble, creating test-specific design matrices:
       \begin{itemize}
           \item 
           If covariates vary between the non-diseased and diseased groups, then users provide a list of two tibbles instead,
           with the first tibble being for the non-diseased group and the second tibble being for the diseased group.
           In our example here, the covariates do not vary between groups,
           hence we're only providing a single tibble.
           \item
           This function filters to studies present in the actual test data
           (i.e., the data \verb|x|,
           which we defined in section \ref{MetaOrdDTA_results_data_format}).
           \item 
           This function also converts categorical variables (with 3+ levels) to binary dummy indicators.
           \item 
           It can also handle missing values
           (continuous missing variables marked for Bayesian/Stan model estimation,
           and binary and/or categorical missing data will treated as a separate missing category).
           \item 
           This function will also create separate design matrices for the diseased and non-diseased groups - 
           even if only a single tibble is inputted - 
           since this structure is needed for the Stan models.
           \item 
           Users can also input subsets of covariates (hence the functions name) - 
           so users do not need to subset their tibble before inputting it into the function.
           Users can also choose to input different subset of covariates for the non-diseased vs. diseased population, 
           or even inputting zero covariates in the non-diseased,
           but some/all covariates in the diseased population (or vice-versa).
           \item 
           Furthermore, for NMA (where there's multiple tests), 
           users can also choose to have different subsets of covariates for each test.
       \end{itemize}
    \item \textbf{Matrix expansion for Stan model compatibility}:
    The
    \newline \verb|R_fn_expand_covariates_to_all_studies()| 
    function ensures all covariate matrices have consistent
    dimensions matching the indicator matrix structure required by the Stan model, 
    padding with zeros for studies without specific test data.
\end{enumerate}
% %%%%
% %%%% -------------------------------------------------------------------------------------------------------------------------------------
% The function automatically handles mixed data types: 
% continuous covariates retain their numeric values 
% (with missings tracked for estimation), binary covariates are coded as 0/1 with optional missing indicators, 
% and categorical covariates are expanded into dummy variables using reference category coding.

%%%%
%%%% ------------------------------------------------------------------------------------------------------
The covariate list (or nested list of lists for NMA) of matrices - \verb|X| -
is defined as:
\textbf{For NMA analyses}, the covariate structure \verb|X| is organized as:
\begin{itemize}
    \item Outer list of length 2: 
    \verb|X[[1]]| for non-diseased group, 
    \verb|X[[2]]| for diseased group.
    \item Inner list of length \verb|n_tests|: 
    Each element contains the covariate matrix for that specific test.
    \item Each matrix has dimensions \verb|n_studies| × \verb|n_covariates|, 
    with rows padded with zeros for studies not containing that test.
\end{itemize}
Note that this structure (for \verb|X|) inverts the nesting of the outcome data \verb|x|, 
which has tests as the outer index.
For example, \verb|X[[1]][[3]]| contains the non-diseased covariate matrix for test 3 (HADS),
whilst \verb|x[[3]][[1]]| contains the non-diseased outcome data for the same test.

%%%%
%%%% ------------------------------------------------------------------------------------------------------
\textbf{For standard (non-NMA) meta-analyses},
the structure simplifies to:
\begin{itemize}
    \item 
    List of length 2: 
    \verb|X[[1]]| for non-diseased, 
    \verb|X[[2]]| for diseased.
    \item 
    Each element is a single matrix of dimensions \verb|n_studies| × \verb|n_covariates|.
\end{itemize}
Furthermore, each covariate matrix includes an intercept column,
as well as columns for all covariates after processing 
(i.e., continuous variables as-is, categorical variables expanded to dummy indicators). 
Studies without data for a particular test have their corresponding rows filled with zeros,
maintaining consistent matrix dimensions as required by Stan's array structure.

%%%%
%%%% ------------------------------------------------------------------------------------------------------
The \verb|create_subsetted_covariates_for_tests()|
function automatically generates this structure from the input covariate tibble,
handling the test-specific subsetting and padding internally.
The \verb|indicator_index_test_in_study()|
matrix ensures proper alignment between the covariate matrices and outcome data.

%%%%
%%%% ------------------------------------------------------------------------------------------------------
\textbf{Baseline case specification:}
In addition to the covariate matrices, 
meta-regression models require specification of baseline cases 
(\verb|baseline_case_nd| and \verb|baseline_case_d|)
for the non-diseased and diseased groups, respectively. 
These vectors define the covariate profile for which baseline sensitivity and specificity are estimated. 
Typically, this is set to the reference categories for all covariates:
\begin{itemize}
    \item Intercept = 1.
    \item Continuous covariates = 0 (if centered) or their mean value (if not centered).
    \item Binary covariates = 0 (reference category).
    \item Categorical dummy variables = 0 (representing the reference level).
\end{itemize}
For example, with our covariates, \verb|baseline_case = c(1, 0, 0, 0, 0, 0, 0)| represents:
intercept, mean prevalence (centered), primary care setting (reference), 
MINI reference test (reference), and high risk of bias (reference).
The \verb|create_subsetted_covariates_for_tests()| function automatically generates 
appropriate baseline cases based on the covariate structure.

%%%%
%%%% ------------------------------------------------------------------------------------------------------
\textbf{Test-specific covariate preparation:}
For NMA analyses where disease prevalence or other covariates may vary between tests, 
the covariate tibble must first be processed to create test-specific versions. 
Table \ref{table:cov_data_per_test} 
shows how the same study can have different 
prevalence values across tests, reflecting the actual data from each test.
%%%%
%%%%
\begin{table}[H]
\centering
\caption{Structure of test-specific covariate data (\texttt{cov\_data\_per\_test})}
\small
\begin{tabular}{cclcccc}
\toprule
\multicolumn{7}{c}{\textbf{Test 1 (GAD-2)}} \\
\midrule
Study & Author & Setting & prev\_GAD & Ref\_test & RoB\_low & logit\_prev \\
\midrule
8 & Study H & 2 & 0.15 & MINI & Yes & -1.73 \\
10 & Study J & 3 & 0.06 & SCID & No & -2.75 \\
... & ... & ... & ... & ... & ... & ... \\
\midrule
\multicolumn{7}{c}{\textbf{Test 2 (GAD-7)}} \\
\midrule
Study & Author & Setting & prev\_GAD & Ref\_test & RoB\_low & logit\_prev \\
\midrule
8 & Study H & 2 & 0.09 & MINI & Yes & -2.31 \\
10 & Study J & 3 & 0.08 & SCID & No & -2.44 \\
... & ... & ... & ... & ... & ... & ... \\
\midrule
\multicolumn{7}{c}{\textbf{Test 3 (HADS)}} \\
\midrule
Study & Author & Setting & prev\_GAD & Ref\_test & RoB\_low & logit\_prev \\
\midrule
8 & Study H & 2 & NA & MINI & Yes & NA \\
10 & Study J & 3 & NA & SCID & No & NA \\
... & ... & ... & ... & ... & ... & ... \\
\midrule
\multicolumn{7}{c}{\textbf{Test 4 (BAI)}} \\
\midrule
Study & Author & Setting & prev\_GAD & Ref\_test & RoB\_low & logit\_prev \\
\midrule
8 & Study H & 2 & 0.12 & MINI & Yes & -2.00 \\
10 & Study J & 3 & NA & SCID & No & NA \\
... & ... & ... & ... & ... & ... & ... \\
\bottomrule
\end{tabular}
\begin{tablenotes}
\footnotesize
\item Note: Study 8 participates in Tests 1, 2, and 4 with different prevalences in each.
\item Study 10 participates in Tests 1 and 2 only (NA values indicate non-participation).
\item The \texttt{cov\_data\_per\_test} structure is a list of 4 tibbles, one for each test.
\end{tablenotes}
\label{table:cov_data_per_test}
\end{table}
%%%%
%%%%
Studies not participating in a particular test have NA values for test-specific covariates 
like prevalence, which are handled appropriately during matrix creation.

%%%%
%%%% ------------------------------------------------------------------------------------------------------
The following code box (code box \ref{code_box_meta_reg_data_prep}) 
demonstrates the complete transformation from the tibble format 
(as shown in table \ref{table:covariate_data_structure}) 
to the nested list structure (i.e., \verb|X|), which are required by the Stan models.

%%%%
%%%% -------------------------------------------------------------------------------------------------------------------------------------
\refstepcounter{codebox} %% increment the counter
\begin{tcolorbox}
[colback=gray!75,colframe=gray!75!black,colupper=black,fonttitle=\color{white}\bfseries,
title=Code box \thecodebox: Meta-regression data preparation (NMA example)]
\label{code_box_meta_reg_data_prep}
\begin{lstlisting}[escapeinside={(*@}{@*)},basicstyle=\scriptsize\ttfamily\color{black},columns=fullflexible]
(*@\textcolor{blue}{\#\# ---- Prepare test data for NMA:}@*)
test_names <- c("GAD-2", "GAD-7", "HADS", "BAI")
max_scores <- c(6, 21, 21, 63)
test_data_list <- list(data_gad2, data_gad7, data_HADS, data_BAI)
(*@\textcolor{blue}{\#\# ---- Use prepare\_NMA\_data function to structure the data:}@*)
NMA_data <- MetaOrdDTA::(*@\textcolor{red}{prepare\_NMA\_data}@*)(
  test_data_list = test_data_list,
  test_names = test_names,
  max_scores = max_scores)
(*@\textcolor{blue}{\#\# ---- Prepare covariates from tibble format: }@*)
(*@\textcolor{blue}{\#\# Start with covariate tibble (as shown in manuscript table) }@*)
continuous_covs <- c("logit_prev_GAD")
binary_covs <- c("RoB_low") 
categorical_covs <- c("study_setting", "Ref_test")
(*@\textcolor{blue}{\#\# ---- Create test-specific covariate data (see manuscript for table example):}@*)
cov_data_per_test <- MetaOrdDTA::(*@\textcolor{red}{update\_covariates\_for\_tests}@*)(
  cov_data = cov_data_Klaus_3,
  test_data_list = test_data_list)
(*@\textcolor{blue}{\#\# ---- Create test-specific covariate matrices:}@*)
covs_output <- MetaOrdDTA::(*@\textcolor{red}{create\_subsetted\_covariates\_for\_tests}@*)(
  cov_data = cov_data_per_test, (*@\textcolor{blue}{\#\# List of test-specific tibbles (see manuscript for more info)}@*)
  indicator_index_test_in_study = NMA_data$indicator_index_test_in_study,
  test_names = test_names,
  continuous_covs = continuous_covs,
  binary_covs = binary_covs,
  categorical_covs = categorical_covs,
  center_and_scale_cts = TRUE) (*@\textcolor{blue}{\#\# Standardize continuous covariates}@*)
(*@\textcolor{blue}{\#\# ---- Extract the covariate matrices X:}@*)
X <- covs_output$X (*@\textcolor{blue}{\#\# List structure: X[[ND/D]][[test]][[matrix]]}@*)
(*@\textcolor{blue}{\#\# ---- Define baseline cases (covariate profile for baseline Se/Sp):}@*)
(*@\textcolor{blue}{\#\# Default: intercept=1, all other covariates=0 (reference categories)}@*)
n_covariates <- ncol(X[[1]][[1]])
baseline_vec <- c(1, rep(0, n_covariates - 1))
(*@\textcolor{blue}{\#\# ---- Create baseline case lists (can vary by test if desired):}@*)
(*@\textcolor{blue}{\#\# ---- NOTE: You can re-compute Se/Sp with new baseline post-hoc (shown in susequent sections):}@*)
baseline_case_nd <- rep(list(baseline_vec), NMA_data$n_index_tests)
baseline_case_d <- rep(list(baseline_vec), NMA_data$n_index_tests)
(*@\textcolor{blue}{\#\# ---- Example: Different baseline for test 3 (HADS):}@*)
(*@\textcolor{blue}{\#\# E.g., set prevalence to non-zero for HADS baseline}@*)
baseline_vec_custom <- baseline_vec
baseline_vec_custom[2] <- 0.5  (*@\textcolor{blue}{\#\# Higher logit-prevalence baseline for HADS}@*)
baseline_case_nd[[3]] <- baseline_vec_custom
baseline_case_d[[3]] <- baseline_vec_custom
(*@\textcolor{blue}{\#\# ---- Expand matrices to match all studies (for Stan compatibility):}@*)
X_expanded <- MetaOrdDTA::(*@\textcolor{red}{expand\_covariates\_to\_all\_studies}@*)(
  X,
  indicator_index_test_in_study = NMA_data$indicator_index_test_in_study)
(*@\textcolor{blue}{\#\# ---- Create covariate data structure for model:}@*)
cov_data <- list(
  X = X_expanded,
  baseline_case_nd = baseline_case_nd,
  baseline_case_d = baseline_case_d)
\end{lstlisting}
\end{tcolorbox}
%%%%
%%%%
%%%%%%%%%%%%%%%%%%%%%%%%%%%%%%%%%%%%%%%%%%%%%%%%%%%%%%%%%%%%%%%%%%%%%%%%%%%%%%%%
\subsubsection{ Model fitting}
\label{MetaOrdDTA_results_meta_reg_model_fitting}
%%%%%%%%%%%%%%%%%%%%%%%%%%%%%%%%%%%%%%%%%%%%%%%%%%%%%%%%%%%%%%%%%%%%%%%%%%%%%%%%
%%%%
%%%% -------------------------------------------------------------------------------------------------------------------------------------
\refstepcounter{codebox} %% increment the counter
%% \begin{tcolorbox}[colback=black!75,colframe=gray!75!black,title=Code box \thecodebox: Plotting sROC plots]
\begin{tcolorbox}
[colback=gray!75,colframe=gray!75!black,colupper=black,fonttitle=\color{white}\bfseries,
title=Code box \thecodebox: Meta-regression model fitting (NMA example)]
\label{code_box_MetaOrdDTA_meta_reg_model_fitting}
\begin{lstlisting}[escapeinside={(*@}{@*)},basicstyle=\scriptsize\ttfamily\color{black},columns=fullflexible]
(*@\textcolor{blue}{\#\# ---- Initialize model WITH meta-regression covariates:}@*) 
model_prep_obj <- MetaOrdDTA::(*@\textcolor{red}{MetaOrd\_model\$new}@*)(  
  x = NMA_data$x,
  indicator_index_test_in_study = NMA_data$indicator_index_test_in_study,
  intercept_only = FALSE, (*@\textcolor{blue}{\#\# FALSE for meta-regression}@*)
  cov_data = cov_data, (*@\textcolor{blue}{\#\# Covariate data from previous code box}@*)
  network = TRUE,
  compound_symmetry = TRUE,
  model_parameterisation = "ord_bivariate",
  random_thresholds = FALSE)

(*@\textcolor{blue}{\#\# ---- Extract and optionally modify priors for regression coefficients:}@*)    
priors <- model_prep_obj$priors
priors$prior_beta_effect_sigma_SD (*@\textcolor{blue}{\#\# Prior SD for regression coefficients}@*)
priors$prior_beta_effect_sigma_SD <- c(1.0, 1.0) (*@\textcolor{blue}{\#\# Wider priors if needed}@*)

(*@\textcolor{blue}{\#\# ---- Sample the model:}@*)
n_chains <- 16
model_samples_obj <- model_prep_obj$(*@\textcolor{red}{sample}@*)(
  n_chains = n_chains,
  n_burnin = 1000,
  n_iter = 1000,
  priors = priors,
  adapt_delta = 0.80)

(*@\textcolor{blue}{\#\# ---- Extract results including regression coefficients:}@*)
model_summary <- model_samples_obj$(*@\textcolor{red}{summary}@*)()

(*@\textcolor{blue}{\#\# ---- Extract parameters, e.g. meta-regression coefficients:}@*)
beta_mu <- model_summary$(*@\textcolor{red}{extract\_params}@*)(
  params = c("beta_mu"")
) %>% print(n = 50)

\end{lstlisting}
\end{tcolorbox}
%%%%
%%%% ------------------------------------------------------------------------------------------------
The model fitting procedure here
(shown in code box \ref{code_box_MetaOrdDTA_meta_reg_model_fitting}) 
is very similar to the non-meta-regression case
(see section \ref{MetaOrdDTA_results_baseline_analysis_model_fitting}), 
except that here, for meta-regression, we must specify 
\verb|intercept_only = FALSE|, 
and we must also provide the 
\verb|cov_data| structure, which we showed how to construct in 
section \ref{MetaOrdDTA_results_meta_reg_data_prep},
and code box \ref{code_box_meta_reg_data_prep}.

%%%%%%%%%%%%%%%%%%%%%%%%%%%%%%%%%%%%%%%%%%%%%%%%%%%%%%%%%%%%%%%%%%%%%%%%%%%%%%%%
\subsubsection{ Covariate selection using K-fold CV}
\label{MetaOrdDTA_results_meta_reg_K_fold}
%%%%%%%%%%%%%%%%%%%%%%%%%%%%%%%%%%%%%%%%%%%%%%%%%%%%%%%%%%%%%%%%%%%%%%%%%%%%%%%%
%%%%
%%%% ------------------------------------------------------------------------------------------------
Code snippet box \ref{code_box_MetaOrdDTA_meta_reg_k_fold_cv}
shows the K-fold cross-validation we conducted for the meta-regression.
This is essentially the same process as the non-meta-regression K-fold CV
(see section \ref{MetaOrdDTA_results_baseline_analysis_model_selection_k_fold},
and code box \ref{code_box_MetaOrdDTA_model_selection_k_fold_cv});
the only difference is that here, rather than comparing 4 models to one another
(fixed/random cutpoints vs. compound-symmetry/unstructured),
we are comparing all possible covariate combinations -
and since we have up to 4 possible covariates,
this means we had 14 different models to run
(15 total models, but one model was intercept-only - which we already did K-fold CV for in
section \ref{MetaOrdDTA_results_baseline_analysis_model_selection_k_fold}).
%%%%
%%%%
\refstepcounter{codebox} %% increment the counter
%% \begin{tcolorbox}[colback=black!75,colframe=gray!75!black,title=Code Snippet \thecodebox: Plotting sROC plots]
\begin{tcolorbox}
[colback=gray!75,colframe=gray!75!black,colupper=black,fonttitle=\color{white}\bfseries,
title=Code box \thecodebox: Model selection using K-fold cross-validation (CV)]
\label{code_box_MetaOrdDTA_meta_reg_k_fold_cv}
%% \begin{lstlisting}[escapeinside={(*@}{@*)},basicstyle=\footnotesize\ttfamily,columns=fullflexible]
\begin{lstlisting}[escapeinside={(*@}{@*)},basicstyle=\scriptsize\ttfamily\color{black},columns=fullflexible]
(*@\textcolor{blue}{\#\# ---- Call "\$make\_folds\_k\_fold\_CV()" R6 method: }@*) 
fold_assignments <- Model_A$model_summary_and_trace_obj$(*@\textcolor{red}{make\_folds\_k\_fold\_CV}@*)( 
                                         seed = 123, (*@\textcolor{blue}{\#\# make sure always use same seed when comparing models }@*) 
                                         K = 5) (*@\textcolor{blue}{\#\# number of folds }@*) 
(*@\textcolor{blue}{\#\# ---- Next, call "\$run\_k\_fold\_CV()" R6 method: }@*) 
outs_kfold <- Model_A$model_summary_and_trace_obj$(*@\textcolor{red}{run\_k\_fold\_CV}@*)(    
                                   seed = 123,
                                   fold_assignments = fold_assignments,
                                   priors = NULL, (*@\textcolor{blue}{\#\# optional (see manuscript text for details) }@*)
                                   init_lists_per_chain = NULL, (*@\textcolor{blue}{\#\# optional (see manuscript text for details) }@*)
                                   n_burnin = 1000,
                                   n_iter = 1000,
                                   n_chains = 2,
                                   adapt_delta = 0.80,
                                   max_treedepth = 10, 
                                   parallel = TRUE, (*@\textcolor{blue}{\#\# Note: No. of threads used = K*n_workers*n_chains }@*) 
                                   n_workers = 5)
(*@\textcolor{blue}{\#\# ---- Save the results somewhere: }@*)     
saveRDS(outs_kfold, file.path("path_to_project_dir", "cv_results", "Model_A_fixed_cutpoints_CS.RDS"))
(*@\textcolor{blue}{\#\# ---- Now, suppose we have conducted K-fold all meta-reg models, all w/ same "fold_assignments" + seed. }@*) 
(*@\textcolor{blue}{\#\# ---- We will now load all of these K-fold objects, and then use the MetaOrdDTA helper functions }@*) 
(*@\textcolor{blue}{\#\# ---- to compare them to one another: }@*) 
kfold_Model_A    <- readRDS(file.path("cv_results", "Model_A_fixed_cutpoints_CS.RDS"))
kfold_MR_Model_1 <- readRDS(file.path("cv_results", "Model_A_MR_model_1.RDS"))
kfold_MR_Model_2 <- readRDS(file.path("cv_results", "Model_A_MR_model_2.RDS"))
(*@\textcolor{green!50!black}{\textit{... load the rest of the previously saved meta-regression K-fold results here ...}}@*)
(*@\textcolor{blue}{\#\# ---- Now we need to make a K-fold results list and a vector of model names: }@*) 
kfold_results_list <- list(kfold_Model_A, 
                           kfold_MR_Model_1, kfold_MR_Model_2, kfold_MR_Model_3, kfold_MR_Model_4,
                           kfold_MR_Model_5_FULL,
                           kfold_MR_Model_6, kfold_MR_Model_7, kfold_MR_Model_8, 
                           kfold_MR_Model_9, kfold_MR_Model_10, kfold_MR_Model_11,
                           kfold_MR_Model_12 kfold_MR_Model_13, kfold_MR_Model_14, kfold_MR_Model_15)
model_names <- c("MR_Model_0 (intercept-only - FIXED-C + CS)",
                 (*@\textcolor{blue}{\#\# ---- 1-covariate models:}@*)
                 "MR_Model_1 (intercept + prev_GAD)", 
                 (*@\textcolor{green!50!black}{\textit{... rest of 1-way models here ...}}@*)
                 (*@\textcolor{blue}{\#\# ---- Full model:}@*)
                 "MR_Model_FULL (int. + prev_GAD + ref_test + study_setting + RoB)",                 
                 (*@\textcolor{blue}{\#\# ---- 2-covariate models:}@*)
                 "MR_Model_6  (intercept + prev_GAD + ref_test)",
                 (*@\textcolor{green!50!black}{\textit{... rest of 2-way models here ...}}@*)
                 (*@\textcolor{blue}{\#\# ---- 3-covariate models:}@*)
                 "MR_Model_12 (intercept + prev_GAD + ref_test + study_setting)",
                 (*@\textcolor{green!50!black}{\textit{... rest of 3-way models here ...}}@*))
(*@\textcolor{blue}{\#\# ---- Then, we call the "compare_models_k_fold_CV" helper function to compare model fit: }@*)      
comparison <- MetaOrdDTA::(*@\textcolor{red}{compare\_models\_k\_fold\_CV}@*)( kfold_results_list = kfold_results_list,
                                                model_names = model_names,
                                                min_ess_threshold = 100)
(*@\textcolor{blue}{\#\# ---- We can also call the "summarise_model_comparison_k_fold_CV" helper function to get a model ranking }@*) 
(*@\textcolor{blue}{\#\# ---- (in order from best -> worst) by ELPD: }@*) 
MetaOrdDTA::(*@\textcolor{red}{summarise\_model\_comparison\_k\_fold\_CV}@*)(comparison)
(*@\textcolor{blue}{\#\# ---- We can also make a LaTeX table of the K-fold comparison results: }@*)
MetaOrdDTA::(*@\textcolor{red}{make\_latex\_table\_k\_fold\_CV}@*)(comparison)
\end{lstlisting}
\end{tcolorbox}
%%%%
%%%% ---------------------------------------------------------------------------------------------------
Table \ref{table_MetaOrdDTA_Meta_reg_kfold_cv_on_dummy_data}
shows the K-fold cross-validation results comparing meta-regression models on the dummy simulated dataset
($80$ studies) on the anxiety screening test data.
Note that model rankings and performance will vary depending
on the specific dataset and covariate relationships present in your data.
We also recommend against only doing univariate meta-regression and then building up from there -
since then one may miss important multi-variable patterns.

%%%%
%%%% ---------------------------------------------------------------------------------------------------
For this demonstration,
we can see from table \ref{table_MetaOrdDTA_Meta_reg_kfold_cv_on_dummy_data}
that model \#6
(i.e., \verb|intercept| + \verb|prev_GAD| + \verb|ref_test|)
achieved the highest ELPD value,
and was hence selected for subsequent examples;
hence, we used this meta-regression model (MR model \#6),
for the remainder of this section
(i.e.,
sub-sections
\ref{MetaOrdDTA_results_meta_reg_parameters} -
\ref{MetaOrdDTA_results_meta_reg_plots_pairwise_diffs}).

%%%%
%%%% ---------------------------------------------------------------------------------------------------
\begin{table}[H]
\centering
\footnotesize
\caption{K-fold cross-validation model comparison: comparing meta-regression models (K = 5)}
\label{tab:kfold_comparison}
\begin{tabular}{clcc}
\hline
Rank & Model & ELPD (SE) & $\Delta$ELPD (SE) \\
\hline
1 & MR\_Model\_6  (intercept + prev\_GAD + ref\_test) & $-20909$ ($1697$) & ---  \textbf{[Best]} \\
2 & MR\_Model\_1 (intercept + prev\_GAD) & $-20938$ ($1645$) & $3943$ ($2559$) \\
3 & MR\_Model\_13 (intercept + prev\_GAD + ref\_test + RoB) & $-20957$ ($1676$) & $3924$ ($2579$) \\
4 & MR\_Model\_8  (intercept + prev\_GAD + RoB) & $-20967$ ($1633$) & $3914$ ($2551$) \\
5 & MR\_Model\_7  (intercept + prev\_GAD + study\_setting) & $-21176$ ($1733$) & $3705$ ($2616$) \\
6 & MR\_Model\_12 (intercept + prev\_GAD + ref\_test + study\_setting) & $-21259$ ($1835$) & $3622$ ($2685$) \\
7 & MR\_Model\_14 (intercept + prev\_GAD + study\_setting + RoB) & $-21358$ ($1793$) & $3523$ ($2656$) \\
8 & MR\_Model\_FULL (int. + prev\_GAD + ref\_test + study\_setting + RoB) & $-21468$ ($1883$) & $3413$ ($2718$) \\
9 & MR\_Model\_10 (intercept + ref\_test + RoB) & $-24437$ ($1801$) & $445$ ($2662$) \\
10 & MR\_Model\_2 (intercept + ref\_test) & $-24591$ ($1930$) & $290$ ($2751$) \\
11 & MR\_Model\_4 (intercept + RoB) & $-24831$ ($1920$) & $50$ ($2744$) \\
12 & MR\_Model\_0 (intercept-only - FIXED-C + CS) & $-24881$ ($1960$) & $NA$ ($NA$)NA \\
13 & MR\_Model\_15 (intercept + ref\_test + study\_setting + RoB) & $-25138$ ($2045$) & $-256$ ($2832$) \\
14 & MR\_Model\_3 (intercept + study\_setting) & $-25147$ ($2060$) & $-266$ ($2844$) \\
15 & MR\_Model\_9  (intercept + ref\_test + study\_setting) & $-25170$ ($2063$) & $-288$ ($2846$) \\
16 & MR\_Model\_11 (intercept + study\_setting + RoB) & $-25315$ ($2092$) & $-433$ ($2867$) \\
\hline
\end{tabular}
\vspace{0.2cm}
\begin{flushleft}
\footnotesize{
ELPD = Expected Log Pointwise Predictive Density (higher is better). \\
All models use fixed cutpoints with compound symmetry correlation structure (i.e. based on Model A from baseline analysis). \\
}
\end{flushleft}
\label{table_MetaOrdDTA_Meta_reg_kfold_cv_on_dummy_data}
\end{table}
%%%%
%%%%

%%%%%%%%%%%%%%%%%%%%%%%%%%%%%%%%%%%%%%%%%%%%%%%%%%%%%%%%%%%%%%%%%%%%%%%%%%%%%%%%
\subsubsection{ Tables: Meta-regression parameters and heterogeneity}
\label{MetaOrdDTA_results_meta_reg_parameters}
%%%%%%%%%%%%%%%%%%%%%%%%%%%%%%%%%%%%%%%%%%%%%%%%%%%%%%%%%%%%%%%%%%%%%%%%%%%%%%%%
%%%%
%%%% ----------------------------------------------------------------------------------------------------------------------------------
Following model selection via K-fold cross-validation 
(section \ref{MetaOrdDTA_results_meta_reg_K_fold})
we examine the fitted meta-regression parameters to understand covariate effects 
and their impact on between-study heterogeneity.
%%%%
%%%%
\refstepcounter{codebox} %% increment the counter
\begin{tcolorbox}
[colback=gray!75,colframe=gray!75!black,colupper=black,fonttitle=\color{white}\bfseries,
title=Code box \thecodebox: Meta-regression: Pairwise accuracy differences plots]
\label{code_box_meta_reg_pairwise_diffs_plots}
\begin{lstlisting}[escapeinside={(*@}{@*)},basicstyle=\scriptsize\ttfamily\color{black},columns=fullflexible]
(*@\textcolor{blue}{ \#\# ---- Load chosen (best-fitting according to K-fold CV) intercept-only model: }@*)
Intercept_only_model_dummy <-  readRDS(file.path("path_to_best_fitting_intercept_only_model.RDS"))  
Intercept_only_model_summary_and_trace_obj <- Intercept_only_model_dummy$model_summary_and_trace_obj
(*@\textcolor{blue}{ \#\# ---- Load chosen (best-fitting according to K-fold CV) meta-reg model: }@*)
MR_model_dummy <- readRDS(file.path("path_to_best_fitting_meta_reg_model.RDS"))
MR_model_summary_and_trace_obj <- MR_model_dummy$model_summary_and_trace_obj

(*@\textcolor{blue}{ \#\# ---- Load "X" from your chosen meta-reg model - the columns of X must be named }@*)
X <- MR_model_dummy$model_summary_and_trace_obj$cov_data$X
covariate_names <- dimnames(X[[1]][[1]])[[2]]
(*@\textcolor{blue}{ \#\# ---- If not, either name them or enter a vector of column names, like this: }@*)
covariate_names <- c("intercept", "logit_prev_GAD", "Ref_test_clean_SCID", "Ref_test_clean_Structured")

(*@\textcolor{blue}{ \#\# ---- Call the "map_beta_coefficients" R function: }@*)
MR_coeff_outs <- MetaOrdDTA::(*@\textcolor{red}{map\_beta\_coefficients}@*)( 
                                beta_tibble = beta_mu, 
                                covariate_names = covariate_names,
                                test_names = test_names, 
                                min_magnitude = 0.125,
                                borderline_threshold = 0.10,
                                show_borderline = TRUE) 
(*@\textcolor{blue}{ \#\# ---- Make the meta-regression coefficients LaTeX table: }@*)
outs_latex <- MetaOrdDTA::(*@\textcolor{red}{create\_metareg\_coef\_table}@*)( 
                             beta_tibble = outs\$full_table, 
                             test_names = test_names)
cat(outs_latex) (*@\textcolor{blue}{ \#\# Print LaTeX code for the meta-regression coefficients table: }@*)

(*@\textcolor{blue}{ \#\# ---- Make the heterogeneity comparison (MR vs. intercept-only) LaTeX table: }@*)
outs_latex <- MetaOrdDTA::(*@\textcolor{red}{create\_heterogeneity\_comparison\_table}@*)( 
                  MR_model_summary_and_trace_obj = MR_model_summary_and_trace_obj,
                  Intercept_only_model_summary_and_trace_obj = Intercept_only_model_summary_and_trace_obj,
                  compound_symmetry = TRUE)
cat(outs_latex) (*@\textcolor{blue}{ \#\# Print LaTeX code for heterogeneity comparison (MR vs. intercept-only) table: }@*)
\end{lstlisting}
\end{tcolorbox}
%%%%
%%%% ----------------------------------------------------------------------------------------------------------------------------------
Table \ref{table_MR_coefficients} presents the meta-regression coefficients on the probit scale, 
with bolded values indicating $95\%$ credible intervals excluding zero. 
These coefficients quantify how covariates modify the latent continuous response 
underlying the ordinal test outcomes.

%%%% ----------------------------------------------------------------------------------------------------------------------------------
Furthermore, table \ref{table_MR_vs_intercept_heterogeneity_components}
compares heterogeneity parameters between the baseline 
and meta-regression models. 
The change in total heterogeneity indicates whether 
the included covariates explain between-study variation. 
In this example, heterogeneity decreased by approximately $24\%$, 
though in practice covariates may increase, decrease, or leave heterogeneity unchanged -
all outcomes provide insights into the data structure.

%%%%
%%%% ----------------------------------------------------------------------------------------------------------------------------------
These parameter estimates inform the subsequent post-hoc exploration 
(section \ref{MetaOrdDTA_results_post_hoc_baseline_exploration}), 
where we systematically vary covariate values to examine their clinical implications.  
%%%%
%%%%
\begin{table}[H]
\centering
\caption{Meta-regression coefficients (probit scale)}
\small
\begin{tabular}{lcc}
\toprule
& \multicolumn{1}{c}{Non-diseased} & \multicolumn{1}{c}{Diseased} \\
\midrule
\multicolumn{3}{l}{\textbf{GAD-2}} \\
Intercept            & $\mathbf{-1.015~(-1.757,~-0.276)}$ & $0.401~(-0.363,~1.124)$ \\
Prevalence (per SD)  & $0.037~(-0.118,~0.192)$ & $\mathbf{0.462~(0.312,~0.604)}$ \\
Ref test: SCID       & $0.050~(-0.211,~0.306)$ & $0.033~(-0.218,~0.281)$ \\
Ref_test_clean_Structured & $0.009~(-1.006,~0.941)$ & $-0.003~(-0.969,~0.951)$ \\
%%%%
\midrule
%%%%
\multicolumn{3}{l}{\textbf{GAD-7}} \\
Intercept            & $\mathbf{-1.484~(-1.917,~-1.037)}$ & $0.039~(-0.418,~0.483)$ \\
Prevalence (per SD)  & $\mathbf{0.154~(0.050,~0.260)}$ & $\mathbf{0.348~(0.237,~0.459)}$ \\
Ref test: SCID       & $\mathbf{0.266~(0.039,~0.487)}$ & $\mathbf{0.382~(0.152,~0.604)}$ \\
Ref_test_clean_Structured & $0.500~(-0.074,~1.065)$ & $0.037~(-0.577,~0.645)$ \\
%%%%
\midrule
%%%%
\multicolumn{3}{l}{\textbf{HADS}} \\
Intercept            & $\mathbf{-1.052~(-1.507,~-0.612)}$ & $0.243~(-0.204,~0.668)$ \\
Prevalence (per SD)  & $-0.002~(-0.117,~0.110)$ & $\mathbf{-0.316~(-0.440,~-0.195)}$ \\
Ref test: SCID       & $-0.067~(-0.292,~0.153)$ & $\mathbf{-0.334~(-0.557,~-0.106)}$ \\
Ref_test_clean_Structured & $-0.144~(-0.504,~0.207)$ & $-0.180~(-0.565,~0.192)$ \\
%%%%
\midrule
%%%%
\multicolumn{3}{l}{\textbf{BAI}} \\
Intercept            & $\mathbf{-1.937~(-2.237,~-1.639)}$ & $\mathbf{-0.386~(-0.700,~-0.083)}$ \\
Prevalence (per SD)  & $\mathbf{0.328~(0.176,~0.480)}$ & $0.056~(-0.113,~0.224)$ \\
Ref test: SCID       & $0.003~(-0.961,~0.925)$ & $-0.001~(-0.966,~0.985)$ \\
Ref_test_clean_Structured & $0.204~(-0.244,~0.649)$ & $-0.184~(-0.702,~0.314)$ \\
%%%%
\bottomrule
%%%%
\end{tabular}
\begin{tablenotes}
\item Bold indicates significant at 95\% level
\end{tablenotes}
\label{table_MR_coefficients}
\end{table}
%%%%
%%%%
%%%%
%%%%
\begin{table}[H]
\centering
\caption{Impact of covariates on heterogeneity parameters}
\small
\begin{tabular}{lcc}
\toprule
\textbf{Parameter}  & \textbf{Model A}          & \textbf{Model A + Covariates} \\
                    & \textit{(intercept-only)} & \textit{(prevalence, ref test, setting)} \\
%%%%
\midrule
%%%%
\multicolumn{3}{l}{\textit{\textbf{NMA shared between-study SD's:}}} \\
$\sigma_{\beta}^{(d-)}$ & $0.241~(0.131,~0.341)$ & $0.192~(0.075,~0.281)$ \\
$\sigma_{\beta}^{(d+)}$ & $0.334~(0.192,~0.460)$ & $0.332~(0.263,~0.411)$ \\
%%%%
\midrule
%%%%
\multicolumn{3}{l}{\textit{\textbf{NMA test-specific deviation SD's (CS structure):}}} \\
$\tau^{(d-)}$ (all tests) & $0.308~(0.248,~0.378)$ & $0.281~(0.226,~0.348)$ \\
%%%%
\midrule
%%%%
\multicolumn{3}{l}{\textit{\textbf{Total within-test, between-study heterogeneity:}}} \\
$\sigma_{total}^{(d-)}$ (all tests) & $0.393~(0.344,~0.455)$ & $0.342~(0.298,~0.398)$ \\
$\sigma_{total}^{(d+)}$ (all tests) & $0.530~(0.464,~0.614)$ & $0.377~(0.323,~0.447)$ \\
%%%%
\midrule
%%%%
\multicolumn{3}{l}{\textit{\textbf{Correlation structure:}}} \\
$\rho_{\beta}$ (NMA shared corr.) & $0.593~(0.171,~0.882)$ & $0.406~(0.005,~0.741)$ \\
$\rho_{12}$ (within-test corr.) & $0.217~(0.051,~0.386)$ & $0.187~(0.002,~0.372)$ \\
%%%%
\bottomrule
%%%%
\end{tabular}
%%%%
\begin{tablenotes}
\footnotesize
\item Both models use fixed cutpoints with compound symmetry correlation structure.
\item CS = compound symmetry; all tests share the same variance components.
\end{tablenotes}
\label{table_MR_vs_intercept_heterogeneity_components}
%%%%
\end{table}
%%%%
%%%%

%%%%%%%%%%%%%%%%%%%%%%%%%%%%%%%%%%%%%%%%%%%%%%%%%%%%%%%%%%%%%%%%%%%%%%%%%%%%%%%%
\subsubsection{ Post-hoc baseline exploration for best-fitting model }
\label{MetaOrdDTA_results_post_hoc_baseline_exploration}
%%%%%%%%%%%%%%%%%%%%%%%%%%%%%%%%%%%%%%%%%%%%%%%%%%%%%%%%%%%%%%%%%%%%%%%%%%%%%%%%
\begin{tcolorbox}
[colback=gray!75,colframe=gray!75!black,colupper=black,fonttitle=\color{white}\bfseries,
title=Code box \thecodebox: Post-hoc baseline exploration]
\label{code_box_meta_reg_baseline_exploration}
\begin{lstlisting}[escapeinside={(*@}{@*)},basicstyle=\scriptsize\ttfamily\color{black},columns=fullflexible]
(*@\textcolor{blue}{\#\# ---- Define scenarios ("all_MR_scenarios") to explore:}@*)
all_MR_scenarios <- list(
  (*@\textcolor{blue}{\#\# For Ref test = MINI:}@*)
  "Low prev (~6%), \nMINI" = list(logit_prev_GAD = -1.0,  Ref_test_SCID = 0, Ref_test_Struct = 0),
  "Med prev (~15%), \nMINI" = list(logit_prev_GAD = 0.0,  Ref_test_SCID = 0, Ref_test_Struct = 0),
  "High prev (~33%), \nMINI" = list(logit_prev_GAD = 1.0, Ref_test_SCID = 0, Ref_test_Struct = 0),
  (*@\textcolor{blue}{\#\# For Ref test = SCID:}@*)
  "Low prev (~6%), \nSCID" = list(logit_prev_GAD = -1.0,  Ref_test_SCID = 1, Ref_test_Struct = 0),
  "Med prev (~15%), \nSCID" = list(logit_prev_GAD = 0.0,  Ref_test_SCID = 1, Ref_test_Struct = 0),
  "High prev (~33%), \nSCID" = list(logit_prev_GAD = 1.0, Ref_test_SCID = 1, Ref_test_Struct = 0),
  (*@\textcolor{blue}{\#\# For Ref test = Structured:}@*)
  "Low prev (~6%), \nStructured" = list(logit_prev_GAD = -1.0,  Ref_test_SCID = 0, Ref_test_Struct = 1),
  "Med prev (~15%), \nStructured" = list(logit_prev_GAD = 0.0,  Ref_test_SCID = 0, Ref_test_Struct = 1),
  "High prev (~33%), \nStructured" = list(logit_prev_GAD = 1.0, Ref_test_SCID = 0, Ref_test_Struct = 1)
)
(*@\textcolor{blue}{\#\# ---- Rapidly update (recompute) Se/Sp/AUC (internally) for each scenario: }@*)
new_baseline_outs <- MR_model_summary_and_trace_obj$(*@\textcolor{red}{recompute\_baseline}@*)(
                                                    MR_scenarios = all_MR_scenarios,
                                                    test_names = test_names)
names(new_baseline_outs) <- names(all_MR_scenarios)

k <- 1 (*@\textcolor{blue}{\#\# to select scenario, in this case k can be 1 -> 9 (3x3 = 9 scenarios total):}@*)
(*@\textcolor{blue}{\#\# ---- Look at/compare AUCs across all_MR_scenarios:}@*)
print(new_baseline_outs[[k]]$AUC_summary)
print(new_baseline_outs[[k]]$AUC_pred_summary)
print(new_baseline_outs[[k]]$AUC_diff_summary)
(*@\textcolor{blue}{\#\# ---- Extract Se/Sp's:}@*)
new_baseline_outs[[k]]$Se_summaries %>% print(n=25)
new_baseline_outs[[k]]$Sp_summaries %>% print(n=25)
(*@\textcolor{blue}{\#\# ---- Extract Se/Sp prediction intervals:}@*)
new_baseline_outs[[k]]$Se_pred_summaries %>% print(n=25)
new_baseline_outs[[k]]$Sp_pred_summaries %>% print(n=25)
\end{lstlisting}
\end{tcolorbox}
%%%%
%%%% -------------------------------------------------------------------------------------------------------------------------------------
Following K-fold cross-validation identifying the optimal meta-regression model
(see section \ref{MetaOrdDTA_results_meta_reg_K_fold}),
the MetaOrdDTA R package also enables efficient post-hoc exploration of how different covariate
profiles affect test performance.
This is crucial for meta-regression analyses, where understanding how results (e.g., AUC, Se, Sp)
vary with covariate values directly informs clinical decision-making.
Critically, this exploration occurs \textit{without refitting the model}, 
using an internal Rcpp
\supercite{rcpp_ref_1_Eddelbuettel_and_Francois_2011, 
rcpp_ref_2_Eddelbuettel_2013, 
rcpp_ref_3_Eddelbuettel_and_Balamuta_2018} 
(C++) implementation that rapidly 
recomputes posterior distributions for new baseline cases.
This functionality allows researchers to explore clinically relevant scenarios -
such as test accuracy across different prevalence levels and reference standards -
as demonstrated in Code Box \ref{code_box_meta_reg_baseline_exploration}.

%%%%
%%%% -------------------------------------------------------------------------------------------------------------------------------------
Furthermore, the \verb|$recompute_baseline()| method 
(see code box \ref{code_box_meta_reg_baseline_exploration})
generates complete posterior distributions for all test 
performance metrics under each scenario, enabling comprehensive comparison of how test 
accuracy varies across clinical contexts. 
The subsequent sections 
(sections
\ref{MetaOrdDTA_results_meta_reg_AUC} - 
\ref{MetaOrdDTA_results_meta_reg_plots_pairwise_diffs})
extract and visualise these results for clinical interpretation.

%%%%%%%%%%%%%%%%%%%%%%%%%%%%%%%%%%%%%%%%%%%%%%%%%%%%%%%%%%%%%%%%%%%%%%%%%%%%%%%%
\subsubsection{ Area under the curve (AUC)}
\label{MetaOrdDTA_results_meta_reg_AUC}
%%%%%%%%%%%%%%%%%%%%%%%%%%%%%%%%%%%%%%%%%%%%%%%%%%%%%%%%%%%%%%%%%%%%%%%%%%%%%%%%
%%%%
%%%% -------------------------------------------------------------------------------------------------------------------------------------
\refstepcounter{codebox} %% increment the counter
\begin{tcolorbox}[colback=gray!75,colframe=gray!75!black,title=Code box \thecodebox: Extracting AUC Results]
\label{code_box_MetaOrdDTA_base_model_AUC_results_for_best_kfold_model}
%% \scriptsize
\begin{lstlisting}[escapeinside={(*@}{@*)},basicstyle=\scriptsize\ttfamily,columns=fullflexible]
(*@\textcolor{blue}{ \#\# ---- AUC: Now, AFTER running "recompute_baseline", can run "\$extract\_AUC()": }@*)
(*@\textcolor{blue}{ \#\# ---- Note: No need to pass either "all_MR_scenarios" or "new_MR_baseline_outs" objects - }@*)
(*@\textcolor{blue}{ \#\# ---- they're automatically stored internally when you run "\$recompute\_baseline()" }@*)
new_MR_AUC_outs <-  MR_Model_dummy$model_summary_and_trace_obj$(*@\textcolor{red}{extract\_AUC}@*)( test_names = test_names)

(*@\textcolor{blue}{ \#\# ---- Output tibbles: }@*)
new_MR_AUC_outs$outs_AUC$auc (*@\textcolor{blue}{\#\# tibble with AUC estimates per test: }@*)
new_MR_AUC_outs$outs_AUC$auc_pred (*@\textcolor{blue}{\#\# tibble with prediction intervals and pairwise diffs: }@*)
new_MR_AUC_outs$outs_AUC$auc_diff (*@\textcolor{blue}{\#\# tibble with prediction pairwise diffs: }@*)

(*@\textcolor{blue}{ \#\# ---- LaTeX table using the general AUC MR table function: }@*)
latex_code <- MetaOrdDTA::(*@\textcolor{red}{create\_MR\_AUC\_latex\_table}@*)(
  baseline_outs_list = new_MR_baseline_outs, (*@\textcolor{blue}{ \#\# Explicitly pass the "new_MR_baseline_outs" object here }@*)
  test_names = test_names,
  group_by = c(1,1,1, 2,2,2, 3,3,3), (*@\textcolor{blue}{ \#\# Groups of 3 }@*)
  group_labels = c("MINI", "SCID", "Structured"), (*@\textcolor{blue}{ \#\# order matching "all_MR_scenarios" }@*)
  row_labels = rep(c("Low (6\\%)", "Med (15\\%)", "High (33\\%)"), 3) (*@\textcolor{blue}{ \#\# order matching "all_MR_scenarios" }@*)
)
cat(latex_code)
(*@\textcolor{blue}{ \#\# ---- LaTeX table using the 2-variable AUC MR table fn: }@*)
latex_code <- MetaOrdDTA::(*@\textcolor{red}{create\_MR\_AUC\_latex\_table\_2vars}@*)(
  baseline_outs_list = new_MR_baseline_outs, (*@\textcolor{blue}{ \#\# Explicitly pass the "new_MR_baseline_outs" object here }@*)
  test_names = test_names,
  var1_levels = c("MINI", "SCID", "Structured"), (*@\textcolor{blue}{ \#\# order matching "all_MR_scenarios" }@*)
  var2_levels = c("Low (6\\%)", "Med (15\\%)", "High (33\\%)"), (*@\textcolor{blue}{ \#\# order matching "all_MR_scenarios" }@*)
  var1_name = "Reference",
  var2_name = "Prevalence"
)
cat(latex_code)
\end{lstlisting}
\end{tcolorbox}
%%%%
%%%% -------------------------------------------------------------------------------------------------------------------------------------
Code Box \ref{code_box_MetaOrdDTA_base_model_AUC_results_for_best_kfold_model} demonstrates 
AUC extraction from the stored scenario results and formatting options for tabular presentation.

%%%%
%%%% -------------------------------------------------------------------------------------------------------------------------------------
\begin{table}[H]
\centering
\caption{AUC comparison}
\begin{tabular}{llcccc}
\hline
Reference & Prevalence & GAD-2 & GAD-7 & HADS & BAI \\
 &  & AUC (95\% CI) & AUC (95\% CI) & AUC (95\% CI) & AUC (95\% CI) \\
\hline
MINI & Low (6\%) & 81.2 (76.8-85.2) & 85.4 (82.0-88.4) & 89.6 (86.9-92.1) & 91.7 (88.2-94.6) \\
 & Med (15\%) & 87.9 (84.2-91.1) & 89.0 (86.5-91.2) & 85.0 (81.6-88.2) & 88.7 (85.3-91.7) \\
 & High (33\%) & 92.4 (87.9-95.7) & 91.4 (88.4-94.0) & 79.0 (72.7-84.6) & 84.6 (78.5-89.9) \\
\hline
SCID & Low (6\%) & 81.0 (75.1-86.1) & 88.0 (83.9-91.3) & 85.7 (81.2-89.5) & 89.1 (65.8-99.1) \\
 & Med (15\%) & 87.7 (83.5-91.2) & 90.5 (87.3-93.3) & 79.9 (75.0-84.4) & 86.0 (59.9-98.5) \\
 & High (33\%) & 92.2 (88.0-95.5) & 92.3 (88.6-95.1) & 72.6 (65.1-79.3) & 81.8 (51.7-97.8) \\
\hline
Structured & Low (6\%) & 78.4 (47.8-96.3) & 76.9 (58.2-90.9) & 88.8 (82.2-93.8) & 86.2 (74.4-94.2) \\
 & Med (15\%) & 85.1 (58.5-98.2) & 80.9 (64.4-92.7) & 83.8 (75.1-90.8) & 82.1 (69.3-91.5) \\
 & High (33\%) & 90.0 (67.9-99.2) & 83.9 (68.7-94.2) & 77.4 (65.2-87.3) & 76.8 (61.9-88.4) \\
\hline
\end{tabular}
\label{table_meta_reg_AUC}
\end{table}
%%%%
%%%% -------------------------------------------------------------------------------------------------------------------------------------
Table \ref{table_meta_reg_AUC} presents AUC values across all nine scenarios, 
generated using the 
\newline \verb|create_MR_AUC_latex_table_2vars()| function
(see bottom of code box 
\ref{code_box_MetaOrdDTA_base_model_AUC_results_for_best_kfold_model}), 
following the
\verb|$recompute_baseline()| computation shown in the previous section
(see section \ref{MetaOrdDTA_results_post_hoc_baseline_exploration}, and
code box \ref{code_box_meta_reg_baseline_exploration}).

%%%%
%%%% -------------------------------------------------------------------------------------------------------------------------------------
The MetaOrdDTA R package also provides two formatting functions: 
\verb|create_MR_AUC_latex_table()| 
for general grouping structures, 
and \verb|create_MR_AUC_latex_table_2vars()| -
specifically for two-covariate designs -
as demonstrated here with reference standard × prevalence. 
Note that the \verb|$extract_AUC()| method automatically uses the internally stored scenario 
results from \verb|$recompute_baseline()| (see section \ref{MetaOrdDTA_results_post_hoc_baseline_exploration}),
eliminating redundant computation; 
however, the LaTeX functions require explicit passing of the baseline outputs, 
as shown in code box \ref{code_box_MetaOrdDTA_base_model_AUC_results_for_best_kfold_model}.

%%%%%%%%%%%%%%%%%%%%%%%%%%%%%%%%%%%%%%%%%%%%%%%%%%%%%%%%%%%%%%%%%%%%%%%%%%%%%%%%
\subsubsection{ Plots: sROC (with custom aesthetics mapping example)}
\label{MetaOrdDTA_results_meta_reg_plots_sROC}
\begin{figure}[H]
    \centering
    \includegraphics[width=14cm]{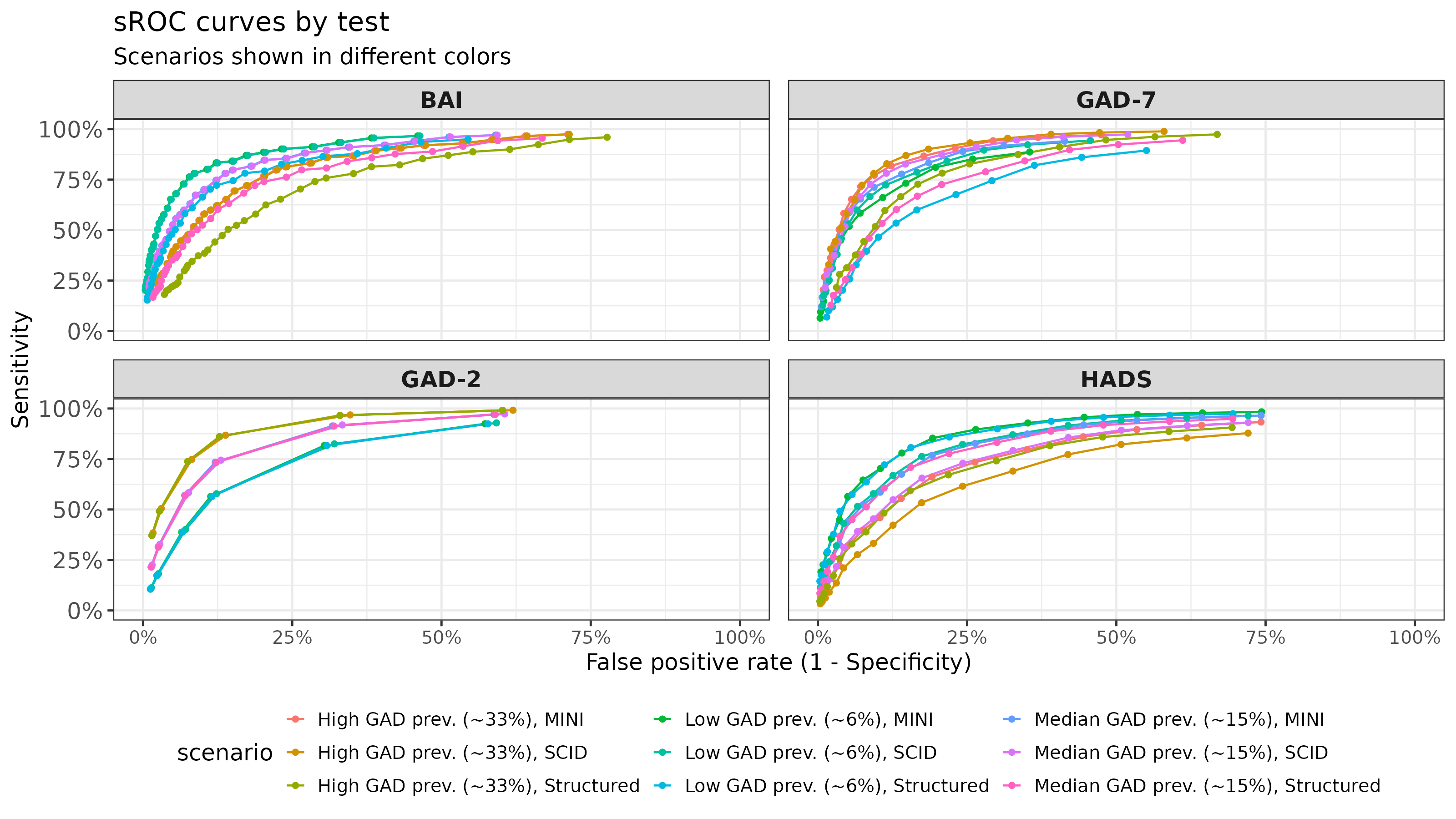}
    \caption{
    \footnotesize{
          Meta-regression sROC Plot; \\
          For best-fitting model according to K-fold cross-validation
          (Model A: Fixed-effect thresholds \& compound-symmetry NMA structure),
          with 2 covariates (logit-transformed GAD prevalence and reference test).
    }
    }
    \label{DUMMY_MetaOrdDTA_MR_sROC_2x2_ALL_TESTS_ALL_SCENARIOS}
\end{figure}

\begin{figure}[H]
    \centering
    \includegraphics[width=14cm]{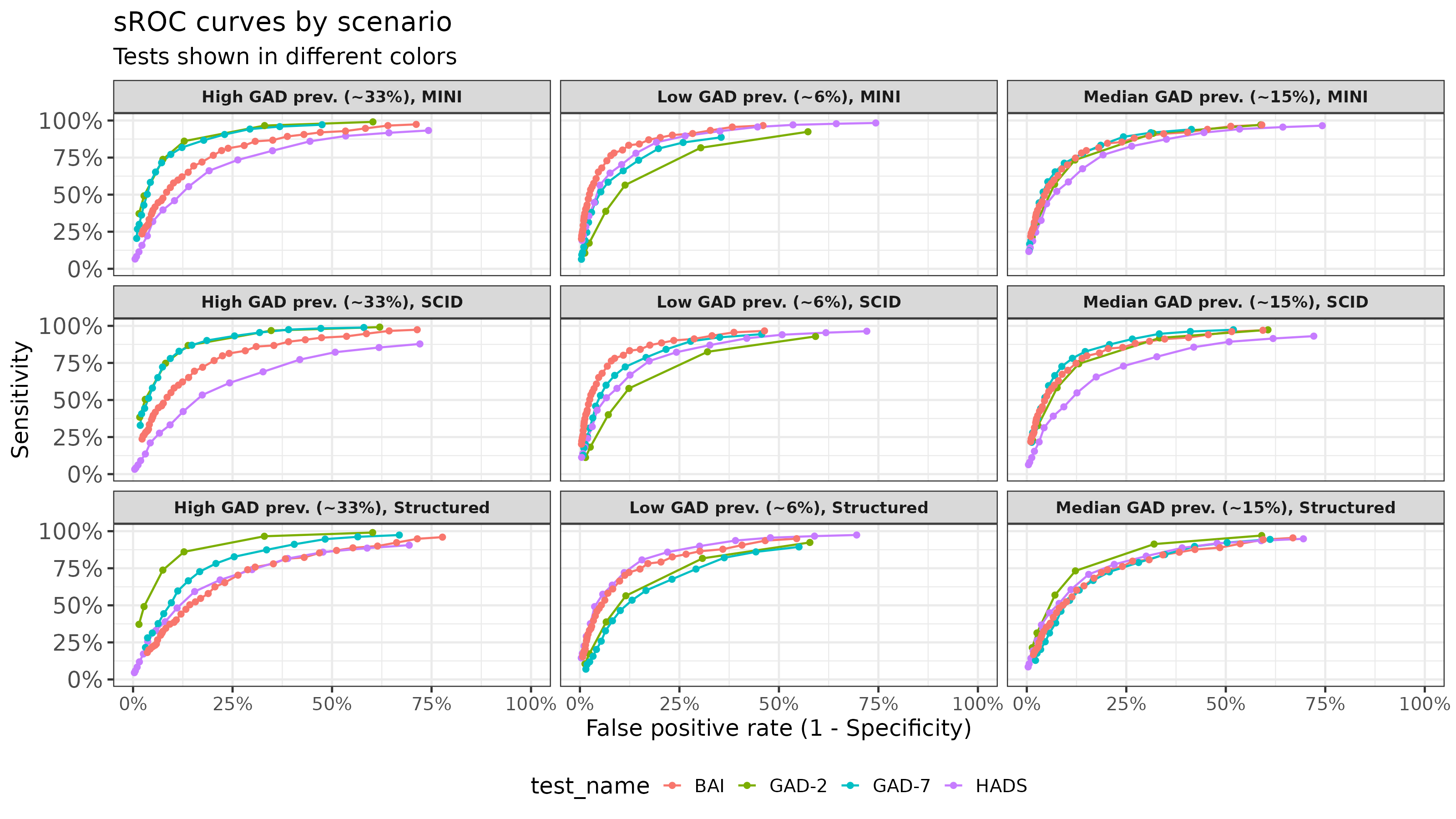}
    \caption{
    \footnotesize{
          Meta-regression sROC plot; \\
          For best-fitting model according to K-fold cross-validation
          (Model A: Fixed-effect thresholds \& compound-symmetry NMA structure),
          with 2 covariates (logit-transformed GAD prevalence and reference test).
    }
    }
    \label{DUMMY_MetaOrdDTA_MR_sROC_3x3_ALL_TESTS_ALL_SCENARIOS}
\end{figure}

\begin{figure}[H]
    \centering
    \includegraphics[width=14cm]{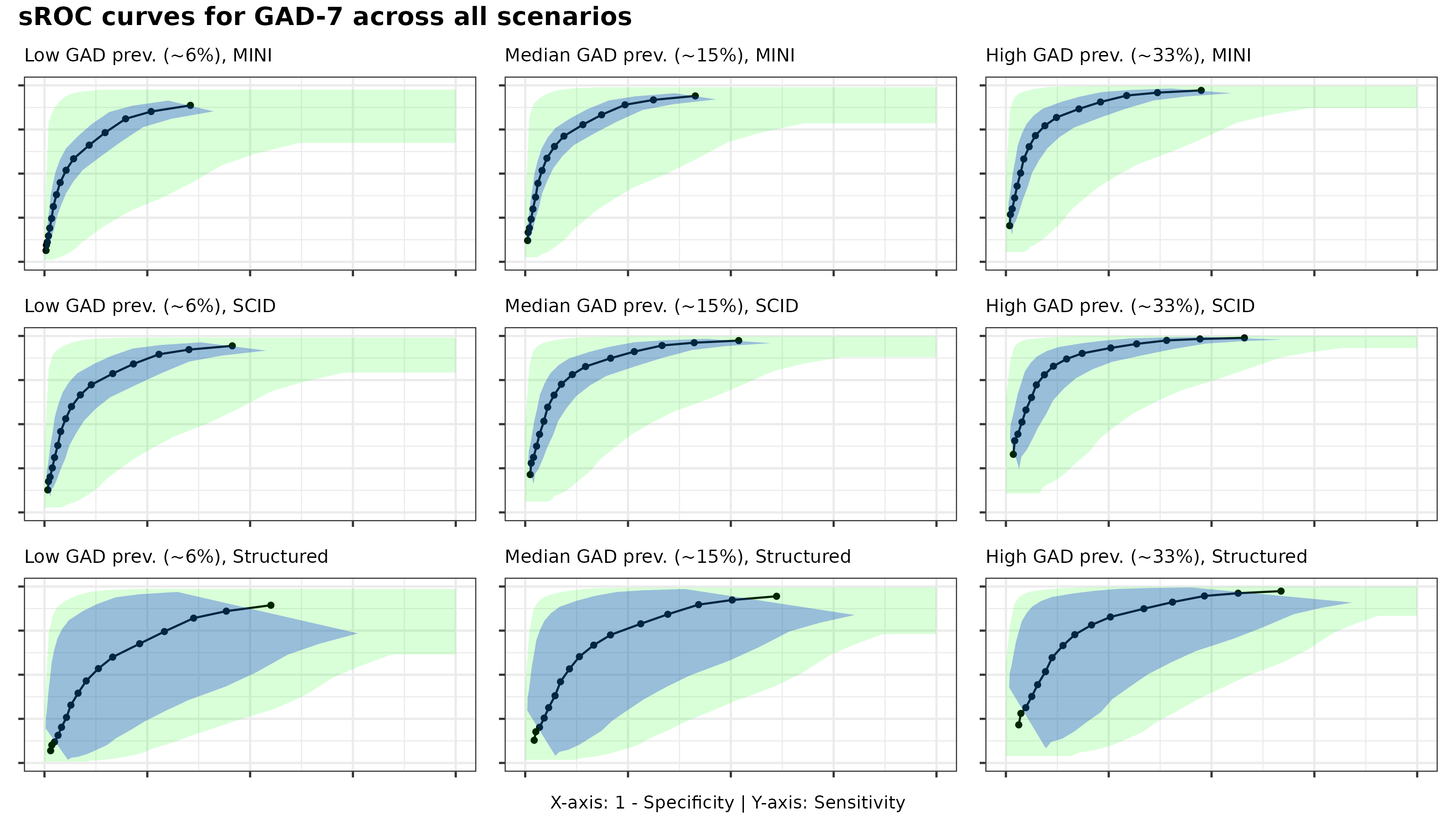}
    \caption{
    \footnotesize{
          Meta-regression sROC plot; \\
          Specifically for GAD-7 test; \\
          With 95\% CrI's and 95\% PrI's;  \\
          For best-fitting model according to K-fold cross-validation
          (Model A: Fixed-effect thresholds \& compound-symmetry NMA structure),
          with 2 covariates (logit-transformed GAD prevalence and reference test).
    }
    }
    \label{DUMMY_MetaOrdDTA_MR_sROC_3x3_GAD_7_w_CrI_PrI}
\end{figure}

%%%%%%%%%%%%%%%%%%%%%%%%%%%%%%%%%%%%%%%%%%%%%%%%%%%%%%%%%%%%%%%%%%%%%%%%%%%%%%%%
%%%%
%%%% -------------------------------------------------------------------------------------------------------------------------------------
The sROC plotting functionality extends naturally to meta-regression, 
displaying how sROC curves vary across covariate scenarios. 
Code Box \ref{code_box_meta_reg_sROC_plots_basic}
demonstrates three layout options for visualizing all nine scenarios simultaneously.
The plots produced from the R code in code box \ref{code_box_meta_reg_sROC_plots_basic} are shown in figures 
\ref{DUMMY_MetaOrdDTA_MR_sROC_2x2_ALL_TESTS_ALL_SCENARIOS},
\ref{DUMMY_MetaOrdDTA_MR_sROC_3x3_ALL_TESTS_ALL_SCENARIOS}, and
\ref{DUMMY_MetaOrdDTA_MR_sROC_3x3_GAD_7_w_CrI_PrI}.
%%%%
%%%% -------------------------------------------------------------------------------------------------------------------------------------
\refstepcounter{codebox} %% increment the counter
\begin{tcolorbox}
[colback=gray!75,colframe=gray!75!black,colupper=black,fonttitle=\color{white}\bfseries,
title=Code box \thecodebox: Meta-regression sROC plots]
\label{code_box_meta_reg_sROC_plots_basic}
\begin{lstlisting}[escapeinside={(*@}{@*)},basicstyle=\scriptsize\ttfamily\color{black},columns=fullflexible]
(*@\textcolor{blue}{\#\# ---- Input relevant thresholds for each test (optional - otherwise shows all thresholds): }@*)  
relevant_thresholds = list( "GAD-2" = c(1:6),  "GAD-7" =  c(3:18), 
                            "HADS" =  c(3:18), "BAI" = c(3:38))                    
(*@\textcolor{blue}{\#\# ---- Note: we can pass optional ggplot customization options }@*) 
by_scenario_grid_sROC_settings <- list()
by_scenario_grid_sROC_settings$ncol <- 3 (*@\textcolor{blue}{\#\# Change to 3, as 3 covariate levels (default = 2): }@*) 
(*@\textcolor{blue}{\#\# ---- Call "plot\_sROC()" R6 method: }@*) 
(*@\textcolor{blue}{\#\# ---- (make sure "\$recompute\_baseline()" is called first, if not already done!): }@*) 
outs_plots_MR_sROC <- MR_model_summary_and_trace_obj$(*@\textcolor{red}{plot\_sROC}@*)( 
                      test_names = test_names,
                      relevant_thresholds = relevant_thresholds,
                      separate_sROC_settings = NULL, (*@\textcolor{blue}{\#\# Meta-reg only }@*) 
                      by_test_sROC_settings = NULL, (*@\textcolor{blue}{\#\# Meta-reg only }@*) 
                      by_scenario_sROC_settings = NULL, (*@\textcolor{blue}{\#\# also for non-meta-reg }@*)
                      by_test_grid_sROC_settings = NULL, (*@\textcolor{blue}{\#\# Meta-reg only }@*) 
                      by_scenario_grid_sROC_settings = by_scenario_grid_sROC_settings,  (*@\textcolor{blue}{\#\# Meta-reg only }@*)
                      by_test_grid_separate_sROC_settings = NULL, (*@\textcolor{blue}{\#\# Meta-reg only }@*) 
                      by_scenario_grid_separate_sROC_settings = NULL) (*@\textcolor{blue}{\#\# also for non-meta-reg }@*)
(*@\textcolor{blue}{\#\# ---- MR sROC plot 1: 2x2 grid - shows all 4 tests + all 9 scenarios. }@*) 
(*@\textcolor{blue}{\#\# ---- Each panel is a test: }@*) 
outs_plots_MR_sROC$grid$grid_by_test
(*@\textcolor{blue}{\#\# ---- MR sROC plot 2: 3x3 grid - shows all 4 tests + all 9 scenarios. }@*)
(*@\textcolor{blue}{\#\# ---- Each panel is a scenario: }@*) 
outs_plots_MR_sROC$grid$grid_by_scenario (*@\textcolor{blue}{ \#\# Note: this is where we changed "ncol" to 3 }@*)
(*@\textcolor{blue}{\#\# ---- MR sROC plot 3: 3x3 grid - but just for GAD-7 test, and all 9 scenarios. }@*) 
(*@\textcolor{blue}{\#\# ---- Also shows the 95\% credible and prediction regions. }@*)
outs_plots_sROC$grid_separate$`grid_separate_test_GAD-7`
\end{lstlisting}
\end{tcolorbox}
%%%%
%%%% -------------------------------------------------------------------------------------------------------------------------------------
A key difference here, compared to the non-meta-regression (i.e., intercept-only) sROC plotting
(see section \ref{MetaOrdDTA_results_baseline_analysis_sROC_plots},
code box \ref{code_box_MetaOrdDTA_base_model_sROC_plots})
is that we can pass more plotting options (as R lists) to the \verb|"$plot_sROC()"| R6 method. 
More specifically, 
just like in section \ref{MetaOrdDTA_results_baseline_analysis_sROC_plots}, 
here we can again pass the 
\verb|by_scenario_sROC_settings| and
\verb|by_scenario_grid_separate_sROC_settings| options.
However, unlike the intercept-only case, 
we can also now pass (for meta-regression only) the following R lists:
\verb|separate_sROC_settings|,
\verb|by_test_sROC_settings|, \newline
\verb|by_test_grid_sROC_settings|, 
\verb|by_scenario_grid_sROC_settings|, \newline
\verb|by_test_grid_separate_sROC_settings|.

%%%% 
%%%%
%%%%
%%%%
%%%%
%%%%
%%%%
%%%% --------------------------------------------------------------------------------------------------------------------------

\begin{figure}[H]
    \centering
    \includegraphics[width=16cm]{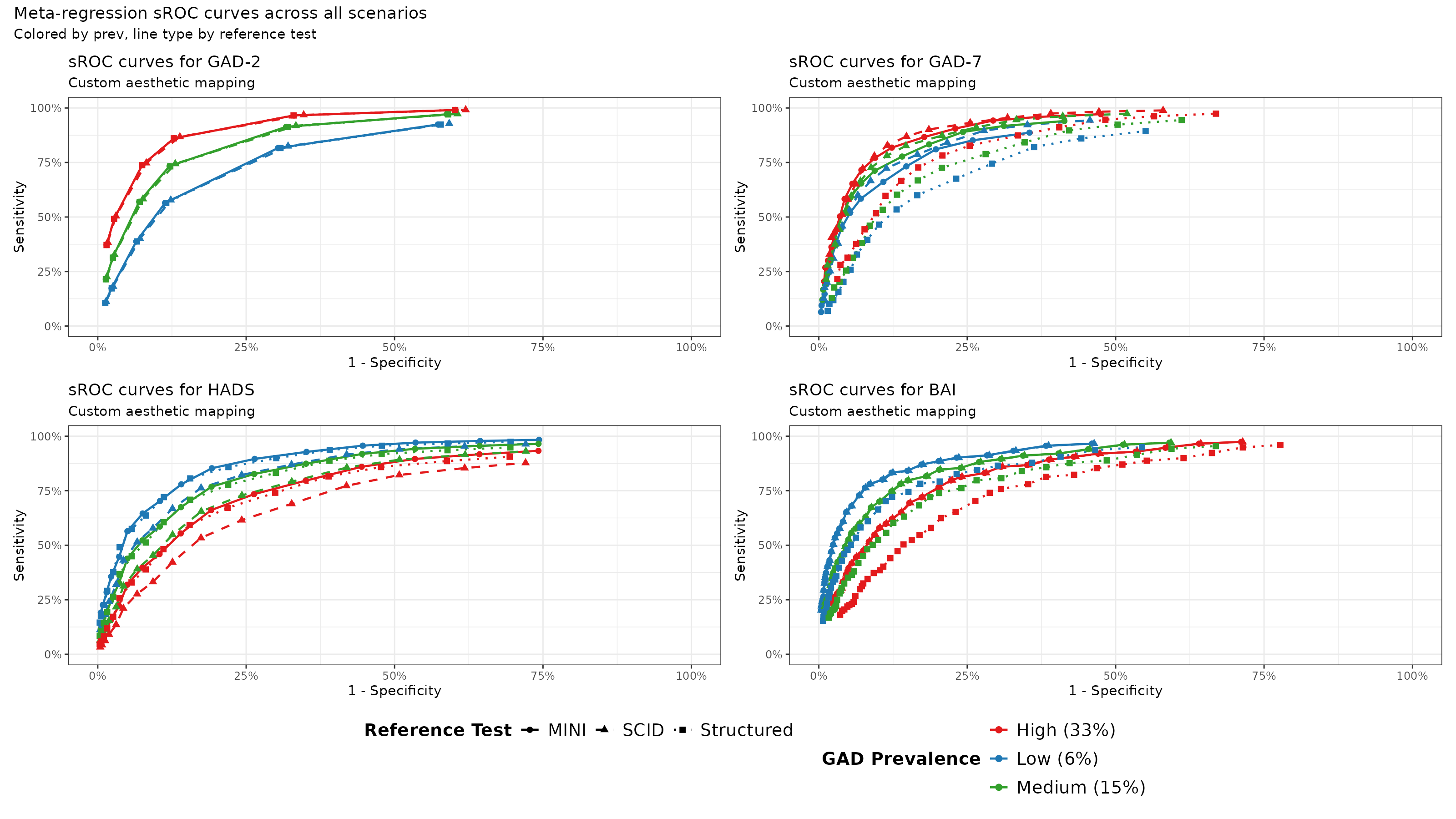}
    \caption{\footnotesize{
          Meta-regression sROC plot; \\
          Note: This is the same as figure 
          \ref{DUMMY_MetaOrdDTA_MR_sROC_2x2_ALL_TESTS_ALL_SCENARIOS},
          but with custom ggplot aesthetic mapping; \\
          For best-fitting model according to K-fold cross-validation \\
          (Model A: Fixed-effect thresholds \& compound-symmetry NMA structure),
          with 2 covariates (logit-transformed GAD prevalence and reference test).
          }}
    \label{DUMMY_MetaOrdDTA_MR_sROC_2x2_ALL_TESTS_ALL_SCENARIOS_custom}
\end{figure}

%%%%%%%%%%%%%%
\paragraph*{\textbf{\underline{Custom aesthetics mapping}}} 
%%%%%%%%%%%%%%
%%%% ------------------------------------------------------------------------------------------------------------------------
For complex meta-regression models with multiple covariates, custom aesthetic mappings
can enhance interpretation by encoding different covariate levels through color, 
line type, and shape. 
Code Box \ref{code_box_meta_reg_sROC_plots_custom_mapping} demonstrates
parsing scenario names into structured variables for systematic visualization.
%%%%
%%%% --------------------------------------------------------------------------------------------------------------------------
\refstepcounter{codebox} %% increment the counter
\begin{tcolorbox}
[colback=gray!75,colframe=gray!75!black,colupper=black,fonttitle=\color{white}\bfseries,
title=Code box \thecodebox: Meta-regression sROC plots with custom aesthetics mapping]
\label{code_box_meta_reg_sROC_plots_custom_mapping}
\begin{lstlisting}[escapeinside={(*@}{@*)},basicstyle=\footnotesize\ttfamily\color{black},columns=fullflexible]
(*@\textcolor{blue}{\#\# ---- Define custom aesthetic mapping: }@*)
my_aesthetic_mapping <- list(
  parse_fn = function(data) { (*@\textcolor{blue}{\#\# Parsing function to create new variables }@*)
        data %>% mutate(    prev_level = case_when(
                              grepl("Low.*prev", scenario) ~ "Low (6%)",
                              grepl("Median.*prev", scenario) ~ "Medium (15%)",
                              grepl("High.*prev", scenario) ~ "High (33%)",
                              TRUE ~ "Unknown"),    
                            ref_test = case_when(
                              grepl("MINI", scenario) ~ "MINI",
                              grepl("SCID", scenario) ~ "SCID",
                              grepl("Structured", scenario) ~ "Structured",
                              TRUE ~ "Unknown"))
  },
  (*@\textcolor{blue}{\#\# ---- Variable mappings: }@*) 
  color_var = "prev_level", linetype_var = "ref_test", shape_var = "ref_test",
  (*@\textcolor{blue}{\#\# ---- Custom scales: }@*) 
  color_scale = scale_color_manual(
    values = c("Low (6%)" = "#1f78b4", "Medium (15%)" = "#33a02c", "High (33%)" = "#e31a1c"),
    name = "GAD Prevalence"),
  linetype_scale = scale_linetype_manual(
    values = c("MINI" = "solid", "SCID" = "dashed", "Structured" = "dotted"),
    name = "Reference Test"),
  shape_scale = scale_shape_manual(
    values = c("MINI" = 16,  "SCID" = 17, "Structured" = 15), name = "Reference Test")
)
(*@\textcolor{blue}{\#\# ---- Run "\$plot\_sROC()" R6 method w/ aesthetic_mapping option: }@*)
plots <- model_summary_and_trace_obj$(*@\textcolor{red}{plot\_sROC}@*)( test_names = test_names, 
                                                layout = "by_test",
                                                relevant_thresholds = relevant_thresholds,
                                                aesthetic_mapping = my_aesthetic_mapping)      
(*@\textcolor{blue}{\#\# ---- Create grid with the enhanced plots: }@*)
custom_grid <- (plots$by_test[["by_test_GAD-2"]] + plots$by_test[["by_test_GAD-7"]]) /
               (plots$by_test[["by_test_HADS"]]  + plots$by_test[["by_test_BAI"]]) +
  plot_layout(guides = "collect") +
  plot_annotation(
    title = "Meta-regression sROC curves across all scenarios",
    subtitle = "Colored by prev, line type by reference test") & 
  theme_bw() + 
  theme(legend.position = "bottom", legend.box = "horizontal",
        legend.title = element_text(size = 14, face = "bold"), 
        legend.text = element_text(size = 14)) 
\end{lstlisting}
\end{tcolorbox}
%%%%
%%%% -------------------------------------------------------------------------------------------------------------------------------------
%%
Figure \ref{DUMMY_MetaOrdDTA_MR_sROC_2x2_ALL_TESTS_ALL_SCENARIOS_custom} 
demonstrates the custom ggplot aesthetics mapping feature,
which allows us to more easily see how test accuracy varies with both prevalence (colour gradient) and reference test (line type).
%%%%
%%%%
%%%%
%%%%

%%%%%%%%%%%%%%%%%%%%%%%%%%%%%%%%%%%%%%%%%%%%%%%%%%%%%%%%%%%%%%%%%%%%%%%%%%%%%%%%
\subsubsection{ Plots: Accuracy vs. threshold }
\label{MetaOrdDTA_results_meta_reg_plots_accuracy_vs_thr}
\begin{figure}[H]
    \centering
    \includegraphics[width=14cm]{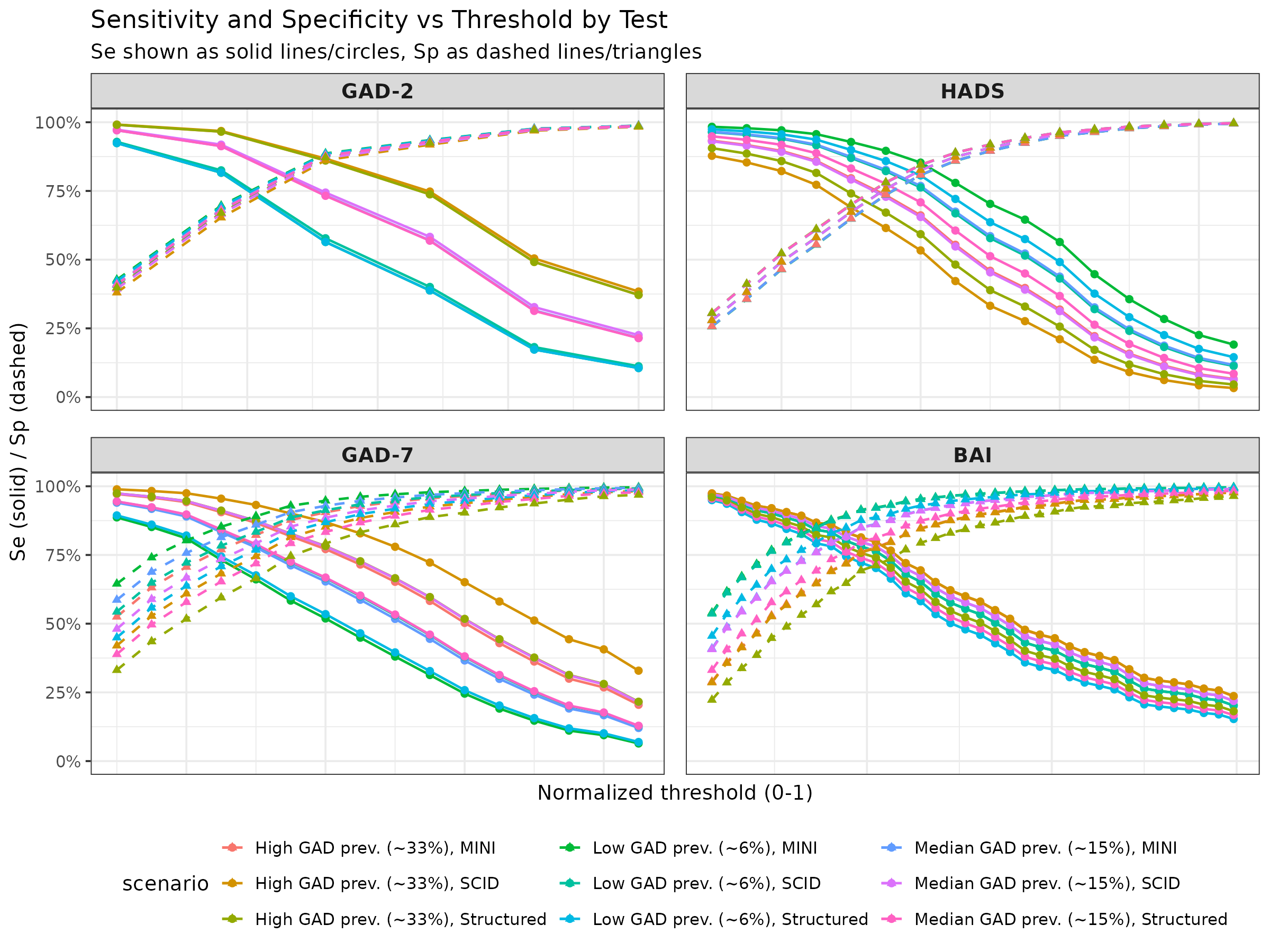}
    \caption{\footnotesize{
          Meta-regression accuracy vs threshold Plot; \\
          For best-fitting model according to K-fold cross-validation \\
          (Model A: Fixed-effect thresholds \& compound-symmetry NMA structure); \\
          With 2 covariates (logit-transformed GAD prevalence and reference test).
    }}
    \label{DUMMY_MetaOrdDTA_MR_accuracy_vs_thr_2x2_ALL_TESTS_ALL_SCENARIOS}
\end{figure}

%%%%%%%%%%%%%%%%%%%%%%%%%%%%%%%%%%%%%%%%%%%%%%%%%%%%%%%%%%%%%%%%%%%%%%%%%%%%%%%%
%%%%
%%%% -------------------------------------------------------------------------------------------------------------------------------------
Code snippet box \ref{code_box_meta_reg_accuracy_vs_thr_plots} 
shows how we can plot accuracy vs. threshold plots for our fitted model, 
by using the "\verb|$plot_threshold()|" R6 class method.
This is very similar to the non-meta-regression procedure 
(see code box \ref{code_box_MetaOrdDTA_base_model_accuracy_vs_thr_plots}, 
section \ref{MetaOrdDTA_results_baseline_analysis_accuracy_vs_thr_plots}).
%%%%
%%%% -------------------------------------------------------------------------------------------------------------------------------------
\refstepcounter{codebox} %% increment the counter
%% \begin{tcolorbox}[colback=black!75,colframe=gray!75!black,title=Code Snippet \thecodebox: Plotting sROC plots]
\begin{tcolorbox}
[colback=gray!75,colframe=gray!75!black,colupper=black,fonttitle=\color{white}\bfseries,
title=Code box \thecodebox: Meta-regression: Accuracy vs. threshold plots]
\label{code_box_meta_reg_accuracy_vs_thr_plots}
%% \begin{lstlisting}[escapeinside={(*@}{@*)},basicstyle=\footnotesize\ttfamily,columns=fullflexible]
\begin{lstlisting}[escapeinside={(*@}{@*)},basicstyle=\scriptsize\ttfamily\color{black},columns=fullflexible]
(*@\textcolor{blue}{\#\# ---- First, make sure recompute_baseline() is run (if not already e.g. for sROC plots, etc): }@*) 
new_MR_baseline_outs <-  MR_model_summary_and_trace_obj$(*@\textcolor{red}{recompute\_baseline}@*)(
                              test_names = test_names,
                              MR_scenarios = all_MR_scenarios)
(*@\textcolor{blue}{\#\# ---- Call "plot\_threshold()" R6 method: }@*) 
outs_plots_thr <- MR_model_summary_and_trace_obj$(*@\textcolor{red}{plot\_threshold}@*)( 
                                test_names = test_names, 
                                plot_type = "both",  (*@\textcolor{blue}{ \#\# either "Se", "Sp", or "both" }@*)
                                show_actual_thresholds = FALSE, 
                                relevant_thresholds = relevant_thresholds)           
(*@\textcolor{blue}{\#\# ---- Grid by test (2x2 grid) [shown in manuscript]: }@*) 
outs_plots_thr$grid$both_joined_grid_by_test
(*@\textcolor{blue}{\#\# ---- Grid by scenario (3x3 grid) [NOT shown in manuscript]: }@*) 
outs_plots_thr$grid$Se_grid_by_scenario
outs_plots_thr$grid$Sp_grid_by_scenario
outs_plots_thr$grid$both_joined_grid_by_scenario
(*@\textcolor{blue}{\#\# ---- Grid by test with 95\% CrI and PrI - GAD-7 only (2x2 grid) [NOT shown in manuscript]: }@*) 
outs_plots_thr$grid_separate$"both_joined_grid_separate_test_GAD-7"
\end{lstlisting}
\end{tcolorbox}
%%%%
%%%% --------------------------------------------------------------------------------------------------
Figure \ref{DUMMY_MetaOrdDTA_MR_accuracy_vs_thr_2x2_ALL_TESTS_ALL_SCENARIOS} 
displays sensitivity and specificity curves across all nine scenarios in a $2 \times 2$ grid. 
Each panel represents one test, with curves colored by prevalence and distinguished by reference standard (line type). 
This visualization reveals how accuracy-threshold relationships shift with covariate values.
Alternative layouts available through \verb|$plot_threshold()| include for example:
scenario-specific grids (\verb|$grid$both_joined_grid_by_scenario|), 
and individual test panels with uncertainty bands (\verb|$grid_separate|) -
useful for detailed examination of specific contexts.

%%%%
%%%% --------------------------------------------------------------------------------------------------
%%%% 
%%%%
%%%%
%%%%
% \begin{figure}[H]
%     \centering
%     \includegraphics[width=14cm]{Figures/DUMMY_MetaOrdDTA_MR_accuracy_vs_thr_3x3_ALL_TESTS_ALL_SCENARIOS.png}
%     \caption{
%       Meta-regression accuracy vs threshold plot; \\
%       For best-fitting model according to K-fold cross-validation (Model A); \\
%       With 2 covariates (logit-transformed GAD prevalence and reference test).
%     }
%     \label{DUMMY_MetaOrdDTA_MR_accuracy_vs_thr_3x3_ALL_TESTS_ALL_SCENARIOS}
% \end{figure}
% %%%%
%%%%
% \begin{figure}[H]
%     \centering
%     \includegraphics[width=14cm]{Figures/DUMMY_MetaOrdDTA_MR_accuracy_vs_thr_3x3_GAD_7_w_CrI_PrI.png}
%     \caption{
%       Meta-regression accuracy vs threshold plot; \\
%       Specifically for GAD-7 test; \\
%       With 95\% CrI's and 95\% PrI's; \\
%       For best-fitting model according to K-fold cross-validation (Model A); \\
%       With 2 covariates (logit-transformed GAD prevalence and reference test).
%     }
%     \label{DUMMY_MetaOrdDTA_MR_accuracy_vs_thr_3x3_GAD_7_w_CrI_PrI}
% \end{figure}
%%%%
%%%% ------------------------------------------------------------------------------------------------------------------------

\begin{figure}[H]
    \centering
    \includegraphics[width=16cm]{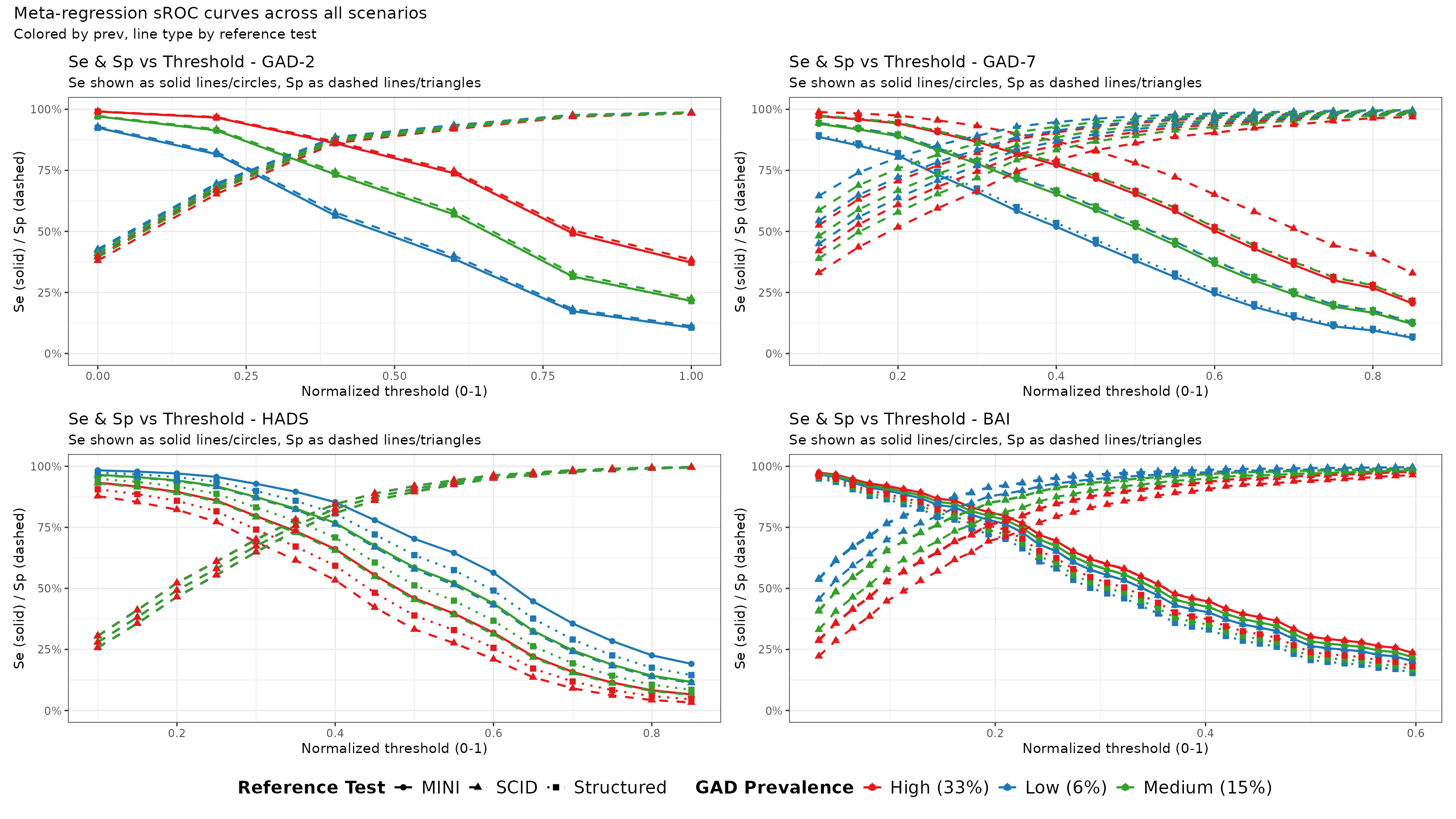}
    \caption{\footnotesize{
          Meta-regression accuracy vs threshold plot; \\
          Note: This is the same as 
          figure \ref{DUMMY_MetaOrdDTA_MR_accuracy_vs_thr_2x2_ALL_TESTS_ALL_SCENARIOS},
          but with custom ggplot aesthetic mapping); \\
          For best-fitting model according to K-fold cross-validation (Model A); \\ 
          With 2 covariates (logit-transformed GAD prevalence and reference test).
    }}
    \label{DUMMY_MetaOrdDTA_MR_accuracy_vs_thr_2x2_ALL_TESTS_ALL_SCENARIOS_custom}
\end{figure}

%%%%%%%%%%%%%%
\paragraph*{\textbf{\underline{Custom aesthetics mapping}}} 
%%%%%%%%%%%%%%
%%%% ------------------------------------------------------------------------------------------------------------------------
For complex meta-regression models with multiple covariates, custom aesthetic mappings
can enhance interpretation by encoding different covariate levels through color, 
line type, and shape. 
Code Box \ref{code_box_meta_reg_accuracy_vs_thr_plots_custom_mapping} demonstrates
parsing scenario names into structured variables for systematic visualization.
%%%%
%%%% ------------------------------------------------------------------------------------------------------------------------
\refstepcounter{codebox} %% increment the counter
\begin{tcolorbox}
[colback=gray!75,colframe=gray!75!black,colupper=black,fonttitle=\color{white}\bfseries,
title=Code box \thecodebox: Meta-regression: Accuracy vs. threshold plots with custom aesthetics mapping]
\label{code_box_meta_reg_accuracy_vs_thr_plots_custom_mapping}
\begin{lstlisting}[escapeinside={(*@}{@*)},basicstyle=\scriptsize\ttfamily\color{black},columns=fullflexible]
(*@\textcolor{blue}{\#\# ---- Run "\$plot\_sROC()" R6 method with "aesthetic_mapping" option: }@*)
(*@\textcolor{blue}{\#\# ---- Use custom aesthetic mapping ("my_aesthetic_mapping") which we defined previously for sROC plots: }@*)
outs_plots_thr <- MR_model_summary_and_trace_obj$(*@\textcolor{red}{plot\_threshold}@*)( 
                                         test_names = test_names, 
                                         plot_type = "both",
                                         show_actual_thresholds = FALSE,
                                         relevant_thresholds = relevant_thresholds, 
                                         aesthetic_mapping = my_aesthetic_mapping)
(*@\textcolor{blue}{\#\# ---- Create grid with the enhanced plots: }@*)
custom_grid <- (outs_plots_thr$by_test[["both_joined_by_test_GAD-2"]] +
                outs_plots_thr$by_test[["both_joined_by_test_GAD-7"]]) /
               (outs_plots_thr$by_test[["both_joined_by_test_HADS"]]  +
                outs_plots_thr$by_test[["both_joined_by_test_BAI"]]) +
  plot_layout(guides = "collect") +
  plot_annotation(
    title = "Meta-regression sROC curves across all scenarios",
    subtitle = "Colored by prev, line type by reference test") & 
  theme_bw() + 
  theme(legend.position = "bottom",
        legend.box = "horizontal",
        # nrow = 1,
        legend.title = element_text(size = 14, face = "bold"),
        legend.text = element_text(size = 14)) 
\end{lstlisting}
\end{tcolorbox}
%%%%
%%%% -------------------------------------------------------------------------------------------------------------------------------
Figure \ref{DUMMY_MetaOrdDTA_MR_accuracy_vs_thr_2x2_ALL_TESTS_ALL_SCENARIOS_custom} 
demonstrates the custom ggplot aesthetics mapping feature,
which allows us to more easily see how test accuracy varies 
with both prevalence (colour gradient) and reference test (line type).
%%%%
%%%%
%%%%
%%%%
%%%%%%%%%%%%%%%%%%%%%%%%%%%%%%%%%%%%%%%%%%%%%%%%%%%%%%%%%%%%%%%%%%%%%%%%%%%%%%%%
\subsubsection{ Plots and tables: Pairwise accuracy differences}
\label{MetaOrdDTA_results_meta_reg_plots_pairwise_diffs}
%%%%%%%%%%%%%%%%%%%%%%%%%%%%%%%%%%%%%%%%%%%%%%%%%%%%%%%%%%%%%%%%%%%%%%%%%%%%%%%%
%%%% -------------------------------------------------------------------------------------------------------------------------------------
Pairwise comparisons in meta-regression reveal how relative test performance varies across clinical contexts. 
The \verb|$compute_pairwise_comparisons()| method 
(see code box \ref{code_box_meta_reg_pairwise_diffs_plots})
efficiently computes these using stored MCMC traces through the internal Rcpp (C++) implementation, avoiding model re-fitting.

\begin{figure}[H]
    \centering
    \includegraphics[width=16cm]{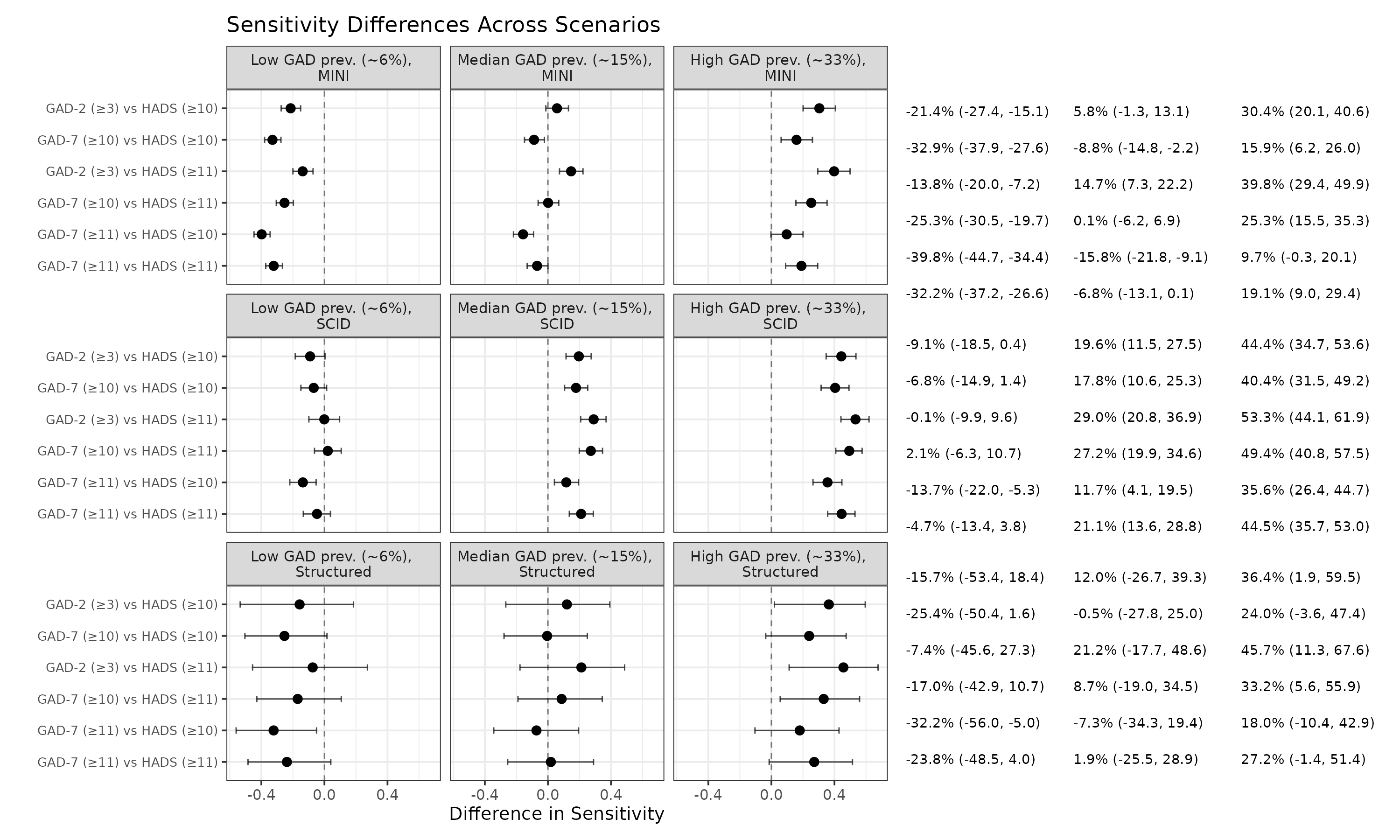}
    \caption{\footnotesize{
      Meta-regression sensitivity pairwise differences with $95\%$ CrI's; \\
      For best-fitting model according to K-fold cross-validation (Model A).
    }}
    \label{DUMMY_MetaOrdDTA_MR_Se_pairwise_diffs_screening_all}
\end{figure}

\begin{figure}[H]
    \centering
    \includegraphics[width=16cm]{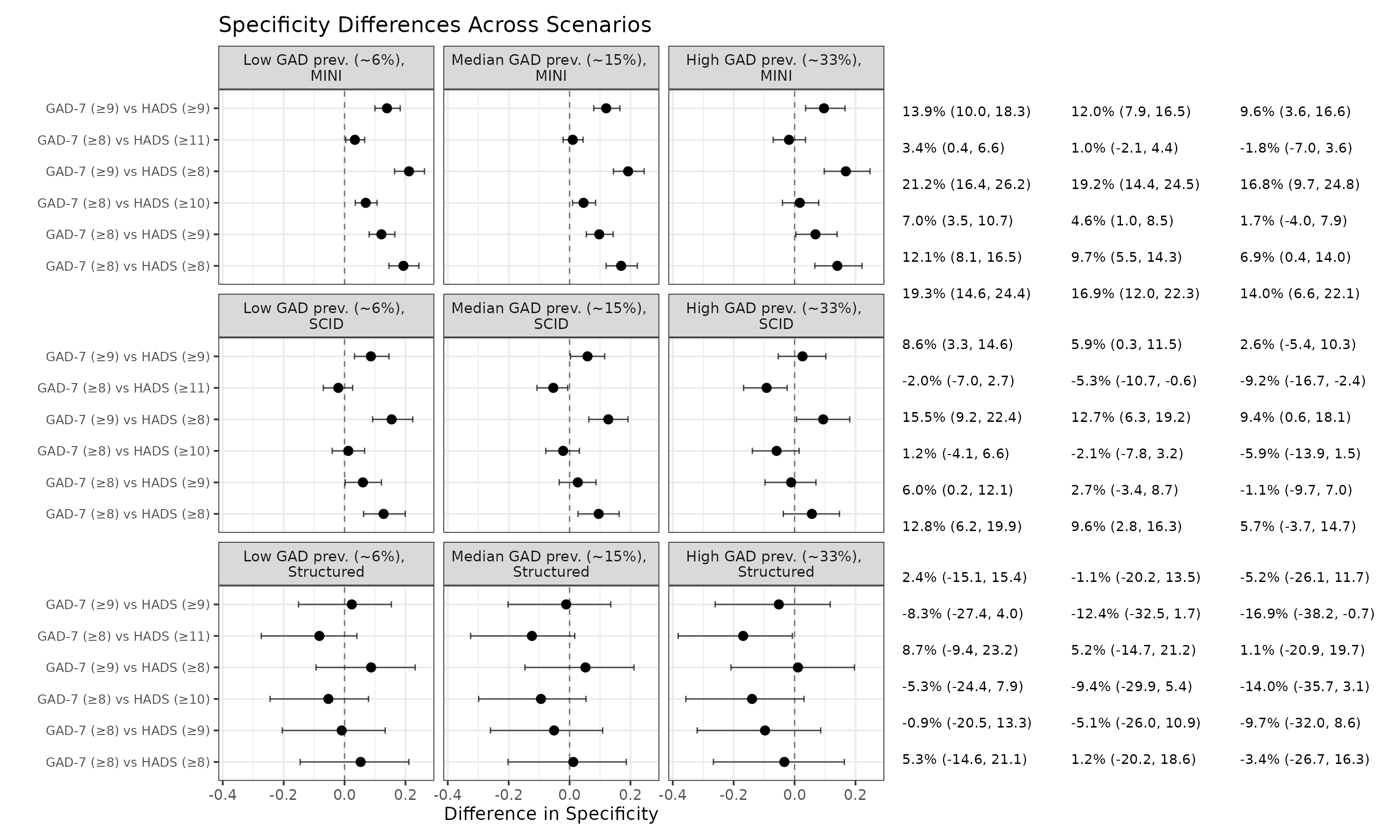}
    \caption{\footnotesize{
      Meta-regression specificity pairwise differences with $95\%$ CrI's; \\
      For best-fitting model according to K-fold cross-validation (Model A).
    }}
    \label{DUMMY_MetaOrdDTA_MR_Sp_pairwise_diffs_screening_all}
\end{figure}

%%%% -------------------------------------------------------------------------------------------------------------------------------------
Note that when we run \verb|$compute_pairwise_comparisons()|,
we can set \verb|"recompute_baseline = FALSE"|
because we ran \verb|$compute_baseline()| first. 
However, if you have not run this first, then switch this to \verb|TRUE|
(or just leave it out - as it will recompute baseline automatically).
%%%%
%%%% -------------------------------------------------------------------------------------------------------------------------
\refstepcounter{codebox} %% increment the counter
\begin{tcolorbox}
[colback=gray!75,colframe=gray!75!black,colupper=black,fonttitle=\color{white}\bfseries,
title=Code box \thecodebox: Meta-regression: Pairwise accuracy differences plots]
\label{code_box_meta_reg_pairwise_diffs_plots}
\begin{lstlisting}[escapeinside={(*@}{@*)},basicstyle=\scriptsize\ttfamily\color{black},columns=fullflexible]
(*@\textcolor{blue}{ \#\# ---- Make R list w/ thresholds of interest for each test (not including BAI for this example): }@*)
thr_screening = list(
      "GAD-2" = c(3),              (*@\textcolor{blue}{ \#\# 3 is typical screening thr. for GAD }@*)
      "GAD-7" = c(8, 9, 10),       (*@\textcolor{blue}{ \#\# 10 is a typical screening thr. }@*)
      "HADS" = c(9, 10, 11))
(*@\textcolor{blue}{ \#\# ---- Make sure "\$recompute\_baseline()" is run first (runs internal fast C++ fn): }@*)
MR_baseline_outs <-  MR_model_summary_and_trace_obj$(*@\textcolor{red}{recompute\_baseline}@*)(
                                                test_names = test_names,
                                                MR_scenarios = all_MR_scenarios)
(*@\textcolor{blue}{ \#\# ---- Then run "\$compute\_pairwise\_comparisons()" for NMA pairwise Se/Sp (runs internal fast C++ fn): }@*)
MR_pairwise_baseline_outs <- MR_model_summary_and_trace_obj$(*@\textcolor{red}{compute\_pairwise\_comparisons}@*)(
                                                                  test_names = test_names,
                                                                  recompute_baseline = FALSE,
                                                                  MR_scenarios = all_MR_scenarios)                                
(*@\textcolor{blue}{ \#\# ---- Modify MR plot settings ("metareg_plot_settings"): }@*)
metareg_plot_settings <- list() (*@\textcolor{blue}{ \#\# do NOT use "baseline_plot_settings" for meta-reg }@*)
metareg_plot_settings$show_values <- TRUE (*@\textcolor{blue}{ \#\# also shows posterior medians and 95\% CrI's for pairwise diffs }@*)
metareg_plot_settings$facet_ncol <- 3 (*@\textcolor{blue}{ \#\#  default is 2 }@*)
metareg_plot_settings$facet_scales <- "fixed"
metareg_plot_settings$ratio_plot <- 3.0 (*@\textcolor{blue}{ \#\# default is 3 }@*)
metareg_plot_settings$ratio_text <- 2.25 (*@\textcolor{blue}{ \#\# default is 1.5 }@*)
(*@\textcolor{blue}{ \#\# ---- Call the "\$plot\_pairwise()" R6 class method: }@*)
outs_plot_pairwise <- MR_model_summary_and_trace_obj$(*@\textcolor{red}{plot\_pairwise}@*)( 
                                comparison_type = "both", (*@\textcolor{blue}{ \#\# either "Se", "Sp", or "both" }@*) 
                                test_names = c("GAD-2", "GAD-7", "HADS"), 
                                relevant_thresholds = thr_screening,                     
                                baseline_plot_settings = NULL, (*@\textcolor{blue}{ \#\# NOT for meta-reg }@*)  
                                metareg_plot_settings = metareg_plot_settings) (*@\textcolor{blue}{ \#\# optional, only for MR }@*)
outs_plot_pairwise$Se  (*@\textcolor{blue}{ \#\# Se pairwise diffs only }@*)    
outs_plot_pairwise$Sp  (*@\textcolor{blue}{ \#\# Sp pairwise diffs only }@*)
outs_plot_pairwise$combined (*@\textcolor{blue}{ \#\# Se and Sp pairwise diffs, side-by-side }@*)
\end{lstlisting}
\end{tcolorbox}
%%%%
%%%% ------------------------------------------------------------------------------------------------------------------------
Figures \ref{DUMMY_MetaOrdDTA_MR_Se_pairwise_diffs_screening_all} and 
\ref{DUMMY_MetaOrdDTA_MR_Sp_pairwise_diffs_screening_all} 
present sensitivity and specificity differences across the nine scenarios.

%%%%
%%%% ------------------------------------------------------------------------------------------------------------------------
Other arguments which can be passed onto the R list \verb|metareg_plot_settings|
(similarly to the \verb|baseline_plot_settings| R list in 
section \ref{MetaOrdDTA_results_baseline_analysis_pairwise_diffs} - 
for the non-meta-regression case) include:
\verb|base_size|,  %% yes
\verb|point_size|, %% yes
%% \verb|line_size|, %% no
\verb|error_bar_width|, %% yes
\verb|error_bar_alpha|, %% yes
%% \verb|legend_position|, %% no
%% \verb|facet_ncol|, %% no
%% \verb|facet_scales|, %% no
\verb|y_axis_text_size_multiplier|, %% yes
\verb|show_values| (as we already mentioned), %% yes
\verb|value_size_multiplier|, %% yes
\verb|ratio_plot|, %% yes
and
\verb|ratio_text|. %% yes
However, there are some additional arguments we can put into this R list  - 
which are not relevant for the non-meta-regression case
(section \ref{MetaOrdDTA_results_baseline_analysis_pairwise_diffs}) - 
these additional meta-regression-only arguments are:
\verb|legend_position| (default = \verb|"bottom"|),
\verb|facet_ncol| (default = $2$),
\verb|facet_scales| (default = \verb|"fixed"|),
\verb|n_auto_select| (default = $6$), and
\verb|strip_text_size_multiplier| (default = $0.8$).

%%%%
%%%% ---------------------------------------------------------------------------------------------------------------------------
%%%%
%%%%
\newpage
\section{ Discussion}
\label{section_discussion}
%%%%%%%%%%%%%%%%%%%%%%%%%%%%%%%%%%%%%%%%%%%%%%%%%%%%%%%%%%%%%%%%%%%%%%%%%%%%%%%%%%%%%%%%%%%%%%%%%%%%%%%%%%%%%%%%%%%%%%%%%
%%%%
%%%%
%%%%%%%%%%%%%%%%%%%%%%%%%%%%%%%%%%%%%%%%%%%%%%%%%%%%%%%%%%%%%%%%%%%%%%%%%%
\subsection{ Discussion of simulation study findings}
\label{discussion_summary_key_findings}
%%%%%%%%%%%%%%%%%%%%%%%%%%%%%%%%%%%%%%%%%%%%%%%%%%%%%%%%%%%%%%%%%%%%%%%%%%
%%%%
%%%% ----------------------------------------------------------------------------------------------------------------------
Our simulation study (see section \ref{Sim_study}) compared five modeling approaches for ordinal diagnostic tests:
\begin{itemize}
    \item[(i)] The three proposed ordinal models (see section \ref{section_model_specs}),
    that properly treat ordinal tests as ordinal
    (O-bivariate-FC, O-bivariate-RC, O-HSROC-RC).
    \item[(ii)] The Jones model (Jones et al, 2019\supercite{Jones2019}) -
    which also handles multiple thresholds -
    but assumes that the tests are continuous (e.g. biomarkers).
    However, it can also be used for ordinal tests and,
    hence we compared our proposed models against this.
    \item[(iii)] The standard stratified-bivariate approach
    (Reitsma et al, 2005\supercite{Reitsma2005}),
    this partitions/dichotomizes the data at each observed test threshold,
    and then applies the standard bivariate model\supercite{Reitsma2005} to that subset of data.
    Hence, this method cannot make use of all available data -
    unless every single study reports accuracy at every single cutoff evaluated -
    which is rarely the case in practice.
\end{itemize}
RMSE served as our primary performance measure, 
because it captures both bias and variance:
RMSE$^2$ = Bias$^2$ + Variance.
A model with low bias but high variance produces wide credible intervals with poor precision,
whilst high bias with low variance yields more narrow intervals that miss the true value.
RMSE appropriately penalizes both problems, making it superior to bias alone for overall model evaluation.
%%%%
%%%%
%%%%%%%%%%%%%%%%%%%%%%%%%%%%%%%%%%%%%%
\subsubsection{ Discussion of simulation study findings: Jones DGM (continuous assumption)}
\label{discussion_jones_dgm}
%%%%%%%%%%%%%%%%%%%%%%%%%%%%%%%%%%%%%%
%%%%
%%%% ----------------------------------------------------------------------------------------------------------------------
%%
%% ---- Key finding: ordinal models robust to continuous assumption
%%
The Jones DGM was included to test whether the proposed ordinal models (see section \ref{section_model_specs})
perform effectively even when the true DGM assumes continuity.
Ordinal models demonstrated strong robustness to this model misspecification.

%%%%%%%%%%%%%%%%%%%%%%%%%%%%%%%%
\paragraph{\underline{ 
Jones DGM: GAD-2
}}
\label{discussion_jones_dgm_GAD_2}
%%%%%%%%%%%%%%%%%%%%%%%%%%%%%%%%
%%%%
%%%% ----------------------------------------------------------------------------------------------------------------------
For the GAD-2 (see section \ref{Sim_study_GAD_2})
with $10$-studies under the Jones DGM (see section \ref{Sim_study_GAD_2_Jones_DGM_1}),
the continuous-assumption Jones model achieved the best RMSE ($9.75$) -
as one would typically expect - given correct model specification.
However, the ordinal O-bivariate-FC performed competitively ($RMSE = 10.25$) -
falling into "Group 2 - worse (stat only)" -
with just $5.2\%$ (less than our $10\%$ "practically significant" threshold)
relative difference compared to the Jones model.
Furthermore, both the O-bivariate-FC and the Jones model substantially outperformed the stratified-bivariate model
(RMSE(strat-biv) = $11.76$ - $20.6\%$ worse than Jones).

%%%%
%%%% ---------------------------------------------------------------------------------------------------------------------
With $50$-studies, the ordinal O-bivariate-FC model joined Jones in the "best" group
(RMSE = $4.81$ vs. $4.55$ for Jones), whilst stratified-bivariate remained in "Group 4 - Worse"
(RMSE = $5.32$, rel. diff = $17.0\%$).
%%%%
%%%%
%%%%%%%%%%%%%%%%%%%%%%%%%%%%%%%%
\paragraph{\underline{ 
Jones DGM: HADS
}}
\label{discussion_jones_dgm_HADS}
%%%%%%%%%%%%%%%%%%%%%%%%%%%%%%%%
%%%%
%%%% ---------------------------------------------------------------------------------------------------------------------
For the HADS (see section \ref{Sim_study_HADS}, sub-section \ref{Sim_study_HADS_Jones_DGM_1})
under the Jones DGM, with $10$ studies,
both the Jones and the ordinal O-bivariate-FC made it into the "Best" group
(see table \ref{Table:summary_table_overall_HADS};
$RMSE = 5.59$ and $5.84$ for Jones and O-bivariate-FC, respectively),
with the stratified-bivariate model being the only model falling into the "Worse" group
($RMSE = 6.98$) - being $24.8\%$ worse than the leading O-bivariate-FC model.

%%%%
%%%% ---------------------------------------------------------------------------------------------------------------------
With $50$ studies, the pattern persisted:
the Jones ($RMSE = 3.22$) and O-bivariate-FC ($RMSE = 3.29$) models remained statistically equivalent -
with the Jones model now being only $2.4\%$ worse than the O-bivariate-FC model.
Furthermore, the stratified-bivariate model ($RMSE = 3.56$) showed $10.8\%$ ($> 10\%$) worse RMSE
than the leading Jones model - being both practically and statistically significantly worse than the
leading Jones model.
%%%%
%%%%
%%%%%%%%%%%%%%%%%%%%%%%%%%%%%%%%
\paragraph{\underline{ 
Jones DGM: BAI
}}
\label{discussion_jones_dgm_BAI}
%%%%%%%%%%%%%%%%%%%%%%%%%%%%%%%%
%%%%
%%%% ---------------------------------------------------------------------------------------------------------------------
For the BAI under the Jones DGM
(see section \ref{Sim_study_BAI}, sub-section \ref{Sim_study_BAI_Jones_DGM_1}),
the continuous-assumption Jones model showed clearer advantages with $10$ studies,
being the only model in the "Best" group ($RMSE = 6.14$).
Furthermore, the ordinal O-bivariate-FC model fell into the "worse (stat only)" group
($RMSE = 6.59\% - 7.4\%$ worse than the leading Jones model),
whilst the stratified-bivariate model performed poorly and fell into the "worse" group
($RMSE = 8.42\% - 37.2\%$ worse than the leading Jones model).

%%%%
%%%% ---------------------------------------------------------------------------------------------------------------------
However, with $50$-studies, both Jones and O-bivariate-FC made it into the "Best" group
($RMSE = 2.64$ and $2.87$ for Jones and O-bivariate-FC models, respectively),
suggesting that the Jones and O-bivariate-FC models perform just as well as one another here with sufficient data.

%%%%
%%%% ---------------------------------------------------------------------------------------------------------------------
It is also important to note that for the BAI simulated datasets,
approximately $55\%$ of the data is missing.
Furthermore, the Jones model is generally more parsimonious than the O-bivariate-FC model,
which might help to explain its clearer performance benefit here for when we only have $10$ studies,
but not when we have $50$ studies.
%%%%
%%%%
%%%%%%%%%%%%%%%%%%%%%%%%%%%%%%%%
\paragraph{\underline{ 
Jones DGM: Summary (between-test comparison) 
}}
\label{discussion_jones_dgm_summary}
%%%%%%%%%%%%%%%%%%%%%%%%%%%%%%%%
%%%%
%%%% ---------------------------------------------------------------------------------------------------------------------
Compared to the GAD-2 data 
($6$ thresholds, $\sim 15\%$ missing data; see section \ref{discussion_jones_dgm_GAD_2} above) -
for the $50$-study case -
the performance differences between the three models were more pronounced than they were for the HADS
($21$ thresholds, $\sim 40\%$ missing data; see section \ref{discussion_jones_dgm_HADS}).

%%%%
%%%% ---------------------------------------------------------------------------------------------------------------------
More specifically, for the GAD-2, we obtained differences
(vs. Jones - which was the best model for both the GAD-2 and HADS)
of:
$5.8\%$ (O-biv-FC vs. Jones) and $17.0\%$ (stratified-bivariate vs. Jones). %% GAD-2
On the other hand, for the HADS, we obtained differences of:
$2.4\%$ (O-biv-FC vs. Jones) and $10.8\%$ (stratified-bivariate vs. Jones). %% HADS
In other words, it may be that more thresholds ($21$ vs. $6$),
and/or more missing data ($\sim 15\%$ vs. $\sim 40\%$),
and/or different between-study heterogeneity (heterogeneity was approximately 2-fold higher for GAD-2 than HADS)
caused the performance gaps between the models to be wider for the GAD-2 data relative to HADS,
where performance between models was somewhat more homogeneous.

%%%%
%%%% ---------------------------------------------------------------------------------------------------------------------
On the other hand, for the BAI ($63$ thresholds; see section \ref{discussion_jones_dgm_BAI}),
we obtained relative differences of:
$8.8\%$ (O-biv-FC vs. Jones) and $52.6\%$ (stratified-bivariate vs. Jones).
These gaps are in fact wider than either the GAD-2 or HADS relative differences discussed in the previous paragraph.
This discrepancy - that is, the apparent paradoxical difference between GAD-2 vs. HADS and GAD-2 vs. BAI -
%%% check the following somehow?!??!: ---- bookmark
could be potentially due to the fact that, even with $50$ total studies, many thresholds
(note we only computed RMSE for the BAI using thresholds $5 - 43$) are only observed in 3 studies (the minimum we required),
or very few (e.g. 4 or 5 studies). This would help explain these results, since the stratified-bivariate model would
perform relatively poorly for these thresholds as it does not borrow information between thresholds like the
other models (e.g. Jones, O-bivariate-FC) do.
%%%%
%%%%
%%%%%%%%%%%%%%%%%%%%%%%%%%%%%%%%%%%%%
\subsubsection{ Discussion of simulation study findings: Ordinal DGMs}
\label{discussion_ordinal_dgms}
%%%%%%%%%%%%%%%%%%%%%%%%%%%%%%%%%%%%%
%%%%
%%%% ---------------------------------------------------------------------------------------------------------------------
When data truly followed ordinal processes -
which we would expect for these tests since they are by definition ordinal anxiety screening questionnaires
(i.e., the GAD-2\supercite{Kroenke_2007_GAD_2}, HADS\supercite{Zigmond_1983_HADS} and BAI\supercite{Beck_1988_BAI}) -
ordinal models generally outperformed both the continuous-assumption Jones\supercite{Jones2019} model,
as well as the standard stratified-bivariate\supercite{Reitsma2005} model.
A more detailed discussion of the results for the ordinal DGMs is provided in 
appendix \ref{appendix_F_additional_discussion_material}, 
section \ref{appendix_F_additional_discussion_material_ordinal_DGMs}.

%%
%% ---- Summary for: GAD-2; 50 studies
%%
For the GAD-2 ($7$ categories, with $\sim 15\%$ missing threshold data), when we have $50$ studies,
overall the O-bivariate-FC (for 4/4 DGMs) and O-bivariate-RC
(for DGMs \#3 and \#4, as we could not run this for DGMs \#1 - \#2)
models performed the best;
namely - 
as discussed in more detail in 
appendix \ref{appendix_F_additional_discussion_material} 
(see section \ref{discussion_jones_dgm}) -
for the Jones DGM, the O-bivariate-FC was statistically equivalent,
and in relative terms its RMSE was only $5.8\%$ ($< 10\%$) worse than the Jones model.
For both the O-bivariate-FC (DGM \#2) and O-bivariate-RC (DGM \#3) DGMs -
as we also discussed above 
(see appendix \ref{appendix_F_additional_discussion_material}, 
sub-sections \ref{discussion_ordinal_dgms_O_biv_FC} 
and \ref{discussion_ordinal_dgms_O_biv_RC}) -
the O-bivariate-FC model came in first place, having the lowest RMSE,
with the Jones model performing $14.6\%$ and $16.1\%$ (both $> 10\%$) worse for DGM \#2 and DGM \#3, respectively.
Finally, for the O-HSROC-RC DGM (DGM \#4), both the O-bivariate-RC ($0.9\%$ worse vs. O-HSROC-RC model),
as well as the O-bivariate-FC ($7.5\%$ worse vs. O-HSROC-RC model) models,
were statistically equivalent (within MCSE) of the leading O-HSROC-RC model,
with the Jones model performing statistically significantly - as well as practically - 
worse than the leading O-HSROC-RC model,
being $10.3\%$ ($> 10\%$) worse than the leading O-HSROC-RC model.

%%%%
%%%% ---------------------------------------------------------------------------------------------------------------------
%%
%% ---- Summary for: GAD-2; 10 studies
%%
On the other hand, in contrast to when we had only $10$ studies,
the Jones model overall actually performed better (compared to when we have $50$ studies).
This is because it was the best model for the Jones DGM itself (DGM \#1);
furthermore, for the other three ordinal DGMs, it performed effectively just as well as the ordinal models themselves did.
More specifically, for the Jones DGM (DGM \#1), it was the only model in the "best" group;
however, it is important to note that the O-bivariate-FC model was equivalent in practical terms
(despite being statistically significantly worse),
being only $5.2\%$ worse then the Jones model (RMSE = $9.75$ vs. $10.25$).
Furthermore, for the other three (i.e., ordinal) DGMs -
as we discuss in more detail in 
appendix \ref{appendix_F_additional_discussion_material}, 
sections
\ref{discussion_ordinal_dgms_O_biv_FC},
\ref{discussion_ordinal_dgms_O_biv_RC}, and
\ref{discussion_ordinal_dgms_O_HSROC_RC}  -
the Jones model was in the "best" group every single time -
being only $0.2\%$, $2.7\%$, and $1.4\%$ worse than the leading model for DGM \#2, \#3, and \#4, respectively.
Also, it is important to note that the leading model for
GAD-2 (for $N_{studies} = 10$) for these three DGMs was the O-bivariate-FC model for DGM \#2,
and the O-bivariate-RC for DGMs \#3 and \#4.
However, the DGMs \#3 and \#4, the O-bivariate-FC model performed essentially the same as the RC variant -
and hence the same as the Jones model -
being only $2.1\%$ and $2.8\%$ worse than the leading O-bivariate-RC models
for DGMs \#3 and \#4, respectively.

%%%%
%%%% ---------------------------------------------------------------------------------------------------------------------
%%
%% ---- Summary for: GAD-2; stratified-bivariate summary (vs. Jones, O-biv, etc).
%%
Regarding the standard stratified-bivariate model, for the $10$-study case, this was:
$20.6\%$, $14.3\%$, $7.6\%$, and $10.5\%$ worse than the leading models
(which were: Jones, O-biv-FC, O-biv-RC, O-biv-RC)
for DGMs \#1, \#2, \#3 and \#4, respectively.
Notably, DGM \#3 (the O-bivariate-RC DGM) was the only DGM for which the
stratified-bivariate model did not end up in the "worse" group.
For the $50$-study case, the stratified-bivariate model performed better overall
(and overall performing better than the Jones model in this case).
More specifically, it was only in the "worse" group for:
DGM \#1 (Jones; $17.0\%$ worse than leading Jones model), and
DGM \#3 (O-bivariate-RC; $11.0\%$ worse than leading O-bivariate-FC model).
In contrast, for DGM \#2 (O-bivariate-FC), it was only statistically significantly worse,
but not practically ($9.2\%$ worse than leading O-bivariate-FC model - just under our $10\%$ threshold),
and for DGM \#4, it was only $7.2\%$ worse than the leading O-HSROC-RC model, and fell into the "best" group.
Overall, the notable but not overwhelming RMSE performance differences between the standard stratified-bivariate model
and the ordinal models we proposed for the GAD-2 is not surprising,
since the amount of missing threshold data is quite low
(only $\sim 15\%$).

%%%%
%%%% ---------------------------------------------------------------------------------------------------------------------
%%
%% ---- Comparison to HADS (GAD-2 vs. HADS; for stratified-bivariate):
%%
On the other hand, for the HADS ($22$ categories, with $\sim 40\%$ missing threshold data),
as well as the BAI ($64$ categories, with $\sim 55\%$ missing threshold data),
the performance differences between the stratified-bivariate model and the
leading models for each DGM become much more obvious and pronounced.
For instance, when we have $10$ studies, the stratified-bivariate model was (for HADS):
$24.8\%$, $37.8\%$, $29.8\%$, and $32.7\%$ worse than the leading models
(which were: Jones, O-biv-FC, O-biv-RC, O-biv-RC)
for DGMs \#1, \#2, \#3 and \#4, respectively.
For the BAI, the corresponding numbers (for $10$ studies) were:
$37.2\%$, $33.9\%$, $33.0\%$, and $38.1\%$.

%%%%
%%%% ---------------------------------------------------------------------------------------------------------------------
This range (for the $10$-study case) we obtained for the HADS ($24.8\% - 37.8\%$) -
as well as the BAI ($33.0\% - 38.1\%$) -
is much larger than the range we saw with the GAD-2 ($7.6\% - 20.6\%$).
Furthermore, for the $50$-study case, we obtained ranges of:
$7.2\% - 17.0\%$, %% GAD-2, 50 studies
$10.2\% - 30.7\%$, and %% HADS, 50 studies
$18.8\% - 52.6\%$ %% BAI, 50 studies
for the GAD-2, HADS, and BAI, respectively.

%%%%
%%%% ---------------------------------------------------------------------------------------------------------------------
This discrepancy (i.e., much larger improvement seen when using multiple threshold models vs. stratified-bivariate
for the HADS and BAI vs. the GAD-2) could be because there is much more missing data with the HADS ($\sim 40\%$ missing)
as well as the BAI ($\sim 55\%$ missing) -
verses only $\sim 15\%$ missing threshold data on average for the GAD-2.
This would affect the estimates, because it means that, for the HADS and BAI data, there will be more thresholds where
we only have a few studies worth of data,
whereas for the GAD-2 most thresholds will only have a few (and also more likely to have zero) missing studies.
This will cause the stratified-bivariate to perform worse for the HADS and BAI (vs. GAD-2),
because unlike the "multiple threshold" models, it cannot borrow strength between thresholds -
hence, those studies must be discarded completely.
%%%%
%%%%

%%%%%%%%%%%%%%%%%%%%%%%%%%%%%%%%%%%%%
\subsubsection{ Coverage and width discussion}
\label{discussion_cvg_and_width_discussion}
%%%%%%%%%%%%%%%%%%%%%%%%%%%%%%%%%%%%%
%%%%
%%%%
We will now discuss the coverage and interval width results
from section \ref{Sim_study_coverage_and_interval_widths} in more detail.
As mentioned in that section, the complete results for coverage and interval widths are contained in
appendix \ref{appendix_E_detailed_results_tables};
more specifically, in:
\ref{appendix_E_detailed_results_tables_GAD_2_test_I},
\ref{appendix_E_detailed_results_tables_HADS_test_II}, and
\ref{appendix_E_detailed_results_tables_BAI_test_III},
for the GAD-2, HADS and the BAI, respectively.
%%
%% ---- Coverage discussion (focusing on correctly specified scenarios):
%%
%%%%%%%%%%%%%%%%%%%%%%%%%%%%%%%%
\paragraph{\underline{ 
GAD-2:
}}
\label{discussion_cvg_and_width_discussion_GAD_2}
%%%%%%%%%%%%%%%%%%%%%%%%%%%%%%%%
%%%%
%%%% --------------------------------------------------------------------------------------------------------------------
All correctly specified scenarios achieved acceptable coverage ($91.1\% - 97.0\%$), 
confirming the models perform as expected under favorable conditions 
($6$ thresholds, $15\%$ missingness).

%%%%%%%%%%%%%%%%%%%%%%%%%%%%%%%%
\paragraph{\underline{ 
HADS:
}}
\label{discussion_cvg_and_width_discussion_HADS}
%%%%%%%%%%%%%%%%%%%%%%%%%%%%%%%%
%%%%
%%%% --------------------------------------------------------------------------------------------------------------------
Two notable patterns emerged for the HADS 
($21$ thresholds, $40\%$ missingness).
Firstly, the Jones model with $50$ studies showed substantial under-coverage
(Se: $67.3\%$, ~Sp: $88.8\%$) paired with narrow intervals (Se: $4.72$, ~Sp: $2.68$).
Despite lowest RMSE $(2.30\%, 0.92\%)$, the continuous-assumption framework
produced overly confident intervals when applied to ordinal data.

%%%%
%%%% --------------------------------------------------------------------------------------------------------------------
Secondly, under the O-Biv-RC DGM with $50$ studies,
the \textit{misspecified} fixed-cutpoint model (O-Biv-FC) achieved
better coverage (Se: $95.8\%$, Sp: $95.2\%$)
than the \textit{correctly specified} random-cutpoint model
(O-Biv-RC: Se: $91.3\%$, Sp: $87.1\%$),
despite similar interval widths
(FC: $5.05/2.62$ vs. RC: $5.27/2.62$).
Given the low cutpoint heterogeneity in this DGM
(table ~\ref{table:cutpoint_heterogeneity_by_dgm} shows SD = $0.010$),
random effects added unnecessary complexity that degraded uncertainty calibration
without improving point estimates.

%%%%%%%%%%%%%%%%%%%%%%%%%%%%%%%%
\paragraph{\underline{
BAI:
}}
\label{discussion_cvg_and_width_discussion_BAI}
%%%%%%%%%%%%%%%%%%%%%%%%%%%%%%%%
%%%%
%%%% --------------------------------------------------------------------------------------------------------------------
For the BAI ($63$ thresholds, $55\%$ missingness),
even with $50$ studies -
a large meta-analysis by conventional standards -
the correctly specified O-Biv-RC model showed under-coverage
(Se: $74.6\%$, Sp: $87.1\%$) with narrow intervals
(Se: $3.99$, Sp: $3.39$).

%%%%
%%%% --------------------------------------------------------------------------------------------------------------------
However, the RMSE ($1.64\%$) and bias ($1.29\%$) remained low,
indicating point estimates converged accurately - despite poor uncertainty quantification.
This reflects severe \textit{threshold-level sparsity}:
with $64$ categories and $55\%$ missingness,
many thresholds were reported by only $\sim 3-5$ studies,
insufficient for reliable variance estimation even when the overall sample size is substantial.
Critically, threshold-level sparsity - not total study count - determined
whether models could properly calibrate uncertainty intervals.

%%%%%%%%%%%%%%%%%%%%%%%%%%%%%%%%%%%%%
\subsubsection{ RMSE-Bias discrepancies reveal precision-accuracy tradeoffs}
\label{discussion_rmse_bias_discrepancy}
%%%%%%%%%%%%%%%%%%%%%%%%%%%%%%%%%%%%%
%%%%
%%%% ---------------------------------------------------------------------------------------------------------------------
%%
%% ---- Why discrepancies occur:
%%
Critical discrepancies between RMSE (primary) and bias (secondary) rankings 
revealed why RMSE better captures overall performance.
Since RMSE$^2$ = Bias$^2$ + Variance, models can achieve good RMSE through different combinations:
low bias with moderate variance, or higher bias with excellent precision.

%%%%
%%%% ---------------------------------------------------------------------------------------------------------------------
%%
%% ---- GAD-2 examples:
%%
For GAD-2 - under Jones DGM with $10$ studies 
(see section \ref{Sim_study_GAD_2_Jones_DGM_1}) -
Jones achieved the best RMSE ($9.75$), 
despite obtaining worse bias ($2.08$)
than the O-bivariate-FC (bias = $1.61$).
However, when ranked by bias alone 
(see section 
\ref{Sim_study_GAD_2_Jones_DGM_1}, 
table 
\ref{Table:summary_table_overall_GAD_2_bias}), 
the O-bivariate-FC became the sole "best" model,
with Jones falling to "Group 3 - Practical only" ($29.0\%$ worse bias).
Similar patterns occurred several times throughout,
which shows that the Jones model achieved superior overall accuracy estimates through better precision,
despite higher systematic error (i.e., bias).

%%%%
%%%% --------------------------------------------------------------------------------------------------------------------
%%
%% ---- HADS examples:
%%
For HADS under the DGM \#3 (i.e., the O-bivariate-RC DGM) with $50$ studies
(see section \ref{Sim_study_HADS_O_biv_RC_DGM_3},
tables
\ref{Table:summary_table_overall_HADS} and
\ref{Table:summary_table_overall_HADS_bias}),
the stratified-bivariate achieved surprisingly good bias ($0.33$),
which was substantially better than Jones ($1.43$), and comparable to O-bivariate-FC ($0.24$).
Yet, for RMSE, stratified-bivariate fell to "Group 4 - Worse" ($2.59$),
whilst O-bivariate-FC (RMSE = $2.03$) and O-bivariate-RC (RMSE = $2.08$) led.
This demonstrates the stratified-bivariate's fundamental limitation:
failing to utilize ordinal structure generally produces high variance -
hence yielding poor RMSE - despite often obtaining good/reasonable bias.

%%%%
%%%% --------------------------------------------------------------------------------------------------------------------
%%
%% ---- BAI examples:
%%
For BAI under DGM \#4 (i.e., the O-bivariate-RC DGM) with $50$-studies
(see section
\ref{Sim_study_BAI_O_biv_RC_DGM_4}),
table \ref{Table:summary_table_overall_BAI},
bias rankings were very inconsistent with the RMSE rankings.
More specifically, the O-HSROC-RC and stratified-bivariate models achieved the best bias
($2.07$ and $2.25$, respectively;
see table \ref{Table:summary_table_overall_BAI_bias}) -
which were within MCSE of one another -
and hence being the only two models in the "best" (by bias) group.
However, when ranking by RMSE
(see table \ref{Table:summary_table_overall_BAI}),
the O-bivariate-RC model had worse bias ($2.38$),
falling to "Group 4 - Worse" for bias.
Yet O-bivariate-RC's superior precision yielded optimal overall accuracy (RMSE = $3.32$),
demonstrating that bias alone provides incomplete performance assessment.
%%%%
%%%%
%%%%%%%%%%%%%%%%%%%%%%%%%%%%%%%%%%%%%%
\subsubsection{ Questioning the continuous assumption}
%%%%%%%%%%%%%%%%%%%%%%%%%%%%%%%%%%%%%%
% %%%%
% %%%% -----------------------------------------------------------------------------------------------------------------
% Perhaps most strikingly, our results with the BAI (64 categories) definitively refute the common assumption that tests with many 
% ordinal categories can be treated as continuous. 
% %%
% Despite having more categories than many researchers' arbitrary threshold for assuming normality (often >20 or >30 categories), 
% the Jones continuous-threshold model still performed significantly worse than our ordinal approach. 
% %%
% This finding validates our hypothesis that the fundamental ordinal nature of these instruments cannot be ignored simply because they 
% have numerous categories - 
% the O-bivariate model achieved 11-17\% lower RMSE than the Jones model even with 64 categories, 
% demonstrating that proper ordinal modeling matters regardless of scale granularity.

%%%%
%%%% --------------------------------------------------------------------------------------------------------------------
Perhaps most strikingly, our simulation study results with the BAI 
($64$ categories; see section \ref{Sim_study_BAI})
contradict the common assumption that outcomes
(in this case, screening/diagnostic tests) with many ordinal categories can be treated as continuous.
Despite having more categories than many researchers' arbitrary threshold for assuming normality
(often only $> 20$ or $> 30$ categories) -
as we discussed in discussion 
section \ref{discussion_ordinal_dgms} -
the O-bivariate model still performed significantly better than or equal to the Jones continuous model across most scenarios.

%%%%
%%%% --------------------------------------------------------------------------------------------------------------------
For example,
For the BAI with $50$ studies, under the O-bivariate-FC DGM 
(see section 
\ref{Sim_study_BAI},
table
\ref{Table:summary_table_overall_BAI}),
the O-bivariate-FC achieved an RMSE of $2.32$,
making it the sole model in the "best" group,
whilst the Jones model obtained a statistically significantly worse RMSE of $3.01$.
This $\sim 0.70$ percentage point difference in RMSE -
combined with the substantially better bias ($0.27$ vs. $1.53$) -
demonstrates that proper ordinal modelling 
(at least, for tests which are truly ordinal, such as ours) 
matters regardless of scale granularity.

%%%%
%%%% --------------------------------------------------------------------------------------------------------------------
The benefits were not limited to RMSE-bias patterns revealed additional advantages.
The O-bivariate models consistently achieved lower bias than the Jones model,
particularly evident in the BAI results where bias differences often exceeded 1 percentage point.
This finding validates our hypothesis that the fundamental ordinal nature of these instruments cannot be ignored
simply because they have numerous categories.

%%%%%%%%%%%%%%%%%%%%%%%%%%%%%%%%%%%%%%%%%%%%%%%%%%%%%%%%%%%%%%%%%%%%%%%%%%
\subsection{ Potential practical \& clinical implications}
\label{section_discussion_practical_and_clinical_implications}
%%%%%%%%%%%%%%%%%%%%%%%%%%%%%%%%%%%%%%%%%%%%%%%%%%%%%%%%%%%%%%%%%%%%%%%%%%
%%%%
%%%%

%%%%%%%%%%%%%%%%%%%%%%%%%%%%%%%%%%%%%%
\subsubsection{ Statistical models \& simulation study}
\label{section_discussion_practical_and_clinical_implications_sim_study}
%%%%%%%%%%%%%%%%%%%%%%%%%%%%%%%%%%%%%%
%%%%
%%%% --------------------------------------------------------------------------------------------------------------------
Our simulation study results (see section \ref{Sim_study}),
which we also discussed in discussion section 
\ref{discussion_summary_key_findings} above,
suggest that our proposed ordinal-bivariate model
(see section \ref{section_model_specs_Cerullo_bivariate_Reitsma_extension}) -
with fixed-effects cutpoint parameters (i.e., the "O-bivariate-FC" model) -
may be an ideal approach for meta-analyzing ordinal screening and/or diagnostic test accuracy data.
This is because of its relatively consistent superiority over the more standard
stratified-bivariate\supercite{Reitsma2005} approach,
as well as over continuous-assumption "multiple threshold" methods
(Jones et al, 2019\supercite{Jones2019}) -
even for $64$-category questionnaires -
suggest these models should perhaps be preferred,
regardless of the number of ordinal categories.

%%%%
%%%% --------------------------------------------------------------------------------------------------------------------
%% ---- Fixed vs. random cutpoints (i.e., O-bivarite-FC vs. O-bivarite-RC models)
Furthermore - perhaps somewhat surprisingly -
the O-bivariate-FC model performed competitively, 
even under random-cutpoint DGMs
(see discussion section \ref{discussion_ordinal_dgms} for more details).
For instance, for the GAD-2 with $50$ studies under the O-bivariate-RC DGM
(see section \ref{discussion_ordinal_dgms_O_biv_RC}),
the O-bivariate-FC paradoxically outperformed the correctly-specified O-bivariate-RC,
although differences were still within MCSE.
This suggests explicit threshold heterogeneity modeling might be unnecessary (at least in some cases),
with fixed-cutpoint models providing adequate flexibility with better precision.

%%%%
%%%% --------------------------------------------------------------------------------------------------------------------
For clinical practice, these findings mean more reliable estimates of test accuracy across different thresholds,
enabling better-informed decisions about optimal cutpoints for specific clinical contexts.
Furthermore, the NMA extensions
(see section \ref{section_model_specs_NMA_extensions})
allow simultaneous comparison of multiple screening and/or diagnostic instruments,
providing comprehensive evidence for test selection.

%%%%
%%%% --------------------------------------------------------------------------------------------------------------------
%%
%% ---- Model selection guidance:
%%
For simple ordinal tests like GAD-2 ($7$ categories, $\sim 15\%$ missing data),
the Jones model performed competitively with $10$ studies under its own DGM,
though O-bivariate-FC was within $5.2\%$ RMSE -
below our practical significance threshold.
With $50$ studies, O-bivariate-FC dominated under ordinal DGMs
(Jones model $14.6\% - 16.1\%$ worse under O-bivariate DGMs).
For moderate complexity tests like HADS 
($22$ categories, $\sim 40\%$ missing data),
O-bivariate-FC and Jones often fell in the same performance group,
whilst the stratified-bivariate showed consistent failure
($24.8\% - 37.8\%$ worse with $10$ studies).

%%%%
%%%% --------------------------------------------------------------------------------------------------------------------
For complex tests like BAI ($64$ categories, $\sim 55\%$ missing data),
ordinal models became essential - O-bivariate-FC was often the sole "best" model,
with Jones struggling due to threshold-level sparsity
(many thresholds observed in only $3 - 5$ studies).
The stratified-bivariate approach showed increasing degradation with test complexity:
$7.6\% - 20.6\%$ worse for GAD-2, $24.8\% - 37.8\%$ worse for HADS,
and $33.0\% - 52.6\%$ worse for BAI.

%%%%
%%%% -------------------------------------------------------------------------------------------------------------------
%%
%% ---- RMSE-bias tradeoffs:
%%
Additionally, critical discrepancies between the RMSE and bias rankings - 
as we discussed in more detail above in discussion 
section \ref{discussion_rmse_bias_discrepancy} - 
revealed important precision-accuracy tradeoffs.
For instance, for the GAD-2 under Jones DGM, 
Jones achieved best RMSE - 
despite worse bias than the O-bivariate-FC -
indicating superior precision compensated for systematic error.
For the HADS under the O-bivariate-RC DGM,
the stratified-bivariate achieved surprisingly good bias
but poor RMSE - due to high variance.

% %%%%
% %%%% -------------------------------------------------------------------------------------------------------------------
% %%
% %% ---- Reporting recommendations:
% %%
% Both RMSE and bias should be reported to reveal precision-accuracy tradeoffs.
% %%
% RMSE should guide primary model selection as it captures total estimation error.
% %%
% When RMSE and bias rankings diverge, this signals that one model achieves accuracy through precision
% whilst another minimizes systematic error - important information for interpretation.
% %%
% Researchers requiring unbiased point estimates for specific applications might reasonably choose bias-optimal models,
% accepting wider confidence intervals as the price for eliminating systematic error.
% %%%%
% %%%%
%%%%%%%%%%%%%%%%%%%%%%%%%%%%%%%%%%%%%%
\subsubsection{ MetaOrdDTA R package}
\label{section_discussion_practical_and_clinical_implications_MetaOrdDTA}
%%%%%%%%%%%%%%%%%%%%%%%%%%%%%%%%%%%%%%
%%%%
%%%% ------------------------------------------------------------------------------------------------------------------
The MetaOrdDTA\supercite{Cerullo_MetaOrdDTA_2025} R package we developed (see section \ref{MetaOrdDTA})
provides a comprehensive implementation of the DTA-MA and DTA-NMA ordinal models - which we developed in this paper in 
section \ref{section_model_specs} - 
the Jones model (Jones et al, 2019\supercite{Jones2019}),
as well as a DTA-NMA extension of the Jones model which we programmed in Stan.
MetaOrdDTA is freely available at \url{https://github.com/CerulloE1996/MetaOrdDTA/}.

%%%%
%%%% ------------------------------------------------------------------------------------------------------------------
%% \paragraph{\underline{Model selection framework}}
A critical innovation in MetaOrdDTA is the integrated K-fold cross-validation capability for model selection
(see section \ref{MetaOrdDTA_results_baseline_analysis_model_selection_k_fold}).
More standard information criteria,
such as the LOO-IC (Vehtari et al, 2017\supercite{Vehtari2017}),
systematically fail for hierarchical NMA models (high Pareto $\hat{k}$ values exceeding $0.7$),
due to the influential nature of individual studies.
The \verb|$run_k_fold_CV()| method
(see e.g., code box \ref{code_box_MetaOrdDTA_model_selection_k_fold_cv})
addresses this through proper out-of-sample validation,
enabling principled comparison of competing model specifications.
In our DTA-NMA application,
this framework identified fixed-cutpoint models with compound symmetry as optimal
(see table \ref{table_MetaOrdDTA_model_selection_k_fold_table}),
avoiding unnecessary model complexity;
however, note that this will not necessarily always be the case for a given user's data/analysis.
This finding - that simpler models often outperform complex alternatives -
emerged only through formal cross-validation,
and would have been obscured by reliance on failing information criteria
(e.g., the DIC\supercite{DIC}, or LOO-IC\supercite{Vehtari2017} and ignoring the pareto-k warnings).

%%%%
%%%% -------------------------------------------------------------------------------------------------------------------
%% \paragraph{\underline{Computational efficiency}}
%%%%
The post-hoc baseline exploration functionality
(section ~\ref{MetaOrdDTA_results_post_hoc_baseline_exploration})
enables rapid scenario analysis without model refitting -
via an internal Rcpp
\supercite{
rcpp_ref_1_Eddelbuettel_and_Francois_2011,
rcpp_ref_2_Eddelbuettel_2013,
rcpp_ref_3_Eddelbuettel_and_Balamuta_2018}
(C++) implementation for speed.
The \verb|$recompute_baseline()| method
(see code box ~\ref{code_box_meta_reg_baseline_exploration})
regenerates complete posterior distributions for sensitivity, specificity,
and AUC across user-specified covariate profiles in seconds rather than minutes or even hours.
This capability proved essential in our meta-regression analysis
(see section ~\ref{MetaOrdDTA_results_meta_reg}),
where we explored nine scenarios
(3 prevalence levels $\times$ 3 reference standards; see table ~\ref{table_meta_reg_AUC})
to understand how test performance varies across clinical contexts.
The ability to iterate rapidly through "what-if" scenarios transforms meta-regression
from a tedious exercise into an exploratory tool for understanding covariate effects.

%%%%
%%%% -------------------------------------------------------------------------------------------------------------------
%% \paragraph{\underline{Visualization and reporting ecosystem}}
%%%%
The R package also generates publication-ready figures through a layered approach that balances automation with customization. 
For instance, the sROC plotting functions
(see section ~\ref{MetaOrdDTA_results_baseline_analysis_sROC_plots},
code box ~\ref{code_box_MetaOrdDTA_base_model_sROC_plots})
produce both overview panels 
(see figure ~\ref{Figure_MetaOrdDTA_baseline_analysis_sROC})
and detailed test-specific plots with credible and prediction regions
(see figure ~\ref{Figure_MetaOrdDTA_baseline_analysis_sROC_panel_w_CrI_PrI}).
%% ---- LaTeX tables:
The automated LaTeX table generation eliminates error-prone manual transcription whilst ensuring reproducibility.
For instance, functions like \verb|create_MR_AUC_latex_table_2vars()|
(code box ~\ref{code_box_MetaOrdDTA_base_model_AUC_results_for_best_kfold_model})
handle arbitrary factorial designs,
automatically formatting results with appropriate precision and statistical annotations.

%%%%
%%%% -------------------------------------------------------------------------------------------------------------------
%% ---- Pairwise comparisons:
For DTA-NMA, the pairwise comparison framework
(section ~\ref{MetaOrdDTA_results_baseline_analysis_pairwise_diffs} for baseline analysis;
section ~\ref{MetaOrdDTA_results_meta_reg_plots_pairwise_diffs} for meta-regression)
directly addresses clinical questions about relative test performance at specific thresholds
(figures ~\ref{Figure_MetaOrdDTA_baseline_analysis_Se_and_Sp_pairwise_diffs_screening_GAD},
\ref{DUMMY_MetaOrdDTA_MR_Se_pairwise_diffs_screening_all},
and ~\ref{DUMMY_MetaOrdDTA_MR_Sp_pairwise_diffs_screening_all}) -
moving beyond isolated accuracy metrics to actionable recommendations.

%%%%
%%%% -------------------------------------------------------------------------------------------------------------------
%% ---- Meta-regression:
Furthermore, the meta-regression implementation (see section ~\ref{MetaOrdDTA_results_meta_reg})
demonstrated how covariates explained approximately $24\%$ of between-study heterogeneity
(see table ~\ref{table_MR_vs_intercept_heterogeneity_components}),
with disease prevalence showing test-specific effects—positive for GAD-2/GAD-7,
negative for HADS, mixed for BAI (see table ~\ref{table_MR_coefficients}).
The covariate preparation workflow
(see section ~\ref{MetaOrdDTA_results_meta_reg_data_prep},
and code box \ref{code_box_meta_reg_data_prep})
handles mixed data types (continuous, binary, categorical) with appropriate missing data treatment,
converting user-friendly tibble formats into the nested list structures required by Stan.
Furthermore, the custom meta-regression aesthetic mapping functionality
(see code boxes ~\ref{code_box_meta_reg_sROC_plots_custom_mapping}
and ~\ref{code_box_meta_reg_accuracy_vs_thr_plots_custom_mapping})
allows systematic visualization of how test performance varies across covariate combinations
(see figures ~\ref{DUMMY_MetaOrdDTA_MR_sROC_2x2_ALL_TESTS_ALL_SCENARIOS_custom}
and ~\ref{DUMMY_MetaOrdDTA_MR_accuracy_vs_thr_2x2_ALL_TESTS_ALL_SCENARIOS_custom}),
effectively communicating complex patterns to help clinicians understand how test accuracy
varies across different patient populations and clinical settings.

%%%%
%%%% -------------------------------------------------------------------------------------------------------------------
To summarise: the MetaOrdDTA package enables rigorous meta-analysis of ordinal diagnostic accuracy tests,
filling a methodological gap that has limited evidence synthesis in this domain.
The demonstrated NMA framework
(see sections \ref{section_model_specs_NMA_extensions} and \ref{MetaOrdDTA})
further enables simultaneous comparison of multiple tests,
essential for informing test selection in practice.
We anticipate this software will facilitate broader adoption of ordinal methods in
screening and/or diagnostic test evaluation,
ultimately improving the evidence base for clinical decision-making.

%%%%%%%%%%%%%%%%%%%%%%%%%%%%%%%%%%%%%%%%%%%%%%%%%%%%%%%%%%%%%%%%%%%%%%%%%%
\subsection{ Limitations \& future work}
\label{section_discussion_limitations_and_future_work}
%%%%%%%%%%%%%%%%%%%%%%%%%%%%%%%%%%%%%%%%%%%%%%%%%%%%%%%%%%%%%%%%%%%%%%%%%%
%%%%
%%%%
%%%%%%%%%%%%%%%%%%%%%%%%%%%%%%%%%%%%%%
\subsubsection{ Statistical models \& simulation study}
\label{section_discussion_limitations_and_future_work_sim_study}
%%%%%%%%%%%%%%%%%%%%%%%%%%%%%%%%%%%%%%
%%%%
%%%% -------------------------------------------------------------------------------------------------------------------
Our simulation study has several limitations that should be considered.
For instance, we focused on three specific data-generating mechanisms based on real (ordinal) screening test data.
Whilst the simulated data based on real data from these tests (GAD-2, HADS, and the BAI)
span a range of characteristics
($7$, $22$ and $64$ categories,
and $15\%$, $40\%$, and $55\%$ missing data, respectively),
other combinations or other patterns of missingness may yield different results,
and may also produce additional insights.

%%%%
%%%% -------------------------------------------------------------------------------------------------------------------
Furthermore, whilst the proposed O-bivariate-FC model performed overall very well -
even under the random-cutpoint DGMs -
we did not explore scenarios with extreme cutpoint heterogeneity
(since the real data itself did not have extremely high cutpoint heterogeneity),
hence in this case the random-effect cutpoint models might show clearer advantages.
However, it is important to note that the true values of the cutpoint heterogeneity (for the RC DGMs)
was still quite notable, and is likely consistent with many other real-life datasets -
see section \ref{Sim_study_design_DGMs} for more details on cutpoint heterogeneity.

%%%%
%%%% ---------------------------------------------------------------------------------------------------------
It is also important to note that we did not adopt a full factorial simulation design,
which would have crossed all combinations of threshold counts and missingness proportions.
More specifically, a full factorial design would have yielded:
$\#\{\text{threshold levels}\} \times \#\{\text{missingness levels}\}$
$= 3 \times 3 = 9$ test configurations per DGM (rather than the 3 we evaluated).
This would have enabled stronger causal statements such as:
"holding threshold count constant at $22$, increasing missingness from $15\%$ to $55\%$
degrades RMSE by $X\%$, isolating the pure effect of data sparsity".
However, our design - where tests with more categories naturally exhibited higher missingness -
likely reflects more realistic patterns in screening/diagnostic test accuracy literature,
where instruments with more categories (e.g., BAI with $64$ categories) likely tend to have studies
reporting less thresholds - and hence more sparse data - compared to instruments with less categories
(e.g., the GAD-2 with $7$ categories).
Furthermore, with $4$ DGMs $\times$ 3 tests $\times$ 2 sample sizes $= 24$ scenarios
(plus 5 models compared per scenario),
resource and time constraints necessitated prioritizing realistic configurations over factorial completeness.
Future simulation work could systematically cross threshold count with missingness
to isolate their independent effects.

%%%%
%%%% -------------------------------------------------------------------------------------------------------------------
Another limitation for our simulation study is that we compared our proposed models
only against the Jones model (Jones et al, 2019\supercite{Jones2019}),
and also to the standard, routinely used stratified-bivariate
(Reitsma et al, 2005\supercite{Reitsma2005}) approach;
however, we did not include other recently proposed "multiple thresholds" methods,
namely the time-to-event based models
(i.e.,
Hoyer et al, 2018\supercite{Hoyer_et_al_2018_mult_thr};
Zapf et al,  2024\supercite{Zapf_et_al_2024_mult_thr}).
Note that these models - just like Jones et al\supercite{Jones2019} -
also rely on continuous latent variable assumptions;
however, without testing them explicitly, we cannot draw any conclusions.

%%%%
%%%% -------------------------------------------------------------------------------------------------------------------
We also only did the simulation study for the DTA-MA models -
not the DTA-NMA model extensions.
However, we expect our findings would apply to the NMA models approximately the same,
especially for larger number of studies ($50$ studies),
and also for the much smaller sample size ($10$ studies) -
assuming priors are set somewhat equivalently.
That being said, future research should investigate this directly.

% %%%%
% %%%% -------------------------------------------------------------------------------------------------------------------
% Several directions merit investigation based on our findings.
% %%
% The surprising robustness of fixed-cutpoint models under random-cutpoint DGMs 
% suggests further research into when random effects for thresholds are truly necessary.
% %%
% The performance degradation for all models with sparse threshold data 
% (as seen with BAI where many thresholds had only 3-5 studies) 
% indicates need for methods explicitly handling threshold-level sparsity.

%%%%
%%%% ---------------------------------------------------------------------------------------------------------
Furthermore, the imperfect gold standard issue remains unaddressed - whilst the methods in this paper 
represent significant advances for DTA-MA and DTA-NMA, they assume a perfect reference standard.
However, truly addressing imperfect gold standards fully (not merely as a meta-regression covariate) 
requires models which require full cross-tabulation tables (essentially IPD-level data), 
which are rarely available. 
This is because these models - such as latent class multivariate probit models 
(LC-MVP; 
Cerullo et al, 2025\supercite{Cerullo_2025_big_sim_study_LC_MVP_and_LT},
Xu et al, 2009 \supercite{Xu2009},
Xu et al, 2013\supercite{Xu2013},
Ueabersax et al, 1999 \supercite{Uebersax},
Chib \& Greenberg, 1998 \supercite{Chib_Greenberg_MVP_1998}) -
operate on individual-level data; hence, one must re-create the individual
data from the full cross-tabulation tables, as well as how results co-vary (between tests) for each patient in order to model
the within-study correlations conditional on the true disease status
(i.e., in order to model conditional dependence,
which is essential for obtaining acceptable accuracy estimates with such models).

%%%%
%%%% ---------------------------------------------------------------------------------------------------------
Unfortunately, our conditional binomial factorization approach
(based on Jones et al, 2019\supercite{Jones2019}) works for the current models,
but cannot extend to latent class models (e.g., LC-MVP) due to their requirement for conditional independence.
The only way to truly address this issue is for authors to report full individual-level response data as standard, 
but unfortunately this is rarely (if ever) done in clinical practice.
Also, it is important to note that this data reporting does not necessarily need to include individual-level covariates -
although this would be useful of course -
as individual-level covariates can easily be incorporated into an IPD-NMA extension of the ordinal LC-MVP model,
which we will be implementing into our BayesMVP R package
(Cerullo et al, 2025\supercite{Cerullo_BayesMVP_2025})
in the future.

%%%%%%%%%%%%%%%%%%%%%%%%%%%%%%%%%%%%%%
\subsubsection{ MetaOrdDTA R package}
\label{section_discussion_limitations_and_future_work_MetaOrdDTA}
%%%%%%%%%%%%%%%%%%%%%%%%%%%%%%%%%%%%%%
%%%%
%%%% -------------------------------------------------------------------------------------------------------------------
The MetaOrdDTA package implements the proposed ordinal models,
Jones model, and their NMA extensions,
but does not currently include time-to-event based "multiple thresholds" approaches
\supercite{Hoyer_et_al_2018_mult_thr, Zapf_et_al_2024_mult_thr}.
While our simulation results suggest ordinal approaches outperform continuous assumptions,
providing implementations of alternative methods would enable direct comparisons
on users' own datasets.

%%%%
%%%% -------------------------------------------------------------------------------------------------------------------
Furthermore, for the MetaOrdDTA R package,
implementing other proposed "multiple threshold" models besides Jones et al
(i.e.,
Hoyer et al, 2018\supercite{Hoyer_et_al_2018_mult_thr};
Zapt et al,  2024\supercite{Zapf_et_al_2024_mult_thr})
would enable an efficient and comprehensive comparison of ordinal verses Jones versus
time-to-event based "multiple thresholds" methods
\supercite{Hoyer_et_al_2018_mult_thr, Zapf_et_al_2024_mult_thr},
which would clarify their relative strengths.
This is important because, whilst all the aforementioned approaches -
besides the standard stratified-bivariate model -
handle multiple thresholds,
they make different distributional assumptions that may favor one or the other
depending on data characteristics.
Integration of these alternative methods into MetaOrdDTA\supercite{Cerullo_MetaOrdDTA_2025}
would facilitate such comparisons,
and provide analysts with a complete toolkit for ordinal test (network) meta-analysis.

%%%%%%%%%%%%%%%%%%%%%%%%%%%%%%%%%%%%%%%%%%%%%%%%%%%%%%%%%%%%%%%%%%%%%%%%%%
\subsection{ Conclusions}
\label{section_discussion_conclusions}
%%%%%%%%%%%%%%%%%%%%%%%%%%%%%%%%%%%%%%%%%%%%%%%%%%%%%%%%%%%%%%%%%%%%%%%%%%
%%%%
%%%% -------------------------------------------------------------------------------------------------------------------
Our simulation study demonstrates that ordinal models -
particularly the O-bivariate-FC -
provide superior performance for meta-analyzing ordinal diagnostic tests across diverse scenarios.
Key findings include:
%%
% sim_data_GAD_2_summary %>% filter(n_studies == 10) %>% pull(avg_RMSE) %>% mean() ## 8.916097
% sim_data_GAD_2_summary %>% filter(n_studies == 50) %>% pull(avg_RMSE) %>% mean() ## 4.206083
% ##
% sim_data_HADS_summary %>% filter(n_studies == 10) %>% pull(avg_RMSE) %>% mean() ## 5.03443
% sim_data_HADS_summary %>% filter(n_studies == 50) %>% pull(avg_RMSE) %>% mean() ## 2.31455
% ##
% sim_data_BAI_summary %>% filter(n_studies == 10) %>% pull(avg_RMSE) %>% mean() ## 6.504302
% sim_data_BAI_summary %>% filter(n_studies == 50) %>% pull(avg_RMSE) %>% mean() ## 3.232993
%%
\begin{itemize}
    \item
    Overall, the proposed ordinal models -
    particularly the ordinal-bivariate-FC model -
    showed remarkable robustness,
    performing excelling under truly ordinal processes (DGMs \#2 - \#4),
    often performing better than the well-performing Jones\supercite{Jones2019} model.
    \item
    Furthermore, the ordinal-bivariate-FC model performed competitively
    even under the continuous-assumption Jones DGM (DGM \#1),
    despite it being designed for continuous screening/diagnostic tests.
    \item
    The standard stratified-bivariate approach showed poor performance overall,
    especially for the HADS and the BAI -
    i.e., the two tests with higher number of thresholds and a higher proportion of missing threshold data -
    with performance penalties of $30\% - 50\%$ worse RMSE
    (vs. the leading model for a given DGM).
    \item 
    Contrary to popular belief, 
    we showed that it may not be ideal or appropriate to assume continuity,
    even when ones data has a large amount of ordinal categories.
    For instance, the Jones model - which assumes continuity - performed consistently worse
    (between $\sim 10.4\% - 29.6\%$ worse) than the ordinal models for the BAI ($64$ categories).
    \item 
    It is important to note that this is not a full simulation study per se,
    since we did not adopt the traditional/standard approach
    of multiplying each set of possible conditions to come up with the
    total number of combinations/scenarios.
    More specifically, if performing a "full" simulation study,
    we would have had 3-fold more scenarios, since:
    $\#\{$thresholds/tests$\}$ $\cdot$ $\#\{$prop. of missing data$\}$
    $= 3 \cdot 3 = 9$ (not $3$).
    \item
    The MetaOrdDTA R package makes these methods accessible,
    enabling researchers to properly analyze ordinal diagnostic accuracy data,
    rather than forcing dichotomization or assuming continuity.
    \item 
    Furthermore, we implemented these models in a user-friendly R package –
    MetaOrdDTA (\url{https://github.com/CerulloE1996/MetaOrdDTA}).
    The package uses Stan
    (or optionally our BayesMVP R package which is often faster),
    and offers a large array of features,
    including:
    producing MCMC summaries,
    sROC plots with credible/prediction regions,
    meta-regression,
    pairwise accuracy differences for the NMA model extensions,
    parallel K-fold cross-validation for appropriate model selection 
    (as well as meta-regression covariate selection),
    and more.
    \item 
    These advances should improve the reliability of diagnostic test evidence synthesis,
    ultimately supporting better clinical decision-making.
\end{itemize}

\newpage
\appendix 
%%
%% \section{Appendices}
%%%%%%%%%%%%%%%%%%%%%%%%%%%%%%%%%%%%%%%%%%%%%%%%%%%%%%%%%%%%%%%%%%%%%%%%%%%%%%%%%%%%%%%%%%
%%%%
%%%%
%%%%%%%%%%%%%%%%%%%%%%%%%%%%%%%%%%%%%%%%%%%%%%%%%%%%%%%%%%%%%%%%%%%%%%%%%%%%%%%%%%%%%%%%%%%%%%%%%%%%%%%%%%%%%%%%%%%%%%%%
\setcounter{figure}{0}
\setcounter{table}{0}
\renewcommand{\thefigure}{A.\arabic{figure}}
\renewcommand{\thetable}{A.\arabic{table}}
% %%%%%%%%%%%%%%%%%%%%%%%%%%%%%%%%%%%%%%%%%%%%%%%%%%%%%%%%%%%%%%%%%%%%%%%%%%%%%%%%%%%%%%%%%%%%%%%%%%%%%%%%%%%%%%%%%%%%%%
% \section{
% Appendix A: The relationship between the proposed ordinal-bivariate and the ordinal-HSROC models
% }
% \label{appendix_A}
% %%%%%%%%%%%%%%%%%%%%%%%%%%%%%%%%%%%%%%%%%%%%%%%%%%%%%%%%%%%%
%%%%
%%%%
%%%%%%%%%%%%%%%%%%%%%%%%%%%%%%%%%%%%%%%%%%%%%%%%%%%%%%%%%%%%%%%%%%%%%%%%%%%%%%%%
\section{
Appendix A: The relationship between the proposed ordinal-bivariate and the ordinal-HSROC models
}
\label{appendix_model_specs_relationship_between_Cerullo_bivariate_and_Cerullo_R_and_G_HSROC_MA_models}
%%%%%%%%%%%%%%%%%%%%%%%%%%%%%%%%%%%%%%%%%%%%%%%%%%%%%%%%%%%%%%%%%%%%%%%%%%%%%%%%
%%%%
%%%% -----------------------------------------------------------------------------------------------------------------------------
In this section, we will show how the two models we proposed in sections
\ref{section_model_specs_Cerullo_R_and_G_HSROC} and
\ref{section_model_specs_Cerullo_bivariate_Reitsma_extension}
are very closely related -
and that the ordinal-HSROC model can in fact be thought of a more restricted,
test-accuracy-specific version of the ordinal-bivariate model.

%%%%
%%%% ------------------------------------------------------------------------------------------------------------------------------
For the ordinal-HSROC model,
for the remainder of this section we will denote all parameters belonging to this model with a H-superscript,
to clearly differentiate them from the parameters of the ordinal-bivariate model.

%%%%%%%%%%%%%%%%%%%%%%%%%%%%%%%%%%%%%%%%%%%%%%%%%%%%%%%%%%%%
\subsection{
Location and cutpoint parameter relationships
}
\label{section_model_specs_relationship_between_Cerullo_bivariate_and_Cerullo_R_and_G_HSROC_MA_models_location_and_scale}
%%%%%%%%%%%%%%%%%%%%%%%%%%%%%%%%%%%%%%%%%%%%%%%%%%%%%%%%%%%%
%%%%
%%%% ------------------------------------------------------------------------------------------------------------------------------
With this in mind, 
recall (from section \ref{section_model_specs_Cerullo_R_and_G_HSROC}) that for the \textbf{Ordinal-HSROC} model:
%%%%
\begin{equation}
\begin{aligned}
\text{Sp}_{s, k} & =     \Phi\left( \frac{\left( C^H_{s, k} - (-1) \beta^H_{s}\right)}{\text{f}( (-1) \gamma^H_{s}) }  \right) \\
\text{Se}_{s, k} & = 1 - \Phi\left( \frac{\left( C^H_{s, k} - (+1) \beta^H_{s}\right)}{\text{f}( (+1) \gamma^H_{s}) }  \right)
\end{aligned}
\end{equation}
%%%%
Which means we have:
%%%%
\begin{equation}
\begin{aligned}
\Phi^{-1}({ \text{Sp}_{s, k} }) & = \frac{\left( C^H_{s, k} + \beta^H_{s}\right)}{\text{f}( -\gamma^H_{s} ) }   \\
\Phi^{-1}({ \text{Fn}_{s, k} }) & = \frac{\left( C^H_{s, k} - \beta^H_{s}\right)}{\text{f}(  \gamma^H_{s} ) } 
\end{aligned}
\label{eqsn_for_relationship_between_ordinal_HSROC_and_ordinal_bivariate_equation_1}
\end{equation}
%%%%
And for the \textbf{ordinal-bivariate} model, 
recall from section \ref{section_model_specs_Cerullo_bivariate_Reitsma_extension} that:
%%%%
\begin{equation}
\begin{aligned}
\text{Sp}_{s, k} & =     \Phi \left( C_{s, k}^{[d-]} - \beta^{[d-]}_{s}\right) \\
\text{Se}_{s, k} & = 1 - \Phi \left( C_{s, k}^{[d+]} - \beta^{[d+]}_{s}\right)
\end{aligned}
\end{equation}
%%%%
Which means we have:
%%%%
\begin{equation}
\begin{aligned}
C_{s, k}^{[d-]} - \beta^{[d-]}_{s} = \Phi^{-1}(\text{Sp}_{s, k})     \\ 
C_{s, k}^{[d+]} - \beta^{[d+]}_{s} = \Phi^{-1}(\text{Fn}_{s, k})
\end{aligned}
\label{eqsn_for_relationship_between_ordinal_HSROC_and_ordinal_bivariate_equation_2}
\end{equation}
%%%%
Hence, putting
\ref{eqsn_for_relationship_between_ordinal_HSROC_and_ordinal_bivariate_equation_1} 
and 
\ref{eqsn_for_relationship_between_ordinal_HSROC_and_ordinal_bivariate_equation_2} 
together, we get:
%%%%
\begin{equation}
\begin{aligned}
C_{s, k}^{[d-]} - \beta^{[d-]}_{s} &= \frac{ \left( C^H_{s, k} + \beta^H_{s} \right)  }{  \text{f}( -\gamma^H_{s} )  } \\
                                   &= \frac{  C^H_{s, k}    }{  \text{f}( -\gamma^H_{s} ) }  +
                                   \frac{ \beta^H_{s}  }{  \text{f}( -\gamma^H_{s} )   }  \\
C_{s, k}^{[d+]} - \beta^{[d+]}_{s} &= \frac{ \left( C^H_{s, k} - \beta^H_{s}\right)  }{  \text{f}(  \gamma^H_{s} )   } \\
                                   &= \frac{  C^H_{s, k}  }{  \text{f}(\gamma^H_{s})  }   -
                                   \frac{  {\beta^H_{s}}  }{\text{f}( \gamma^H_{s} ) }  \\
\end{aligned}
\label{eqsn_for_relationship_between_ordinal_HSROC_and_ordinal_bivariate_equation_3}
\end{equation}
%%%%
From \ref{eqsn_for_relationship_between_ordinal_HSROC_and_ordinal_bivariate_equation_3},
we can see that we can parameterize the ordinal-HSROC model as an ordinal-bivariate model,
by selecting the ordinal-bivariate cutpoint parameters as:
%%%%
\begin{equation}
\begin{aligned}
C_{s, k}^{[d+]} & = \frac{  C^H_{s, k}   }{   \text{f}(   \gamma^H_{s})   }, ~
C_{s, k}^{[d-]}   = \frac{  C^H_{s, k}   }{   \text{f}( - \gamma^H_{s})   }
\end{aligned}
\label{eqsn_for_relationship_between_ordinal_HSROC_and_ordinal_bivariate_equation_4}
\end{equation}
%%%%
And then selecting the ordinal-bivariate locations as:
%%%%
\begin{equation}
\begin{aligned}
\beta_{s}^{[d+]} & = \frac{    \beta^H_{s}   }{   \text{f}(   \gamma^H_{s})   }, ~ ~ 
\beta_{s}^{[d-]}   = \frac{  - \beta^H_{s}   }{   \text{f}( - \gamma^H_{s})   }
\end{aligned}
\label{eqsn_for_relationship_between_ordinal_HSROC_and_ordinal_bivariate_equation_5}
\end{equation}
%%%%
For now, we will focus on the case where $ \text{f}(\cdot) = \exp(\cdot) $. 
In this case, the cutpoints from equation 
\ref{eqsn_for_relationship_between_ordinal_HSROC_and_ordinal_bivariate_equation_4} become:
%%%%
\begin{equation}
\begin{aligned}
C_{s, k}^{[d+]} & =     C^H_{s, k}      \cdot \text{exp}\left( - \gamma^H_{s}  \right), \\ ~ ~ 
C_{s, k}^{[d-]} & =     C^H_{s, k}      \cdot \text{exp}\left(   \gamma^H_{s}  \right)  
\end{aligned}
\label{eqsn_for_relationship_between_ordinal_HSROC_and_ordinal_bivariate_equation_6}
\end{equation}
%%%%
And the locations from equation \ref{eqsn_for_relationship_between_ordinal_HSROC_and_ordinal_bivariate_equation_5} then become:
%%%%
\begin{equation}
\begin{aligned}
\beta_{s}^{[d+]} & =   \beta^H_{s}  \cdot \text{exp}\left( - \gamma^H_{s} \right), \\ ~ ~
\beta_{s}^{[d-]} & = - \beta^H_{s}  \cdot \text{exp}\left(   \gamma^H_{s} \right)
\end{aligned}
\label{eqsn_for_relationship_between_ordinal_HSROC_and_ordinal_bivariate_equation_7}
\end{equation}
%%%%
Evaluating \ref{eqsn_for_relationship_between_ordinal_HSROC_and_ordinal_bivariate_equation_6} at the means (i.e. pooled estimates), we obtain the following expression for the \textbf{average cutpoints}:
%%%%
\begin{equation}
\begin{aligned}
\mu_{C_{k}}^{[d+]} & =     \mu_{C_{k}}^H      \cdot \text{exp}\left( -  \mu_{\gamma}^H \right), \\ ~ ~
\mu_{C_{k}}^{[d-]} & =     \mu_{C_{k}}^H      \cdot \text{exp}\left(    \mu_{\gamma}^H \right)
\end{aligned}
\label{eqsn_for_relationship_between_ordinal_HSROC_and_ordinal_bivariate_equation_6_pt_2_evaluation_at_means}
\end{equation}
%%%%
And then evaluating \ref{eqsn_for_relationship_between_ordinal_HSROC_and_ordinal_bivariate_equation_7} at the means (i.e. pooled estimates), we obtain the following expression for the \textbf{average locations}:
%%%%
\begin{equation}
\begin{aligned}
\mu_{\beta}^{[d+]} & =   \mu_{\beta}^H  \cdot \text{exp}\left( - \mu_{\gamma}^H \right), \\ ~ ~
\mu_{\beta}^{[d-]} & = - \mu_{\beta}^H  \cdot \text{exp}\left(   \mu_{\gamma}^H \right)
\end{aligned}
\label{eqsn_for_relationship_between_ordinal_HSROC_and_ordinal_bivariate_equation_7_pt_2_evaluation_at_means}
\end{equation}
%%%%
Hence, from equations
\ref{eqsn_for_relationship_between_ordinal_HSROC_and_ordinal_bivariate_equation_6} and
\ref{eqsn_for_relationship_between_ordinal_HSROC_and_ordinal_bivariate_equation_7},
we have shown that both the cutpoints of the ordinal-bivariate model
(i.e., $ \{ C^{[d+]}_{s, k}, C^{[d-]}_{s, k}  \} $) and
the locations of the ordinal-bivariate model (i.e., $ \{ \beta^{[d+]}_{s}, ~ \beta^{[d-]}_{s} \} $)
can be expressed in terms of the locations and scale parameters of the ordinal-HSROC model
(i.e., in terms of $ \{ \beta^H_{s}, \gamma^H_{s}  \} $).
%%%%
In other words, both the cutpoints and location parameters of the ordinal-bivariate model can be written 
as functions of the parameters of the ordinal-HSROC model.
%%%%%%%%%%%%%%%%%%%%%%%%%%%%%%%%%%%%%%%%%%%%%%%%%%%%%%%%%%%%%%%%%%%%%%%%
\subsection{
Between-study heterogeneity relationships
}
\label{section_model_specs_relationship_between_Cerullo_bivariate_and_Cerullo_R_and_G_HSROC_MA_models_between_study_hetero}
%%%%%%%%%%%%%%%%%%%%%%%%%%%%%%%%%%%%%%%%%%%%%%%%%%%%%%%%%%%%%%%%%%%%%%%%
%%%%
%%%% ------------------------------------------------------------------------------------------------------------------------------
Next, we will explore the relationships between the ordinal-bivariate between-study heterogeneity and correlation parameters 
(i.e., $ \{ \sigma^{[d+]}_{\beta}, \sigma^{[d-]}_{\beta}, \rho_{\beta} \} $) and the ordinal-HSROC model parameters 
(i.e., $ \{ \beta^H_{s}, \gamma^H_{s}, \sigma_{\beta^H}, \sigma_{\gamma^H}) \} $).
%%%%
We will begin this subsection by first recalling
(from section \ref{section_model_specs_Cerullo_R_and_G_HSROC})
that the between-study model of the ordinal-HSROC model is characterized by:
%%%%
\begin{equation}
\begin{aligned}
\beta^H_s  & \sim \text{normal}\left(  \mu_{\beta}^H , \sigma^H_{\beta}   \right) \\
\gamma^H_s & \sim \text{normal}\left(  \mu_{\gamma}^H, \sigma^H_{\gamma}  \right) \\
 & \text{Cov}(\beta^H_s, \gamma^H_s) = 0
\end{aligned}
\label{eqsn_for_relationship_between_ordinal_HSROC_and_ordinal_bivariate_equation_8}
\end{equation}
%%%%%%%%%%%%%%%%%%%%%%%%%%%%%%%%%%%%%%%%%%%%%%%%%%%%%%%%%%%%%%%%%%%%%%%%
\subsubsection{\underline{
The delta method (Version I)
}}
%%%%%%%%%%%%%%%%%%%%%%%%%%%%%%%%%%%%%%%%%%%%%%%%%%%%%%%%%%%%%%%%%%%%%%%%
%%%%
%%%% ------------------------------------------------------------------------------------------------------------------------------
Before we do the next step, we will first have to define the \textbf{delta method} -
specifically the delta method applied to a single function ($g(\cdot)$)
which is a function of two \textbf{independent} random variables $X$ and $Y$,
which each have a univariate normal distribution.
%%%%
In other words, we can write $g$ as:
$g(\cdot) = g(X, Y)$.
%%%%
And we can write the independent random variables $X$ and $Y$ as:
%%%%
\begin{equation}
\begin{aligned}
& X \sim \text{normal}\left( \mu_{X}, \sigma^2_{X} ~ \right), ~  \\
& Y \sim \text{normal}\left( \mu_{Y}, \sigma^2_{Y} ~ \right)
\end{aligned}
\end{equation}
%%%%
Then, the delta method says that the variance of the function $g = g(X, Y)$ can be approximated to be:
%%%%
\begin{equation}
\begin{aligned}
\text{Var}[g(X,Y)] \approx \left(  {\left[ \frac{  \partial   g(\mu_{X}, \mu_{Y})   }{  \partial X  } \right]}^2 \cdot \left[ \sigma^2_{X}    \right]   \right) & +
                           \left(  {\left[ \frac{  \partial   g(\mu_{X}, \mu_{Y})   }{  \partial Y  } \right]}^2 \cdot \left[ \sigma^2_{Y}    \right]   \right)
\end{aligned}
\label{eqsn_for_relationship_between_ordinal_HSROC_and_ordinal_bivariate_equation_9_delta_method_single_fn_two_indep_RVs}
\end{equation}
%%%%
%%%%
%%%%%%%%%%%%%%%%%%%%%%%%%%%%%%%%%%%%%%%%%%%%%%%%%%%%%%%%%%%%%%%%%%%%%%%%
\subsubsection{\underline{  
Variance relationships ($ \{ ( \sigma^{[d+]}_{\beta}  )^2, ( \sigma^{[d-]}_{\beta} )^2 \} $)  
}}
%%%%%%%%%%%%%%%%%%%%%%%%%%%%%%%%%%%%%%%%%%%%%%%%%%%%%%%%%%%%%%%%%%%%%%%%
%%%%
%%%% ------------------------------------------------------------------------------------------------------------------------------
Hence, using the delta method defined in
equation \ref{eqsn_for_relationship_between_ordinal_HSROC_and_ordinal_bivariate_equation_9_delta_method_single_fn_two_indep_RVs} above, 
we can write $ \left( \sigma^{[d+]}_{\beta} \right)^2 $ (i.e., the variance of $\beta_s^{[D+]}$) as:
%%%%
%%%%
\begin{equation}
\begin{aligned}
\left( \sigma^{[d+]}_{\beta} \right)^2  & = \left[  \frac{  \partial \mu_{\beta}^{[D+]}  }{  \partial \mu_{\beta}^H    }  \right]^2  \cdot \left[ (\sigma^H_{\beta})^2  \right] + 
                                            \left[  \frac{  \partial \mu_{\beta}^{[D+]}  }{  \partial \mu_{\gamma}^H   }  \right]^2  \cdot \left[ (\sigma^H_{\gamma})^2 \right]
\end{aligned}
\label{eqsn_for_relationship_between_ordinal_HSROC_and_ordinal_bivariate_equation_10_var_of_beta_diseased_initial_form}
\end{equation}
%%%%
Then, since $ \mu_{\beta}^{[D+]} = \frac{  \mu_{\beta}^H }{ \exp\left( \mu_{\gamma}^H \right)  } $, we can compute the two partial derivatives needed:
%%%%
\begin{align*}
\frac{  \partial \mu_{\beta}^{[D+]}  }{  \partial \mu_{\beta}^H  }  & = \exp( - \mu_{\gamma}^H ) \\
\frac{  \partial \mu_{\beta}^{[D+]}  }{  \partial \mu_{\gamma}^H  } & = - \mu_{\beta}^H \cdot \exp( - \mu_{\gamma}^H )
\end{align*}
%%%%
And then their squares:
%%%%
\begin{align*}
\left[  \frac{  \partial \mu_{\beta}^{[D+]}  }{  \partial \mu^H_{\beta}    }  \right]^2  & = \exp\left(  -2 \cdot \mu_{\gamma}^H  \right) \\
\left[  \frac{  \partial \mu_{\beta}^{[D+]}  }{  \partial \mu^H_{\gamma}   }  \right]^2  & = \left(  \mu_{\beta}^H  \right)^2 \cdot \exp\left( -2 \cdot \mu_{\gamma}^H  \right)
\end{align*}
%%%%
Then, substituting these squared partial-derivatives into equation
\ref{eqsn_for_relationship_between_ordinal_HSROC_and_ordinal_bivariate_equation_10_var_of_beta_diseased_initial_form} given above, 
we get (after factorizing out the exponential term):
%%%%
\begin{equation}
\begin{aligned}
\left( \sigma^{[d+]}_{\beta} \right)^2   = \exp\left( - 2 \cdot \mu_{\gamma}^H \right) \cdot \left[  \left(  \sigma^H_{\beta} \right)^2 ~ + ~ \left( \mu_{\beta}^H  \right)^2 \left( \sigma^H_{\gamma} \right)^2   \right] 
\end{aligned}
\label{eqsn_for_relationship_between_ordinal_HSROC_and_ordinal_bivariate_equation_11_var_of_beta_diseased_final_form}
\end{equation}
%%%%
%%%%
%%%%
%%%%
%%%% ------------------------------------------------------------------------------------------------------------------------------
Following the same steps as above,
we get a very similar expression for the variance in the non-diseased population ($\beta_s^{[D-]}$), 
by first applying the delta method again:
%%%%
%%%%
\begin{equation}
\begin{aligned}
\left( \sigma^{[d-]}_{\beta} \right)^2  & = \left[  \frac{  \partial \mu_{\beta}^{[d-]}  }{  \partial \mu_{\beta}^H    }  \right]^2  \cdot \left[ ( \sigma^H_{\beta} )^2  \right] + 
                                            \left[  \frac{  \partial \mu_{\beta}^{[d-]}  }{  \partial \mu_{\gamma}^H   }  \right]^2  \cdot \left[ ( \sigma^H_{\gamma} )^2 \right]
\end{aligned}
\label{eqsn_for_relationship_between_ordinal_HSROC_and_ordinal_bivariate_equation_12_var_of_beta_non_diseased_initial_form}
\end{equation}
%%%%
Then, since $ \mu_{\beta}^{[D-]} = - \frac{ \mu_{\beta}^H }{ \exp\left( - \mu_{\gamma}^H \right) } $, we can compute the two partial derivatives as well as their squares:
%%%%
\begin{align*}
\frac{  \partial \mu_{\beta}^{[D-]}  }{   \partial \mu_{\beta}^H  }   = - \exp( \mu_{\gamma}^H )                       \implies &
        \left[ \frac{\partial \mu_{\beta}^{[D-]}}{\partial \mu_{\beta}^H}   \right]^2 = \exp( 2 \cdot \mu_{\gamma}^H )              \\
\frac{  \partial \mu_{\beta}^{[D-]}  }{   \partial \mu_{\gamma}^H  }  = - \mu_{\beta}^H \cdot \exp( \mu_{\gamma}^H )   \implies &
        \left[ \frac{  \partial \mu_{\beta}^{[D-]}  }{  \partial \mu_{\gamma}^H  }  \right]^2 = \left(   \mu_{\beta}^H \right)^2 \cdot   \exp( 2 \cdot \mu_{\gamma}^H )
\end{align*}
%%%%
Then, substituting these squared partial derivatives into equation 
\ref{eqsn_for_relationship_between_ordinal_HSROC_and_ordinal_bivariate_equation_12_var_of_beta_non_diseased_initial_form} above, 
we obtain:
%%%%
\begin{equation}
\begin{aligned}
\left( \sigma^{[d-]}_{\beta} \right)^2 & =
\exp\left( 2 \cdot \mu_{\gamma}^H \right) \cdot
\left[ \left( \sigma^H_{\beta} \right)^2 ~ + ~ \left( \mu_{\beta}^H \right)^2 \left( \sigma^H_{\gamma} \right)^2 \right]
\end{aligned}
\label{eqsn_for_relationship_between_ordinal_HSROC_and_ordinal_bivariate_equation_13_var_of_beta_non_diseased_final_form}
\end{equation}
%%%%
Furthermore, from equations 
\ref{eqsn_for_relationship_between_ordinal_HSROC_and_ordinal_bivariate_equation_13_var_of_beta_non_diseased_final_form} and
\ref{eqsn_for_relationship_between_ordinal_HSROC_and_ordinal_bivariate_equation_11_var_of_beta_diseased_final_form}
derived above, 
we can see that using the ordinal-HSROC model is equivalent to using an ordinal-bivariate model, 
but one where the ratio of the ratio of the between-study SD's (of the ordinal-bivariate model) 
is determined solely by the mean of the scale parameter ($ \mu_{\gamma}^H $):
%%%%
\begin{equation}
\begin{aligned}
\frac{    \sigma^{[d-]}_{\beta}    }{    \sigma^{[d+]}_{\beta}     } = \exp\left( 2 \cdot \mu_{\gamma}^H \right)
\end{aligned}
\label{eqsn_for_relationship_between_ordinal_HSROC_and_ordinal_bivariate_equation_14_variance_ratio_relationship}
\end{equation}
%%%%
%%%%
%%%%%%%%%%%%%%%%%%%%%%%%%%%%%%%%%%%%%%%%%%%%%%%%%%%%%%%%%%%%%%%%%%%%%%%%
\subsection{
Between-study correlation parameter relationships
}
\label{section_model_specs_relationship_between_Cerullo_bivariate_and_Cerullo_R_and_G_HSROC_MA_models_between_study_corr}
%%%%%%%%%%%%%%%%%%%%%%%%%%%%%%%%%%%%%%%%%%%%%%%%%%%%%%%%%%%%%%%%%%%%%%%%
%%%%
%%%% ------------------------------------------------------------------------------------------------------------------------------
In this section, we will derive the relationship between the between-study correlation parameter of the ordinal-bivariate model 
($ \rho_{\beta} $) and the ordinal-HSROC model parameters. 
%%%%%%%%%%%%%%%%%%%%%%%%%%%%%%%%%%%%%%%%%%%%%%%%%%%%%%%%%%%%%%%%%%%%%%%%
\subsubsection{\underline{
The delta method (Version II)
}}
%%%%%%%%%%%%%%%%%%%%%%%%%%%%%%%%%%%%%%%%%%%%%%%%%%%%%%%%%%%%%%%%%%%%%%%%
%%%%
%%%% ------------------------------------------------------------------------------------------------------------------------------
Before we find the relationship between the between-study correlation parameter of the ordinal-bivariate model
(i.e. $\rho_{\beta}$) and the ordinal-HSROC parameters, 
we will first need to define a second variation of the delta method -
specifically the delta method applied to two functions ($f$ and $g$), 
where both $f$ and $g$ are functions of two random variables $X$ and $Y$ which are jointly distributed using a bivariate normal distribution.
%%%%
In other words, we can write the functions $f$ and $g$ as:
$f(\cdot) = f(X, Y)$ and $g(\cdot) = g(X, Y)$.
%%%%
Furthermore, we can write the two-dimensional random variable $\{ X, Y \} $ as:
%%%%
\begin{equation}
\begin{aligned} 
          \begin{bmatrix}    
                 X \\
                 Y   
            \end{bmatrix}  
\sim
\text{bivariate_normal} \left(    
            \begin{bmatrix} 
                  \mu_{X} \\
                  \mu_{Y} 
            \end{bmatrix}, ~~
            \begin{bmatrix} 
                 \   \sigma^2_{X} &
                 \rho  \sigma_{X}  \sigma_{Y} \\
                 \rho  \sigma_{X}  \sigma_{Y} & 
                \ \sigma^2_{Y}
             \end{bmatrix}
 \right),
\end{aligned}
\end{equation}
%%%%
Then, the delta method (two-dimensional version) - 
applied to two functions where each is a function of two random variables which have a bivariate normal distribution -
says that the covariance between the two functions can be approximated to be:
%%%%
\begin{equation}
\begin{aligned}
\text{Cov}[f(X,Y), g(X,Y)] \approx \left( \frac{\partial f}{\partial X} \cdot \frac{\partial g}{\partial X} \cdot \left[ \sigma^2_{X}    \right]   \right) & +  
                                   \left( \frac{\partial f}{\partial Y} \cdot \frac{\partial g}{\partial Y} \cdot \left[ \sigma^2_{Y}    \right]   \right)   ~ ~ +  \\ 
                                   \left( \frac{\partial f}{\partial X} \cdot \frac{\partial g}{\partial Y} \cdot \left[ \rho \sigma_{X} \sigma_{Y} \right]  \right) & +  
                                   \left( \frac{\partial f}{\partial Y} \cdot \frac{\partial g}{\partial X} \cdot\left[  \rho \sigma_{X} \sigma_{Y} \right] \right)
\end{aligned}
\label{eqsn_for_relationship_between_ordinal_HSROC_and_ordinal_bivariate_equation_15_delta_method_two_dim_version}
\end{equation}
%%%%
%%%%
%%%%%%%%%%%%%%%%%%%%%%%%%%%%%%%%%%%%%%%%%%%%%%%%%%%%%%%%%%%%%%%%%%%%%%%%
\subsubsection{\underline{
Correlation relationships ($ \rho_{\beta} $)
}}
%%%%%%%%%%%%%%%%%%%%%%%%%%%%%%%%%%%%%%%%%%%%%%%%%%%%%%%%%%%%%%%%%%%%%%%%
%%%%
%%%% ------------------------------------------------------------------------------------------------------------------------------
Now, using the delta method defined in 
equation \ref{eqsn_for_relationship_between_ordinal_HSROC_and_ordinal_bivariate_equation_9_delta_method_two_dim_version},
we can write the covariance between the location parameters of the ordinal-bivariate model in terms of the ordinal-HSROC model as:
%%%%
%%%%
\begin{equation}
\begin{aligned}
\text{Cov}[ \beta_{s}^{[d+]}, \beta_{s}^{[d-]} ] \approx 
                                   \left( \frac{\partial \mu_{\beta}^{[d+]}  }{\partial \mu_{\beta}^H    } \cdot \frac{\partial \mu_{\beta}^{[d-]}  }{  \partial \mu_{\beta}^H       } \cdot \left[ (\sigma^H_{\beta})^2    \right]   \right) & +  
                                   \left( \frac{\partial \mu_{\beta}^{[d+]}  }{\partial \gamma_{\beta}^H } \cdot \frac{\partial \mu_{\beta}^{[d-]}  }{  \partial \gamma_{\beta}^H    } \cdot \left[ (\sigma^H_{\gamma})^2    \right]   \right)   ~ ~ +  \\ 
                                   \left( \frac{\partial \mu_{\beta}^{[d+]}  }{\partial \mu_{\beta}^H    } \cdot \frac{\partial \mu_{\beta}^{[d-]}  }{  \partial \gamma_{\beta}^H    } \cdot \left[ (\rho_{\beta, \gamma}^H) \sigma_{\beta}^H \sigma_{\gamma}^H \right]  \right) & +  
                                   \left( \frac{\partial \mu_{\beta}^{[d+]}  }{\partial\gamma_{\beta}^H  } \cdot \frac{\partial \mu_{\beta}^{[d-]}  }{  \partial \mu_{\beta}^H       } \cdot\left[  (\rho_{\beta, \gamma}^H) \sigma_{\beta}^H \sigma_{\gamma}^H \right] \right)
\end{aligned}
\label{eqsn_for_relationship_between_ordinal_HSROC_and_ordinal_bivariate_equation_16_covariance_using_2nd_delta_method}
\end{equation}
%%%%
Where: 
$ \rho_{\beta, \gamma}^H  $ is the correlation between $ \beta_{s}^H $ and $ \gamma_{s}^H $.
However, by the definition of the ordinal-HSROC model, this correlation is zero (i.e. the location and scale parameters are independent), 
and therefore equation 
\ref{eqsn_for_relationship_between_ordinal_HSROC_and_ordinal_bivariate_equation_16_covariance_using_2nd_delta_method} 
simplifies to:
%%%%
\begin{equation}
\begin{aligned}
\text{Cov}[ \beta_{s}^{[d+]}, \beta_{s}^{[d-]} ] \approx 
                                   \left( \frac{\partial \mu_{\beta}^{[d+]}  }{\partial \mu_{\beta}^H    } \cdot \frac{\partial \mu_{\beta}^{[d-]}  }{  \partial \mu_{\beta}^H       } \cdot \left[ (\sigma^H_{\beta})^2    \right]   \right) & +  
                                   \left( \frac{\partial \mu_{\beta}^{[d+]}  }{\partial \gamma_{\beta}^H } \cdot \frac{\partial \mu_{\beta}^{[d-]}  }{  \partial \gamma_{\beta}^H    } \cdot \left[ (\sigma^H_{\gamma})^2    \right]   \right)
\end{aligned}
\label{eqsn_for_relationship_between_ordinal_HSROC_and_ordinal_bivariate_equation_17_covariance_using_2nd_delta_method_simplified}
\end{equation}
%%%%
%%%%
The partial derivatives we need to compute in equation \ref{eqsn_for_relationship_between_ordinal_HSROC_and_ordinal_bivariate_equation_17_covariance_using_2nd_delta_method_simplified}
have already been computed as part of the computations needed to work out the relationship in the variance parameters
(see section \ref{section_model_specs_relationship_between_Cerullo_bivariate_and_Cerullo_R_and_G_HSROC_MA_models_____between_study_hetero}). 
%%%%
Hence, we can write this as:
%%%%
\begin{equation*}
\begin{aligned}
\text{Cov}[ \beta_{s}^{[d+]}, \beta_{s}^{[d-]} ]  \approx 
                                     \left(  \left[   \exp\left( - \mu_{\gamma}^H \right)  \right]  \cdot  \left[ -  \exp\left(  \mu_{\gamma}^H \right)   \right]   \cdot \left[ (\sigma^H_{\beta})^2    \right]   \right) & \\ +  
                                     \left(  \left[ \left( - \mu_{\beta}^H  \right)  \exp\left( - \mu_{\gamma}^H \right)  \right]  \cdot  \left[   \left( - \mu_{\beta}^H  \right) \exp\left(  \mu_{\gamma}^H \right)   \right]   \cdot \left[ (\sigma^H_{\gamma})^2    \right]   \right)
\end{aligned}
\end{equation*}
%%%%
Which simplifies to:
%%%%
\begin{equation}
\begin{aligned}
\text{Cov}[ \beta_{s}^{[d+]}, \beta_{s}^{[d-]} ]  \approx      \left( \sigma_{\gamma}^H \right)^2  \left( \mu_{\beta}^H \right)^2  ~  -  ~   \left(  \sigma_{\beta}^H  \right)^2                 
\end{aligned}
\label{eqsn_for_relationship_between_ordinal_HSROC_and_ordinal_bivariate_equation_18_covariance_using_2nd_delta_method_final_form}
\end{equation}
%%%%
Hence the correlation between the location parameters in the diseased and non-diseased populations can be found by computing:
%%%%
\begin{equation}
\begin{aligned}
\rho_{\beta} = \frac{      \left( \sigma_{\gamma}^H \right)^2  \left( \mu_{\beta}^H \right)^2  ~  -  ~   \left(  \sigma_{\beta}^H  \right)^2       }{   \sigma_{\beta}^{[d+]}  \cdot \sigma_{\beta}^{[d-]}   }
\end{aligned}
\label{eqsn_for_relationship_between_ordinal_HSROC_and_ordinal_bivariate_equation_19_correlation_using_2nd_delta_method_final_form}
\end{equation}
%%%%
Where 
$  \sigma_{\beta}^{[d+]} $ and  $ \sigma_{\beta}^{[d-]}   $
were found previously in section 
\ref{section_model_specs_relationship_between_Cerullo_bivariate_and_Cerullo_R_and_G_HSROC_MA_models_____between_study_hetero}. 
%%%%
%%%%
%%%%%%%%%%%%%%%%%%%%%%%%%%%%%%%%%%%%%%%%%%%%%%%%%%%%%%%%%%%%%%%%%%%%%%%%%%%%%%%%%%%%%%%%%%%%%%%%%%%%%%%%%%%%%%%%%%%%%%%%
\setcounter{figure}{0}
\setcounter{table}{0}
\renewcommand{\thefigure}{B.\arabic{figure}}
\renewcommand{\thetable}{B.\arabic{table}}
%%%%%%%%%%%%%%%%%%%%%%%%%%%%%%%%%%%%%%%%%%%%%%%%%%%%%%%%%%%%%%%%%%%%%%%%%%%%%%%%%%%%%%%%%%%%%%%%%%%%%%%%%%%%%%%%%%%%%%%%
%%%%%%%%%%%%%%%%%%%%%%%%%%%%%%%%%%%%%%%%%%%%%%%%%%%%%%%%%%%%%%%%%%%%%%%%%%%%%%%%%%%%%%%%%%%%%%%%%%%%%%%%%%%%%%%%%%%%%%%%
\section{ Appendix B: RMSE plots}
\label{appendix_B_RMSE_plots}
%%%%%%%%%%%%%%%%%%%%%%%%%%%%%%%%%%%%%%%%%%%%%%%%%%%%%%%%%%%%%%%%%%%%%%%%%%%%%%%%%%%%%%%%%%%%%%%%%%%%%%%%%%%%%%%%%%%%%%%%
%%%%%%%%%%%%%%%%%%%%%%%%%%%%%%%%%%%%%%%%%%%%%%%%%%%%%%%%%%%%%%%%%%%%%%%%%%%%%%%%%%%%%%%%%%%%%%%%%%%%%%%%%%%%%%%%%%%%%%%%
\subsection{ Appendix B: RMSE plots; GAD-2 (test I)}
\label{appendix_B_RMSE_plots_GAD_2_test_I}
%%%%%%%%%%%%%%%%%%%%%%%%%%%%%%%%%%%%%%%%%%%%%%%%%%%%%%%%%%%%%%%%%%%%%%%%%%%%%%%%%%%%%%%%%%%%%%%%%%%%%%%%%%%%%%%%%%%%%%%%
%%%%
%%%%
%%%%%%%%%%%%%%%%%%%%%%%%%%%%%%%%%%%%%%%%
\begin{figure}[H]
    \centering
    \includegraphics[width=15cm]{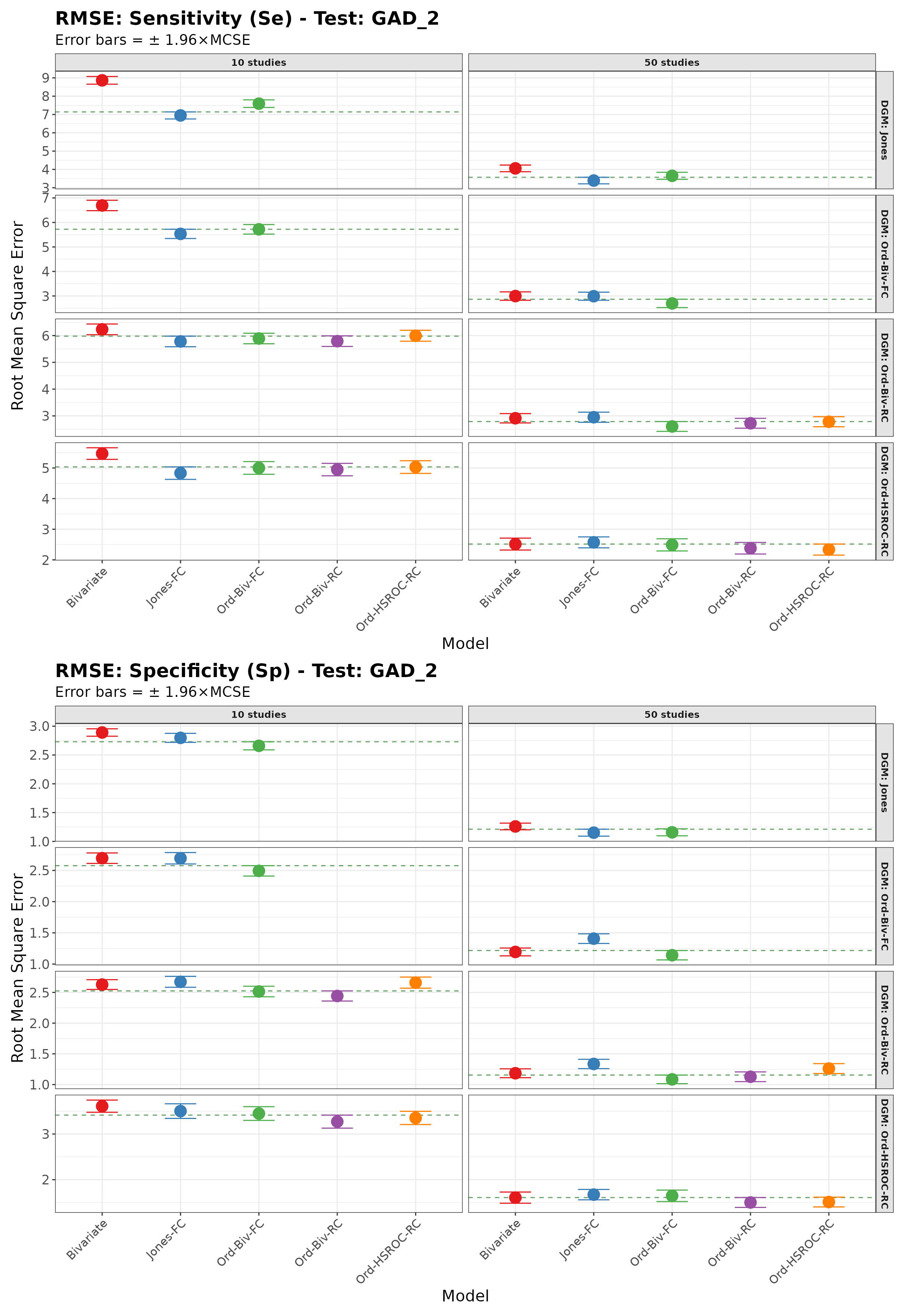}
    \caption{\footnotesize{
        Plot of RMSE (with $95\%$ MCSE intervals) for simulation study - for GAD-2 test.
    }}
    \label{Sim_study_RMSE_GAD_2}
\end{figure}
%%%%%%%%%%%%%%%%%%%%%%%%%%%%%%%%%%%%%%%%
%%%%
%%%%
%%%%%%%%%%%%%%%%%%%%%%%%%%%%%%
\subsection{ Appendix B: RMSE plots; HADS (test II)}
\label{appendix_B_RMSE_plots_HADS_test_II}
%%%%%%%%%%%%%%%%%%%%%%%%%%%%%%
%%%%
%%%%
%%%%%%%%%%%%%%%%%%%%%%%%%%%%%%%%%%%%%%%%
\begin{figure}[H]
    \centering
    \includegraphics[width=15cm]{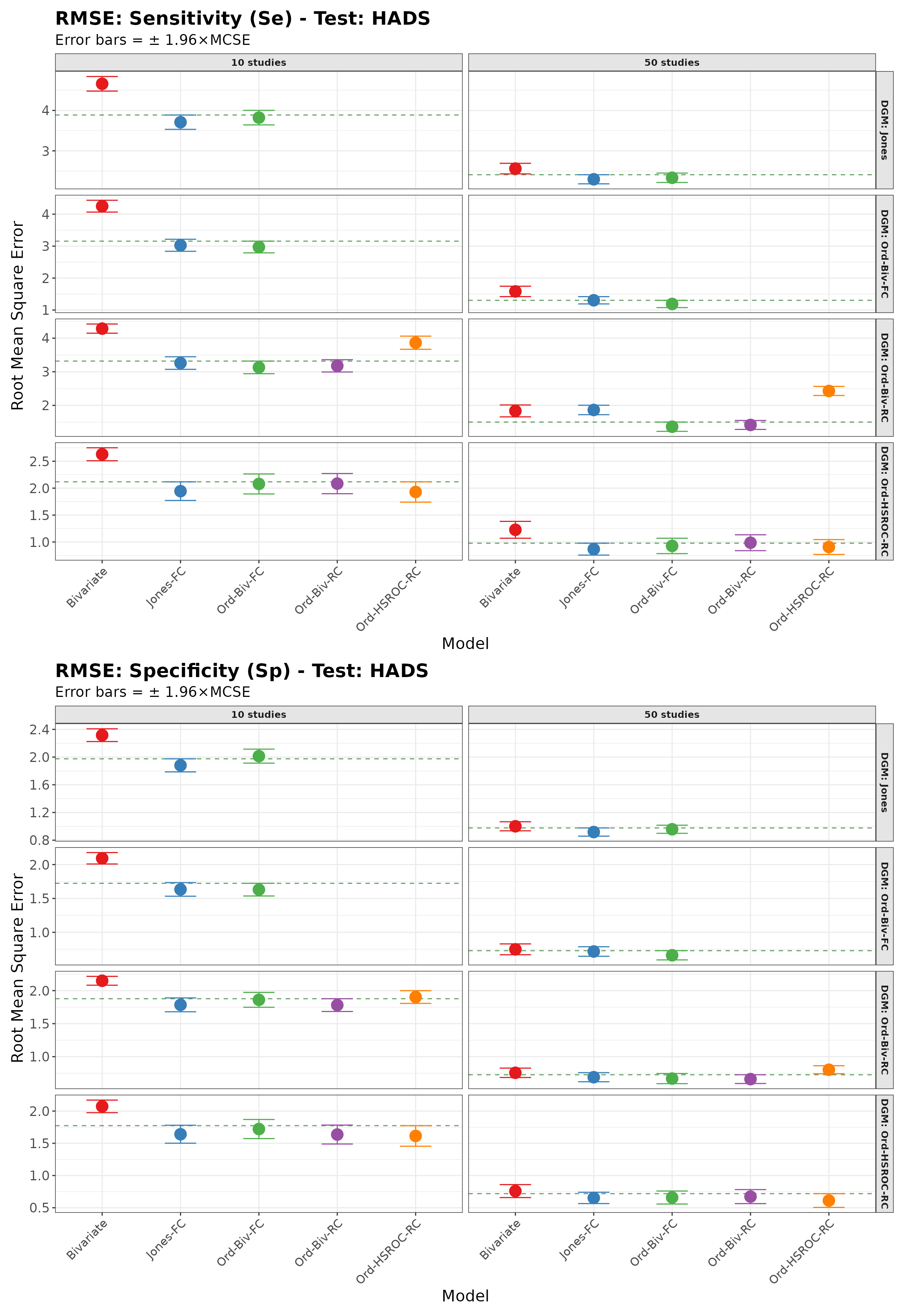}
    \caption{\footnotesize{
        Plot of RMSE (with $95\%$ MCSE intervals) for simulation study - for HADS test.
    }}
    \label{Sim_study_RMSE_HADS}
\end{figure}
%%%%%%%%%%%%%%%%%%%%%%%%%%%%%%%%%%%%%%%%
%%%%
%%%%
%%%%%%%%%%%%%%%%%%%%%%%%%%%%%%
\subsection{ Appendix B: RMSE plots; BAI (test III)}
\label{appendix_B_RMSE_plots_BAI_test_III}
%%%%%%%%%%%%%%%%%%%%%%%%%%%%%%
%%%%
%%%%
%%%%%%%%%%%%%%%%%%%%%%%%%%%%%%%%%%%%%%%%
\begin{figure}[H]
    \centering
    \includegraphics[width=15cm]{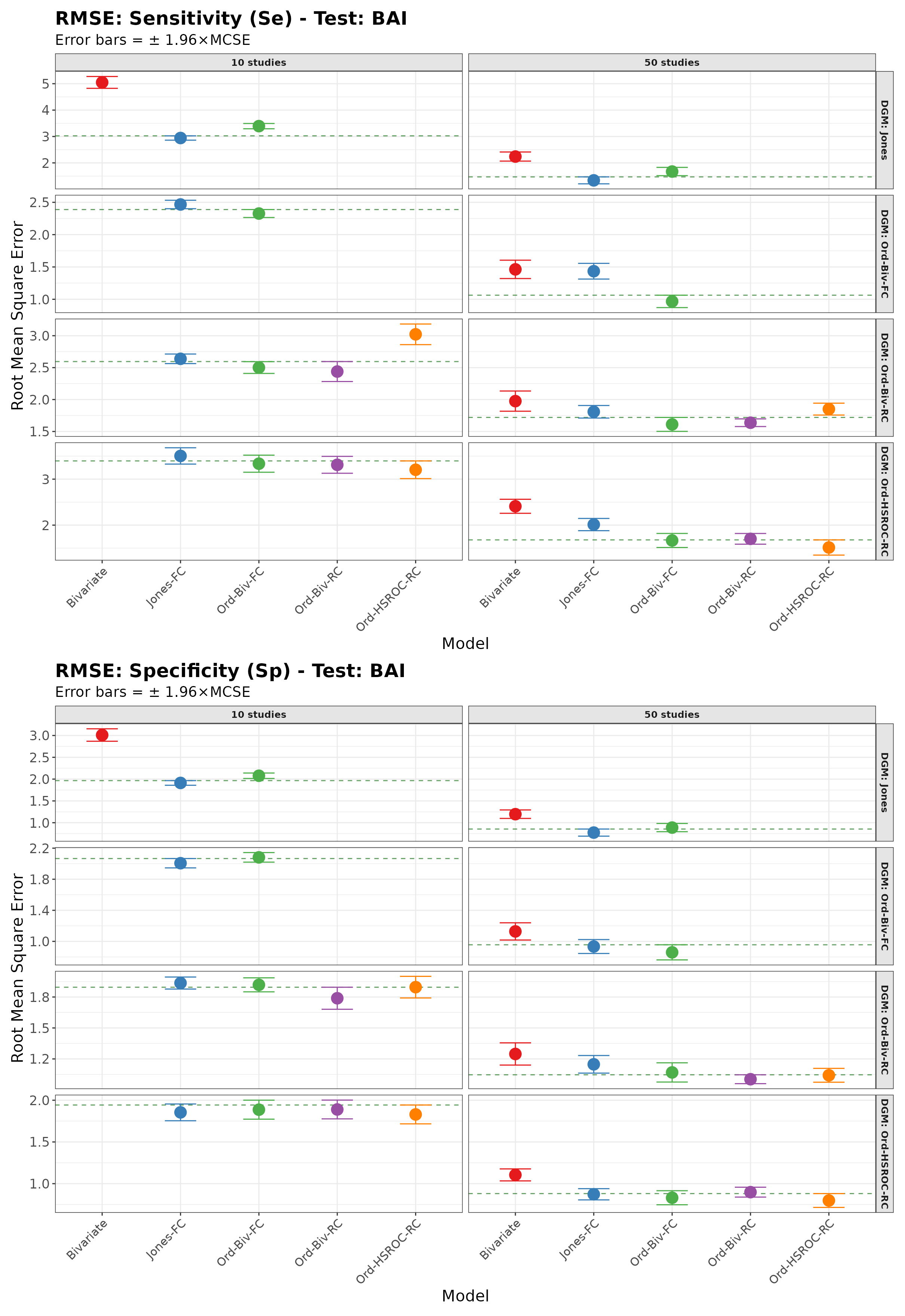}
    \caption{\footnotesize{
        Plot of RMSE (with $95\%$ MCSE intervals) for simulation study - for BAI test.
    }}
    \label{Sim_study_RMSE_BAI}
\end{figure}
%%%%%%%%%%%%%%%%%%%%%%%%%%%%%%%%%%%%%%%%
%%%%
%%%%
%%%%
%%%%
%%%%%%%%%%%%%%%%%%%%%%%%%%%%%%%%%%%%%%%%%%%%%%%%%%%%%%%%%%%%%%%%%%%%%%%%%%%%%%%%%%%%%%%%%%%%%%%%%%%%%%%%%%%%%%%%%%%%%%%%
\setcounter{figure}{0}
\setcounter{table}{0}
\renewcommand{\thefigure}{C.\arabic{figure}}
\renewcommand{\thetable}{C.\arabic{table}}
%%%%%%%%%%%%%%%%%%%%%%%%%%%%%%%%%%%%%%%%%%%%%%%%%%%%%%%%%%%%%%%%%%%%%%%%%%%%%%%%%%%%%%%%%%%%%%%%%%%%%%%%%%%%%%%%%%%%%%%%
%%%%%%%%%%%%%%%%%%%%%%%%%%%%%%%%%%%%%%%%%%%%%%%%%%%%%%%%%%%%%%%%%%%%%%%%%%%%%%%%%%%%%%%%%%%%%%%%%%%%%%%%%%%%%%%%%%%%%%%%
\section{ Appendix C: Bias plots}
\label{appendix_C_Bias_plots}
%%%%%%%%%%%%%%%%%%%%%%%%%%%%%%%%%%%%%%%%%%%%%%%%%%%%%%%%%%%%%%%%%%%%%%%%%%%%%%%%%%%%%%%%%%%%%%%%%%%%%%%%%%%%%%%%%%%%%%%%
%%%%%%%%%%%%%%%%%%%%%%%%%%%%%%
\subsection{ Appendix C: Bias plots; GAD-2 (test I)}
\label{appendix_C_Bias_plots_GAD_2_test_I}
%%%%%%%%%%%%%%%%%%%%%%%%%%%%%%
%%%%
%%%%
%%%%%%%%%%%%%%%%%%%%%%%%%%%%%%%%%%%%%%%%
\begin{figure}[H]
    \centering
    \includegraphics[width=15cm]{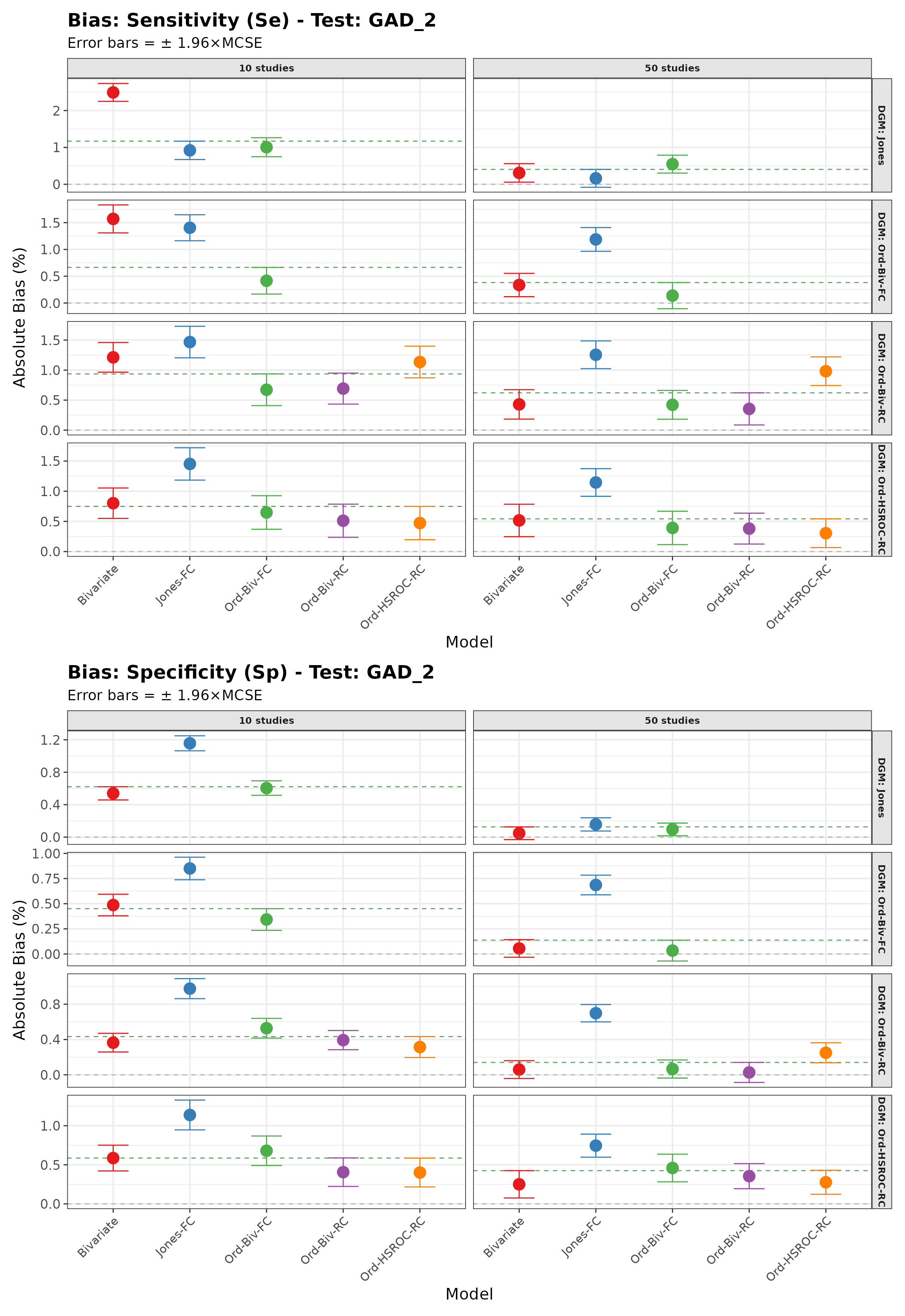}
    \caption{\footnotesize{
        Plot of Bias (with $95\%$ MCSE intervals) for simulation study - for GAD-2 test.
    }}
    \label{Sim_study_Bias_GAD_2}
\end{figure}
%%%%%%%%%%%%%%%%%%%%%%%%%%%%%%%%%%%%%%%%
%%%%
%%%%
%%%%%%%%%%%%%%%%%%%%%%%%%%%%%%
\subsection{ Appendix C: Bias plots; HADS (test II)}
\label{appendix_C_Bias_plots_HADS_test_II}
%%%%%%%%%%%%%%%%%%%%%%%%%%%%%%
%%%%
%%%%
%%%%%%%%%%%%%%%%%%%%%%%%%%%%%%%%%%%%%%%%
\begin{figure}[H]
    \centering
    \includegraphics[width=15cm]{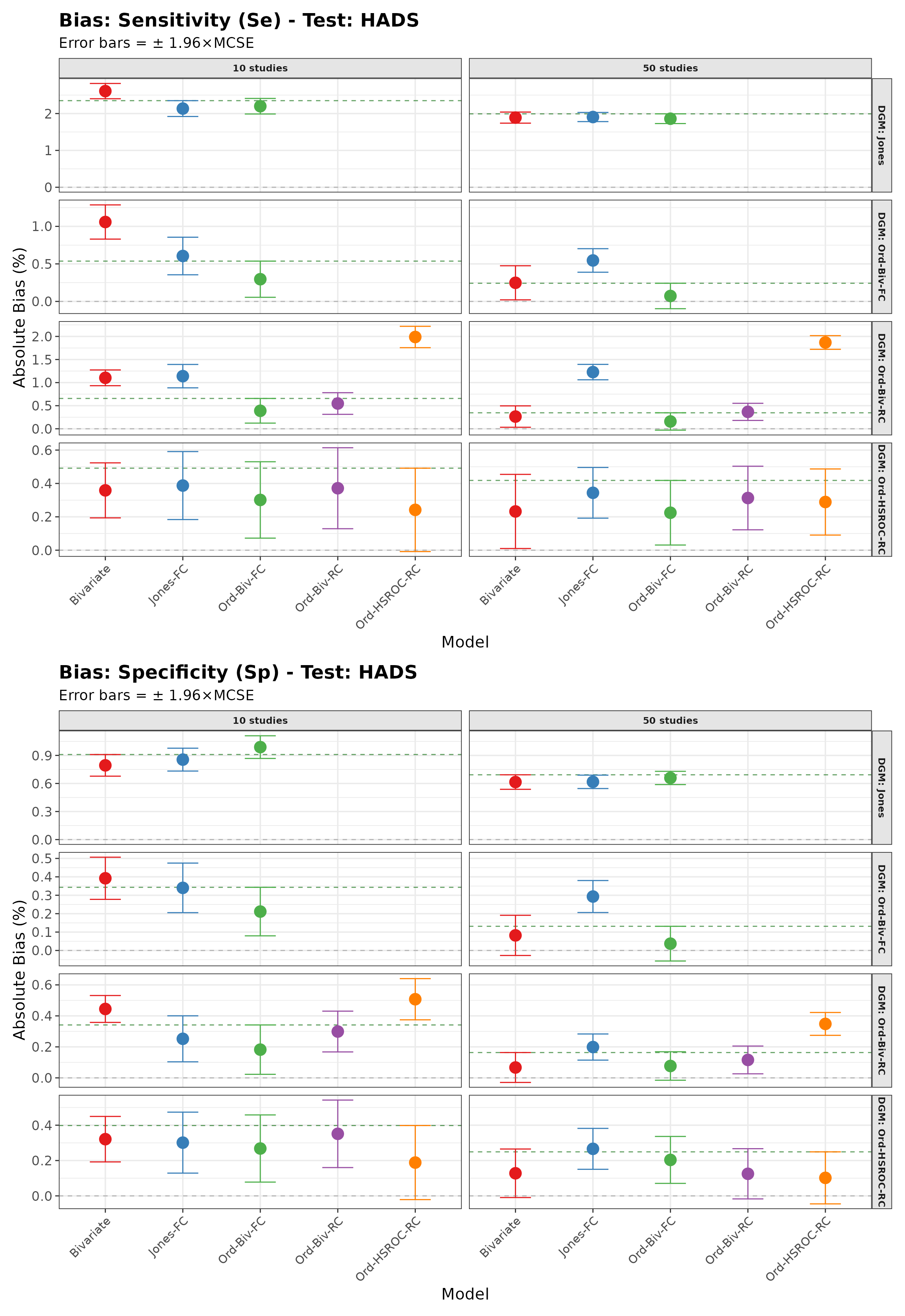}
    \caption{\footnotesize{
        Plot of Bias (with $95\%$ MCSE intervals) for simulation study - for HADS test.
    }}
    \label{Sim_study_Bias_HADS}
\end{figure}
%%%%%%%%%%%%%%%%%%%%%%%%%%%%%%%%%%%%%%%%
%%%%
%%%%
%%%%%%%%%%%%%%%%%%%%%%%%%%%%%%
\subsection{ Appendix C: Bias plots; BAI (test III)}
\label{appendix_C_Bias_plots_BAI_test_III}
%%%%%%%%%%%%%%%%%%%%%%%%%%%%%%
%%%%
%%%%
%%%%%%%%%%%%%%%%%%%%%%%%%%%%%%%%%%%%%%%%
\begin{figure}[H]
    \centering
    \includegraphics[width=15cm]{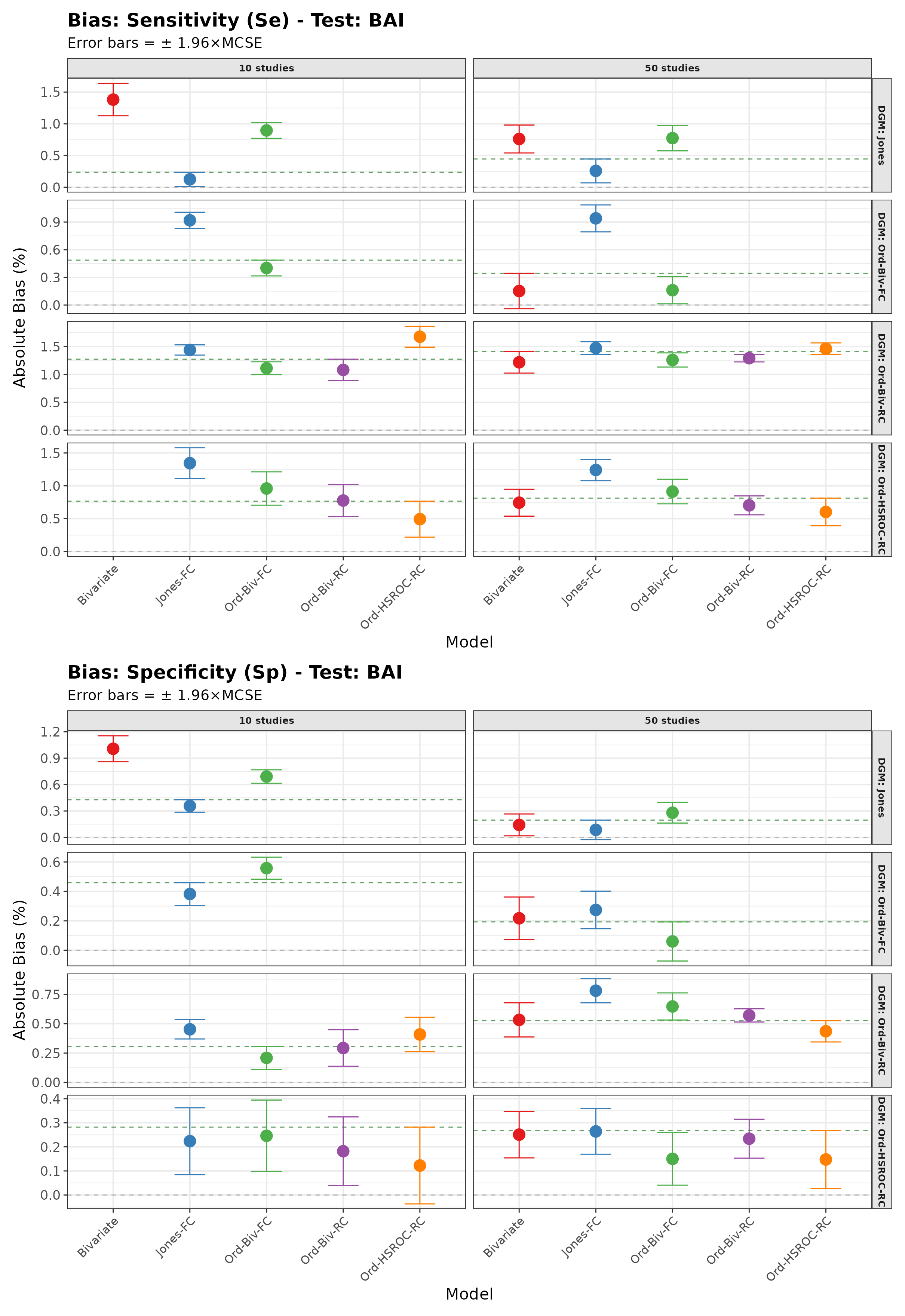}
    \caption{\footnotesize{
        Plot of Bias (with $95\%$ MCSE intervals) for simulation study - for BAI test.
    }}
    \label{Sim_study_Bias_BAI}
\end{figure}
%%%%%%%%%%%%%%%%%%%%%%%%%%%%%%%%%%%%%%%%
%%%%
%%%%
%%%%%%%%%%%%%%%%%%%%%%%%%%%%%%%%%%%%%%%%%%%%%%%%%%%%%%%%%%%%%%%%%%%%%%%%%%%%%%%%%%%%%%%%%%%%%%%%%%%%%%%%%%%%%%%%%%%%%%%%
\setcounter{figure}{0}
\setcounter{table}{0}
\renewcommand{\thefigure}{D.\arabic{figure}}
\renewcommand{\thetable}{D.\arabic{table}}
%%%%%%%%%%%%%%%%%%%%%%%%%%%%%%%%%%%%%%%%%%%%%%%%%%%%%%%%%%%%%%%%%%%%%%%%%%%%%%%%%%%%%%%%%%%%%%%%%%%%%%%%%%%%%%%%%%%%%%%%
%%%%%%%%%%%%%%%%%%%%%%%%%%%%%%%%%%%%%%%%%%%%%%%%%%%%%%%%%%%%%%%%%%%%%%%%%%%%%%%%%%%%%%%%%%%%%%%%%%%%%%%%%%%%%%%%%%%%%%%%
\section{ Appendix D: Coverage plots}
\label{appendix_D_Coverage_plots}
%%%%%%%%%%%%%%%%%%%%%%%%%%%%%%%%%%%%%%%%%%%%%%%%%%%%%%%%%%%%%%%%%%%%%%%%%%%%%%%%%%%%%%%%%%%%%%%%%%%%%%%%%%%%%%%%%%%%%%%%
%%%%%%%%%%%%%%%%%%%%%%%%%%%%%%
\subsection{ Appendix D: Coverage plots; GAD-2 (test I)}
\label{appendix_D_Coverage_plots_GAD_2_test_I}
%%%%%%%%%%%%%%%%%%%%%%%%%%%%%%
%%%%
%%%%
%%%%%%%%%%%%%%%%%%%%%%%%%%%%%%%%%%%%%%%%
\begin{figure}[H]
    \centering
    \includegraphics[width=15cm]{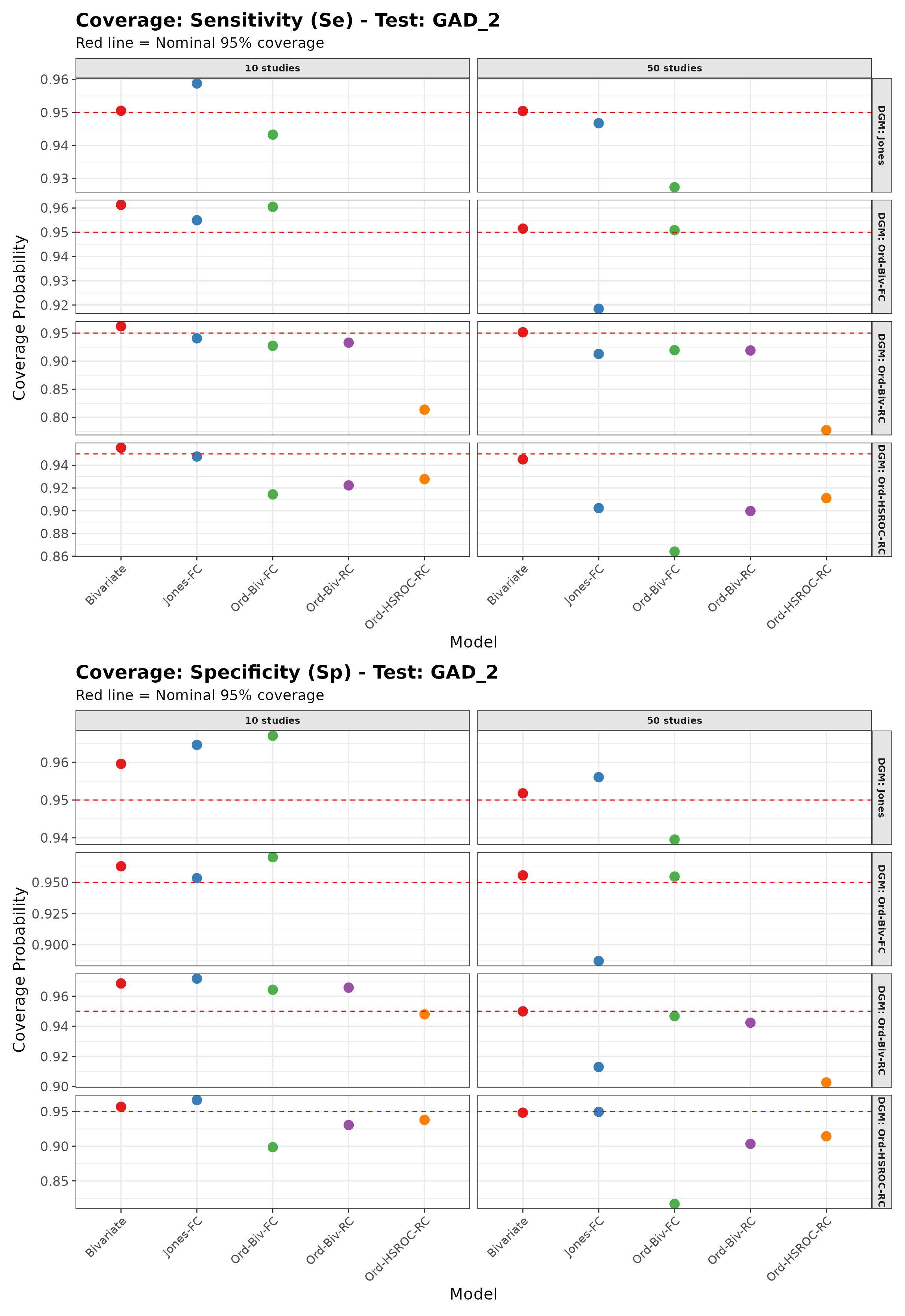}
    \caption{\footnotesize{
            Plot of Coverage for simulation study - for GAD-2 test.
    }}
    \label{Sim_study_Coverage_GAD_2}
\end{figure}
%%%%%%%%%%%%%%%%%%%%%%%%%%%%%%%%%%%%%%%%
%%%%
%%%%
%%%%%%%%%%%%%%%%%%%%%%%%%%%%%%
\subsection{ Appendix D: Coverage plots; HADS (test II)}
\label{appendix_D_Coverage_plots_HADS_test_II}
%%%%%%%%%%%%%%%%%%%%%%%%%%%%%%
%%%%
%%%%
%%%%%%%%%%%%%%%%%%%%%%%%%%%%%%%%%%%%%%%%
\begin{figure}[H]
    \centering
    \includegraphics[width=15cm]{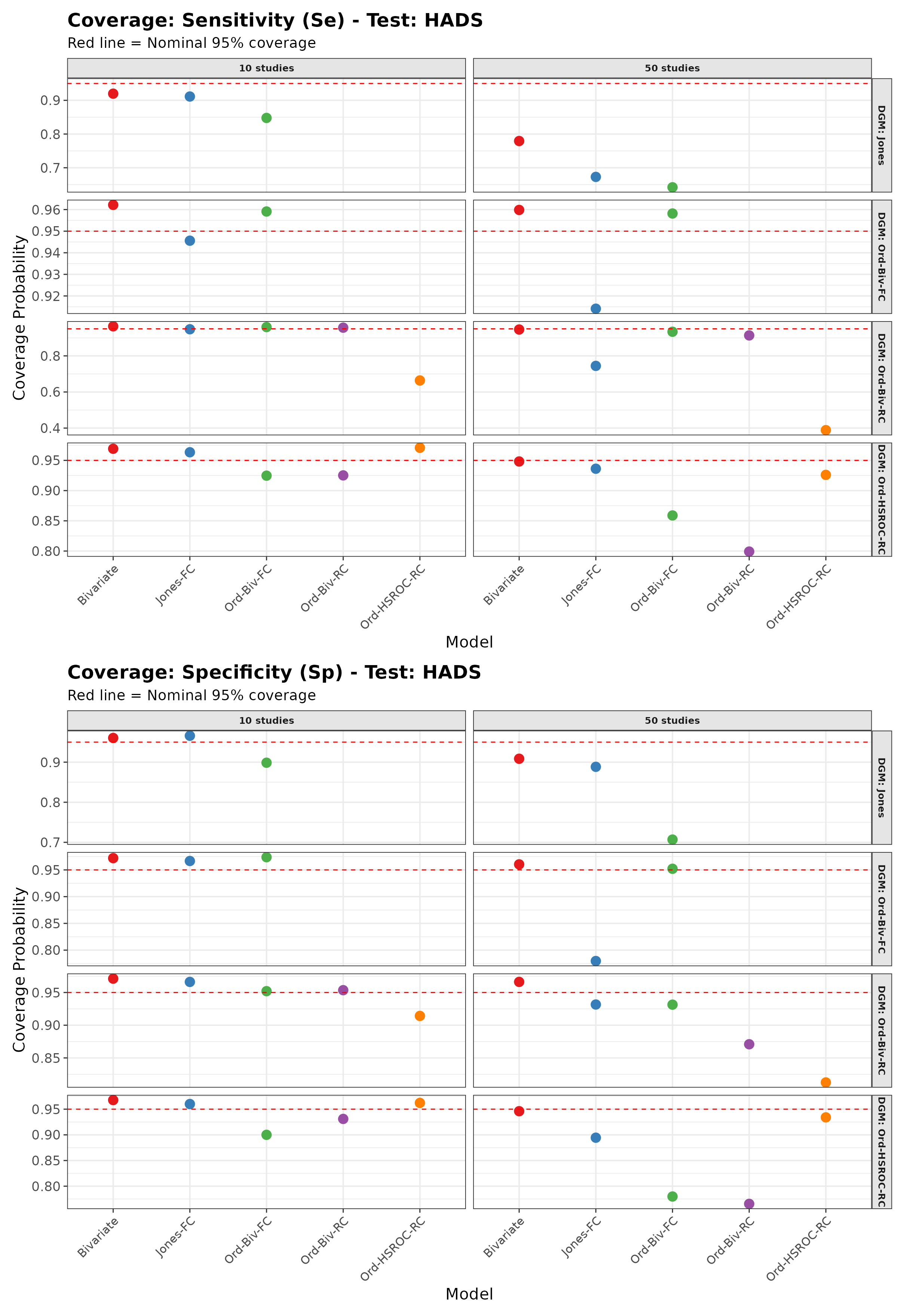}
    \caption{\footnotesize{
            Plot of Coverage for simulation study - for HADS test.
    }}
    \label{Sim_study_Coverage_HADS}
\end{figure}
%%%%%%%%%%%%%%%%%%%%%%%%%%%%%%%%%%%%%%%%
%%%%
%%%%
%%%%%%%%%%%%%%%%%%%%%%%%%%%%%%
\subsection{ Appendix D: Coverage plots; BAI (test III)}
\label{appendix_D_Coverage_plots_BAI_test_III}
%%%%%%%%%%%%%%%%%%%%%%%%%%%%%%
%%%%
%%%%
%%%%%%%%%%%%%%%%%%%%%%%%%%%%%%%%%%%%%%%%
\begin{figure}[H]
    \centering
    \includegraphics[width=15cm]{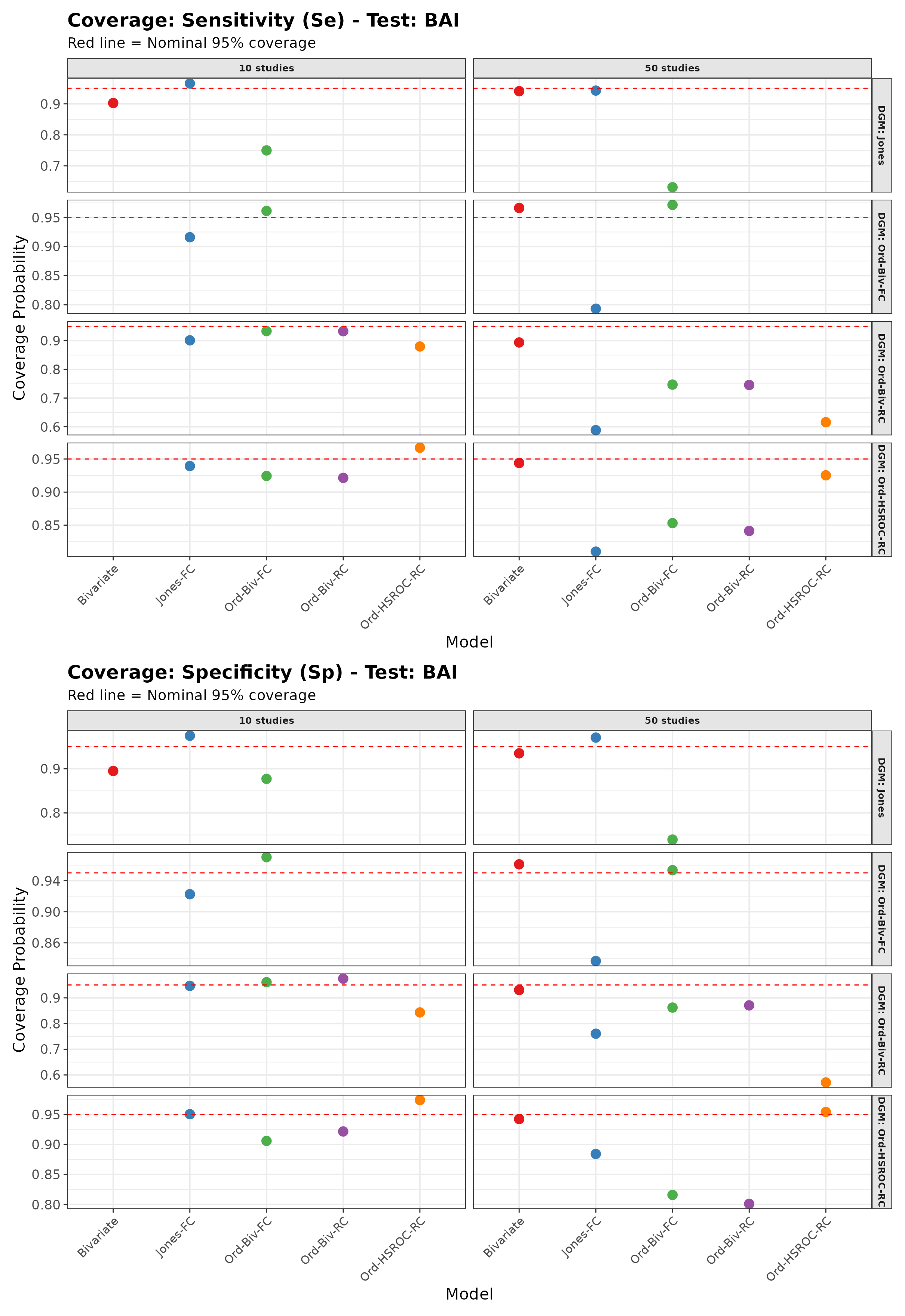}
    \caption{\footnotesize{
            Plot of Coverage for simulation study - for BAI test.
    }}
    \label{Sim_study_Coverage_BAI}
\end{figure}
%%%%%%%%%%%%%%%%%%%%%%%%%%%%%%%%%%%%%%%%

%%%%%%%%%%%%%%%%%%%%%%%%%%%%%%%%%%%%%%%%%%%%%%%%%%%%%%%%%%%%%%%%%%%%%%%%%%%%%%%%%%%%%%%%%%%%%%%%%%%%%%%%%%%%%%%%%%%%%%%%
\setcounter{figure}{0}
\setcounter{table}{0}
\renewcommand{\thefigure}{E.\arabic{figure}}
\renewcommand{\thetable}{E.\arabic{table}}
%%%%%%%%%%%%%%%%%%%%%%%%%%%%%%%%%%%%%%%%%%%%%%%%%%%%%%%%%%%%%%%%%%%%%%%%%%%%%%%%%%%%%%%%%%%%%%%%%%%%%%%%%%%%%%%%%%%%%%%%
%%%%
%%%%
%%%%%%%%%%%%%%%%%%%%%%%%%%%%%%%%%%%%%%%%%%%%%%%%%%%%%%%%%%%%%%%%%%%%%%%%%%%%%%%%%%%%%%%%%%%%%%%%%%%%%%%%%%%%%%%%%%%%%%%%
\section{ Appendix E: Detailed results tables}
\label{appendix_E_detailed_results_tables}
%%%%%%%%%%%%%%%%%%%%%%%%%%%%%%%%%%%%%%%%%%%%%%%%%%%%%%%%%%%%%%%%%%%%%%%%%%%%%%%%%%%%%%%%%%%%%%%%%%%%%%%%%%%%%%%%%%%%%%%%
%%%%
%%%%
%%%%%%%%%%%%%%%%%%%%%%%%%%%%%%
\subsection{ Appendix E: Detailed results tables; GAD-2 (test I)}
\label{appendix_E_detailed_results_tables_GAD_2_test_I}
%%%%%%%%%%%%%%%%%%%%%%%%%%%%%%
%%%%
%%%%
\begin{table}[H]
\centering
\caption{GAD-2 simulation results: Model performance across DGMs and sample sizes}
%% \tiny
\setlength{\tabcolsep}{2pt}
\begin{tabular}{llc|cc|cc|cc|cc}
\toprule
\multirow{2}{*}{\textbf{DGM}} & 
\multirow{2}{*}{\textbf{Model}} &
\multirow{2}{*}{\textbf{$n$}} & 
\multicolumn{8}{c}{\textbf{Performance Metrics}} \\
\cmidrule(lr){4-11}
& & & \multicolumn{2}{c|}{\textbf{RMSE (\%)}} & 
\multicolumn{2}{c|}{\textbf{$|$Bias$|$ (\%)}} &
\multicolumn{2}{c|}{\textbf{Cvg. (\%)}} & 
\multicolumn{2}{c}{\textbf{Width}} \\
\cmidrule(lr){4-5} \cmidrule(lr){6-7} \cmidrule(lr){8-9} \cmidrule(lr){10-11}
& & & \textbf{Se} & \textbf{Sp} & \textbf{Se} & \textbf{Sp} & \textbf{Se} & \textbf{Sp} & \textbf{Se} & \textbf{Sp} \\
%%%%
\midrule
%%%%
Jones & Bivariate & 10 & 8.87 {\scriptsize(0.11)} & 2.89 {\scriptsize(0.03)} & 2.49 {\scriptsize(0.12)} & 0.54 {\scriptsize(0.04)} &
95.0 & 96.0 & 37.7 & 14.1 \\
Jones & Jones-FC & 10 & \textbf{6.95} {\scriptsize(0.10)} & \textbf{2.80} {\scriptsize(0.04)} & \textbf{0.92} {\scriptsize(0.13)} & \textbf{1.16} {\scriptsize(0.05)} & \textbf{95.9} & \textbf{96.5} & \textbf{28.7} & \textbf{12.8} \\
Jones & Ord-Biv-FC & 10 & 7.59 {\scriptsize(0.11)} & 2.66 {\scriptsize(0.04)} & 1.01 {\scriptsize(0.13)} & 0.61 {\scriptsize(0.05)} & 94.3 & 96.7 & 30.0 & 12.9 \\
%%%%
\cmidrule(lr){2-11}
%%%%
Jones & Bivariate & 50 & 4.06 {\scriptsize(0.09)} & 1.26 {\scriptsize(0.03)} & 0.31 {\scriptsize(0.13)} & 0.05 {\scriptsize(0.04)} & 95.0 & 95.2 & 16.1 & 5.09 \\
Jones & Jones-FC & 50 & \textbf{3.39} {\scriptsize(0.09)} & \textbf{1.15} {\scriptsize(0.03)} & \textbf{0.16} {\scriptsize(0.12)} & \textbf{0.16} {\scriptsize(0.04)} & \textbf{94.7} & \textbf{95.6} & \textbf{13.7} & \textbf{4.74} \\
Jones & Ord-Biv-FC & 50 & 3.65 {\scriptsize(0.10)} & 1.16 {\scriptsize(0.03)} & 0.55 {\scriptsize(0.12)} & 0.10 {\scriptsize(0.04)} & 92.7 & 94.0 & 13.5 & 4.69 \\
%%%%
\midrule
%%%%
O-Biv-FC & Bivariate & 10 & 6.69 {\scriptsize(0.11)} & 2.70 {\scriptsize(0.04)} & 1.57 {\scriptsize(0.13)} & 0.49 {\scriptsize(0.05)} & 96.1 & 96.3 & 30.5 & 13.7 \\
O-Biv-FC & Jones-FC & 10 & 5.53 {\scriptsize(0.10)} & 2.69 {\scriptsize(0.05)} & 1.41 {\scriptsize(0.12)} & 0.85 {\scriptsize(0.06)} & 95.5 & 95.4 & 22.9 & 12.0 \\
O-Biv-FC & Ord-Biv-FC & 10 & \textbf{5.72} {\scriptsize(0.10)} & \textbf{2.49} {\scriptsize(0.04)} & \textbf{0.42} {\scriptsize(0.13)} & \textbf{0.34} {\scriptsize(0.06)} & \textbf{96.0} & \textbf{97.0} & \textbf{25.0} & \textbf{12.4} \\
\cmidrule(lr){2-11}
O-Biv-FC & Bivariate & 50 & 3.00 {\scriptsize(0.09)} & 1.19 {\scriptsize(0.03)} & 0.34 {\scriptsize(0.11)} & 0.06 {\scriptsize(0.04)} & 95.2 & 95.6 & 12.1 & 4.86 \\
O-Biv-FC & Jones-FC & 50 & 2.99 {\scriptsize(0.08)} & 1.41 {\scriptsize(0.04)} & 1.19 {\scriptsize(0.11)} & 0.69 {\scriptsize(0.05)} & 91.8 & 88.7 & 10.5 & 4.52 \\
O-Biv-FC & Ord-Biv-FC & 50 & \textbf{2.70} {\scriptsize(0.09)} & \textbf{1.14} {\scriptsize(0.04)} & \textbf{0.14} {\scriptsize(0.13)} & \textbf{0.03} {\scriptsize(0.05)} & \textbf{95.1} & \textbf{95.5} & \textbf{10.6} & \textbf{4.56} \\
%%%%
\midrule
%%%%
O-Biv-RC & Bivariate & 10 & 6.23 {\scriptsize(0.10)} & 2.63 {\scriptsize(0.04)} & 1.21 {\scriptsize(0.13)} & 0.36 {\scriptsize(0.05)} & 96.2 & 96.8 & 29.1 & 13.3 \\
O-Biv-RC & Jones-FC & 10 & 5.78 {\scriptsize(0.10)} & 2.67 {\scriptsize(0.05)} & 1.47 {\scriptsize(0.13)} & 0.98 {\scriptsize(0.06)} & 94.1 & 97.2 & 22.8 & 12.3 \\
O-Biv-RC & Ord-Biv-FC & 10 & 5.89 {\scriptsize(0.10)} & 2.52 {\scriptsize(0.04)} & 0.67 {\scriptsize(0.14)} & 0.53 {\scriptsize(0.06)} & 92.7 & 96.4 & 23.2 & 11.9 \\
O-Biv-RC & Ord-Biv-RC & 10 & \textbf{5.79} {\scriptsize(0.10)} & \textbf{2.44} {\scriptsize(0.04)} & \textbf{0.69} {\scriptsize(0.13)} & \textbf{0.39} {\scriptsize(0.06)} & \textbf{93.3} & \textbf{96.6} & \textbf{22.8} & \textbf{11.5} \\
O-Biv-RC & Ord-HSROC-RC & 10 & 5.99 {\scriptsize(0.10)} & 2.66 {\scriptsize(0.05)} & 1.13 {\scriptsize(0.13)} & 0.31 {\scriptsize(0.06)} & 81.4 & 94.8 & 16.0 & 10.5 \\
%%%%
\cmidrule(lr){2-11}
%%%%
O-Biv-RC & Bivariate & 50 & 2.91 {\scriptsize(0.09)} & 1.18 {\scriptsize(0.04)} & 0.43 {\scriptsize(0.13)} & 0.06 {\scriptsize(0.05)} & 95.2 & 95.0 & 11.5 & 4.78 \\
O-Biv-RC & Jones-FC & 50 & 2.95 {\scriptsize(0.10)} & 1.34 {\scriptsize(0.04)} & 1.26 {\scriptsize(0.12)} & 0.70 {\scriptsize(0.05)} & 91.3 & 91.3 & 10.2 & 4.55 \\
O-Biv-RC & Ord-Biv-FC & 50 & 2.61 {\scriptsize(0.09)} & 1.08 {\scriptsize(0.04)} & 0.42 {\scriptsize(0.12)} & 0.07 {\scriptsize(0.05)} & 92.0 & 94.7 & 9.76 & 4.32 \\
O-Biv-RC & Ord-Biv-RC & 50 & \textbf{2.73} {\scriptsize(0.09)} & \textbf{1.13} {\scriptsize(0.04)} & \textbf{0.35} {\scriptsize(0.14)} & \textbf{0.03} {\scriptsize(0.06)} & \textbf{91.9} & \textbf{94.2} & \textbf{9.81} & \textbf{4.33} \\
O-Biv-RC & Ord-HSROC-RC & 50 & 2.78 {\scriptsize(0.10)} & 1.26 {\scriptsize(0.04)} & 0.98 {\scriptsize(0.12)} & 0.25 {\scriptsize(0.06)} & 77.7 & 90.3 & 6.85 & 4.36 \\
%%%%
\midrule
%%%%
O-HSROC-RC & Bivariate & 10 & 5.47 {\scriptsize(0.10)} & 3.61 {\scriptsize(0.07)} & 0.80 {\scriptsize(0.13)} & 0.59 {\scriptsize(0.08)} & 95.5 & 95.7 & 25.6 & 17.8 \\
O-HSROC-RC & Jones-FC & 10 & 4.83 {\scriptsize(0.10)} & 3.50 {\scriptsize(0.08)} & 1.45 {\scriptsize(0.14)} & 1.14 {\scriptsize(0.10)} & 94.8 & 96.7 & 20.3 & 15.9 \\
O-HSROC-RC & Ord-Biv-FC & 10 & 5.00 {\scriptsize(0.11)} & 3.45 {\scriptsize(0.08)} & 0.65 {\scriptsize(0.14)} & 0.68 {\scriptsize(0.10)} & 91.4 & 89.9 & 19.1 & 13.8 \\
O-HSROC-RC & Ord-Biv-RC & 10 & 4.95 {\scriptsize(0.10)} & 3.27 {\scriptsize(0.07)} & 0.51 {\scriptsize(0.14)} & 0.41 {\scriptsize(0.09)} & 92.2 & 93.1 & 18.8 & 13.5 \\
O-HSROC-RC & Ord-HSROC-RC & 10 & \textbf{5.03} {\scriptsize(0.11)} & \textbf{3.35} {\scriptsize(0.07)} & \textbf{0.47} {\scriptsize(0.14)} & \textbf{0.40} {\scriptsize(0.09)} & \textbf{92.8} & \textbf{93.8} & \textbf{18.9} & \textbf{12.8} \\
%%%%
\cmidrule(lr){2-11}
%%%%
O-HSROC-RC & Bivariate & 50 & 2.52 {\scriptsize(0.10)} & 1.61 {\scriptsize(0.06)} & 0.52 {\scriptsize(0.14)} & 0.25 {\scriptsize(0.09)} & 94.5 & 94.8 & 9.98 & 6.30 \\
O-HSROC-RC & Jones-FC & 50 & 2.57 {\scriptsize(0.09)} & 1.67 {\scriptsize(0.06)} & 1.14 {\scriptsize(0.12)} & 0.75 {\scriptsize(0.08)} & 90.2 & 95.0 & 8.67 & 6.25 \\
O-HSROC-RC & Ord-Biv-FC & 50 & 2.49 {\scriptsize(0.10)} & 1.65 {\scriptsize(0.06)} & 0.39 {\scriptsize(0.14)} & 0.46 {\scriptsize(0.09)} & 86.4 & 81.7 & 7.84 & 5.31 \\
O-HSROC-RC & Ord-Biv-RC & 50 & 2.38 {\scriptsize(0.10)} & 1.50 {\scriptsize(0.06)} & 0.38 {\scriptsize(0.13)} & 0.36 {\scriptsize(0.08)} & 90.0 & 90.3 & 8.12 & 5.39 \\
O-HSROC-RC & Ord-HSROC-RC & 50 & \textbf{2.34} {\scriptsize(0.09)} & \textbf{1.51} {\scriptsize(0.05)} & \textbf{0.30} {\scriptsize(0.12)} & \textbf{0.28} {\scriptsize(0.08)} & \textbf{91.1} & \textbf{91.4} & \textbf{7.98} & \textbf{5.26} \\
%%%%
\bottomrule
%%%%
\end{tabular}
\begin{tablenotes}
\scriptsize
\item RMSE and Bias values are percentages; Values in parentheses are Monte Carlo Standard Errors (MCSE).
\item Cvg. = Coverage probability (\%); Width = Mean 95\% credible interval width.
\item \textbf{Bold values} indicate correctly specified scenarios (DGM matches fitted model).
\end{tablenotes}
\label{table:sim_results_GAD2}
\end{table}
%%%%
%%%%
%%%%%%%%%%%%%%%%%%%%%%%%%%%%%%
\subsection{ Appendix E: Detailed results tables; HADS (test II)}
\label{appendix_E_detailed_results_tables_HADS_test_II}
%%%%%%%%%%%%%%%%%%%%%%%%%%%%%%
%%%%
%%%% HADS Results
%%%%
\begin{table}[H]
\centering
\caption{HADS simulation results: Model performance across DGMs and sample sizes}
%% \tiny
\setlength{\tabcolsep}{2pt}
\begin{tabular}{llc|cc|cc|cc|cc}
\toprule
\multirow{2}{*}{\textbf{DGM}} & 
\multirow{2}{*}{\textbf{Model}} &
\multirow{2}{*}{\textbf{$n$}} & 
\multicolumn{8}{c}{\textbf{Performance Metrics}} \\
\cmidrule(lr){4-11}
& & & \multicolumn{2}{c|}{\textbf{RMSE (\%)}} & 
\multicolumn{2}{c|}{\textbf{$|$Bias$|$ (\%)}} &
\multicolumn{2}{c|}{\textbf{Cvg. (\%)}} & 
\multicolumn{2}{c}{\textbf{Width}} \\
\cmidrule(lr){4-5} \cmidrule(lr){6-7} \cmidrule(lr){8-9} \cmidrule(lr){10-11}
& & & \textbf{Se} & \textbf{Sp} & \textbf{Se} & \textbf{Sp} & \textbf{Se} & \textbf{Sp} & \textbf{Se} & \textbf{Sp} \\
\midrule
Jones & Bivariate & 10 & 4.66 {\scriptsize(0.09)} & 2.32 {\scriptsize(0.05)} & 2.61 {\scriptsize(0.11)} & 0.79 {\scriptsize(0.06)} & 92.0 & 96.0 & 22.5 & 14.8 \\
Jones & Jones-FC & 10 & \textbf{3.71} {\scriptsize(0.09)} & \textbf{1.88} {\scriptsize(0.05)} & \textbf{2.14} {\scriptsize(0.11)} & \textbf{0.86} {\scriptsize(0.06)} & \textbf{91.1} & \textbf{96.6} & \textbf{13.0} & \textbf{8.71} \\
Jones & Ord-Biv-FC & 10 & 3.82 {\scriptsize(0.09)} & 2.01 {\scriptsize(0.05)} & 2.20 {\scriptsize(0.11)} & 0.99 {\scriptsize(0.06)} & 84.8 & 89.9 & 12.1 & 8.13 \\
\cmidrule(lr){2-11}
Jones & Bivariate & 50 & 2.56 {\scriptsize(0.07)} & 1.00 {\scriptsize(0.03)} & 1.89 {\scriptsize(0.08)} & 0.62 {\scriptsize(0.04)} & 77.9 & 90.9 & 6.86 & 3.38 \\
Jones & Jones-FC & 50 & \textbf{2.30} {\scriptsize(0.06)} & \textbf{0.92} {\scriptsize(0.03)} & \textbf{1.91} {\scriptsize(0.06)} & \textbf{0.62} {\scriptsize(0.04)} & \textbf{67.3} & \textbf{88.8} & \textbf{4.72} & \textbf{2.68} \\
Jones & Ord-Biv-FC & 50 & 2.33 {\scriptsize(0.06)} & 0.96 {\scriptsize(0.03)} & 1.86 {\scriptsize(0.07)} & 0.66 {\scriptsize(0.04)} & 64.2 & 70.7 & 4.68 & 2.52 \\
\midrule
O-Biv-FC & Bivariate & 10 & 4.25 {\scriptsize(0.09)} & 2.09 {\scriptsize(0.04)} & 1.06 {\scriptsize(0.12)} & 0.39 {\scriptsize(0.06)} & 96.2 & 97.2 & 22.5 & 13.3 \\
O-Biv-FC & Jones-FC & 10 & 3.03 {\scriptsize(0.10)} & 1.63 {\scriptsize(0.05)} & 0.61 {\scriptsize(0.13)} & 0.34 {\scriptsize(0.07)} & 94.6 & 96.7 & 13.4 & 8.26 \\
O-Biv-FC & Ord-Biv-FC & 10 & \textbf{2.97} {\scriptsize(0.09)} & \textbf{1.63} {\scriptsize(0.05)} & \textbf{0.30} {\scriptsize(0.12)} & \textbf{0.21} {\scriptsize(0.07)} & \textbf{95.9} & \textbf{97.4} & \textbf{13.7} & \textbf{8.32} \\
\cmidrule(lr){2-11}
O-Biv-FC & Bivariate & 50 & 1.58 {\scriptsize(0.08)} & 0.75 {\scriptsize(0.04)} & 0.25 {\scriptsize(0.12)} & 0.08 {\scriptsize(0.06)} & 96.0 & 96.0 & 7.01 & 3.18 \\
O-Biv-FC & Jones-FC & 50 & 1.30 {\scriptsize(0.06)} & 0.72 {\scriptsize(0.04)} & 0.55 {\scriptsize(0.08)} & 0.29 {\scriptsize(0.04)} & 91.4 & 78.0 & 4.92 & 2.59 \\
O-Biv-FC & Ord-Biv-FC & 50 & \textbf{1.19} {\scriptsize(0.06)} & \textbf{0.66} {\scriptsize(0.03)} & \textbf{0.07} {\scriptsize(0.09)} & \textbf{0.04} {\scriptsize(0.05)} & \textbf{95.8} & \textbf{95.2} & \textbf{5.05} & \textbf{2.62} \\
\midrule
O-Biv-RC & Bivariate & 10 & 4.28 {\scriptsize(0.07)} & 2.15 {\scriptsize(0.03)} & 1.10 {\scriptsize(0.09)} & 0.45 {\scriptsize(0.04)} & 96.4 & 97.1 & 22.7 & 13.3 \\
O-Biv-RC & Jones-FC & 10 & 3.26 {\scriptsize(0.10)} & 1.78 {\scriptsize(0.05)} & 1.14 {\scriptsize(0.13)} & 0.25 {\scriptsize(0.08)} & 94.7 & 96.6 & 14.1 & 8.36 \\
O-Biv-RC & Ord-Biv-FC & 10 & 3.13 {\scriptsize(0.10)} & 1.86 {\scriptsize(0.06)} & 0.39 {\scriptsize(0.14)} & 0.18 {\scriptsize(0.08)} & 95.9 & 95.2 & 13.9 & 8.22 \\
O-Biv-RC & Ord-Biv-RC & 10 & \textbf{3.18} {\scriptsize(0.09)} & \textbf{1.78} {\scriptsize(0.05)} & \textbf{0.55} {\scriptsize(0.12)} & \textbf{0.30} {\scriptsize(0.07)} & \textbf{95.6} & \textbf{95.4} & \textbf{13.8} & \textbf{8.20} \\
O-Biv-RC & Ord-HSROC-RC & 10 & 3.86 {\scriptsize(0.10)} & 1.90 {\scriptsize(0.05)} & 1.99 {\scriptsize(0.12)} & 0.51 {\scriptsize(0.07)} & 66.4 & 91.4 & 8.38 & 6.82 \\
\cmidrule(lr){2-11}
O-Biv-RC & Bivariate & 50 & 1.84 {\scriptsize(0.09)} & 0.76 {\scriptsize(0.04)} & 0.26 {\scriptsize(0.12)} & 0.07 {\scriptsize(0.05)} & 94.6 & 96.6 & 7.66 & 3.32 \\
O-Biv-RC & Jones-FC & 50 & 1.86 {\scriptsize(0.07)} & 0.69 {\scriptsize(0.04)} & 1.23 {\scriptsize(0.09)} & 0.20 {\scriptsize(0.04)} & 74.5 & 93.2 & 5.14 & 2.59 \\
O-Biv-RC & Ord-Biv-FC & 50 & 1.37 {\scriptsize(0.07)} & 0.67 {\scriptsize(0.04)} & 0.16 {\scriptsize(0.10)} & 0.08 {\scriptsize(0.05)} & 93.4 & 93.1 & 5.23 & 2.58 \\
O-Biv-RC & Ord-Biv-RC & 50 & \textbf{1.42} {\scriptsize(0.07)} & \textbf{0.66} {\scriptsize(0.03)} & \textbf{0.37} {\scriptsize(0.09)} & \textbf{0.12} {\scriptsize(0.05)} & \textbf{91.3} & \textbf{87.1} & \textbf{5.27} & \textbf{2.62} \\
O-Biv-RC & Ord-HSROC-RC & 50 & 2.43 {\scriptsize(0.07)} & 0.80 {\scriptsize(0.03)} & 1.87 {\scriptsize(0.08)} & 0.35 {\scriptsize(0.04)} & 38.9 & 81.3 & 3.43 & 2.41 \\
\midrule
O-HSROC-RC & Bivariate & 10 & 2.63 {\scriptsize(0.06)} & 2.07 {\scriptsize(0.05)} & 0.36 {\scriptsize(0.08)} & 0.32 {\scriptsize(0.07)} & 96.9 & 96.8 & 15.5 & 13.2 \\
O-HSROC-RC & Jones-FC & 10 & 1.94 {\scriptsize(0.09)} & 1.64 {\scriptsize(0.07)} & 0.39 {\scriptsize(0.10)} & 0.30 {\scriptsize(0.09)} & 96.3 & 96.0 & 9.35 & 7.66 \\
O-HSROC-RC & Ord-Biv-FC & 10 & 2.08 {\scriptsize(0.09)} & 1.72 {\scriptsize(0.08)} & 0.30 {\scriptsize(0.12)} & 0.27 {\scriptsize(0.10)} & 92.5 & 90.0 & 8.82 & 6.92 \\
O-HSROC-RC & Ord-Biv-RC & 10 & 2.08 {\scriptsize(0.10)} & 1.64 {\scriptsize(0.07)} & 0.37 {\scriptsize(0.12)} & 0.35 {\scriptsize(0.10)} & 92.5 & 93.1 & 9.00 & 7.36 \\
O-HSROC-RC & Ord-HSROC-RC & 10 & \textbf{1.93} {\scriptsize(0.10)} & \textbf{1.61} {\scriptsize(0.08)} & \textbf{0.24} {\scriptsize(0.13)} & \textbf{0.19} {\scriptsize(0.11)} & \textbf{97.1} & \textbf{96.2} & \textbf{8.89} & \textbf{7.56} \\
\cmidrule(lr){2-11}
O-HSROC-RC & Bivariate & 50 & 1.23 {\scriptsize(0.08)} & 0.76 {\scriptsize(0.05)} & 0.23 {\scriptsize(0.11)} & 0.13 {\scriptsize(0.07)} & 94.8 & 94.6 & 5.20 & 3.15 \\
O-HSROC-RC & Jones-FC & 50 & 0.87 {\scriptsize(0.06)} & 0.65 {\scriptsize(0.04)} & 0.34 {\scriptsize(0.08)} & 0.27 {\scriptsize(0.06)} & 93.6 & 89.4 & 3.31 & 2.49 \\
O-HSROC-RC & Ord-Biv-FC & 50 & 0.93 {\scriptsize(0.07)} & 0.66 {\scriptsize(0.05)} & 0.23 {\scriptsize(0.10)} & 0.20 {\scriptsize(0.07)} & 85.9 & 78.0 & 3.25 & 2.26 \\
O-HSROC-RC & Ord-Biv-RC & 50 & 0.99 {\scriptsize(0.08)} & 0.67 {\scriptsize(0.06)} & 0.31 {\scriptsize(0.10)} & 0.13 {\scriptsize(0.07)} & 79.9 & 76.6 & 3.44 & 2.44 \\
O-HSROC-RC & Ord-HSROC-RC & 50 & \textbf{0.91} {\scriptsize(0.07)} & \textbf{0.61} {\scriptsize(0.05)} & \textbf{0.29} {\scriptsize(0.10)} & \textbf{0.10} {\scriptsize(0.08)} & \textbf{92.6} & \textbf{93.4} & \textbf{3.47} & \textbf{2.51} \\
\bottomrule
\end{tabular}
\begin{tablenotes}
\scriptsize
\item RMSE and Bias values are percentages; Values in parentheses are Monte Carlo Standard Errors (MCSE).
\item Cvg. = Coverage probability (\%); Width = Mean 95\% credible interval width.
\item \textbf{Bold values} indicate correctly specified scenarios (DGM matches fitted model).
\end{tablenotes}
\label{table:sim_results_HADS}
\end{table}
%%%%
%%%%
%%%%%%%%%%%%%%%%%%%%%%%%%%%%%%
\subsection{ Appendix E: Detailed results tables; BAI (test III)}
\label{appendix_E_detailed_results_tables_BAI_test_III}
%%%%%%%%%%%%%%%%%%%%%%%%%%%%%%
%%%%
%%%% BAI Results
%%%%
\begin{table}[H]
\centering
\caption{BAI simulation results: Model performance across DGMs and sample sizes}
%% \tiny
\setlength{\tabcolsep}{2pt}
\begin{tabular}{llc|cc|cc|cc|cc}
\toprule
\multirow{2}{*}{\textbf{DGM}} & 
\multirow{2}{*}{\textbf{Model}} &
\multirow{2}{*}{\textbf{$n$}} & 
\multicolumn{8}{c}{\textbf{Performance Metrics}} \\
\cmidrule(lr){4-11}
& & & \multicolumn{2}{c|}{\textbf{RMSE (\%)}} & 
\multicolumn{2}{c|}{\textbf{$|$Bias$|$ (\%)}} &
\multicolumn{2}{c|}{\textbf{Cvg. (\%)}} & 
\multicolumn{2}{c}{\textbf{Width}} \\
\cmidrule(lr){4-5} \cmidrule(lr){6-7} \cmidrule(lr){8-9} \cmidrule(lr){10-11}
& & & \textbf{Se} & \textbf{Sp} & \textbf{Se} & \textbf{Sp} & \textbf{Se} & \textbf{Sp} & \textbf{Se} & \textbf{Sp} \\
\midrule
Jones & Bivariate & 10 & 5.05 {\scriptsize(0.11)} & 3.01 {\scriptsize(0.07)} & 1.38 {\scriptsize(0.13)} & 1.01 {\scriptsize(0.08)} & 90.3 & 89.5 & 26.5 & 20.7 \\
Jones & Jones-FC & 10 & \textbf{2.94} {\scriptsize(0.04)} & \textbf{1.91} {\scriptsize(0.03)} & \textbf{0.13} {\scriptsize(0.06)} & \textbf{0.36} {\scriptsize(0.04)} & \textbf{96.6} & \textbf{97.5} & \textbf{13.2} & \textbf{9.27} \\
Jones & Ord-Biv-FC & 10 & 3.39 {\scriptsize(0.05)} & 2.08 {\scriptsize(0.03)} & 0.90 {\scriptsize(0.06)} & 0.69 {\scriptsize(0.04)} & 75.0 & 87.7 & 9.46 & 8.66 \\
\cmidrule(lr){2-11}
Jones & Bivariate & 50 & 2.24 {\scriptsize(0.09)} & 1.20 {\scriptsize(0.05)} & 0.76 {\scriptsize(0.11)} & 0.14 {\scriptsize(0.06)} & 94.1 & 93.5 & 9.63 & 5.12 \\
Jones & Jones-FC & 50 & \textbf{1.34} {\scriptsize(0.07)} & \textbf{0.77} {\scriptsize(0.04)} & \textbf{0.26} {\scriptsize(0.10)} & \textbf{0.09} {\scriptsize(0.06)} & \textbf{94.3} & \textbf{97.1} & \textbf{5.19} & \textbf{3.45} \\
Jones & Ord-Biv-FC & 50 & 1.67 {\scriptsize(0.08)} & 0.89 {\scriptsize(0.05)} & 0.77 {\scriptsize(0.10)} & 0.28 {\scriptsize(0.06)} & 63.1 & 74.0 & 3.83 & 3.18 \\
\midrule
O-Biv-FC & Bivariate & 50 & 1.46 {\scriptsize(0.07)} & 1.13 {\scriptsize(0.06)} & 0.15 {\scriptsize(0.10)} & 0.22 {\scriptsize(0.07)} & 96.6 & 96.1 & 6.89 & 5.09 \\
O-Biv-FC & Jones-FC & 10 & 2.47 {\scriptsize(0.03)} & 2.01 {\scriptsize(0.03)} & 0.92 {\scriptsize(0.04)} & 0.38 {\scriptsize(0.04)} & 91.6 & 92.3 & 10.1 & 9.41 \\
O-Biv-FC & Ord-Biv-FC & 10 & \textbf{2.33} {\scriptsize(0.03)} & \textbf{2.08} {\scriptsize(0.03)} & \textbf{0.40} {\scriptsize(0.04)} & \textbf{0.56} {\scriptsize(0.04)} & \textbf{96.1} & \textbf{97.0} & \textbf{10.3} & \textbf{9.72} \\
\cmidrule(lr){2-11}
O-Biv-FC & Jones-FC & 50 & 1.43 {\scriptsize(0.06)} & 0.93 {\scriptsize(0.05)} & 0.94 {\scriptsize(0.07)} & 0.27 {\scriptsize(0.07)} & 79.3 & 83.7 & 3.94 & 3.55 \\
O-Biv-FC & Ord-Biv-FC & 50 & \textbf{0.97} {\scriptsize(0.05)} & \textbf{0.86} {\scriptsize(0.05)} & \textbf{0.16} {\scriptsize(0.08)} & \textbf{0.06} {\scriptsize(0.07)} & \textbf{97.2} & \textbf{95.3} & \textbf{4.07} & \textbf{3.58} \\
\midrule
O-Biv-RC & Bivariate & 10 & 1.98 {\scriptsize(0.08)} & 1.25 {\scriptsize(0.06)} & 1.22 {\scriptsize(0.10)} & 0.53 {\scriptsize(0.07)} & 89.4 & 93.0 & 6.70 & 4.81 \\
O-Biv-RC & Jones-FC & 10 & 2.64 {\scriptsize(0.04)} & 1.94 {\scriptsize(0.03)} & 1.44 {\scriptsize(0.05)} & 0.45 {\scriptsize(0.04)} & 90.1 & 94.6 & 9.67 & 9.18 \\
O-Biv-RC & Ord-Biv-FC & 10 & 2.50 {\scriptsize(0.05)} & 1.92 {\scriptsize(0.03)} & 1.11 {\scriptsize(0.06)} & 0.21 {\scriptsize(0.05)} & 93.3 & 96.1 & 9.71 & 9.05 \\
O-Biv-RC & Ord-Biv-RC & 10 & \textbf{2.44} {\scriptsize(0.08)} & \textbf{1.79} {\scriptsize(0.05)} & \textbf{1.08} {\scriptsize(0.10)} & \textbf{0.29} {\scriptsize(0.08)} & \textbf{93.3} & \textbf{97.5} & \textbf{9.71} & \textbf{9.02} \\
O-Biv-RC & Ord-HSROC-RC & 10 & 3.02 {\scriptsize(0.08)} & 1.90 {\scriptsize(0.05)} & 1.68 {\scriptsize(0.10)} & 0.41 {\scriptsize(0.07)} & 88.0 & 84.3 & 10.8 & 6.12 \\
\cmidrule(lr){2-11}
O-Biv-RC & Jones-FC & 50 & 1.81 {\scriptsize(0.05)} & 1.15 {\scriptsize(0.04)} & 1.47 {\scriptsize(0.06)} & 0.78 {\scriptsize(0.05)} & 58.8 & 76.0 & 3.74 & 3.41 \\
O-Biv-RC & Ord-Biv-FC & 50 & 1.61 {\scriptsize(0.06)} & 1.07 {\scriptsize(0.05)} & 1.26 {\scriptsize(0.07)} & 0.65 {\scriptsize(0.06)} & 74.7 & 86.2 & 3.88 & 3.40 \\
O-Biv-RC & Ord-Biv-RC & 50 & \textbf{1.64} {\scriptsize(0.03)} & \textbf{1.00} {\scriptsize(0.02)} & \textbf{1.29} {\scriptsize(0.03)} & \textbf{0.57} {\scriptsize(0.03)} & \textbf{74.6} & \textbf{87.1} & \textbf{3.99} & \textbf{3.39} \\
O-Biv-RC & Ord-HSROC-RC & 50 & 1.85 {\scriptsize(0.05)} & 1.04 {\scriptsize(0.03)} & 1.46 {\scriptsize(0.05)} & 0.44 {\scriptsize(0.05)} & 61.6 & 57.1 & 4.40 & 2.44 \\
\midrule
O-HSROC-RC & Bivariate & 50 & 2.41 {\scriptsize(0.08)} & 1.10 {\scriptsize(0.04)} & 0.74 {\scriptsize(0.11)} & 0.25 {\scriptsize(0.05)} & 94.4 & 94.2 & 9.79 & 4.51 \\
O-HSROC-RC & Jones-FC & 10 & 3.50 {\scriptsize(0.09)} & 1.85 {\scriptsize(0.05)} & 1.34 {\scriptsize(0.12)} & 0.22 {\scriptsize(0.07)} & 93.9 & 95.0 & 14.3 & 8.28 \\
O-HSROC-RC & Ord-Biv-FC & 10 & 3.34 {\scriptsize(0.09)} & 1.89 {\scriptsize(0.06)} & 0.96 {\scriptsize(0.13)} & 0.25 {\scriptsize(0.08)} & 92.4 & 90.6 & 13.6 & 7.59 \\
O-HSROC-RC & Ord-Biv-RC & 10 & 3.31 {\scriptsize(0.09)} & 1.89 {\scriptsize(0.06)} & 0.78 {\scriptsize(0.12)} & 0.18 {\scriptsize(0.07)} & 92.2 & 92.2 & 13.6 & 7.80 \\
O-HSROC-RC & Ord-HSROC-RC & 10 & \textbf{3.20} {\scriptsize(0.10)} & \textbf{1.83} {\scriptsize(0.06)} & \textbf{0.49} {\scriptsize(0.14)} & \textbf{0.12} {\scriptsize(0.08)} & \textbf{96.7} & \textbf{97.4} & \textbf{14.3} & \textbf{8.47} \\
\cmidrule(lr){2-11}
O-HSROC-RC & Jones-FC & 50 & 2.01 {\scriptsize(0.07)} & 0.87 {\scriptsize(0.03)} & 1.24 {\scriptsize(0.08)} & 0.26 {\scriptsize(0.05)} & 81.0 & 88.4 & 5.55 & 3.20 \\
O-HSROC-RC & Ord-Biv-FC & 50 & 1.67 {\scriptsize(0.08)} & 0.83 {\scriptsize(0.04)} & 0.91 {\scriptsize(0.10)} & 0.15 {\scriptsize(0.06)} & 85.3 & 81.6 & 5.36 & 2.94 \\
O-HSROC-RC & Ord-Biv-RC & 50 & 1.70 {\scriptsize(0.06)} & 0.90 {\scriptsize(0.03)} & 0.70 {\scriptsize(0.07)} & 0.23 {\scriptsize(0.04)} & 84.1 & 80.1 & 5.43 & 3.06 \\
O-HSROC-RC & Ord-HSROC-RC & 50 & \textbf{1.51} {\scriptsize(0.08)} & \textbf{0.80} {\scriptsize(0.04)} & \textbf{0.60} {\scriptsize(0.11)} & \textbf{0.15} {\scriptsize(0.06)} & \textbf{92.5} & \textbf{95.4} & \textbf{5.62} & \textbf{3.21} \\
\bottomrule
\end{tabular}
\begin{tablenotes}
\scriptsize
\item RMSE and Bias values are percentages; Values in parentheses are Monte Carlo Standard Errors (MCSE).
\item Cvg. = Coverage probability (\%); Width = Mean 95\% credible interval width.
\item \textbf{Bold values} indicate correctly specified scenarios (DGM matches fitted model).
\end{tablenotes}
\label{table:sim_results_BAI}
\end{table}
%%%%
%%%%

%%%%%%%%%%%%%%%%%%%%%%%%%%%%%%%%%%%%%%%%%%%%%%%%%%%%%%%%%%%%%%%%%%%%%%%%%%%%%%%%%%%%%%%%%%%%%%%%%%%%%%%%%%%%%%%%%%%%%%%%
\setcounter{figure}{0}
\setcounter{table}{0}
\renewcommand{\thefigure}{F.\arabic{figure}}
\renewcommand{\thetable}{F.\arabic{table}}
%%%%%%%%%%%%%%%%%%%%%%%%%%%%%%%%%%%%%%%%%%%%%%%%%%%%%%%%%%%%%%%%%%%%%%%%%%%%%%%%%%%%%%%%%%%%%%%%%%%%%%%%%%%%%%%%%%%%%%%%
%%%%
%%%%
%%%%%%%%%%%%%%%%%%%%%%%%%%%%%%%%%%%%%%%%%%%%%%%%%%%%%%%%%%%%%%%%%%%%%%%%%%%%%%%%%%%%%%%%%%%%%%%%%%%%%%%%%%%%%%%%%%%%%%%%
\section{ Appendix F: Additional discussion material}
\label{appendix_F_additional_discussion_material}
%%%%%%%%%%%%%%%%%%%%%%%%%%%%%%%%%%%%%%%%%%%%%%%%%%%%%%%%%%%%%%%%%%%%%%%%%%%%%%%%%%%%%%%%%%%%%%%%%%%%%%%%%%%%%%%%%%%%%%%%
%%%%
%%%%
%%%%%%%%%%%%%%%%%%%%%%%%%%%%%%
\subsection{Discussion of simulation study findings: Ordinal DGMs}
\label{appendix_F_additional_discussion_material_ordinal_DGMs} 
%%%%%%%%%%%%%%%%%%%%%%%%%%%%%%
%%%%
%%%% ---------------------------------------------------------------------------------------------------------------------
When data truly followed ordinal processes -
which we would expect for these tests since they are by definition ordinal anxiety screening questionnaires
(i.e., the GAD-2\supercite{Kroenke_2007_GAD_2}, HADS\supercite{Zigmond_1983_HADS} and BAI\supercite{Beck_1988_BAI}) -
ordinal models generally outperformed both the continuous-assumption Jones\supercite{Jones2019} model,
as well as the standard stratified-bivariate\supercite{Reitsma2005} model.
%%%%
%%%%
%%%%%%%%%%%%%%%%%%%%%%%%%%%%%%%%
\paragraph{\underline{ 
Ordinal DGMs: O-bivariate-FC (DGM \#2)
}}
\label{discussion_ordinal_dgms_O_biv_FC}
%%%%%%%%%%%%%%%%%%%%%%%%%%%%%%%%
%%%%
%%%% ---------------------------------------------------------------------------------------------------------------------
Under O-bivariate-FC DGM, the correctly specified ordinal model excelled -
especially for when we had more data ($50$ studies) -
where for two out of three tests (the GAD-2 and BAI; see sections \ref{Sim_study_GAD_2} and \ref{Sim_study_BAI})
the O-bivariate-FC model was the only model which made it into the "best" group
(with the Jones model being in the "worse" group both times).
Furthermore, even when the Jones model did make it into the best group for $N_{studies} = 50$ -
which was for the HADS (see results, section \ref{Sim_study_HADS}) -
it was a borderline case
(relative difference for O-bivariate-FC vs. Jones model = $9.2\%$ - just below our $10\%$ threshold).

%%%%
%%%% ---------------------------------------------------------------------------------------------------------------------
On the other hand, for the $10$-study case,
the difference between the O-biv-FC and Jones model was consistently negligible;
more specifically, the relative improvement in using the Jones model -
as opposed to the O-bivariate-FC model -
was $0.2\%$, $0.06\%$ and $1.0\%$ for the GAD-2, HADS, and BAI, respectively
(see tables \ref{Table:summary_table_overall_GAD_2},
\ref{Table:summary_table_overall_HADS},
and \ref{Table:summary_table_overall_BAI}).
%%%%
%%%%
%%%%%%%%%%%%%%%%%%%%%%%%%%%%%%%%
\paragraph{\underline{ 
Ordinal DGMs: O-bivariate-RC (DGM \#3)
}}
\label{discussion_ordinal_dgms_O_biv_RC}
%%%%%%%%%%%%%%%%%%%%%%%%%%%%%%%%
%%%%
%%%% ---------------------------------------------------------------------------------------------------------------------
Under the O-bivariate-RC DGM, interesting patterns emerged.
For example, the fixed-cutpoint O-bivariate-FC performed approximately the same as its much more complex
random-cutpoint counterpart (i.e., the O-bivariate-RC); 
this suggests that explicit threshold heterogeneity modeling might be unnecessary.

%%%%
%%%% ---------------------------------------------------------------------------------------------------------------------
More specifically - under the O-bivariate-RC DGM and for the more informative $50$-study case -
for the GAD-2 and HADS, the O-bivariate-FC model paradoxically came out ahead of the O-bivariate-RC model
(by $4.4\%$ and $2.2\%$, respectively) -
although it is important to note that these differences did not reach statistical nor practical significance.
That being said, one might expect the RC variant to come out ahead (and perhaps by a notable margin),
given that the true DGM itself was from the RC variant.
Meanwhile, for the BAI, the O-bivariate-RC and O-bivariate-FC performed essentially identically,
with an RMSE difference of just $0.04$ and the O-bivariate-FC being only $1.0\%$ worse than the O-bivariate-RC model.
Furthermore - again specifically for the $50$-study case - 
both the FC and RC variants of our proposed O-bivariate model performed better than the Jones model
(by $16.1\%$, $25.5\%$, and $10.4\%$ for the GAD-2, HADS and BAI, respectively).

%%%%
%%%% ---------------------------------------------------------------------------------------------------------------------
For the less informative $10$-study case - unlike the $50$-study case discussed above -
the performance differences between the O-bivariate and the Jones models did not reach statistical
nor practical significance.
More specifically, the Jones model was only $2.7\%$, $1.8\%$,
and $6.5\%$ worse than the leading O-biv-RC model for the GAD-2,
HADS and BAI, respectively (when $N_{studies} = 50$).
Furthermore - in contrast to the $50$-study case - 
the RC variant consistently came out on top for all three tests,
although similarly to the $50$-study case,
the differences between the FC and RC variants of our proposed O-bivariate model
were not statistically nor practically significant
(with the O-bivariate-FC model being $2.1\%$, $0.7\%$ and $4.8\%$ 
worse than the O-bivariate-RC model for the GAD-2, HADS and BAI tests, respectively).
%%%%
%%%%
%%%%%%%%%%%%%%%%%%%%%%%%%%%%%%%%
\paragraph{\underline{ 
Ordinal DGMs: O-HSROC-RC (DGM \#4)
}}
\label{discussion_ordinal_dgms_O_HSROC_RC}
%%%%%%%%%%%%%%%%%%%%%%%%%%%%%%%%
%%%%
%%%% ---------------------------------------------------------------------------------------------------------------------
For the random-cutpoint O-HSROC-RC DGM, for the GAD-2 and HADS, the performance differences between the five models
tested was generally more homogeneous than it was for the O-bivariate-RC DGM
(discussed above in section \ref{discussion_ordinal_dgms_O_biv_RC}).
However, for the BAI, this pattern reversed -
with the performance differences between the models actually
varying more than they did compared to the O-bivariate-RC model.

%%%%
%%%% ---------------------------------------------------------------------------------------------------------------------
For example, for the $50$-study case,
for both the GAD-2 and HADS four out of five models made it into the "best" group.
More specifically, for the GAD-2 (see section \ref{Sim_study_GAD_2}), the only model in the "worse" group
was the Jones model (RMSE = $4.24$ vs. $3.85 - 4.14$ for the other four models),
being $10.3\%$ ($> 10\%$) worse than the leading O-HSROC-RC model.
For the HADS (see section \ref{Sim_study_HADS}),
it was instead all models besides the stratified-bivariate which made it into the "best" group
(RMSE range: $1.52 - 1.66$), with the stratified-bivariate being $30.7\%$ worse (RMSE = $1.99$)
than the leading O-HSROC-RC model (RMSE = $1.52$).
However, for the BAI (see section \ref{Sim_study_BAI}), it was only the O-HSROC-RC (RMSE = $2.73$) and -
somewhat paradoxically - the fixed-cutpoints variant of the O-bivariate model (RMSE = $2.94$)
which made it into the "best" group, being only $7.5$ ($< 10\%$) worse than the leading O-HSROC-RC model.
This was followed by the O-bivariate-RC model, 
which fell into the "worse (practical only)" group,
being $11.6\%$ ($> 10\%$) worse than the leading O-HSROC-RC model,
and finally both the Jones ($24.9\%$ worse vs. O-HSROC-RC)
and the stratified-bivariate ($43.5\%$ worse vs. O-HSROC-RC) models
which fell into the "worse" group.

%%%%
%%%% ---------------------------------------------------------------------------------------------------------------------
For the $10$-study case, the GAD-2 (see section \ref{Sim_study_GAD_2}) pattern was quite different:
it was the stratified-bivariate model which was the only model (not Jones like the $50$-study case) 
which fell into the "worse" group, 
being $10.5\%$ ($> 10\%$) worse than the leading O-bivariate-RC model.
On the other hand, for the HADS \ref{Sim_study_HADS}),
there was very little difference between the model rankings/groupings
regardless of whether we have $10$ studies or $50$; for instance, in both cases the O-HSROC-RC model was in the lead,
with the Jones model performing essentially the same
(just $1.5\%$ and $0.1\%$ worse than the O-HSROC-RC model for $10$ and $50$ studies, respectively).
Furthermore - for both sample sizes - the stratified-bivariate model was the only model in the "worse" group
(being $32.7\%$ and $30.7\%$ worse than the leading O-HSROC-RC model for $10$ and $50$ studies, respectively).
For the BAI (see section \ref{Sim_study_BAI}), the pattern was very different when we had only $10$ studies
compared to when we had $50$ studies (which we discussed in the previous paragraph above).
More specifically, for the $10$-study case, all models except for the stratified-bivariate model fell into the "best" group, 
with the stratified-bivariate model being $38.1\%$ worse than the leading O-HSROC-RC model.
%%%%
%%%%

% %%%%%%%%%%%%%%%%%%%%%%%%%%%%%%%%%%%%%%%%%%%%%%%%%%%%%%%%
% \section{References and appendices}
% %%%%%%%%%%%%%%%%%%%%%%%%%%%%%%%%%%%%%%%%%%%%%%%%%%%%%%%%

%% References
%\newpage
\printbibliography

\end{document}